\documentclass[hyper]{JHEP3}

\input{epsf}
\usepackage{epsfig}
\usepackage{amssymb}
\usepackage{amsfonts}
\usepackage{amsbsy}

\def\tn{\tilde{n}}
\def\tb{\tilde{\beta}}

\def\mod{{\rm mod}}

\def\IC{\mathbb{C}}
\def\IN{\mathbb{N}}
\def\IZ{{\mathbb{Z}}}
\def\IR{{\mathbb{R}}}
\def\IP{\mathbb{P}}
\def\IQ{\mathbb{Q}}
\def\ICP{\mathbb{CP}}

\def\CI {{\cal I}}
\def\CC {{\cal C}}
\def\CM {{\cal M}}

\def\CN {{\cal N}}
\def\CR {{\cal R}}
\def\CD {{\cal D}}
\def\CF {{\cal F}}

\def\CL {{\cal L}}

\def\CO {{\cal O}}
\def\CZ {{\cal Z}}

\def\CH {{\cal H}}
\def\CB {{\cal B}}
\def\CS {{\cal S}}

\def\CZ{{\cal Z}}

\def\half{\frac{1}{2}}

\renewcommand{\Im}{{\rm Im }}
\renewcommand{\Re}{{\rm Re }}

\def\one{{\hbox{ 1\kern-.8mm l}}}

\def\p{\partial}

\def\bz{\bar{z}}

\def\bPi{\bar{\Pi}}

\def\bpartial{\bar{\partial}}
\def\bj{{\bar{j}}}
\def\bi{{\bar{i}}}
\def\bk{{\bar{k}}}
\def\bl{{\bar{l}}}

\def\btheta{\bar{\theta}}
\def\bpsi{\bar{\psi}}

\def\bz{\bar{z}}

\def\bnabla{{\bar \nabla}}

\def\vx{{\vec{x}}}

\newcommand{\vd}[2]{|\vec{x}_{#1} - \vec{x}_{#2}|}

\title{Split States, Entropy Enigmas, Holes and Halos}

\author{Frederik~Denef$^1$ and Gregory~W.~Moore$^2$\\
$^1$ Instituut voor Theoretische Fysica, KU Leuven, \\
Celestijnenlaan 200D, B-3001 Leuven, Belgium \\
\\
$^2$ NHETC and Department of Physics and Astronomy,
Rutgers University,\\
Piscataway, NJ 08855--0849, USA\\
\\
{\tt frederik.denef@fys.kuleuven.be, gmoore@physics.rutgers.edu} }

\abstract{  We investigate degeneracies of BPS states of D-branes on
compact Calabi-Yau manifolds.  We develop a factorization formula
for BPS indices using attractor flow trees associated to
multicentered black hole bound states. This enables us to study
background dependence of the BPS spectrum, to compute explicitly
  exact indices of various nontrivial D-brane systems, and to
clarify the subtle relation of Donaldson-Thomas invariants to BPS
indices of stable D6-D2-D0 states, realized in supergravity as
``hole halos.'' We introduce a convergent generating function for D4
indices in the large CY volume limit, and prove it can be written as
a modular average of its polar part, generalizing the fareytail
expansion of the elliptic genus. We show polar states are ``split''
D6-anti-D6 bound states, and that the partition function factorizes
accordingly, leading to a refined version of the OSV conjecture.
This differs from the original conjecture in several aspects. In
particular we obtain a nontrivial measure factor $g_{\rm top}^{-2}
\, e^{-K}$ and find factorization requires a cutoff. We show that
the main factor determining the cutoff and therefore the error is
the existence of ``swing states''
--- D6 states which exist at large radius but do not form stable
D6-anti-D6 bound states. We point out a likely breakdown of the OSV
conjecture at small $g_{\rm top}$ (in the large background CY volume
limit), due to the surprising phenomenon that for sufficiently large
background K\"ahler moduli, a charge $\Lambda \, \Gamma$ supporting
single centered black holes of entropy $\sim \Lambda^2 S(\Gamma)$
also admits two-centered BPS black hole realizations whose entropy
grows like $\Lambda^3$ when $\Lambda\to \infty$.
}

\begin{document}

\section{Introduction}

String theory has been spectacularly successful in microscopically
reproducing the entropy of certain classes of black holes, in
particular of supersymmetric charged black holes. What made this
possible is the fact that for these black holes, the entropy can be
identified with the logarithm of the Witten index of the system,
which is independent of the string coupling constant, enabling one
to count states in the zero coupling limit where the D-brane
description becomes accurate \cite{Strominger:1996sh}. Alternatively
one can use the M-theory description, as was done in
\cite{Maldacena:1997de} for four dimensional D4-D2-D0 BPS black
holes in IIA Calabi-Yau compactifications. In this case the relevant
weak coupling limit is the limit of large Calabi-Yau volume and
large M-theory circle radius.

However, until recently all these derivations were limited to cases
dual to systems in regimes in which some form of the Cardy formula
could be applied. For example the computation of
\cite{Maldacena:1997de} was limited to zero D6-brane charge and
large D0-charge $N$. The restriction to large $N$ followed from the
use of the Cardy formula. The parallel derivation in the D4-D0
picture in string theory \cite{Vafa:1997gr} is restricted to the
same large D0-charge regime. Thus the standard treatments are in
fact valid only in  a very small subset of charge space. In
particular, in this regime none of the IIA worldsheet instanton
contributions to the supergravity entropy are visible.

Further significant progress on the microscopic accounting of
entropy  only came after the supergravity prediction for the entropy,
based on the attractor mechanism \cite{Ferrara:1995ih,Strominger:1996kf},
was refined in \cite{LopesCardoso:1998wt,LopesCardoso:1999cv,LopesCardoso:1999xn,
 Mohaupt:2000mj}, leading in turn to the
formulation of a famous conjecture by Ooguri, Strominger and Vafa
(OSV) \cite{Ooguri:2004zv}. The OSV conjecture predicts  a
far-reaching generalization of the correspondence between
supergravity and statistical entropies, refining it to all orders in
a $1/Q$ expansion, $Q$ being some measure of the charge. One way of
stating the original conjecture is
\begin{equation} \label{OSVintformintro}
 \Omega(p,q) \sim \int d\phi \, e^{- 2 \pi \phi^\Lambda q_\Lambda
 } \, |{\cal Z}_{\rm top}(g_{\rm top}, t)|^2,
\end{equation}
where   $\Omega(p,q)$ is a suitable index of BPS states of given
charge $\Gamma=(p,q)$, defined below in (\ref{indexdef}), and
$\CZ_{\rm top}(g_{\rm top}, t)$ is the topological string partition
function with certain $(p,\phi)$ dependent substitutions for the
topological string coupling $g_{\rm top}$ and K\"ahler moduli $t$,
also detailed below in section \ref{sec:prelim}. By construction,
the leading saddle point approximation to (\ref{OSVintformintro}) is
$e^{S_{\rm BHW}(p,q)}$, where $S_{\rm BHW}$ is the
Bekenstein-Hawking-Wald entropy obtained from the standard $\CN=2$
low energy two derivative action plus F-term $R^2$ corrections,
governed by topological string amplitudes.

A version of the conjecture counting BPS states on \emph{noncompact}
Calabi-Yau manifolds  was subsequently investigated in many
examples. The reason for considering noncompact Calabi-Yau manifolds
is that much more is known about the D-brane systems they carry and
the counting of their BPS states. On the other hand the immediate
black hole interpretation is lost, and one should be wary of drawing
conclusions for the compact case from the noncompact case. In this
paper, we will strictly limit ourselves to the compact case. Direct
tests for this case have been more limited
\cite{Dabholkar:2005by,Dabholkar:2005dt,Shih:2005he}, mainly because
little is known about the behavior of various curve counting
invariants at large degree in this compact setting. However, more
recently progress was made towards model independent derivations of
the conjecture
\cite{talks,Gaiotto:2006ns,deBoer:2006vg,Beasley:2006us}. As a
byproduct, these studies have opened the window to extensions of
microscopic derivations of black hole entropy beyond the ``Cardy
regime,'' and in particular to give an explanation for the
appearance of IIA worldsheet instanton corrections to the entropy.

Nevertheless, the situation is still far from being completely
understood, and several problems were left open in these recent
studies, some explicitly, some only implicitly.

One of the problems in the first category is the fact that there is
still no general derivation for the case with nonzero D6-brane
charge. Another one is the need to keep the black hole attractor
point within some sufficiently small neighborhood of the infinite
radius limit, requiring in particular the   magnetic D4 charge $P$
to lie within the K\"ahler cone (excluding in particular the
so-called ``small'' black holes). These limitations also hold for
the present work.

The more subtle problems on the other hand are related to the
intrinsic ambiguities present in (\ref{OSVintformintro}), some of
which were already pointed out in \cite{Ooguri:2004zv,
Dabholkar:2005by,Dabholkar:2005dt}. The most serious ones are:
\begin{itemize}
 \item The indices $\Omega(p,q)$ in fact depend on the boundary conditions
 of the scalar fields at infinity. Denoting the background by $t_\infty$
 we should, and henceforth will, denote the indices by $\Omega(p,q;t_\infty)$.
 The $t_\infty$ dependence is  due to jumps at walls of marginal stability. On the other hand
the right hand side of (\ref{OSVintformintro}) does not have this
dependence. This raises the question: For which value of $t_\infty$ is the
conjecture is supposed to hold?
 \item Since $\CZ_{\rm top}$ is divergent and  only makes sense
 as an asymptotic perturbative expansion, it is clear that the conjecture can at
 most hold approximately. However, it is  not   clear \textit{a priori} what the regime
 of validity should be, nor what the order of the error is, nor even  how to  define
 properly   the integral (\ref{OSVintformintro}). In other words,
 it is not clear what ``$\sim$'' means.
 \item It is not clear whether there should be an additional integration measure factor in
 (\ref{OSVintformintro}).
\end{itemize}
The first issue has been sidestepped in most studies of the OSV
conjecture so far. Upon closer inspection though, one sees that
typically an implicit choice of $t_\infty$ is made. For example if
one counts states in a classical geometric brane picture, one is
implicitly working in the infinite radius limit, i.e.\
$t_\infty=i\infty$. This will also be the value of $t_\infty$
considered in this paper. We should stress that this is different
from working with local Calabi-Yau manifolds. First, we are taking a
well defined limit $t_{\infty} \to i \infty$ of the full, compact
degeneracies, without making truncations of degrees of freedom as
one does in local models. Moreover, by simultaneously tuning the IIA
string coupling one could actually keep the M-theory CY volume $V_M
\sim V_{\rm IIA}/g_{\rm IIA}^2$ finite in this limit.\footnote{This
indicates that M-theory is perhaps the most natural framework to
consider this limit. We will develop most of our picture in IIA, but
the fundamental building blocks we will use, namely attractor flow
trees, are universal and can equally well be interpreted in IIA, IIB
or M-theory, or even microscopically, as we will see.}

The second issue has been largely ignored by keeping derivations
formal and not worrying about issues of convergence or how to define
the right hand side of (\ref{OSVintformintro}) such that it makes
sense as an integral. In \cite{deBoer:2006vg} the need for a cutoff
in $\CZ_{\rm top}$ and the existence of corrections to the OSV
formula were emphasized but the analysis was not sufficiently
detailed to provide a precise description of either one of those.
For certain $\CN \geq 4$ models
\cite{Dabholkar:2005by,Dabholkar:2005dt,Shih:2005he,LopesCardoso:2004xf,LopesCardoso:2006bg},
where cutoffs are not needed due to the simplicity of the
topological string partition function, explicit exponential
corrections were found.

Finally, the need for an additional measure factor was was pointed
out in \cite{Dabholkar:2005dt} for small black holes, in
\cite{Shih:2005he} for $T^6$ and $T^2 \times K3$ compactifications,
in \cite{LopesCardoso:2006bg} for $\CN = 4$ models, and on general
theoretical grounds in \cite{Verlinde:2004ck}. On the other hand,
none of the derivations \cite{Gaiotto:2006ns,deBoer:2006vg} detected
an additional measure factor.

Our analysis will tackle all these problems head-on for arbitrary
compact (proper) Calabi-Yau manifolds, resulting in a refined,
unambiguous version of the conjecture at $t_\infty=i \infty$, with a
precise cutoff prescription for $\CZ_{\rm top}$. Moreover, we will
in fact find a nontrivial extra measure factor
\begin{equation}
 \mu \sim g_{\rm top}^{-2} \, e^{-K}
\end{equation}
in agreement with previous special case studies
\cite{Dabholkar:2005by,Dabholkar:2005dt,Shih:2005he,LopesCardoso:2004xf}.
The presence of a similar nontrivial measure factor has been
previously discussed from a different point of view in
\cite{LopesCardoso:2006bg}. Finally we will formulate detailed
constraints on $(p,q)$ for the conjecture to hold to exponential
accuracy, and give a concrete error estimate within this domain of
validity somewhat larger than that found in previous special case
studies \cite{Shih:2005he}.

A more or less self-contained technical summary of our final result
--- the refined OSV formula ---  can be found in section
\ref{sec:discussion}.

The main challenge, to which much of the paper is devoted, is
essentially controlling the error and ensuring it does not swamp the
effects of interest. We outline the issues involved a bit further on
in section \ref{sec:challenges}, and discuss a number of unresolved
problems in this context.

\subsection{Outline}

Besides a precise version of the OSV conjecture as outlined above,
we will obtain several results of independent interest:
\begin{enumerate}

 \item In section \ref{sec:fareytail} we show that the ``black hole partition
  function'' of OSV for $p^0=0$  may be obtained from  a
  well-defined, convergent (topologically
 twisted) D4 partition function $\CZ_{D4}$ through a formal substitution of arguments.
 We then demonstrate,
  using
 TST duality, that $\CZ_{D4}$ transforms as a generalized
 multi-variable Jacobi form under modular transformations. From this, we rigorously establish a
 ``fareytail expansion'' \cite{Dijkgraaf:2000fq,Moore:2004fg} for the fareytail-transform
 $\widehat{\CZ}_{D4}$, as well as one  for the
 original $\CZ_{D4}$. This expresses the D4D2D0 indices of arbitrary charges in terms of
 those of a distinguished, finite set of ``polar'' charges, which have reduced D0-charge
 $\hat{q}_0 > 0$. The use of the fareytail expansion in this problem  was suggested in
 \cite{Dabholkar:2005by} and is  dual to the M-theory
 derivations of \cite{deBoer:2006vg,Kraus:2006nb,Gaiotto:2006wm}.
 The result obtained here refines and extends these results. This section is logically quite independent
 of the remainder of the paper, although the final result (\ref{ZPoincSer})  will be used in
 the derivation in section \ref{sec:OSVderivation}.
 %

 \item In section \ref{sec:boundBH}, we give a review of the
 four dimensional supergravity picture of BPS bound states as multicentered black
 hole ``molecules'' and of the phenomenon of decay at marginal stability in this
 setting. We emphasize the power of attractor flow \emph{trees} in establishing
 the existence of such solutions, and argue in general that
 these flow trees give a useful partition of the classical BPS configuration moduli space and quantum BPS Hilbert space.
 We give very concrete explicit examples of D6-anti-D6 two centered bound states, halos, Sun-Earth-Moon
 systems, and iterations of those.

 Within the class of two-centered black hole examples, we encounter a rather surprising phenomenon:
 when one uniformly scales up a generic ($P>0$) D4-D2-D0 charge $\Gamma$ as $\Gamma \to \Lambda \Gamma$, one finds
that for $\Lambda$ sufficiently large and in a background with $\Im
\, t_\infty$ sufficiently large, there always exist two-centered
black hole BPS bound states whose horizon entropy is parametrically
larger than the single centered horizon entropy. More precisely the
horizon entropy of these two centered solutions grows as
$\Lambda^3$, while the single centered entropy only grows as
$\Lambda^2$. Although this is easily seen to be fully compatible
with holography (as all distances scale as $\Lambda^{3/2}$), it is
still quite unexpected,  and appears at odds with the OSV formula
(at $t_{\infty} \to i \infty$) in this limit, as that formula
predicts $\log |\Omega| \sim \Lambda^2$. We refer to this phenomenon
as the ``entropy enigma''.

 We also demonstrate the existence of an interesting class of
 multiparticle
 ``scaling'' solutions, which are characterized by a configuration
 scale modulus $\lambda$ such that in the limit $\lambda \to 0$,
 the solution becomes indistinguishable from a single centered black
 hole to a distant observer, while a near observer keeps on
 seeing nontrivial microstructure.

 Finally, we show that the polar states
 forming the basis of the fareytail expansion correspond to charges which do
 not have a single centered black hole description. Instead they are
 realized as BPS black hole configurations consisting of
 two (or more) clusters of nonzero opposite D6 charges.
 In this sense polar states are ``split states''. This
 split nature will translate into approximately factorized degeneracies,
 which we show in later sections to give rise eventually
 to the factorization $\CZ_{\rm BH} \sim \CZ_{\rm top} \overline{\CZ_{\rm
 top}}$, i.e.\ the OSV conjecture.

 \item In section \ref{sec:microdescription} we briefly review the
 microscopic counterparts of these multicentered configurations in terms of stretched open strings
 and tachyon condensation. We
 also exhibit how to a certain extent the split nature of polar states is mirrored even
 in the large radius geometrical description of these D-brane
 states, by matching charges and moduli spaces.

 \item In section \ref{sec:quivfact}, we get to the actual counting
 of BPS states and describe perhaps the most important result in the
 paper.  We give physical arguments for a wall crossing
 formula giving the jump $\Delta \Omega(\Gamma;t)$ of the index at a
 wall of marginal stability $t=t_{\rm ms}$ corresponding to a decay
 $\Gamma\to \Gamma_1 + \Gamma_2$. The index changes by
 \begin{equation}
 \Delta \Omega(\Gamma;t)=
 (-1)^{\langle \Gamma_1,\Gamma_2
 \rangle-1} |\langle
 \Gamma_1,\Gamma_2
 \rangle| \,\, \Omega(\Gamma_1;t_{\rm ms} ) \,\,
 \Omega(\Gamma_2; t_{\rm ms} )
\end{equation}
when $\Gamma_1$ and $\Gamma_2$ are primitive. Of course, for fixed
$\Gamma_1, \Gamma_2$ this is
also the wall of marginal stability for other charges
 $\Gamma_{N_1,N_2} \to N_1 \Gamma_1 +
N_2 \Gamma_2$, $N_1,N_2>0$. We show that the formula for $N_1=1$ but
arbtirary $N_2$  is
  most conveniently
 given in terms of a generating function:
 \begin{equation}
 \sum_{N_2>0} \Delta \Omega(\Gamma_1 + N_2 \Gamma_2) \, q^N =
 \Omega(\Gamma_1) \prod_{k>0} \biggl( 1-(-1)^{k \langle \Gamma_1,\Gamma_2
 \rangle}\,  q^k \biggr)^{k |\langle \Gamma_1,\Gamma_2 \rangle| \, \Omega(k \Gamma_2)}
\end{equation}
where all indices are understood to be evaluated at $t=t_{\rm ms}$.
These formulae in turn give rise to a powerful factorization formula
for BPS indices based on attractor flow trees.

We verify these formulas explicitly for a number of nontrivial bound
states of branes described by quivers and/or large radius sheaves,
including a three node quiver with a closed
 loop and a generic cubic superpotential. For this case we find
 an intriguing exact formula for the indices in terms of an integral of the product
 of three Laguerre functions.   This shows
 a phase transition (as a function of the charges)
  in the growth of the degeneracies, going from
 polynomial to exponential exactly at the transition point where
 the black hole-like scaling solutions mentioned above come
 into existence. Moreover in this regime we find the rather suggestive
 asymptotics $\Omega \sim 2^{I_{12}} \, 2^{I_{23}} \, 2^{I_{31}}$, where
 $I_{ij}$ denotes the number of arrows between the respective nodes
 in the quiver (note that these grow quadratically with uniform charge scalings, making this a
 macroscopic entropy).

 \item In section \ref{sec:counting} we turn to the counting of the
 BPS states specifically relevant for our derivation of the OSV
 conjecture.

 In section \ref{sec:D6D4D2D0}, we analyze the spectrum of
 D6-D2-D0 BPS bound states with unit D6 charge and the generating function $\CZ_{\rm
 D6-D2-D0}|_{t_{\infty}}$ for their indices in a given background $t_{\infty}=B+iJ$.
 We discuss the relation of these
 generating functions
 with the Donaldson-Thomas(DT) / Gopakumar-Vafa (GV) partition functions. It turns out
 that for D6D2D0 states realized in supergravity as D2D0 ``halos'' around
 a core with nonzero D6 charge,
 there are walls of marginal stability which run all the
 way out to infinite K\"ahler class, leading to jumps in $\CZ_{\rm D6-D2-D0}$ when the $B$-field is
 varied, and hence to explicit deviations of physical stability from
 $\mu$-stability even in the infinite CY volume limit.
 Thus one can only potentially identify $\CZ_{\rm D6-D2-D0}$ with the
 DT partition function in certain
 limits of the background. This includes particular limits
 $B \to \infty$, for which we show that the contribution of all stable halo states to $\CZ_{\rm
 D6-D2-D0}$ is given exactly by the genus $r=0$ factor of the GV/DT infinite product.
 We argue that in such limits we can indeed identify $\CZ_{DT} = \CZ_{\rm D6-D2-D0}$, refining the
 arguments and result of \cite{Dijkgraaf:2006um}. Under this
 identification, the genus $r>0$ part of the DT infinite product counts
 ``core'' states, which are stable for any value of the $B$-field at infinite K\"ahler
 class.

 Sections \ref{sec:D6antiD6deg} and \ref{sec:dilutegaspf} contain
 the most subtle steps in our derivation of the OSV conjecture.
 We will outline the origin of the complications and discuss the
 unresolved issues that arose separately in section
 \ref{sec:challenges} below. In section \ref{sec:D6antiD6deg} we analyze the
 D6-anti-D6 type bound states giving rise to polar
 D4-D2-D0 states, and address in particular the question of which D6
 and anti-D6 states correspond to ``extreme'' polar states, i.e.\
 states whose reduced D0-charge $\hat{q}_0$ is near-extremal.
 Restriction to these states in the generating function is necessary to obtain exact factorization, which we further
 discuss in section \ref{sec:dilutegaspf}. Finally, in section \ref{sec:D4D2D0},  by
 combining the results of the previous sections, this leads to to a derivation of a refined
 version of the OSV formula.


 \item In section \ref{sec:discussion}, we give a thorough
 discussion of our final result.

 \item The seven appendices contain numerous technical details and several additional
 results. In appendix \ref{app:defconv}, we collect some definitions and
 conventions, and in appendix \ref{sec:alggeom} we summarize a number of results
 in algebraic geometry we use. In appendix \ref{app:finiteness}, we prove a
 partial result regarding the finiteness of the number of split
 attractor flows. In appendix \ref{app:attnumerics} we outline
 an efficient algorithm for  checking numerically the existence of attractor flow
 trees. These algorithms helped in checking the extreme polar state conjecture.
In appendix \ref{app:eulerint} we
 give the details of the computation of the closed loop three node quiver
 index mentioned above.  In appendix \ref{app:entropy5dBH}, we clarify some confusion
 which existed in the literature regarding whether one should compare the index
 or total degeneracy computed at zero string coupling to the black
 hole entropy (in particular in five dimensions, where it seemed that the former gave
 wrong results), and show it is in fact the index, if one uses the
 proper one. Finally, in  \ref{app:largeq}, we give an
 independent derivation of our version of the OSV formula in the $g_{\rm top}
 \to \infty$ regime, using techniques originally developed in
 \cite{Ashok:2003gk,Denef:2004ze} to count closed string flux vacua.
\end{enumerate}

\subsection{Challenges for a complete proof and unresolved issues}
\label{sec:challenges}

 Although the basic idea underlying
our derivation of the OSV conjecture is quite simple, turning it
into a complete proof proved to be a rather complex task, and we
have only been partially successful. This is not a shortcoming of
the IIA picture we work in
--- the same would be true if one wanted to turn the ideas of
\cite{Gaiotto:2006ns,deBoer:2006vg} into an actual proof, and the
complications outlined below have all direct equivalents in the
M-theory picture used there. We elaborate on this in section
\ref{sec:comparison}.

At the core of the complexity lies the fact that in order to obtain
the factorized form of the integrand in (\ref{OSVintformintro}) it
is necessary to introduce a cutoff. The factorization ultimately
comes from the fact that, through the fareytail series of section
\ref{sec:fareytail}, all D4 indices can be expressed in terms of the
indices of a finite number of polar D4 states, which as we mentioned
above do not form single centered black holes but can be described
as D6-anti-D6 bound states.\footnote{Here and in the following, by
D4 and D6 states, we mean states with arbitrary induced lower
dimensional charges.} In a suitable limit of the background, single
D6-states are counted by DT invariants, which determine $\CZ_{\rm
top}$. If there were a one-to-one map between all polar D4 states
and all possible pairs of single D6 and single anti-D6 states in the
background in which they are counted by DT invariants, we would thus
get exact factorization and a strong version of the OSV
conjecture.\footnote{Actually there would still be a series of
fareytail corrections, but these are under exact control, and each
of these corrections would again be factorized.}

Unfortunately, this is not the case. Not all polar states are
\emph{single} D6 - anti-D6 bound states, and moreover the subset of
single D6-anti-D6 pairs giving rise to actual bound states is rather
limited and complicated. In particular, even within the set of pairs
which do form bound states, the two elements of the pair cannot in
general be chosen independently, ruining exact factorization.

However, we will argue that for sufficiently polar states, i.e.\ D4
states with charges sufficiently close to those of a pure D4, the
desired one-to-one correspondence does indeed hold. \footnote{By a
``pure D4'' we mean a D4-brane BPS state with $N=0$,$F=0$ in the
notation of section \ref{sec:Dbranemodel} below. It has charge
$\Gamma = P +  (P^3+c_2\cdot P)/24$. Similarly, by a ``pure D6'' we
mean a rank 1 D6 brane with no lower charges, i.e. $\Gamma=1$. A
``pure fluxed D6'' means $\Gamma = e^S $. } ``Sufficiently close''
is measured by a parameter $\eta \geq 0$, defined in section
\ref{sec:extremepolar}, with $\eta=0$ corresponding to the pure D4,
which can be realized as a bound state of a pure D6 and a pure
anti-D6 with suitable fluxes turned on on their worldvolumes. More
precisely $\eta:=\frac{\hat{q}_0 - (\hat{q}_0)_{\rm
max}}{(\hat{q}_0)_{\rm max}}$.

Dropping all polar states with $\eta > \eta_*$ with $\eta_*$
sufficiently small allows us to obtain an approximate factorized
formula. The error introduced in this way turns out to amount to a
multiplicative correction to the integrand in
(\ref{OSVintformintro}) of order
\begin{equation} \label{errorexp}
 \exp \left[ \CO(e^{-\eta_* g_{\rm top} P^3}) \right].
\end{equation}
 Now the
essence of the OSV conjecture is that D-brane BPS degeneracies are
expressed in terms of the data of worldsheet instanton effects.
These effects contribute factors $\exp\left[ \CO(e^{- g_{\rm top}
P}) \right]$ to the integrand, where we used the relation $\Im\, t
\sim g_{\rm top} P$ detailed below in (\ref{osvsubst}). Thus, to
keep the error smaller than the effects of interest, we need to keep
$\eta_* \gg 1/P^2$.

Much of our work is aimed at investigating how large one can take
$\eta_*$ without ruining factorization. This requires getting
sufficient control on the very difficult problem of stability of BPS
bound states. For this purpose, we spend some effort extending the
development of the theory of attractor flow trees
\cite{Denef:2000ar,Denef:2000nb,Denef:2001xn,Bates:2003vx}. This
framework gives in principle a way to  study systematically
 stability issues. We successfully used it  to get large parts of
the problem under control, but regrettably a few gaps remain.

We make these gaps explicit by formulating a number of precise
conjectures. The first one is the ``split attractor flow
conjecture,'' formulated in section \ref{sec:flowtrees}. This states
essentially that we can classify BPS states by attractor flow trees,
and that the number of such trees of a given total charge is finite.
There are very good physical arguments for this, and the analysis of
this paper gives ample evidence for it. We have no reasons to doubt
it. The second one is the ``extreme polar state conjecture,''
formulated in section \ref{sec:extremepolar}. This states that all
polar states with $\eta < \eta_*$ sufficiently small can be realized
as single D6-anti-D6 bound states with the charges of the
constituents being close to those of the pure fluxed D6 and anti-D6
branes describing the pure D4. ``Close'' in this case is measured by
a parameter $\epsilon$ defined in section \ref{sec:D6antiD6deg}, eq.
(\ref{CCdef}),  with $\epsilon \sim \eta_*$. We give some physical
arguments and considerable numerical and analytical evidence for the
extreme polar state conjecture, and we strongly believe it to be
true.

Modulo one assumption, these conjectures then allow us to show that
we get the required factorization for a sufficiently small but
$P$-independent value of $\eta_*$, in which case indeed $\eta_* \gg
1/P^2$ in the large $P$ limit, making the error exponentially
smaller than the instanton contributions. That assumption is that
the BPS indices of the D6 and anti-D6 charges restricted by the
cutoff $\epsilon$ do not jump between the region in moduli space
where they equal DT invariants and the region in moduli space where
the central charges of the two constituents line up.

Unfortunately, this assumption turns out to be \emph{wrong} at large
$P$ with fixed $\epsilon$ ! There can be D6 or anti-D6 states within
the $\epsilon$ bound which \emph{do} decay between these loci in
moduli space. Such pairs cannot combine into D6-anti-D6 BPS bound
states, again spoiling the desired factorization, and moreover
spoiling the identification of the D6 indices with DT invariants and
the topological string.

Such states, which we call ``swing states,'' in fact do exist, at
least when $\eta_* \sim \epsilon
> \CO(1/P)$, as we discuss in section \ref{innocuouscoredump}. When
$\epsilon < \CO(1/P^3)$, we can prove in general that swing states
are absent at sufficiently large $P$. Let $\xi_{cd}$ be the minimal
value of $\xi$ such that taking $\epsilon < \delta/P^{\xi}$, swing
states are absent at sufficiently large $P$ for some fixed $\delta$
(for reasons which will become clear in section
\ref{innocuouscoredump}, we call this the ``core dump exponent'').
From what we just said, we know that $1 \leq \xi_{cd} \leq 3$. But
given the relation $\eta_* \sim \epsilon$, we need $\xi_{cd} \leq 2$
for the error not to be parametrically larger than the instanton
contributions. We suspect that in fact $\xi_{cd}=1$, and give some
circumstantial evidence for this claim, but are not fully confident,
so we consider this to be an unresolved issue. Note that in this
case, the corrections are of order $e^{- g_{\rm top} P^2} \sim e^{-
({\rm Im} \, t)^2/g_{\rm top}}$, suggestive of D4/M5 corrections to
the topological free energy. Indeed, the D6D2D0 swing states we find
are realized in supergravity as two-centered D6-D4 bound states,
which lift to M5 rings circling the center of Taub-NUT in M-theory.

Assuming $\xi_{cd} \leq 2$, there is no further obstacle to  proving
our refined OSV formula for $t_\infty = i \infty$, at least for
$g_{\rm top} > \CO(1)$, that is, at \emph{strong} topological string
coupling. The restriction to strong coupling might seem odd at
first, as this is opposite to the regime for which the OSV
conjecture was intended to be valid, but it becomes less so when one
realizes that all the old successes of microscopically reproducing
the Bekenstein-Hawking entropy, such as \cite{Maldacena:1997de}, in
fact have saddle point values $g_{\rm top} \gg 1$, this being
essentially equivalent to being in the regime of applicability of
the Cardy formula. Also, all of the other recent studies based on
reduction to a weakly coupled brane system
\cite{Gaiotto:2006ns,deBoer:2006vg}, although going beyond the
$g_{\rm top} \gg 1$ limit, are upon closer inspection implicitly
equally limited to the strong topological coupling regime $g_{\rm
top} > \CO(1)$. Moreover, the checks for  small black holes
\cite{Dabholkar:2005by,Dabholkar:2005dt} were in general only valid
in the region of strong topological string coupling.

Technically, the reason for the restriction to $g_{\rm top} >
\CO(1)$ comes from the fact that the error estimate
(\ref{errorexp}), which arises from dropping all polar states with
$\eta > \eta_*$, is actually only manifestly valid for $g_{\rm top}$
sufficiently large. In this case, the most important contributions
to the error come from polar states with $\eta$ near the cutoff, as
the ones with large values of $\eta$ are exponentially suppressed.
However, when $g_{\rm top}$ becomes smaller, the exponential
suppression becomes weaker and at a certain point, the bulk of the
polar terms (order 1 values of $\eta$) will start to dominate the
original partition function and therefore the error produced by
dropping them, because they have more entropy than the extreme polar
ones. So something like a phase transition occurs, with $g_{\rm
top}$ playing the role of inverse temperature. Simple estimates
suggest that the degeneracies of the polar states at fixed $\eta$
grow with $P$ as $e^{\eta P^3}$, and the transition occurs at the
value of $g_{\rm top}$ where this starts to dominate over the
suppression factor $e^{-\eta \, g_{\rm top} P^3}$, hence at some
$\CO(1)$ value of $g_{\rm top}$. If the growth estimate is correct,
then for values of $g_{\rm top}$ less than this, factorization
breaks down.

This is not a failure of the derivation itself, but is in fact very
closely related to the entropy enigma mentioned earlier --- indeed,
the supergravity configurations dominating the entropies of the
polar states are precisely of the same kind as those giving rise to
the entropy enigma. Moreover, since the saddle point value of
$g_{\rm top}$ in (\ref{OSVintformintro}) scales as $1/\Lambda$ when
$(p,q) \to \Lambda (p,q)$, the large $\Lambda$ regime is equivalent
to the small $g_{\rm top}$ regime, and thus the appearance of the
enigmatic configurations with entropy scaling as $\Lambda^3$ is
consistent with the potential failure of the conjecture at weak
$g_{\rm top}$ (and $t_\infty = i\infty$), which predicts $\log
|\Omega| \sim \Lambda^2$

We can only say there is a ``potential failure'' here because there
is  a possible loophole which might still save the conjecture even
in this large $\Lambda$ regime. This loophole is discussed in detail
in section \ref{sec:discussion}. It is based on the fact that since
$\Omega(p,q;t_\infty)$ is an \emph{index}, it receives contributions
of different signs from many multicentered black hole
configurations, so there might in principle be miraculous
cancelations altering the exponential growth from $e^{c \Lambda^3}$
down to $e^{c \Lambda^2}$.   We argue this is very unlikely, but
might still have a (remote) chance of being true if a similar
cancelation happens for the DT invariants approximately building up
our polar indices. The problem for DT invariants can be phrased in a
mathematically precise way. The upshot is that if $N_{DT}(\beta,n)$
is a DT invariant for curve class $\beta$ with $D0$ charge $n$ then
we should study the large $\lambda$ asymptotics of $\log
N_{DT}(\lambda^2 \beta,\lambda^3 n) \sim \lambda^k$. (See eq.
(\ref{preciseasympts}) for a more precise version.) The
straightforward estimate based on the entropy of the corresponding
D6D2D0 black holes suggests $k=3$, which would invalidate weak
coupling OSV at $t_{\infty} = i \infty$. If, due to miraculous
cancelations between different contributions to the DT invariants,
we get $k \leq 2$, this would be suggestive of cancelations between
the related actual indices, and perhaps the extension to weak
coupling might still be possible (but this is by no means
guaranteed). Although such cancelations might seem like ludicrous
wishful thinking, we discuss a number of heuristic arguments pro
(but also contra) this hypothesis.

Of course, it might also be that one should not take $t_\infty = i
\infty$. Other natural prescriptions might be to take $t_\infty$ to
be at the attractor point $t_*(p,q)$ in (\ref{OSVintformintro}) or
to take $t$ finite and fixed while sending $P \to \infty$. Both of
these turn out   automatically to
  eliminate the enigmatic $\Lambda^3$ configurations when $\Lambda \to \infty$, but would
also spoil some of the interesting interpretations of the conjecture
as an example of large radius D-brane gauge theory - gravity
duality. Moreover, they would also push direct microscopic
verification into a quantum geometric regime (due to the importance
of $\alpha'$ correction at finite values of $\Im \, t_\infty$) which
so far has proven intractable. We again refer to section
\ref{sec:discussion} for more details.

\vskip5mm \noindent \emph{Note added in version 2:} \vskip2mm

\begin{enumerate}
 \item After version 1 of this paper appeared on the arXiv, the paper \cite{Huang:2007sb} appeared,
 in which the growth of DT invariants
 $\log N_{DT}(\lambda^2 \beta, \lambda^3 n) \sim \lambda^k$ was
 numerically studied, based on available data sets of DT invariants of a number of compact
 CY manifolds. Although these data sets are too limited to directly extract
 asymptotics, one can get predictions for $k$ by using
 Richardson transforms. Surprisingly, the results
 suggest $k=2$, exactly the critical value for
 the OSV conjecture at $t=i\infty$ to have a chance of being correct even at weak topological string
 coupling, and implying the ``miraculous cancelations'' do indeed occur!
 Although as mentioned above and discussed at length in section
 \ref{sec:rov}, such cancelations at the level of DT
 invariants are not quite enough to make OSV work at weak coupling (for this one also needs cancelations
 at a more detailed level of different contributions to the D6D4D2D0 indices),
 it is clear that if the numerical results of \cite{Huang:2007sb}
 indeed correctly capture the $\lambda \to \infty$ asymptotics of
 the DT invariants, the unknown mechanism underlying these
 cancelations might conceivably also imply the more general
 cancelations required for weak coupling OSV. It would be extremely
 interesting to settle this issue.
 \item We would like to stress that the implications of such miraculous cancelations could be
 enormous, going well beyond the issue of the range of validity of
 the OSV conjecture. In particular, if $k=2$, then this
 raises the possibility that the
 Donaldson-Thomas partition function (\ref{ZDTdefinition}) gives a convergent,
 nonperturbative completion of the topological string partition
 function, of which (\ref{gvprodexp}) is a divergent asymptotic
 expansion. Indeed if $k=2$ the sum over $\beta$ will have a nonzero
 radius of convergence for fixed $n$.
 However proper convergence would actually also require $N_{DT}(\lambda^2
\beta,\lambda^3
 n)$ to vanish identically at sufficiently large $\lambda$
 for all strictly negative $n$, which requires an even more miraculous,
 exact
 cancelation to occur. Further work is needed to determine whether this
 might be the case or not.
 It is  perhaps also worth noting that our equations (\ref{EPfact1})-(\ref{EPfact2}) below in fact
 can be interpreted as defining a nonperturbative completion of the
 norm squared $|\CZ_{\rm top}|^2$ of the topological string wave function, starting
 from the \emph{polar part} $\CZ^-$ of the D4 partition function, since when $P \to
 \infty$,
 $\CZ^\epsilon_{DT}$ becomes $\CZ_{DT}$.

\end{enumerate}

\subsection{Preliminaries} \label{sec:prelim}

Let us conclude this section by reviewing some basic definitions
which will be used in the text.

We will be studying type IIA D-branes wrapping cycles in a
nonsingular compact Calabi-Yau 3-fold $X$ of generic holonomy. The
Hilbert space of type IIA on $\IR^{1,3}\times X$ is graded by  RR
charge $\Gamma \in K^0(X)$. We ignore possible torsion subgroups and
 identify the charge group  with $H^{\rm even}(X;\IZ)$, modulo torsion.
 Near   a large radius limit of
$X$ there is a canonical electromagnetic decomposition
$\Gamma=(p,q)$ with magnetic charges $p\in H^0(X;\IR)\oplus
H^2(X;\IR)$ and electric charges $q\in H^4(X;\IR)\oplus H^6(X;\IR)$.
Picking a basis $\{D_A \}_A$, $A=1,\ldots,h:=h^{1,1}(X)$ of
$H^2(X,\IZ)$, the charges can be written in components as $p =: p^0
+ P^A D_A$, $Q_A := \int_X D_A \wedge q$, $q_0 := \int_X q$. We also
introduce an index $\Lambda$ running over $0,1,\dots, h$, and denote
components of $p$ by $p^\Lambda$ and $q$ by $q_\Lambda$.
\footnote{Sometimes $q$ refers just to the $D2$ charge and not the
total electric charge. In this case it should be clear from context
which one is meant.}   The crucial boundary conditions on the fields
at spatial infinity are those for the vectormultiplet scalar fields
of the effective $\CN=2$ supergravity defined on $\IR^{1,3}$. For
IIA compactifications these are the complexified K\"ahler moduli $t
:=t^A D_A:= B+iJ $. In the superselection sector $(p,q)$ there is a
central charge $Z(p,q;t)$ of the $\CN=2$ supersymmetry algebra and
there is is a well-defined finite-dimensional space of BPS states
$\CH(p,q;t)$. These are the states at rest transforming in the small
representations of the little $\CN=2$ superalgebra. Alternatively,
they are the 1-particle states satisfying the energy bound $E=\vert
Z(p,q;t)\vert$.

As pointed out in \cite{Dabholkar:2005by} the appropriate index to
use in this context
 is the second helicity supertrace. We
will denote it as:\footnote{The normalization factor $-2$ is chosen
such that one (half) hypermultiplet gives $\Omega=+1$.}
\begin{equation} \label{indexdef}
 \Omega(p,q;t):=- 2 \, {\rm Tr}_{\CH(p,q;t)} \, (-1)^{2 J_3} J_3^2.
\end{equation}
Here   $J_3$ is the 3-component of spatial angular momentum. Since
every BPS particle in an $\CN=2$ theory has a universal
half-hypermultiplet factor $({\bf 0},{\bf 0};{\bf \half})$ obtained
from quantizing the fermionic degrees of freedom associated to its
center of mass in $\IR^3$, we can also write $\CH(p,q;t)= ({\bf
0},{\bf 0};{\bf \half})\otimes \CH'(p,q;t)$ and
\begin{equation} \label{indexred}
 \Omega(p,q;t) = -2 \, {\rm Tr}_{\CH'(p,q;t)}\,
 \left( 2 (J_3')^2 - (J_3'-\frac{1}{2})^2 - (J_3'+\frac{1}{2})^2  \right) (-1)^{2
 J'_3} = {\rm Tr}_{\CH'(p,q;t)} \, (-1)^{2 J_3'}.
\end{equation}
Here   $J_3'$ is the   reduced angular momentum.

We include the $t$-dependence since even though $\Omega$ is an
index, it \emph{does} depend on the background complexified K\"ahler
moduli $t^A$, through jumping phenomena at walls of marginal
stability. These are walls where the phases of the central charges
of the constituents of a bound state line up, so decay into them is
no longer energetically obstructed. As we will see this is not just
a minor nuisance; it affects the regime of validity of the OSV
conjecture in a significant way and moreover associated wall
crossing formulae for the index will prove a powerful tool, central
in our derivation.

With these indices one can define a (formal) partition sum at fixed
magnetic charge $p$ by summing over electric charges
$q$:\footnote{Our normalization conventions differ from
\cite{Ooguri:2004zv,Dabholkar:2005by,Dabholkar:2005dt}: $\phi$(here)
= $-\phi$(\cite{Ooguri:2004zv})$/2\pi$ =
$-\phi$(\cite{Dabholkar:2005by,Dabholkar:2005dt})$/2$.}
\begin{equation} \label{Zosv}
 {\cal Z}_{\rm BH}(\phi;t_\infty) := \sum_{q} \Omega(p,q;t_\infty)
 \, e^{2 \pi \phi^\Lambda q_\Lambda }.
\end{equation}
In terms of this generating function the conjecture
\cite{Ooguri:2004zv} states that in a suitable parameter regime,
\begin{equation} \label{osvconj}
 {\cal Z}_{\rm BH}(\phi;t_\infty) \sim |{\cal Z}_{\rm top}(g_{\rm top},t)|^2
\end{equation}
where ${\cal Z}_{\rm top}$ is the topological string partition
function and the following substitutions are understood:
\begin{equation} \label{osvsubst}
  g_{\rm top} = \frac{4\pi i}{X^0}=
  \frac{4\pi}{2 {I^0}_\Lambda \phi^\Lambda + i \, p^0 }, \qquad
  t^A = \frac{2 {I^A}_\Lambda \phi^\Lambda + i \, p^A}{2 {I^0}_\Lambda \phi^\Lambda + i \, p^0}.
\end{equation}
Here  ${I^{\Lambda_1}}_{\Lambda_2}$ is the inverse symplectic
intersection form between magnetic and electric charges, where
intersection products are assumed to equal zero between magnetic
charges and between electric charges. The presence of this
intersection form can be deduced for example from the results of
\cite{Bates:2003vx}. In a canonical symplectic charge basis (which
was assumed in \cite{Ooguri:2004zv}), one has
${I^{\Lambda_1}}_{\Lambda_2}=\delta^{\Lambda_1}_{\Lambda_2}$, but
more generally it is sometimes more natural to work in a basis with
a different intersection form. For example, as we review in appendix
\ref{app:defconv}, in type IIA at large radius, where RR charges are
given by $\Gamma = {\rm ch}(F) \wedge \sqrt{\widehat{A}}$, the
natural choice of basis gives
${I^{\Lambda_1}}_{\Lambda_2}=\sigma_{\Lambda_2} \,
\delta^{\Lambda_1}_{\Lambda_2}$ where $\sigma_0=1$, $\sigma_A = -1$.
We trust the reader will not confuse the background moduli
$t_\infty$ with the $t^A$ substituted on the RHS of the OSV
conjecture.

As mentioned already in the introduction, an alternative way of
writing (\ref{osvconj}) is
\begin{equation} \label{OSVintform}
 \Omega(p,q;t_\infty) \sim \int d\phi \, e^{- 2 \pi \phi^\Lambda q_\Lambda
 } \, |{\cal Z}_{\rm top}|^2,
\end{equation}
where again the substitutions (\ref{osvsubst}) are understood on the
right hand side.

In this text, we will define the topological string partition
function $\CZ_{\rm top}$ associated to  a Calabi-Yau threefold $X$
as follows:
\begin{eqnarray}
 {\cal Z}_{\rm top}(g,t) &:=& {\cal Z}_{\rm pol}(g,t) \,
 {\cal Z}_{GW}^{0}(g) \, {\cal Z}'_{\rm GW}(g,t) \\
 {\cal Z}_{\rm pol}(g,t) &:=& \exp \biggl( -\frac{(2\pi i)^3}{6 g^2} D_{ABC} t^A t^B t^C
 - \frac{2 \pi i}{24} c_{2A} t^A \biggr) \\
 {\cal Z}_{GW}^0(g) &:=& \biggl( \prod_{n} (1-e^{-g n})^n
 \biggr)^{-\chi(X)/2} \label{GWdeg0} \\
 {\cal Z}'_{GW}(g,t) &:=& \exp \biggl( \sum_{\beta \neq 0} \sum_h
 N_{h,\beta} \,
 (-g^2)^{h-1} \, e^{2 \pi i \beta_A t^A} \biggr).
 \label{GWdegnon0}
\end{eqnarray}
 Here $D_{ABC}$ are the triple intersection numbers of the
basis $\{ D_A \}_A$, $c_2$ is the second Chern class of $X$,
$\chi(X)$ is the Euler characteristic of $X$, and $N_{h,\beta}$ are
the Gromov-Witten invariants counting the ``number'' of holomorphic
maps of genus $h$ into class $\beta \in H_2(X,\IZ)$. Henceforth we
will usually drop the subscript on $g_{\rm top}$ and simply write
$g$.

In (\ref{GWdeg0}), we resummed the contributions of the degree zero
Gromov-Witten invariants $N_{h,0}$ into the McMahon form
(\ref{GWdeg0}). At small $g$, it is more suitable to use the
asymptotic expansion given by
\begin{equation} \label{hzergw}
\CZ_{GW}^0(g) \approx K \left( \frac{g}{2 \pi}
\right)^{\frac{\chi(X)}{24}} \exp \left(
   \frac{\chi(X)}{2} \frac{\zeta(3)}{g^{2}} +
   \sum_{h=1}^\infty N_{h,0} \, (-g^2)^{h-1} \right).
\end{equation}
where $K$ is a constant. (See Appendix E of \cite{Dabholkar:2005dt},
or eq.(4.34) et. seq. of \cite{Pioline:2006ni} for a careful
derivation.)

Similarly, one can rewrite $\CZ_{GW}'(g,t)$ as an infinite product
using M2 BPS invariants \footnote{These are also known as
Gopakumar-Vafa invariants.}
\cite{Gopakumar:1998ii,Gopakumar:1998jq,Klemm:2004km}. Denoting the
BPS invariants as $n^r_q$ we have $\CZ_{GW}'(g,t) =
\CZ'_{GV}(e^{-g},e^{2 \pi i t})$, where
\begin{eqnarray}\label{gvprodexp}
 \CZ_{GV}'(e^{-g},e^{2 \pi i t}) &=& \prod_{q>0,k>0} (1-e^{-g k + 2 \pi i q \cdot t})^{k
 n_q^0} \\
 && \quad \times \prod_{q>0,r>0} \prod_{\ell=0}^{2r-2}
 \left( 1 - e^{-g(r-\ell-1) + 2 \pi i q \cdot t} \right)^{(-1)^{r+\ell} {2r-2 \choose \ell} \,
 n^r_q}.
\end{eqnarray}
Finally, $\CZ_{\rm top}$ is conjecturally related
\cite{Iqbal:2003ds,MNOP1,MNOP2,Dijkgraaf:2006um} to the
Donaldson-Thomas partition function for ideal sheaves, as
\begin{equation} \label{DTGVcorrespondence}
 \CZ_{DT}'(u,v) = \CZ_{GV}'(-u,v).
\end{equation}
Here
\begin{eqnarray}
 \CZ_{DT}(u,v) &:=& \sum_{n,\beta} N_{DT}( \beta,n) \,
 u^n \, v^\beta := \CZ^0_{DT}(u) \, \CZ'_{DT}(u,v) \label{ZDTdefinition} \\
 \CZ^0_{DT}(u) &:=& \prod_n (1-(-u)^n)^{-n \chi(X)} = \left( \CZ^0_{\rm GW} \right)^2 \label{ZDT0MacMahon},
\end{eqnarray}
where $v^{\beta} := \prod_A (v_A)^{\beta_A}$, and $N_{DT}( \beta,n)$
are the ideal sheaf DT invariants, defined in \cite{DT1,DT2,DT3}.
Physically they can be thought of as counting D6-D2-D0 BPS states
with D0-charge $n$ and D2-charge $-\beta$, \emph{ignoring} stability

(i.e.\ ignoring D-term constraints on the D6-D2-D0 moduli space).

The conjecture (\ref{DTGVcorrespondence}) has been confirmed by many
case studies, partially proved, and is physically well supported
\cite{Iqbal:2003ds,Dijkgraaf:2006um}. We will assume it is true.

We conclude with two remarks:

\begin{enumerate}

\item Our definition of $\CZ_{\rm top}$ is slightly nonstandard
because of the way we handled the $\beta=0$ invariants. From
(\ref{hzergw}) we see that there is an extra summand
$\frac{\chi}{24}\log\frac{g}{2\pi} $ in the definition of $F_{\rm
top}$. This is large for $g$ both small and large, and has the
important property that the expansion of $F_{\rm top}$ is \emph{not}
analytic in $g$.

\item  For many Calabi-Yau
manifolds one can use known asymptotic growth estimates of the
Gromov-Witten invariants following from the results of
\cite{Candelas:1990rm,Harvey:1995fq} to show that the first line of
(\ref{gvprodexp}) indeed converges as an analytic product for
sufficiently small $u$ and sufficiently large K\"ahler classes. It
will, however, have interesting singularities. On the other hand, it
already follows from the results of \cite{Klemm:2004km} that
$\CZ_{DT}^{r>0}$ has zero radius of convergence, and must be
considered a formal product. In section \ref{sec:D6D4D2D0} we will
give a nice physical interpretation of this second product as a
product over ``core states.''

\end{enumerate}

\section{A fareytail expansion for the D4-D2-D0 partition function}
\label{sec:fareytail}

In this section we will show that $\CZ_{\rm BH}$ for $p^0=0$ and
$t_{\infty}=i \infty$ can in general be expressed as a fareytail (or
Rademacher-Jacobi) series built from its polar part, generalizing
the results of \cite{Dijkgraaf:2000fq,Moore:2004fg}. The derivation
given here is dual to the derivations
\cite{deBoer:2006vg,Kraus:2006nb} which appeared while this paper
was being written. We include it nevertheless for completeness and
because we fill in some of the gaps in those proofs and clarify some
issues which were left open, e.g.\ how to define a fareytail series
for the actual partition function instead of for the fareytail
transform of it.

\subsection{D-brane model, physical interpretation, and
S-duality}\label{sec:Dbranemodel}

Consider a single D4-brane wrapped on a smooth holomorphic surface
in $X$. This surface will be in an ample divisor class $P=P^A D_A$
and we  often (somewhat sloppily) refer to the surface also as $P$.
We assume the surface has  $N$ $\overline{D0}$-branes\footnote{In
our conventions, which follow \cite{Brunner:1999jq} and are natural
from a geometric point of view, D4 branes form bound states with
\emph{anti}-D0 branes at large radius, and D6-branes with anti-D2
branes.} bound to it and $U(1)$ flux $F\in H^2(P)$ turned
on.\footnote{$H^2(P)$ and $H^2(X)$ will refer to the
\textit{integral} cohomology modulo its torsion subgroup. }

Defining electric charges as the quantities coupling to the
RR-potentials, the D2-brane charges are
\begin{equation} \label{q2val}
  q_A = Q_A = D_A \cdot F.
\end{equation}
Here and in what follows the scalar product between two 2-forms in
$H^2(P)$ is the intersection product:
\begin{equation}
  \alpha_1 \cdot \alpha_2 := \int_P \alpha_1 \wedge \alpha_2
\end{equation}
If the 2-forms are pulled back from $H^2(X)$, i.e.\ $\alpha_i =
\iota_P^* \hat{\alpha}_i$, this can also be written as
\begin{equation}
 \alpha_1 \cdot \alpha_2 = P \cdot \hat{\alpha}_1 \cdot
 \hat{\alpha}_2 = D_{ABC} \, P^A \hat{\alpha}_1^B \hat{\alpha}_2^C.
\end{equation}
To avoid cluttering, in the following we usually will not
notationally distinguish between $\hat{\alpha}_i$ and its pullback
$\alpha_i$, hoping that this will not cause confusion.

The total D0-brane charge is
\begin{equation} \label{q0val}
  q_0 = -N + \frac{1}{2} F^2 + \frac{\chi(P)}{24}
\end{equation}
where
\begin{equation}\label{eulercharp}
 \chi(P)=P^3 + c_2(X) \cdot P := D_{ABC} P^A P^B P^C +
 c_{2,A} P^A
\end{equation}
is the Euler characteristic of $P$. The last term in (\ref{q0val})
represents the curvature induced D0-brane charge on the D4-brane.

To have a supersymmetric configuration, one needs
$F^{(2,0)}=0(=F^{(0,2)})$. (The $B$-field does not appear here
because it is always of $(1,1)$ type for flat $B$.) For generic
fluxes at generic points in the D4-brane moduli space, this
condition will not be satisfied. Exceptions are fluxes $F$ which are
pulled back from $H^2(X)=H^{1,1}(X)$: for these, $F^{(2,0)} = 0$
identically. However, in general there will be (many) elements of
$H^2(P)$ which are not pulled back from $H^2(X)$. For these, the
condition $F^{(2,0)}=0$ imposes $h^{2,0}(P)$ equations on the
$h^{2,0}(P)$ geometric moduli of $P$, which will generically
restrict the divisor moduli to a set of isolated points, and more
generally to a subvariety of the original moduli space
\cite{Gomis:2005wc,Gaiotto:2005rp}. The $N$ $\overline{D0}$-branes
bound to the D4 are pointlike\footnote{When a suitable nonzero
$B$-field is turned on, they can alternatively be considered to be
smooth noncommutative $U(1)$ instantons.} and not obstructed by the
fluxes.

Thus we can rewrite (\ref{Zosv}) as
\begin{equation}
 \CZ_{\rm BH}(\phi^0,\Phi) = \sum_{F,N} d(F,N)
 \, e^{2 \pi \phi^0[-N + \frac{1}{2} F^2
 + \frac{\chi(P)}{24}] + 2 \pi \Phi \cdot F 
 }
\end{equation}
where the sum is over $U(1)$ worldvolume fluxes $F$  on $P$ and the
number $N \geq 0$ of bound pointlike anti-D0 branes. The flux
lattice is $L = \sigma/2 + H^2(P)$, with $\sigma/2 := c_1(P)/2 =
\iota_P^* P/2$ mod 1, where $\iota_P$ is the embedding map of the
surface $P$. We will choose the natural representative $\sigma =
\iota_P^* P$, which we will also denote simply as $P$. The
half-integral shift follows from the K-theoretic formulation of RR
charges and is needed to cancel anomalies, both on the brane
worldvolume \cite{Minasian:1997mm} and on the  fundamental string
worldsheet \cite{Freed:1999vc}.

The index $d(F,N)$ is defined similarly to
(\ref{indexdef})-(\ref{indexred}) but now with the trace in a sector
of fixed $(F,N)$, for $J \to \infty$. The angular momentum can be
identified with the Lefshetz $SU(2)$ action on the moduli space, so
in particular the 3-component of the spin of a $p$-form is $J_3' =
(p-\dim)/2$, where $\dim$ is the complex dimension of moduli space
\cite{Witten:1996qb}. This identifies our index $d(F,N)$ up to a
sign with the Euler characteristic of the moduli space $\CM_{F,N}$
of BPS configurations in the sector labeled by $(F,N)$:
\begin{equation} \label{indextoeuler}
 d(F,N)=(-1)^{\dim \CM_{F,N}} \, \chi(\CM_{F,N}).
\end{equation}
Note that $d(F,N)$ is independent of the $B$-field, because $B$ does
not appear in the (large radius) BPS conditions for D4-D2-D0 bound
states and does not affect the moduli spaces.

Typically $\CM_{F,N}$ has singularities, and as a result it is not
directly clear what the proper mathematical definition is of the
Euler characteristic $\chi(\CM_{F,N})$ to get the correct physical
index. It would be worthwhile to have a precise mathematical
definition of the invariants $d(F,N)$. Quite possibly they are DT
invariants for torsion sheaves \cite{DT1,DT2,DT3}. Some discussion
of the subtleties involved can be found in \cite{Gaiotto:2006wm}.

It was noted e.g.\ in section 6 of \cite{Dabholkar:2005dt} that this
partition function is everywhere divergent, but that this can be
cured in a natural way by adding a Boltzmann weight $e^{-\beta H}$
with $H$ the BPS energy of the state. The resulting partition sum is
then naturally interpreted as the BPS partition
function\footnote{This should be given by the partition function of
a suitably topologically twisted D4 DBI theory, possibly the theory
constructed in \cite{Spence:1999xb}. It would be interesting to make
this precise.} of a single D4-brane wrapping $P$ and a Euclidean
time circle of circumference $\beta$, in a background with flat RR
potentials\footnote{Flat $RR$ potentials are properly described by
the compact $K$-group $K^{-1}(X;\IR/\IZ)$. This determines the
proper periodicities for these fields. We will ignore such
subtleties in this paper.}
\begin{equation}
 C_3 =: C \wedge \frac{dt}{\beta}, \qquad C_1 =: C_0 \wedge
 \frac{dt}{\beta}.
\end{equation}
Here $C\in H^2(X,\IC)$.  The BPS partition function is roughly ${\rm
Tr}\, (-1)^{2 J_3'} \, e^{-\beta H - 2\pi i q_\Lambda \cdot
C^\Lambda}$ where the trace sums over all D4 states including all
sectors $(F,N)$. More precisely (in units with $\ell_s := 2 \pi
\sqrt{\alpha'} = 1$):
\begin{eqnarray} \label{ZD4def}
 {\cal Z}_{D4}(\frac{\beta}{g_{\rm IIA}},C_0,C;B+iJ)
 &:=& {\rm Tr} \, (-1)^{2 J_3'}
 e^{-\beta H - 2 \pi
 i [-N + \frac{1}{2} \CF^2 + \frac{\chi(P)}{24}] C_0 -2\pi i \CF \cdot (C + \frac{P}{2})} \\
 &=& \sum_{F,N} d(F,N) \,
 e^{-\frac{2 \pi \beta}{g_{\rm IIA}} |Z(F,N;B+iJ)|} \nonumber \times \\
 && \quad \times \, e^{- 2 \pi
 i [-N + \frac{1}{2} \CF^2 + \frac{\chi(P)}{24}] C_0 -2\pi i \CF \cdot (C  + \frac{ P}{2})}
\end{eqnarray}
where $\CF := F-B$. The ``extra'' factor $e^{-\pi i \CF \cdot P}$
must be there for S-duality to work properly
\cite{Belov:2004ht,Kapustin:2006pk,Nikita}, as we will confirm
below. The string coupling constant $g_{\rm IIA}$ is the physical
IIA coupling, not to be confused with the topological string
coupling. The quantity $Z(F,N;B+iJ)$ denotes the holomorphic central
charge of the D4-D2-D0 system, which in our conventions with charge
$\Gamma \in H^{\rm even}(X)$ at large $J$ is given by
(\ref{centralchargedef}):
\begin{equation} \label{centralcharge}
 Z(F,N;B+iJ) = -\int_X e^{-(B+iJ)} \Gamma = \frac{1}{2} J^2 + i \CF \cdot
 J + [N-\frac{1}{2} \CF^2 - \frac{\chi(P)}{24}].
\end{equation}
Recall that $J^2/2 \equiv \int_P J^2/2$ is the volume of $P$. The
absolute value of $Z$ is proportional to the DBI energy evaluated on
BPS configurations. Note that equation (\ref{ZD4def})  does depend
on $\beta$ and the background metric, but in a quasi-topological
way, fixed by charges and background K\"ahler moduli.

In the limit $J \to \infty$   we have
\begin{equation}
 |Z| = \frac{1}{2} J^2 + \frac{(\CF \cdot J)^2}{J^2} +  [N-\frac{1}{2} \CF^2 - \frac{\chi(P)}{24}]
 + O(1/J^2).
\end{equation}
Since $\CF$ is of type $(1,1)$ on supersymmetric solutions, we can
use the Hodge index theorem, which states that the lattice of
$(1,1)$ classes has Lorentzian signature, to decompose $\CF$ in
self-dual and anti-selfdual parts as:
\begin{equation}
 \CF = \CF_+ + \CF_-, \qquad \CF_+ = \frac{\CF \cdot J}{J^2} \, J,
 \qquad \CF_- \cdot J = 0.
\end{equation}
We refer to appendix \ref{sec:alggeom} for more details. With this
we can redefine $\CZ_{D4}$, dropping an irrelevant (S-duality
invariant) overall factor $e^{- 2 \pi \Im \tau {\rm Vol}}$ as:
\begin{equation} \label{ZD4tautaubar}
 \CZ_{D4}(\tau,C,B) := \sum_{F,N} d(F,N) \,
 e^{2 \pi i \tau [N - \frac{1}{2} \CF_-^2 - \frac{\chi(P)}{24}]
 - 2 \pi i \bar{\tau} \frac{1}{2} \CF_+^2
 -2\pi i \CF \cdot (C + \frac{P}{2})},
\end{equation}
where
\begin{equation}
 \tau = C_0 + \frac{\beta}{g_{\rm IIA}} \, i.
\end{equation}
Since $\CF_-^2<0$ and $\CF_+^2 > 0$, the sum over fluxes is well
behaved in (\ref{ZD4tautaubar}).

Alternatively we can write (\ref{ZD4def}) as
\begin{equation}
 \CZ_{D4} = e^{-2 \pi i \tau \frac{\chi(P)}{24}} \sum_{F,N} d(F,N) \,
 e^{-2\pi [\Im \tau \int \frac{1}{2} \CF \wedge *\CF + i \, \Re \tau \int \frac{1}{2} \CF \wedge
 \CF -i \tau N   + i \CF \cdot (C + \frac{P}{2})]}.
\end{equation}
Not surprisingly, part of the exponent has the form of a $U(1)$
Yang-Mills energy with complexified coupling constant $\tau$.
Nevertheless,  the coefficients $d(F,N)$ are nontrivial and the
expression is not proportional to the standard theta function of a
$U(1)$ gauge theory. The reason for this is that the theory we are
considering is more than just standard topologically twisted $\CN=4$
$U(1)$ Yang-Mills, since we consider arbitrarily large deformations
of the D4-brane, which are only properly described by the full DBI
theory. Moreover, we include  pointlike bound
$\overline{D0}$-branes, which do not correspond to standard smooth
Yang-Mills instantons. Note also that the curvature term
proportional to $\chi(P)$, which is crucial for modular invariance,
appears naturally here.

The OSV black hole partition function ${\cal Z}_{\rm BH}$ is
obtained by making a formal substitution of arguments in  ${\cal
Z}_{D4}$: \footnote{This is sometimes referred to as the ``OSV
limit.'' However it is not in any sense a well-defined limit.}
\begin{equation} \label{osvlimit}
 {\cal Z}_{\rm BH}(\phi^0,\phi^A)
 := {\cal Z}_{\rm D4}(\beta = 0,B = 0,C_0 = i \phi^0,C = i \Phi - \frac{P}{2}).
\end{equation}
We put $B = 0$ because we do not want to introduce explicit
$B$-dependence in ${\cal Z}_{\rm BH}$. Here $\phi^0, \Phi$ are real.

Unlike the partition function ${\cal Z}_{D4}$, which from
(\ref{ZD4tautaubar}) and the large $N$ asymptotics $d(F,N) \sim e^{k
\sqrt{N}}$ \cite{Maldacena:1997de,Vafa:1997gr} is easily seen to
converge for any $\beta
> 0$, the partition function ${\cal Z}_{\rm BH}$ diverges
everywhere. We can nevertheless make sense of it and justify formal
manipulations by turning on $\beta$ at intermediate steps. For
example we can write
\begin{equation} \label{Opqint}
 \Omega(p,q) = \lim_{\beta \to 0} \oint d C_0 \, d C \,
 \CZ_{D4}(\mbox{$C_0 + \frac{\beta}{g_{\rm IIA}} i$},C,B=0) \, e^{2 \pi i q_0 C_0
 + 2 \pi i Q \cdot (C + \frac{P}{2})}.
\end{equation}
The integrals run over one period of all RR potentials.
\footnote{Again, the $K$-theoretic interpretation can modify the
proper periods. This will at most result in a modest numerical
factor in \ref{Opqint}. We will ignore this possibility.}
  They are well defined, and
produce $\Omega(p,q) \, e^{-\beta |Z|/g}$, which in the limit $\beta
\to 0$ reduces to $\Omega(p,q)$. Often however it is not necessary
to perform the regularization explicitly; for example if one is only
interested in a saddle point evaluation of the integral, one can
proceed formally.

Besides providing a good regularization, this ``physical''
interpretation of the OSV partition function also allows applying
the usual T- and S-dualities one expects to be symmetries of
$\CZ_{D4}$. In particular performing a TST duality transforms this
into a form getting significantly closer to what one needs to derive
the conjecture.

The T-duality goes along the time circle, and trivially preserves
${\cal Z}$ but gives it the interpretation of a partition sum over
supersymmetric configurations of a Euclidean D3-brane wrapped on
$P$, with Euclidean time circumference $1/\beta$, IIB coupling
$g_{\rm IIB}=g_{\rm IIA}/\beta$ and RR potentials $C_0$ and $C$.

Next, we S-dualize. The D3-brane is self-dual under S-duality
\cite{Tseytlin:1996it}, which maps $\tau = C_0 + i/g_{\rm IIB}$ to
$-1/\tau$ while acting as electric-magnetic duality on the $U(1)$
gauge fields. The background fields transform as
\begin{equation}
 \tau'=-1/\tau, \quad \quad C'=-B, \quad B'=C,
 \quad J'=\sqrt{C_0^2+{g_{\rm IIB}}^{-2}} \, J.
\end{equation}
Note that the transformation of the K\"ahler form $J$ leaves the
background $J = \infty$ we are considering invariant. The partition
function must transform as a modular form with some weights
$(w,\bar{w})$, that is
\begin{equation}
 \CZ_{D3}' =  \omega_S \, \tau^{w}  \bar{\tau}^{\bar{w}} \CZ_{D3}.
\end{equation}
where $\omega_S$ is a phase and, for fractional $w,\bar w$ we use
the principal branch of the logarithm.  The sum over fluxes $F$ in
$\CZ_{D3}'$ is now over the dual lattice, but since $H^2(P)$ is
self-dual on a compact surface, this is the same as the original
lattice. In examples which can be checked explicitly this equality
of partition sums essentially amounts to a Poisson resummation (see
also appendix \ref{app:largeq}). Finally we can do another T-duality
along the time circle to go back to IIA, but this is again trivial.

We thus have
\begin{eqnarray}
 \CZ(\tau,C) &:=& \CZ_{D4}(\tau,C,B=0) \\
 &=& \omega_S^{-1} \tau^{-w}\bar{\tau}^{-\bar{w}} \,
 \CZ_{D4}(-\frac{1}{\tau},0,C) \\
 &=& \omega_S^{-1} \tau^{-w}\bar{\tau}^{-\bar{w}} \sum_{F,N} d(F,N) \,
 e^{\frac{2 \pi i}{\tau}[-N + \frac{1}{2} (F_- - C_-)^2 +
 \frac{\chi(P)}{24}] + \frac{\pi i}{\bar{\tau}} (F_+ - C_+)^2
 - \pi i F \cdot P} \\
 &=& \omega_S^{-1} \tau^{-w}\bar{\tau}^{-\bar{w}} \, e^{\pi i
 (\frac{1}{\tau} C_-^2 + \frac{1}{\bar{\tau}} C_+^2)} \, \\
 && \quad \times \sum_{F,N} d(F,N) \,
 e^{\frac{2 \pi i}{\tau}[-N + \frac{1}{2} F_-^2 +
 \frac{\chi(P)}{24}] + \frac{\pi i}{\bar{\tau}} F_+^2
 - 2 \pi i (F_- \cdot \frac{C}{\tau} + F_+ \cdot \frac{C}{\bar{\tau}} )
 - \pi i F \cdot P}
 \label{Zosvresummed} \\
 &=& \omega_S^{-1} \tau^{-w}\bar{\tau}^{-\bar{w}} \, e^{\pi i
 (\frac{1}{\tau} C_-^2 + \frac{1}{\bar{\tau}} C_+^2)} \,
 \CZ_{D4}(-\frac{1}{\tau},\frac{C}{\tau},B=0) \\
 &=& \omega_S^{-1} \tau^{-w}\bar{\tau}^{-\bar{w}} \, E[-\frac{C^2}{2
 \tau}] \, \CZ(-\frac{1}{\tau},\frac{C}{\tau}). \label{ZSdual}
\end{eqnarray}
In the last line we introduced the shorthand notation
\begin{equation} \label{shortnot}
 E[f(\tau) X \cdot Y] := e^{-2 \pi i \, f(\tau) \, X_- \cdot Y_-
- 2 \pi i \, f(\bar{\tau}) \, X_+ \cdot Y_+}, \qquad E[A+B] :=
E[A]\,E[B].
\end{equation}
To summarize, if we define
\begin{equation} \label{ZtauCdef}
\CZ(\tau,C) := \sum_{F,N} d(F,N) \,
 e^{ 2 \pi i\tau [N - \frac{1}{2} (F_-)^2 -
 \frac{\chi(P)}{24}]  -2\pi i \bar{\tau} \frac{1}{2} (F_+  )^2
 - 2\pi i F \cdot (C + \frac{P}{2}) }
\end{equation}
with $F \in H^2(P)+P/2$ then we have the following modular
representation. For $A  \in \Gamma := SL(2,\IZ)$ denote
\begin{equation}
A\cdot (\tau,C_+, C_-): = (\frac{a \tau + b}{c \tau + d},
\frac{C_+}{c\bar\tau +d}, \frac{C_-}{c \tau +d} )
\end{equation}
(Sometimes we will abbreviate this equation to $A\cdot(\tau,C) =
(\frac{a \tau + b}{c \tau + d},\frac{C}{c\tau+d})$. Also, note  that
this action does not factor through $PSL(2,\IZ)$, indeed, $S^2 \cdot
(\tau, C) = (\tau, -C)$.) Then
\begin{equation} \label{SdualityZ}
 \CZ(A \cdot(\tau,C)) = \omega_A \, (c\tau +d)^{w}
 (c\bar{\tau}+d)^{\bar w} \, E[\frac{c}{c \tau+d} \frac{C^2}{2}] \,
\CZ(\tau,C).
\end{equation}
Here $\omega_A$ is a phase depending on the $SL(2,\IZ)$ element. For
\begin{equation}
A=T= \left(
\begin{array}{cc}
 1 & 1 \\
 0 & 1
\end{array}
\right)
\end{equation}
direct computation leads to
\begin{equation}\label{omegaTformula}
\omega_T = e^{-2\pi i \frac{P^3}{8}- 2\pi i \frac{\chi}{24}} =
e^{2\pi i \frac{c_2\cdot P}{24}}
\end{equation}
In the second equality we used   the index theorem which says that
\begin{equation} \label{IPdef}
I_P:= \frac{P^3}{6} + \frac{c_2\cdot P}{12}
\end{equation}
is an integer. To verify consistency of the modular representation
of
\begin{equation}
A=S= \left(
\begin{array}{cc}
 0 & -1 \\
 1 & 0
\end{array}
\right)
\end{equation}
 we need to use $d(-F,N)=d(F,N)$.  In this case consistency of
the modular representation and the theta function decomposition
described below leads to
\begin{equation} \label{omegaSformula}
\omega_S = - e^{i \pi I_P}  e^{i \frac{\pi}{2}(\bar w-w)} e^{i
\frac{\pi}{2}P^2}
\end{equation}
We will find later that $\bar w=1/2$ and, for $b_1(X)=0$, $w=-3/2$.
Using this one checks that indeed $\omega_T^3= \omega_S^{-1}$. We
will not need an explicit formula for $\omega_A$ for all $A$.

What we are ultimately interested in are saddle point evaluations of
integrals like (\ref{Opqint}). For saddle points at small $\tau$,
which arise for large $Q_0$, the resummed expansion
(\ref{Zosvresummed}) is particularly useful, since when $\tau \to
0$, the subleading terms in this expression are exponentially
suppressed. We will make this more precise in section
\ref{sec:smalltau}.

\subsection{Theta function decomposition} \label{sec:theta}

It it is useful and instructive to decompose $\CZ$ as a sum of theta
functions. To do this, we decompose $F$ in parts according to the
distinction between fluxes which are pulled back from $X$, and
fluxes orthogonal to these. As before, let $\iota_P$ be the
embedding map for our divisor $P$. Then $L_X := \iota_P^* H^2(X)$ is
the lattice of fluxes pulled back from the ambient space $X$, a
basis of which is formed by $\iota_P^* D_A$. This has metric $D_{AB}
:= D_{ABC} P^C$. Because in general $\det D_{AB} \neq 1$, the
lattice is not unimodular, while $H^2(P)$ is. This implies that the
lattice $L_X \oplus L_X^{\bot}$ is only a sublattice of $H^2(P)$.
The quotient $\CD$ of the latter by the former is a finite group,
parametrized by ``glue vectors'' $\gamma \in \CD$. Taking into
account the half-integral shift $P/2$ of the flux mentioned earlier,
we thus get the following decomposition for fluxes $F \in H^2(P)$:
\begin{equation} \label{Fdecomp}
 F = \frac{P}{2} + f^{\|} + \gamma + f^{\bot},
\end{equation}
where $f^{\|} \in L_X$, $f^{\bot} \in L_X^{\bot}$. We can further
decompose $\gamma$ along $\IQ \otimes L_X$ and its orthogonal
complement $\IQ \otimes L_X^{\bot}$:
\begin{equation}
 \gamma = \gamma^{\|} + \gamma^{\bot}.
\end{equation}
Any nontrivial $\gamma$ must have simultaneously $\gamma^{\|} \neq
0$ and $\gamma^{\bot} \neq 0$. That it must have a nonzero
$\gamma^{\|}$ is clear: otherwise $\gamma$ is an integral flux
orthogonal to $L_X$, which by definition is in $L_X^\bot$ and hence
trivial in $\CD$. That it must have nonzero $\gamma^\bot$ as well is
more subtle. If $\gamma^\bot = 0$, then we can write $\gamma = r^A
\iota_P^* D_A$. Now because $P$ is very ample, by the Lefshetz
hyperplane theorem (see appendix \ref{sec:alggeom}), the map
$\iota_P:H_2(P,\IZ) \to H_2(X,\IZ)$ is surjective, that is, every
2-cycle in $X$ can be realized as a 2-cycle in $P$. Hence there is
in particular a set $\sigma^A$ of 2-cycles on $P$ such that
$\iota_P(\sigma^A)$ is a basis of $H_2(X,\IZ)$ dual to the $D_A$.
Because $\gamma$ is integral, we moreover have $\int_{\sigma^A}
\gamma \in \IZ$. But by construction, this equals $r^A$. Therefore
$\gamma \in L_X$, so it is trivial as an element of $\CD$.

We can also identify $\CD$ with the discriminant group: $ \CD =
L_X^*/L_X$. Since $H^2(P)$ is unimodular, the embedding is specified
by an isomorphism with $(L_X^{\bot})^*/L_X^{\bot}$ preserving
quadratic forms, by the Nikulin primitive embedding theorem
\cite{Nikulin} (the embedding is primitive again because of the
Lefshetz hyperplane theorem). Similarly, $\gamma^\bot\in
(L_X^\bot)^*$ represents $\gamma$ under this isomorphism.

Note that $d(F,N)$ does not depend on the $L_X$ part of $F$, since
this is automatically of type $(1,1)$ and hence does not affect the
BPS condition or the moduli space of supersymmetric configurations.
Using this, the partition function (\ref{ZtauCdef}) can be written
as \footnote{This decomposition can be understood in the $AdS/CFT$
correspondence as the decomposition of the partition function
obtained by factoring out the singleton modes, as in
\cite{Gukov:2004id}. The analogous singleton decomposition for the
M5-brane partition function was used in \cite{deBoer:2006vg}. The
general singleton decomposition of the M5-brane partition function
was described in \cite{Moore:2004jv}. }
\begin{equation} \label{PsiHsplitdef}
 \CZ(\tau,C) = \sum_{\gamma} \Psi_\gamma(\tau,\bar{\tau},C) \,
 H_{\gamma}(\tau).
\end{equation}
Here we defined, using the shorthand notation (\ref{shortnot})
\begin{equation}
 \Psi_{\gamma}(\tau,\bar{\tau},C) := \sum_{f^\|} E[\frac{\tau}{2}
 (\frac{P}{2} + \gamma^\| + f^\|)^2 +
 (\frac{P}{2} + \gamma^\|+ f^\| ) \cdot (C + \frac{P}{2})],
\end{equation}
which is a nonholomorphic Siegel-Narain theta function of signature
$(1,h-1)$, implicitly depending on the K\"ahler form $J$ because
this determines the $(C_+,C_-)$-split. We furthermore defined the
holomorphic
\begin{equation} \label{Hgamdef}
 H_{\gamma}(\tau)  := \sum_{f^{\bot},N} d(\frac{P}{2} + \gamma +
 f^{\bot},N) \, e^{-2 \pi i \tau \hat q_0(F,N) }
 \end{equation}
where
\begin{equation} \label{q0hatH}
 \widehat{q}_0(F,N)=\frac{\chi(P)}{24} +
 \frac{1}{2}(f^{\bot}+\gamma^{\bot})^2 -N=
  q_0 - \frac{1}{2}(\frac{P}{2} + \gamma^\| + f^\|)^2 = q_0 -
\frac{Q^2}{2}
\end{equation}
Note that $H_{\gamma}(\tau) = H_{-\gamma}(\tau)$.

 All nontrivial information about the degeneracies is captured
by the holomorphic $H_\gamma(\tau)$. For example we have for the
degeneracies \footnote{We abbreviate $\Omega(0,P,Q,q_0)$ by
$\Omega(P,Q,q_0)$, and of course $t_\infty=i\infty$ is understood.}
\begin{equation} \label{HOPQ}
 \Omega(P,Q,q_0) = \oint d\tau H_{\gamma_Q}(\tau) \, e^{2 \pi i \tau
 \widehat{q}_0}
\end{equation}
where $\gamma_Q$ is uniquely determined by $(\gamma_Q)_A = Q_A -
D_{ABC} P^B P^C/2$ mod $D_{ABC} P^B n^C$, $n^C \in \IZ$, and
\begin{equation}
 \widehat{q}_0 := q_0 - \frac{Q^2}{2}, \qquad Q^2 :=
 D^{AB}Q_A Q_B.
\end{equation}
The proof of (\ref{HOPQ}) proceeds as follows: First,   note that
fixing the D2-charge $Q_A$ fixes $D_A \cdot F$, which puts
$P/2+\gamma^\|+f^\| = Q$. This determines $\gamma$ (and $f^\|$)
uniquely as stated above, because the difference of two different
$\gamma$'s satisfying this equation would give a nontrivial element
of $\CD$ with zero $\|$-component, which as we saw does not exist.
Put differently,  for each $\gamma$ and $\hat q_0$ we have an
equivalence class
\begin{equation}
 [\gamma,\hat q_0] := \{ (0,P,Q,q_0) \, | \,  q_0 -
\frac{Q^2}{2} = \hat q_0 \quad \mbox{and} \quad Q \in
\iota_{P,*}(L_X + \frac{P}{2} + \gamma) \}.
\end{equation}
As noted above, shifts of $F$ by elements of $L_X$ do not change the
index of BPS states, hence the index
$\Omega([\gamma,\hat{q}_0]):=\Omega(0,P,Q,q_0)$ only depends on the
  equivalence class $  [\gamma,\hat q_0]$.
Another way of phrasing this is that the D4-D2-D0 BPS spectrum at $J
\to \infty$ is invariant under integral $B$-shift monodromy, in
accord with the absence of walls of marginal stability running off
to $J = \infty$ for this system.

Grouping terms in (\ref{Hgamdef}) with fixed $\hat q_0$ we can thus
also write
\begin{equation} \label{HagamOm}
H_{\gamma}(\tau)  := \sum_{\hat q_0 } \Omega([\gamma,\hat q_0])  \,
e^{-2 \pi i \tau \hat q_0 }.
\end{equation}

Now, using the S-duality transformation (\ref{ZSdual}) of $\CZ$ and
integrating both sides with respect to $C$ ranging over $H^2(X,\IR)$
one gets $\bar{w}=1/2$ and
\begin{equation} \label{modtransfH}
 H_\gamma(\tau) = |\CD|^{-1/2} (-i \tau)^{-w +
 \frac{h-1}{2}}(-1)^{I_P+1}
 \sum_{\delta \in \CD} e^{2 \pi i \gamma^\| \cdot
 \delta^\|} H_\delta(-\frac{1}{\tau}).
\end{equation}
where $|\CD| = \# \CD = \det (D_{ABC} P^C)$. Furthermore,
\begin{equation}
 H_\gamma(\tau+n)= e^{-2\pi i n (\frac{(\gamma^\bot)^2}{2} + \frac{\chi}{24})} \,
 H_\gamma(\tau).
\end{equation}
Thus we see that the $H_\gamma$ form a modular vector. One can check
consistency of the modular representation using the Gauss-Milgram
sum formula \cite{Milnor}
\begin{equation}
\frac{1}{\sqrt{\vert \CD\vert }}\sum_{\gamma} e^{-2\pi i
\frac{1}{2}(\gamma^\bot)^2} = e^{-2\pi i \, {\rm sig}(L_X^\bot)/8}.
\end{equation}

\subsection{$\tau \to 0$ limit} \label{sec:smalltau}

When $\widehat{q}_0 \to -\infty$, the saddle point of (\ref{HOPQ})
will be at $\tau \to 0$. To evaluate the integral, it is therefore
useful to   perform first the modular transformation
(\ref{modtransfH}). Indeed, when $\tau \to 0$, the only surviving
term in the resummed series has $\gamma = N = f^{\bot} = 0$ since on
supersymmetric configurations $(\gamma^{\bot}+f^{\bot})^2 \leq 0$
with equality iff $f^\bot=0$ and $\gamma^\bot=0$, which as we saw in
section (\ref{sec:theta}) implies $\gamma=0$. Hence in this limit
\begin{equation}
 \Omega(P,Q,q_0) = d(\frac{P}{2},0) \, |\CD|^{-1/2} \,
 (-1)^{I_P+1}
 \oint d\tau \, e^{2 \pi i \widehat{q}_0 \tau} \,
 (-i \tau)^{-w + \frac{h-1}{2}} \,
 e^{2 \pi i \frac{\chi}{24 \, \tau}}.
\end{equation}
The saddle point of this integral lies at
\begin{equation} \label{tausaddle}
 \tau_* = i \sqrt{-\frac{\chi(P)}{24 \widehat{q}_0}}
\end{equation}
which is indeed small when $-\widehat{q}_0 \gg \chi(P)$, and
\begin{equation} \label{largeq0entropy}
 \ln \Omega(P,Q,q_0) = \frac{4 \pi i \chi(P)}{\tau_*}
 = 2 \pi \sqrt{-\frac{1}{6} \widehat{q}_0 \, \chi(P)}.
\end{equation}
Since $\chi(P) = P^3 + c_2 \cdot P$, this reproduces the well-known
result for the Bekenstein-Hawking-Wald entropy in this limit
\cite{Maldacena:1997de,LopesCardoso:1998wt}. One can do better
however. The $\tau$-integral can be done exactly, resulting in a
Bessel function, as detailed in
\cite{Dabholkar:2005by,Dabholkar:2005dt}. This gives an explicit
formula for $\Omega(P, Q,q_0)$ to all orders in a $1/\widehat{q}_0$
expansion, up to determination of $w$ and $d(P/2,0)$.

In fact, comparison with an independent computation of $\CZ_{\rm
BH}$ in this regime using techniques developed in
\cite{Ashok:2003gk,Denef:2004ze} for counting closed string flux
vacua, which we give in appendix \ref{app:largeq}, fixes $w=-3/2$
for $X$ a proper $SU(3)$ holonomy Calabi-Yau manifold.\footnote{More
generally $w=-3/2+b_1(X)$.} Very roughly, the reason for this is
that in the small $\phi^0$ regime, the OSV partition function
approximately factorizes in a factor $1/\eta(\phi^0)^{\chi(P)/24}$
counting the Euler characteristics of the D0-brane moduli spaces
${\rm Sym}^N P$, and a factor of the form $\int dF \, e^{c \phi^0
F^2} \cdots$, approximately counting flux vacua. The first factor
gives a $(\phi^0)^{\chi(P)/2}$ after modular transformation, and the
second factor a $(\phi^0)^{-b_2(P)/2}$ from integrating out $F$.
Since $\chi(P)=b_2(P)+2$ on an ample divisor in a proper Calabi-Yau,
we thus get a net factor $\phi^0$, corresponding to having
$w+\bar{w}=-1$ in (\ref{ZSdual}). We found in section
\ref{sec:theta} that $\bar{w}=1/2$ (alternatively this can be
directly deduced from the modular transformation properties of the
theta functions $\Psi_{\gamma}$), hence $w=-3/2$ as claimed. We
refer to appendix \ref{app:largeq} for more details.

Furthermore, $d(P/2,0)$ is just the index of BPS states of the pure
D4-brane without any deformation obstructing fluxes, which by
(\ref{indextoeuler}) equals $(-1)^{\dim \CM_P} \chi(\CM_P)$ where
$\CM_P$ is the divisor deformation moduli space. It is  not clear
\emph{a priori} what the physically relevant Euler characteristic of
the divisor moduli space is, since this space has singularities
where the divisor degenerates. It has an obvious compactification
however, namely the corresponding linear system, which is the
projectivization of the space of sections of the line bundle
corresponding to $P$, which, because $P$ is very ample, is just
$\CM_P = \IC\IP^{I_P-1}$. Using this compactification, we thus have
$d(P/2,0)=(-1)^{I_P-1} I_P$. Below we will give more evidence that
this is the correct definition of $\chi(\CM_P)$.

Rephrasing all of this in terms of the original OSV partition
function, we conclude that for the purpose of computing $
\widehat{q}_0 \to -\infty$ degeneracies, we can take
\begin{equation} \label{Zsmalltau}
 \CZ(\tau,C) \approx (-1)^{I_P-1} I_P \, \omega_S^{-1} \tau^{3/2} \bar{\tau}^{-1/2} \,
 e^{2 \pi i \frac{\chi(P)}{24 \tau}} \, E[-\frac{C^2}{2
 \tau}] \, \Psi_0(-\frac{1}{\tau},\frac{C}{\tau}).
\end{equation}
Making the OSV substitution $\tau = \bar{\tau} = i \phi^0$, and
using (\ref{omegaSformula}), this formally becomes:
\begin{equation} \label{resultsmallphi0}
 \CZ_{\rm BH}(\phi^0,\Phi) \approx i \, I_P \, \phi^0 \,
 %
 \sum_{S \in
 H^2(X,\IZ)}
  e^{\frac{2\pi}{\phi^0}
 [ \frac{\chi(P)}{24} - \frac{1}{2} (\Phi + i S)^2] + \pi i P \cdot S}.
\end{equation}
This is in rough agreement with the OSV formula (\ref{osvconj}),
restricted to the polynomial part of the topological string
partition function. The additional sum over shifts of $\Phi$ can be
seen to be necessary to give the right hand side of (\ref{osvconj})
the same periodicity as the left hand side. In the integral
formulation (\ref{OSVintform}) of the conjecture, this sum can be
traded for an extension of the periodic integration contours to the
entire imaginary axis. A similar sum over shifts of $\phi^0$ is
absent here, but this is consistent with the small $\phi^0$
approximation as the shifted terms are exponentially suppressed. We
also find an additional measure factor $i I_P \phi^0$. Finally in
this small $\phi^0$ limit, the nonpolynomial corrections to
$\CZ_{\rm top}$ after the OSV substitutions are all exponentially
small. Hence the above formula is in satisfactory agreement with the
original OSV conjecture at small $\phi^0$ (and in perfect agreement
with our refinement of it which we will derive in the remainder of
the paper).

\subsection{A Rademacher-Jacobi formula} \label{sec:radjacobi}

For larger values of $\tau$, which is the regime relevant to the
full OSV conjecture including instanton corrections, it is no longer
sufficient to do a $\tau \to -1/\tau$ modular transformation to
extract approximate expressions for the degeneracies, because the
subleading terms in the $q$-expansion are no longer sufficiently
suppressed in this limit to justify throwing them away. The key
observation which will allow us to make progress is that because of
its modular properties, $\CZ$ can be entirely expressed in terms of
a some kind of ``$SL(2,\IZ)$ average'' of a finite subset of terms,
analogous to the Rademacher-Jacobi or fareytail expansion of
\cite{Dijkgraaf:2000fq}.

At the end of section \ref{sec:theta} we saw that $H_\gamma$
transforms as a modular vector with weight\footnote{We noted in
section \ref{sec:smalltau} that $w=-3/2$ for proper Calabi-Yau
manifolds but the following works for any value of $w$, so we will
leave this an arbitrary parameter for now.}
\begin{equation}
 w_H := w - \frac{h-1}{2}
\end{equation}
The theta function vector $\Psi_\gamma$ transforms in a conjugate
way to ensure the transformation (\ref{SdualityZ}) of $\CZ$. This
can also verified directly using general properties of theta
functions or by Poisson resummation. Thus, under general $SL(2,\IZ)$
transformations $A$, using the notation introduced above
(\ref{SdualityZ}):
\begin{eqnarray}
\CZ(A \cdot (\tau,C)) &=& \omega_A \, (c \tau + d)^{w} (c \bar{\tau}
+ d)^{\frac{1}{2}} \, E[\frac{c}{c \tau+d} \frac{C^2}{2}] \,
\CZ(\tau,C) \\\label{psitrmn}
 \Psi_{\gamma}(A\cdot(\tau,C)) &=& (c \tau +
d)^{\frac{h-1}{2}} (c \bar{\tau} + d)^{\frac{1}{2}} \, E[\frac{c}{c
\tau+d} \frac{C^2}{2}] \, M(A)_{\gamma \delta} \,
\Psi_{\delta}(\tau,C) \\ \label{Htmn}
 H_\gamma(A \cdot \tau) &=& \omega_A \,(c \tau +d)^{w_H}
 \,
 M(A)^{-1}_{\delta \gamma} \, H_{\delta}(\tau)
\end{eqnarray}
where $\omega_A^{-1}M(A)$ is a  representation of $SL(2,\IZ)$
generated by
\begin{eqnarray}\label{sl2rep}
M(T)_{\gamma\delta} & = & \delta_{\gamma,\delta} \, e^{-i \pi
(\gamma^{\|}+ \frac{P}{2})^2} \\
M(S)_{\gamma \delta} & = & |\CD|^{-1/2} e^{-2 \pi i (\gamma^\| \cdot
 \delta^\| +\frac{P^3}{4} )} e^{-i\frac{\pi}{4}(h-2)}
\end{eqnarray}
The phases $\omega_T, \omega_S$ are given in (\ref{omegaTformula}),
(\ref{omegaSformula})  above.  It is worth noting that it is crucial
to have the extra phase $e^{i \pi P \cdot F}$ in the partition
function in order for the vector of functions $\Psi_{\gamma}$ to
transform into themselves.

Now we would like to write a Poincar\'e series for $H_\gamma$. Since
the modular weight $w_H = w - \frac{h-1}{2} = -1 - h/2$ is negative
we should in fact first define the ``dual'' modular vector
\begin{equation}\label{fareytrmn}
\widetilde{H}_\gamma(\tau) := L^{1-w_H} H_\gamma(\tau), \qquad
\mbox{where } L f(\tau) := \frac{1}{2 \pi i} \frac{\partial
}{\partial \tau} \, f(\tau),
\end{equation}
which transforms according to (\ref{Htmn}) but with   modular weight
$w_H \to 2-w_H>2$. The reason for this is the following nontrivial
 identity, which can be verified by elementary
means and holds for any differentiable function $f$:
\begin{equation} \label{coolidentity}
 L^n \left[ (c \tau + d)^{-1+n} f\left( \frac{a \tau + b}{c \tau + d} \right)
 \right] = (c\tau+d)^{-1-n} (L^n f)\left( \frac{a \tau + b}{c \tau + d}
 \right).
\end{equation}
Next, it is convenient to define $j(A,\tau):=c \tau + d$ so that
\begin{equation}
j(A_1 A_2,\tau) = j(A_1, A_2 \tau)j(A_2,\tau) .
\end{equation}
Finally, let   $\Gamma_\infty$ be the subgroup of $\Gamma$ generated
by $\tau \to \tau + 1$. Then we claim
\begin{equation}\label{HPoinSer}
\widetilde H_{\gamma}(\tau) = \sum_{A \in \Gamma_\infty \backslash
\Gamma} (j(A,\tau))^{w_H-2} \, \omega_A^{-1} M(A)_{\delta\gamma}
\widetilde H_{\delta}^-(A \cdot \tau)
\end{equation}
Here $H_{\gamma}^-( \tau)$ is the \textit{polar part} of
$H_{\gamma}( \tau)$, namely, the terms in the sum (\ref{Hgamdef})
with negative powers of $e^{2 \pi i \tau}$.\footnote{Note that $L$
commutes with taking the polar part, so the notation
$\widetilde{H}^-$ is unambiguous.} Equivalently, because of
(\ref{Hgamdef}), these are the terms with positive $\widehat{q}_0$.
Note that there is a finite number of such terms. Their physical
interpretation will be given in the next section. The quotient by
$\Gamma_\infty$ is necessary because the factor $\omega_A^{-1}
M(A)_{\delta\gamma}$  (\ref{HPoinSer}) cancels the transformation
law of  $\widetilde{H}_\gamma^-(\tau)$ for any $A=\tau \to \tau+b$,
$b \in \IZ$. The proof of (\ref{HPoinSer}) proceeds by noting that
it is in the orthogonal complement of cusp forms, since it is in the
image of the operator (\ref{fareytrmn}), but then it is completely
determined by its Poincar\'e series. See \cite{Dijkgraaf:2000fq,
Moore:2004fg} for more details.

Again using (\ref{coolidentity}), one can formally  pull out the
$L^{1-w_H}$ operation on the right hand side so, formally at least,
we have for the original $H_\gamma$:
\begin{equation} \label{HPoinSer2}
 H_{\gamma}(\tau) = h_\gamma + \sum_{A \in \Gamma_\infty \backslash
\Gamma} (j(A,\tau))^{-w_H} \, \omega_A^{-1} M(A)_{\delta\gamma}
H_{\delta}^-(A \cdot \tau)
\end{equation}
where $h_\gamma(\tau)$ is some function such that $L^{1-w_H}
h_\gamma(\tau) = 0$, i.e., $h_\gamma$ is a polynomial in $\tau$ of
order at most $\vert w_H\vert$. (We assume here that $|w_H|$ is
integral. The case where $|w_H|$ is half-integral is more
complicated and we do not fully understand it.)

Since $w_H<0$ the series (\ref{HPoinSer2}) is in fact not
convergent. We can regularize it as follows. Using $A\tau =
\frac{a}{c} - \frac{1}{c(c\tau+d)}$ we define the notation:
\begin{equation}
\left[ e^{2\pi i k A\tau} \right]_N:= e^{2\pi i k \frac{a}{c} }
\left( e^{-2\pi i k \frac{1}{c(c\tau+d)} } - \sum_{j=0}^N
\frac{1}{j!}\left(\frac{-2\pi i k}{c(c\tau+d)}\right)^j\right)
\end{equation}
 and then, writing
 \begin{equation}
 H_\gamma(\tau) := \sum_{k} \hat H_\gamma(k) e^{2\pi i k \tau}
 \end{equation}
 where $k$ runs over $\frac{1}{M}\IZ$ for some integer $M$,
 we replace the formal
expression (\ref{HPoinSer2}) by
\begin{equation} \label{RegHPoinSer2}
 H_{\gamma}(\tau) = h_\gamma + \sum_{A \in \Gamma_\infty \backslash
\Gamma} (j(A,\tau))^{-w_H} \, \omega_A^{-1} M(A)_{\delta\gamma}
\sum_{k<0} \hat H_{\delta}(k) \left[ e^{2\pi i k A\tau}
\right]_{\vert w_H\vert}
\end{equation}
where $h_\gamma$ is a polynomial of order $\vert w_H\vert$. We claim
(\ref{RegHPoinSer2}) transforms like a form of weight $w_H$, and
extracting the degeneracies from the contour integral proceeds as in
the case where we use the formal expression (\ref{HPoinSer2}).

The Poincar\'e series representation of $H_\gamma(\tau)$ can be
lifted to a Poincar\'e series representation of the partition sum
$\CZ(\tau,C)$ itself. Define the polar part of $\CZ$ as
\begin{equation}
 \CZ^-(\tau,C) := \sum_\gamma \Psi_\gamma(\tau,C) \,
 H_\gamma^-(\tau).
\end{equation}
Equivalently, this is $\CZ$ truncated to the terms for which
$\widehat{q}_0 > 0$. We can now substitute (\ref{RegHPoinSer2}) into
(\ref{PsiHsplitdef}) and use (\ref{psitrmn}) to get a Poincar\'e
series for $\CZ$. Introducing the slash operator
\begin{equation}\label{slashop}
 f\vert^A_{\nu,\bar{\nu}}(\tau,C):= (j(A,\tau))^{-\nu} (j(A,\bar
\tau))^{-\bar \nu} \, \omega_A^{-1} E[- \frac{c}{c\tau + d}
\frac{C^2}{2} ] \, f(A \cdot (\tau,C)),
\end{equation}
on arbitrary function $f(\tau,C)$, where $E[\ldots]$ was defined in
(\ref{shortnot}), this can be written as
\begin{equation} \label{ZPoincSer}
 \CZ = \sum_{A \in \Gamma_\infty \backslash
\Gamma} \CZ^-|^A_{\nu,\bar \nu}
\end{equation}
where $\nu=w=-3/2$ and $\bar \nu=\bar w=1/2$, and we dropped the
divergent $\widehat{q}_0=0$ ``counterterms'' lifted from the
$h_\gamma$, which are not important for the purpose of extracting
$\widehat{q}_0 \not= 0$ degeneracies.

While (\ref{ZPoincSer}) will be our main formula it is perhaps worth
remarking that one could define   a convergent Poincar\'e series for
the quantity  $\widetilde{\CZ}$ analogous to $\widetilde{H}_\gamma$.
To define $\widetilde{\CZ}$, let us extend $L$ as
\begin{eqnarray}
 L_- &:=& \frac{1}{2 \pi i} \frac{\p}{\p \tau} - \frac{1}{8
\pi^2}\frac{\p^2}{\p C_-^2} \\
 L_+ &:=& \frac{1}{2 \pi i} \frac{\p}{\p \bar{\tau}} - \frac{1}{8
\pi^2}\frac{\p^2}{\p C_+^2} \\
L &:=& L_- + L_+ = \frac{1}{2 \pi i} \frac{\p}{\p C_0} - \frac{1}{8
\pi^2}\frac{\p^2}{\p C^2}.
\end{eqnarray}
Plainly,   $L_\pm$   annihilate $\Psi_{\gamma}$ and hence
\begin{equation}\label{Otilde}
 \widetilde{\CZ}:=L^{1-w_H} \CZ = \sum_{\gamma} \Psi_\gamma(\tau,C) \widetilde
H_\gamma(\tau).
\end{equation}
Repeating the same steps as before, but now using (\ref{HPoinSer}),
we get
\begin{equation}\label{RadJacobi}
 \widetilde \CZ =
\sum_{A \in \Gamma_\infty \backslash \Gamma} \widetilde
\CZ^-\vert^{A}_{\nu,\bar{\nu}}
\end{equation}
with $\nu=2-w_H+\frac{h-1}{2}=-w+h+1=h+5/2$ and $\bar \nu = 1/2$.

\section{BPS bound states in supergravity} \label{sec:boundBH}

\subsection{Basic idea} \label{sec:basicidea}

In this section we will argue that the BPS states corresponding to
the polar part of the partition function, i.e.\ D4-D2-D0 states with
$\widehat{q}_0 > 0$, can be concretely thought of as bound states of
D6 and anti-D6 branes (each with lower degree charges turned on),
which moreover can be made to split into those two constituents by
moving the background moduli $t_\infty$  to a wall of marginal
stability, implying in particular that the degeneracies of these
states factorize accordingly. In a suitable asymptotic regime, this
factorization of degeneracies translates in a factorization of the
partition function roughly of the form
\begin{equation} \label{ZtopZtopbarfact}
 \CZ \sim \CZ_{\rm top} \, \overline{\CZ_{\rm top}},
\end{equation}
in other words, to the OSV conjecture.

The starting point to derive this factorization is the observation,
detailed below, that the polar charges do not have single centered
black hole realizations in four dimensions, but instead are realized
as two (or more) centered ``molecular'' bound states with
nonparallel charges at the centers. This structure is mirrored to a
certain extent in the microscopic D-brane description of these
states, which we will develop in section \ref{sec:microdescription}.

Even without getting into any of the detailed descriptions, there is
a simple physical argument for why polar states always ``split,'' in
the sense that they can be made to decay in constituents at some
wall of marginal stability. This goes as follows.

The holomorphic central charge of the D4-D2-D0 system in the large
radius approximation is given by (\ref{centralcharge}):
\begin{equation}
 Z = - \frac{1}{2} P^A D_{ABC}(B+iJ)^B (B+iJ)^C + Q_A
 (B+iJ)^A - q_0.
\end{equation}
We claim that this has a zero in the interior of moduli space if and
only if $\widehat{q}_0 := q_0 - D^{AB} Q_A Q_B > 0$, where we recall
that $D_{AB} := D_{ABC} P^C$, $D^{AB} D_{BC} := \delta^A_C$. To see
this, first make the change of variables $B \to \tilde{B}$:
\begin{equation} \label{covBtrick}
 B = \tilde{B} + D^{AB} Q_B.
\end{equation}
Then
\begin{equation} \label{Zholaftercov}
 Z = - \frac{1}{2} D_{AB}(\tilde{B}+iJ)^A (\tilde{B}+iJ)^B - \widehat{q}_0.
\end{equation}
Requiring $Z=0$, leads to
\begin{equation}
 \tilde{B} \cdot J = 0, \qquad \frac{1}{2} (J^2 - \tilde{B}^2) =
 \widehat{q}_0,
\end{equation}
where as before the dot product is defined using the metric
$D_{AB}$. Recall that $J$ has positive norm squared and all vectors
(in $L_X$) perpendicular to $J$ have negative norm squared. Because
of the first equation, $\tilde{B}$ is of this kind, hence the left
hand side of the second equation is strictly positive in the
interior of moduli space, so we need $\widehat{q}_0 > 0$.
Conversely, when $\widehat{q}_0 > 0$, we can take for example
$\tilde{B}=0$, $J_0 = \sqrt{2 \widehat{q}_0/P^3} \, P$ and obtain
$Z=0$. This proves our claim. Note that at large $P$, this result is
guaranteed to be robust under adding instanton corrections as long
as $J_0^A \gg 1$. In particular this is true for the ``most polar''
states, that is states with $\widehat{q}_0$ near $(P^3 + c_2 P)/24$,
which will be of main interest in the derivation of the OSV
conjecture.

Now when the background moduli are at the zero locus at sufficiently
large $J$, a BPS state of the given charge $\Gamma=(p,q)$ cannot
exist; if it did, the charge would correspond to a massless BPS
particle at this locus, which would cause a singularity of the
moduli space metric
\cite{Seiberg:1994rs,Strominger:1995cz,Moore:1998pn}. Such
singularities exist at conifold points of the mirror complex
structure moduli space, but are (more or less by definition) absent
in the large $J$ region. Since by assumption the state does exist
when $J \to \infty$, there must be a wall of marginal stability
separating the zero locus from $J=\infty$. When crossing this wall
of marginal stability coming from $J=\infty$, the state decays in
two BPS states with charges $\Gamma_1$ and $\Gamma_2$, $\Gamma =
\Gamma_1 + \Gamma_2$, whose central charges are aligned on the wall:
$\alpha_1 \equiv \arg Z(\Gamma_1) = \alpha_2 \equiv \arg Z(\Gamma_2)
= \alpha \equiv \arg Z(\Gamma)$.\footnote{Recall that because of the
BPS condition, decay is only energetically possible when the phases
of the constituents align.}

As we will review below, decay at marginal stability is realized in
the supergravity picture by two (clusters of) centers moving
infinitely far away from each other. In this infinite separation
limit, one physically expects the degeneracies to factorize. In
particular the Witten index $\Omega$ of this configuration, which is
independent of the background moduli as long as the wall of marginal
stability is not crossed, can be expected to have a factorized form
$\Omega(\Gamma) = \Omega(\Gamma_2)\, \Omega(\Gamma_2)$. There is a
slight subtlety however, in that quantizing the position degrees of
freedom of the two parts produces an additional lowest Landau level
degeneracy $|\langle \Gamma_1,\Gamma_2 \rangle|$, where $\langle
\Gamma_1,\Gamma_2 \rangle$ is the Dirac-Schwinger-Zwanziger
symplectic intersection product (\ref{intprodform}) on charge space.
As we review under (\ref{spinformula}), this is most easily
understood by noting that a 2-centered BPS bound state carries an
intrinsic angular momentum $J = \frac{1}{2} (|\langle
\Gamma_1,\Gamma_2 \rangle| - 1)$, leading to an additional
degeneracy $2 J + 1$. Moreover, this intrinsic spin changes the
fermion parity by a factor $(-1)^{2J}=(-1)^{\langle
\Gamma_1,\Gamma_2 \rangle - 1}$, which appears in the index
$\Omega$, defined as in (\ref{indexdef})-(\ref{indexred}). Thus,
summing over different possible splits of $\Gamma$, we arrive at a
factorization formula for polar states  of the form
\begin{equation} \label{Omfact}
 \Omega(\Gamma) = \sum_{\Gamma_1,\Gamma_2} (-1)^{\langle \Gamma_1,\Gamma_2 \rangle-1} \,
  |\langle \Gamma_1,\Gamma_2 \rangle| \, \Omega(\Gamma_1) \, \Omega(\Gamma_2)
\end{equation}
where the sum runs over allowed charge splits $(p,q) \equiv
\Gamma=\Gamma_1 + \Gamma_2$ (with $\Gamma_1,\Gamma_2$ primitive).
We have been sloppy here in the sense
that we did not specify at which $t$ the indices should be
evaluated. We will make this precise in section
\ref{sec:physargfact}.

Precisely which splits can be realized by decays of actual bound
states and therefore have to be summed over is a highly nontrivial
question, and analyzing this as well as to what extent it leads to
the factorization (\ref{ZtopZtopbarfact}) will in fact take up much
of the remainder of this paper. As a byproduct of this analysis
however, we will obtain several new insights in the structure of BPS
states which are of independent interest.

\subsection{Review of BPS black hole bound states and attractor
flow trees} \label{sec:reviewtrees}

In many cases, BPS D-brane states at $g_s |\Gamma| \ll 1$ correspond
to single centered black holes in four dimensional supergravity at
$g_s |\Gamma| \gg 1$. However, this is not always the case
\cite{Douglas:1999vm,Denef:2000nb}. It might even happen that a
single centered BPS solution of the given charge does not exist at
all. This is the case when the attractor flow corresponding to this
charge terminates on a zero of the central charge $Z$ at a regular
point in moduli space \cite{Moore:1998pn,Moore:1998zu}. In such cases, it is
necessary to consider more general multicentered BPS black hole
bound states
\cite{Denef:2000nb,Denef:2000ar,Denef:2001xn,Bates:2003vx}. These
are stationary but in general non-static BPS solutions of
supergravity \cite{Behrndt:1997ny,LopesCardoso:2000qm}, with
\emph{non}-parallel charges at the centers. The distances between
the centers are constrained by equations depending on the charges
and the moduli at spatial infinity, and there is a potential energy
exceeding the BPS bound when going off the constraint locus. Hence
unlike the usual parallel charge multicentered BPS solutions, these
are genuine bound states. Moreover, although time independent, they
carry an intrinsic, quantized angular momentum, due to the Poynting
vector field produced by the simultaneous presence of electric and
magnetic charges.

As we saw above, the BPS states corresponding to the polar part of
$\CZ$ have regular zeros and therefore are of this type: they do not
have single centered solutions, so they must have realizations as
multicentered  bound states. On the other hand D4-D2-D0 states with
$\widehat{q}_0<0$ do have single centered solutions, but here we
will find a surprise (described in section \ref{sec:entropyenigma}):
  when all charges are linearly scaled up by some
sufficiently large $\Lambda$, in addition to the usual single centered
solutions, there are always two-centered BPS configurations whose
Bekenstein-Hawking entropy is parametrically larger than that of the
single centered solution, growing as $\Lambda^3$ instead of the
single centered growth $\Lambda^2$. Clearly, this creates some
tension with the OSV conjecture, which predicts to leading order the
single centered entropy growth $\log \Omega(\Lambda \Gamma) \sim
\Lambda^2$. To what extent this is a problem for the conjecture will
be discussed in detail in section \ref{sec:discussion}.

\subsubsection{General stationary BPS solutions}

Let us now review the description of these solutions in more detail.
The metric of a BPS solution is always of the form
\begin{equation} \label{metric}
 ds^2 = -e^{2U} (dt+\omega)^2 + e^{-2 U} d \vec{x}^2
\end{equation}
satisfying the BPS equations of motion\footnote{We neglect $R^2$
corrections \cite{LopesCardoso:2000qm}, which is justified in the
large charge limit}:
\begin{eqnarray} \label{attfloweq}
 2 \, e^{-U} \, \Im(e^{-i \alpha} \Omega_{\rm nrm}) &=& - H \\
 \label{omeq}
 *_3 d\omega &=& \langle dH , H \rangle.
\end{eqnarray}
where $*_3$ is the Hodge star on flat $\IR^3$,   $\Omega_{\rm nrm}$
and $\langle \cdot,\cdot \rangle$ are defined in appendix
\ref{app:defconv}, and $e^{i \alpha}$ is the phase of $Z(\Gamma;t)$.
The function $H:\IR^3 \to H^{\rm even}(X,\IR)$ is harmonic with
poles at the centers. For an $n$-centered configuration with charges
$\Gamma_i$ in asymptotically flat space:
\begin{equation}
 H(\vec{x}) = \sum_i \frac{\Gamma_i}{|\vx-\vx_i|} - 2 \, \Im(e^{-i\alpha}
 \Omega_{\rm nrm})|_{r=\infty}.
\end{equation}
The phase field $\alpha(\vec{x})$ satisfies the boundary condition
$\alpha|_{r=\infty}=\arg Z(\Gamma)|_{r=\infty}$, with $Z$ given by
(\ref{centralchargedef}).

In this subsection the period vector and central charge will always
be the normalized versions, so to avoid cluttering the formulae
 we will henceforth not explicitly write the subscripts indicating this.

For a single center (\ref{attfloweq}) reduces to the attractor flow
equation
\begin{equation} \label{singlefloweq}
 2 \, e^{-U} \, \Im(e^{-i \alpha} \Omega) = - \Gamma \, \tau + {\rm const.}, \qquad \tau \equiv 1/r.
\end{equation}
The moduli at the horizon $\tau=\infty$ are fixed by the attractor
equation
\begin{equation} \label{attreq}
 2 \, \Im(\overline{Z(\Gamma;t_*(\Gamma))}  \Omega) = - \Gamma,
\end{equation}
and the Bekenstein-Hawking entropy is given by
\begin{equation} \label{entropyZ}
 S(\Gamma) = \pi |Z(\Gamma;t_*(\Gamma))|^2
\end{equation}
evaluated at the attractor point. Attractor flows are gradient flows
of $\log |Z|^2$ \cite{Moore:1998pn,Moore:1998zu}, hence the right
hand side of this expression is minimized at the attractor point
\cite{Ferrara:1997tw}. Equations (\ref{attreq})-(\ref{entropyZ})
hold in the multicentered case for each center separately; in
particular the attractor point and horizon area for each constituent
black hole is not affected by the presence of the other centers.

Under the substitutions (\ref{osvsubst}) and identifying
$\Gamma=(p,q)$, the attractor equations can alternatively be written
as \cite{Ooguri:2004zv}
\begin{equation} \label{legendreattreq}
 2 \pi q_\Lambda = \frac{\partial}{\partial \phi^\Lambda} \CF_0(p,\phi).
\end{equation}
where $\CF_0 = \log |\CZ_{\rm top}^{(h=0)}|^2$ is the genus
zero\footnote{The restriction to genus zero is due to the fact that
we are neglecting $R^2$ corrections.} free energy
\cite{Ooguri:2004zv}, again with the substitutions (\ref{osvsubst}).
The entropy is then obtained as the Legendre transform of the free
energy:
\begin{equation} \label{Slegform}
 S(p,q) = \CF_0(p,\phi) - 2 \pi q_\Lambda \phi^\Lambda.
\end{equation}
It was shown in \cite{Bates:2003vx} that
(\ref{attfloweq})-(\ref{omeq}) (as well as the equations giving the
electromagnetic field) can be solved completely explicitly given
just this entropy function. For example,
\begin{equation} \label{metricfunct}
 e^{-2U(\vec x)} = S(H(\vx))/\pi,
\end{equation}
while the moduli fields $t^A(\vx)$ in our conventions are obtained
as
\begin{equation}
 t^A(\vec x) = \left.
 \frac{\frac{\partial S}{\partial q_A} + \pi i p^A}{\frac{\partial S}{\partial q_0}-\pi i
 p^0} \right|_{(p,q)=H(\vec x)}.
\end{equation}
Depending on the model and the charges, (approximate) expressions
for $S$ may or may not be obtainable analytically. In the large
radius approximation, the general attractor solution was derived in
\cite{Shmakova:1996nz} (see also \cite{Moore:1998pn}, sec. 9).
Parametrizing a general charge $\Gamma \in H^{\rm even}(X,\IR)$ as
\begin{equation}
 \Gamma=r e^S(1-\beta + n \, \omega)
\end{equation}
where $r \in \IR$, $S \in H^2(X,\IR)$, $\beta \in H^4(X,\IR)$, $n \,
\omega \in H^6(X,\IR)$ (with $\int_X \omega \equiv 1$), the
condition to have an attractor point is:
\begin{equation} \label{attrexistence}
 \CD:=8(Y^3)^2 - 9 \, n^2 \geq 0, \quad Y^2:=\beta, \quad Y \in \mbox{ K\"ahler
 cone},
\end{equation}
with entropy
\begin{equation}\label{entropydisc}
 S=\frac{\pi}{3} r^2 \sqrt{\CD}.
\end{equation}
In this case, the region in $H^3(X,\IR)$ for which $S$ is real and
positive is the region for which the discriminant $\CD$ is positive;
we denote this region in general by ${\rm dom} \, S$.

Returning to the general multicentered case, we note that equation
(\ref{omeq}) has nonsingular solutions if and only if the following
integrability condition, obtained by acting with $d*_3$ on both
sides, is satisfied for all centers $i$:
\begin{equation} \label{centerconstraints}
\label{distconstr} \sum_{j=1(\neq i)}^N \frac{\langle
\Gamma_i,\Gamma_j \rangle}{|\vec{x}_i-\vec{x}_j|} = 2 \,
\Im\left(e^{-i \alpha} Z(\Gamma_i) \right)_{\infty}.
\end{equation}
In the case of just two charges $\Gamma_1$ and $\Gamma_2$, this
simplifies to
\begin{equation} \label{eqsep}
|\vec{x}_1-\vec{x}_2| = \frac{\langle \Gamma_1,\Gamma_2 \rangle}{2
\, \Im(e^{-i \alpha} Z_1)_{\infty}} = \left. \frac{\langle
\Gamma_1,\Gamma_2 \rangle}{2} \, \frac{|Z_1+Z_2|}{\Im(Z_1
\bar{Z}_2)} \right|_\infty.
\end{equation}
Since distances are positive, a necessary condition for existence in
this case is
\begin{equation} \label{eqsepcond}
 \langle \Gamma_1,\Gamma_2 \rangle \Im(Z_1 \bar{Z}_2)_\infty > 0.
\end{equation}
From (\ref{eqsep}) it follows that the separation of the centers
diverges when such a wall is approached from the side where the
above inequality is satisfied. Thus, this process is the 4d
supergravity realization of decay at marginal stability.

A crucial property of these multicentered solutions is that despite
being time-independent, they carry intrinsic angular momentum
\cite{Denef:2000nb}, stored in the electromagnetic field, much as
for electron-monopole pairs. This is given by
\begin{equation} \label{spinformula}
 \vec{J} = \sum_{i<j} \frac{1}{2} \langle \Gamma_i,\Gamma_j \rangle
 \, \frac{\vec{x}_i-\vec{x}_j}{|\vec{x}_i - \vec{x}_j|}.
\end{equation}
In particular for a two centered configuration, the angular momentum
stored in the electromagnetic field equals $J = \frac{1}{2} |\langle
\Gamma_1,\Gamma_2 \rangle|$. The presence of this angular momentum
implies that quantizing this 2-particle ``monopole-electron'' system
will give rise to a ground state degeneracy. The quantization of
this system was studied in great detail in \cite{Denef:2002ru}, also
for more complicated multiparticle systems. One subtlety that was
uncovered there was that the position hypermultiplet degrees of
freedom of the particles arrange themselves such that the effective
total angular momentum of the BPS ground state is lowered by $1/2$,
to a total of $J= \frac{1}{2} (|\langle \Gamma_1,\Gamma_2 \rangle| -
1) $. This was derived explicitly in \cite{Denef:2002ru} by
constructing the ground state wave functions, but if we think of the
system as a light electron $\Gamma_1$ moving in the background field
of a heavy monopole $\Gamma_2$, this can physically be understood as
follows. If the electron $\Gamma_1$ lived in empty space, it would
have a half-hypermultiplet $({\bf 0},{\bf 0},{\bf \half})$ of BPS
states associated to its position degrees of freedom in $\IR^3$.
However in the case at hand it is moving in the magnetic field of
the monopole $\Gamma_2$, and the interaction between this radial
magnetic field and the hypermultiplet spin degrees of freedom in
fact selects out a single energy minimizing state in the
hypermultiplet, essentially spin 1/2 down in the radial direction.
As a result, the total angular momentum is lowered by $1/2$, as
claimed, and the total ground state degeneracy (factoring out the
decoupled center of mass half-hyper) equals $2 J + 1=|\langle
\Gamma_1,\Gamma_2 \rangle|$. This can also be interpreted as the
lowest Landau level degeneracy of an electron confined on a sphere
surrounding a magnetic monopole.

\subsubsection{Existence criteria and attractor flow trees} \label{sec:flowtrees}

Whether or not multicentered BPS solutions of given charges
$\Gamma_i$ and positions $\vec x_i$ actually exist is in
general a rather nontrivial problem.
Necessary and sufficient conditions are:
\begin{enumerate}
\item The integrability conditions (\ref{centerconstraints}).
\item To keep the metric warp factor real in (\ref{metricfunct}),
 $H(\vx)$ must have positive discriminant $\CD(H(\vec x))>0$,
 that is, must lie  in ${\rm dom} \, S$ for all $\vx\in \IR^3$.
\item The fields $t^A(\vec x)$ must remain within the physical
moduli space for all $\vx \in \IR^3$.
\end{enumerate}
In particular, conditions such as  $\Gamma_i \in {\rm dom} \, S$ for
all $i$, and (\ref{centerconstraints})  are necessary but, in
general not sufficient for existence of a BPS solution.

\EPSFIGURE{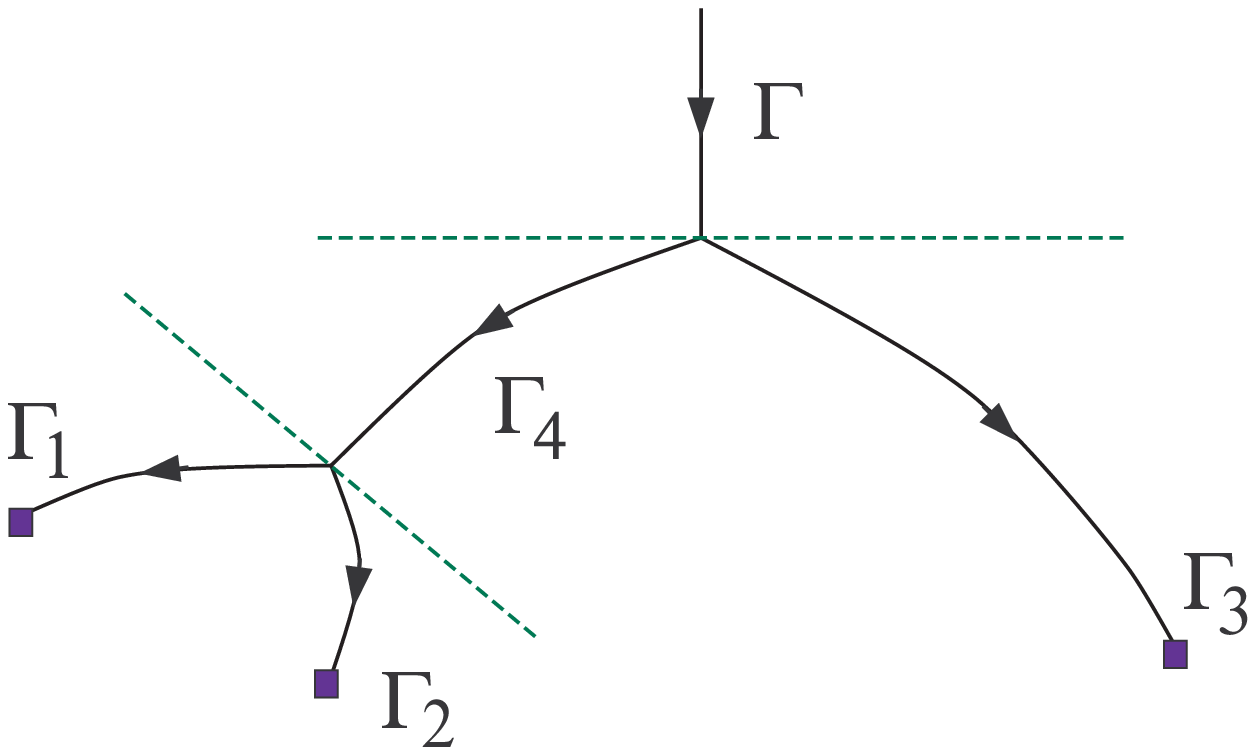,height=5cm,angle=0,trim=0 0 0 0}%
{Sketch of an attractor flow tree. The dotted lines are lines of
marginal stability and the squares are attractor points.
  \label{flowtree}}

Now ideally, one would like to have a local necessary and sufficient
existence criterion in terms of the charge and the background moduli
only. The first condition above is local and easy to evaluate, but
not sufficient, and the second and third conditions are not local,
as they require information about fields at all $\vec x$.
Unfortunately, a local necessary and sufficient existence criterion
is not known, and given the intrinsic mathematical complexity of
stability conditions in the theory of derived categories (see e.g.\
\cite{Douglas:2000gi,bridgeland,Joyce:2006pf}), this is probably too
much to hope for.

In \cite{Denef:2000nb}, an existence criterion was proposed in terms
of \emph{attractor flow trees}, also called \emph{split attractor flows}:
a solution exists iff an attractor
flow tree exists in moduli space starting at the background value of
the moduli and terminating at the $\Gamma_i$ attractor points. Each
edge $E$ of an attractor flow tree is given by a single charge
attractor flow for some charge $\Gamma_E$, charge and energy is
conserved at the vertices, i.e.\ for each vertex $E \to (E_1,E_2)$,
$\Gamma_E = \Gamma_{E_1} + \Gamma_{E_2}$ and
$|Z(\Gamma_E)|=|Z(\Gamma_{E_1})| + |Z(\Gamma_{E_2})|$. The last
condition is equivalent to requiring the vertices to lie on a line
of marginal stability: $\arg Z(\Gamma_{E_1}) = \arg
Z(\Gamma_{E_2})$.
A number of arguments in favor of the equivalence with the full
existence problem were given in \cite{Denef:2000ar}, and a practical
approach for computing split flows on the quintic was developed in
\cite{Denef:2001xn}.

The split flow approach gives a reasonably practical criterion in
sufficiently simple examples, but it often requires case by case
analysis, and is therefore perhaps not as powerful as one would wish
as a general systematic test. Nevertheless, it will be quite useful
in our analysis below. Note that it is not always necessary to
construct the precise flow tree to argue for its existence; for
example to argue for existence of a split flow with two endpoints,
it is sufficient to establish the existence of the two attractor
points and the existence of a wall of marginal stability between the
starting point and the endpoint of the single flow with charge
$\Gamma$ (either a zero of $Z$ or an attractor point).

In appendix \ref{app:attnumerics} we outline an efficient algorithm
for numerically checking existence of flow trees.

The general uplift of arbitrary multicentered IIA/CY$_3$ solutions
to M-theory has been discussed in
\cite{Gaiotto:2005gf,Gaiotto:2005xt,Behrndt:2005he,Cheng:2006yq},
generalizing
\cite{Bena:2004tk,Bena:2005ay,Bena:2005ni,Elvang:2005sa,Berglund:2005vb,Balasubramanian:2006gi}
and relating some of these solutions in four dimensions to the
multi-black hole/black ring and ``bubbling'' solutions in five
dimensions studied e.g.\ in
\cite{Bena:2005va,Berglund:2005vb,Balasubramanian:2006gi}; see
\cite{Bena:2007kg} for a recent review. In
\cite{Bena:2005ni,Berglund:2005vb,Balasubramanian:2006gi,Cheng:2006yq}
it was pointed out that the 4d condition of having positive
discriminant for $H(\vec x)$ everywhere is equivalent to the 5d
condition of having no closed timelike curves, which is similarly
nontrivial to verify directly. Through the correspondence given in
those works, the flow tree picture reviewed here is thus directly
applicable to existence and classification of 5d solutions as well.

In any case, what emerges from examples is that existence of a
certain $\Gamma \to \sum_i \Gamma_i$ bound state realization is
highly constrained; in particular, although a priori there is an
infinite number of ways of splitting up a given charge, only a
finite number of those turn out to correspond to a flow tree.
Physically this is as expected, since an infinite number would imply
an infinite degeneracy of BPS states of a given charge.
\footnote{Even if some states associated with different flow trees
mixed and were lifted quantum mechanically to near-BPS states, the
number of states below any finite energy scale should be finite. }
Some more direct general arguments, based on the monotonic decrease
of $|Z|$ along attractor flows, were given in appendix A of
\cite{Denef:2001xn}. Part of this argument is made more precise in
appendix \ref{app:finiteness}.

To summarize: Throughout this paper we will assume the truth of the
following \textit{split attractor flow conjecture}, which we
consider to be very well-founded:

\bigskip
\textbf{Split Attractor Flow Conjecture}:

\begin{enumerate}

\item[a)] The components of the moduli spaces (in $\vec x_i$) of the
multicentered BPS solutions with constituent charges $\Gamma_i$ and
background $t_\infty$,
 are in 1-1 correspondence
with the attractor flow trees beginning at $t_\infty$ and
terminating on
  attractor points for $\Gamma_i$.

\item[b)] For a fixed $t_\infty$ and total charge $\Gamma$ there are only
a finite number of attractor flow trees.

\end{enumerate}

\noindent We note the following subtleties:
\begin{itemize}

\item Finiteness is of course only valid when
charge quantization is imposed, and hence is not strictly speaking a
property of the classical theory.

\item It is useful to distinguish between attractor flows terminating on
regular points in moduli space and those associated to pure electric
or pure magnetic charges which flow off to infinity. Since the
latter case leads to (mildly) singular solutions, some argument that
transcends supergravity is strictly speaking required to establish
the existence of the corresponding BPS states.

\item It is \emph{not} true that a single flow only
 corresponds to a single centered solution. Indeed, as we will see
 in section \ref{sec:scalingsol}, there exist multicentered solutions
 which are in some sense continuously connected to a single centered
 solution. These are the so-called ``scaling solutions'', first
 identified in \cite{Denef:2002ru}, which develop a capped off ${\rm AdS}_2 \times S^2$ throat
 with a scale modulus $\lambda$ parametrizing the depth of the throat, and which asymptotically
 for $\lambda \to 0$ become indistinguishable from a single centered black hole for a
 distant observer. Such configurations cannot be forced to decay at
 a wall of marginal stability, and thus are not described by a split attractor flow (since a split flow
 can always be made to decay by crossing the wall of marginal stability on which the split occurs).

\end{itemize}

\subsubsection{Attractor flow trees and the Hilbert space of quantum BPS
states} \label{sec:quantumflowtrees}

So far we discussed the relation between attractor flow trees and
classical BPS solutions of four dimensional supergravity. However,
the attractor flow criterion for existence of BPS states can be
argued to be valid beyond the classical four dimensional
supergravity picture. As we mentioned above, even for purely
electric charges, which lead to singular 4d gravity solutions and
are better described as probe particles, the flow tree picture
continues to hold. Moreover, flow trees can be given a purely
microscopic interpretation. This was done for the IIB description of
BPS D-brane as special Lagrangians in \cite{Denef:2001ix}, and we
will sketch a more general argument based on tachyon condensation of
open stretched strings in section \ref{sec:microDbrane}. Finally,
after quantization of the four dimensional BPS configuration moduli
space of a given charge $\Gamma$ with moduli at infinity
$t_{\infty}$, the partitioning of this moduli space by attractor
flow trees leads to a partitioning of the corresponding BPS Hilbert
space $\CH(\Gamma;t_{\infty})$. This suggests that attractor flow
trees provide a classification of BPS states independent of any
particular picture, more refined than classification by total charge
only, but coarser than distinguishing individual states.

In fact a general physical argument for this idea can be given, as
follows. The starting point is the physical expectation that at an
attractor point for charge $\Gamma_i$, (irreducible) BPS states
exist with that charge, while at a zero of the central charge at a
nonsingular point of the moduli space, there cannot be any BPS
states (since zero mass BPS states lead to singularities). The
crucial second ingredient is the observation that if a BPS state of
some charge $\Gamma$ exists at a certain point in moduli space, it
will continue to exist at all points along the attractor flow for
that charge $\Gamma$ passing through this point, when one follows
the flow in the \emph{inverse} direction, that is \emph{decreasing}
$\tau$ in (\ref{singlefloweq}). This can be seen as follows. A BPS
state can only disappear when it decays at a wall of marginal
stability, when crossing the wall from the side where the stability
condition (\ref{eqsepcond}) is satisfied to the side where it is
not. However, an inverted attractor flow will always cross any such
wall in the opposite direction, i.e.\ from unstable to stable.
Indeed, say we are near a wall of $\Gamma \to \Gamma_1 + \Gamma_2$
marginal stability. By taking the intersection product of
(\ref{singlefloweq}) with $\Gamma_1$, it follows that $2 \, e^{-U}
\Im(e^{-i \alpha} Z_1) = - \langle
 \Gamma_1,\Gamma \rangle \, \tau + {\rm const.}$, which can also be
written as
\begin{equation} \label{stabchangeflow}
 2\, e^{-U} \, \Im(Z_1 \, \bar{Z}_2) = - \langle \Gamma_1,\Gamma_2
 \rangle \, \vert Z\vert \tau + {\rm const.}
\end{equation}
From this it is clear that $\langle \Gamma_1,\Gamma_2 \rangle
\Im(Z_1 \, \bar{Z}_2)$ can only increase along an inverted attractor
flow, which proves our claim. Thus, if we have a split attractor
tree, we can start with BPS states of charge $\Gamma_i$ at the
attractor points, let them flow up along the tree edges, ``glue''
them together (microscopically through tachyon condensation as will
be reviewed in section \ref{sec:microDbrane}, macroscopically by
creating multicentered configurations, initially with infinitely
separated centers) at the MS vertices, as described in section
\ref{sec:basicidea}, and then continue to flow up with this newly
formed BPS state, all the way to the starting point of the tree,
where we end up with a BPS state of the required total charge
$\Gamma$.

Conversely, we can start with a charge $\Gamma$ and some point $t$
in moduli space, and consider the Hilbert space $\CH(\Gamma;t)$ of
BPS states with charge $\Gamma$ at $t$. When flowing down along the
attractor flow starting at $t$, some states might decay by splitting
in two BPS states at walls of marginal stability, reducing the
Hilbert space in size. Whenever such a decay occurs, we can
associate to this event a flow split in the obvious way. The
procedure can be repeated for each of the constituents separately
starting from the split point, and so on, until each flow branch
terminates in an attractor point. This algorithm decomposes
$\CH(\Gamma;t)$ in sectors labeled by different flow trees,
according to the decay pattern under the procedure just described.

Thus we arrive at the picture that flow trees label different
sectors of the Hilbert space of BPS states of a given charge in a
given background, independent of the description of these states.

We note the following subtleties:

\begin{itemize}

\item Although every flow tree is associated to a component of the moduli space of classical
 BPS solutions, not all of these components survive quantization.
 The reason for this is the Pauli exclusion principle; for example,
 even if classically we can form a bound state of some charge
 $\Gamma_1$ with an arbitrary number of charges $\Gamma_2$, if the
 $\Gamma_2$ particles happen to be fermions and their number is
 larger than the number of available one-particle BPS ground states,
 the exclusion principle forbids a BPS bound state. This was
 discussed in detail in \cite{Denef:2002ru}.

\item The different sectors of $\CH(\Gamma;t_{\infty})$ labeled by different flow trees
are not necessarily superselection sectors, as quantum tunneling
might occur between different configurations with the same charge.
For the same reason, part of the BPS states obtained say by
quantizing different classical components of moduli space might in
fact be lifted due to quantum tunneling. Presumably these tunneling
amplitudes are exponentially small in some measure of the charges
involved. Similarly, tunneling phenomena may occur when starting
from the microscopic D-brane picture of these states at zero string
coupling. In this case the suppression can be expected to be
exponentially small in the inverse string coupling. However, the
index is of course not affected by this, and can be computed in any
semiclassical picture. It would be interesting to investigate these
tunneling phenomena in more detail.

\end{itemize}

\subsection{Symmetries} \label{sec:symmetries}

Scaling symmetries will be a powerful tool in the following, so we
describe these in detail here.

The BPS equations of motion (neglecting $R^2$ corrections) always
have the following scaling symmetry
\begin{equation} \label{unifscaling}
 \Gamma \to \mu \Gamma, \qquad
 t^A \to t^A, \qquad g_{\rm top} \to g_{\rm
 top}/\mu, \qquad \vec{x} \to \mu \vec{x},
\end{equation}
with $g_{\rm top}$ defined as in (\ref{osvsubst}), while the OSV
potentials scale as $\phi \to \mu \phi$. Under this scaling, the
leading order entropy (i.e.\ without $R^2$ corrections governed by
the higher genus topological string amplitudes) scales as $S \to
\mu^2 S$. Moreover in the large $\mu$ limit, $R^2$ corrections can
be consistently neglected, so this scaling becomes exact.

In the large radius regime, dropping all instanton corrections,
there is a less trivial additional scaling symmetry:
\begin{equation} \label{lefshetzscaling}
 (p^0,P,Q,q_0) \to (p^0,\lambda P,\lambda^2 Q,\lambda^3 q_0), \qquad
 t^A \to \lambda t^A, \qquad g_{\rm top} \to g_{\rm
 top}, \qquad \vec{x} \to \lambda^{3/2} \vec{x},
\end{equation}
with the OSV potentials remaining invariant. The corresponding
leading order entropy scales as $S \to \lambda^3 S$. Moreover in the
large $\lambda$ limit, instanton corrections can be consistently
neglected, so this scaling becomes exact.

There are also two useful discrete symmetries. The first one is
simply charge conjugation $\Gamma \to -\Gamma$ with everything else
invariant. This is valid in all regimes. The second is only valid in
the large radius regime and given by
\begin{equation}
 \Gamma \to \Gamma^*, \quad \mbox{i.e.\ $(p^0,P,Q,q_0) \to (p^0,-P,Q,-q_0)$}, \qquad B \to
 -B.
\end{equation}
  This    leaves the entropy
invariant but inverts intersection products: $\langle
\Gamma_1^*,\Gamma_2^*\rangle = - \langle \Gamma_1,\Gamma_2\rangle $.
Furthermore, the  central charges transform according to
\begin{equation}\label{zeedualcharge}
Z(\Gamma^*;t) = - \overline{Z(\Gamma;-\bar t)}.
\end{equation}
Microscopically it corresponds to taking the dual of the
object in the derived category. In the case of objects described by
vector bundles, this simply amounts to inverting the curvature, $F
\to -F$.

Finally, there is a gauge symmetry
\begin{equation}
 \Gamma \to e^S \Gamma, \qquad B \to B + S.
\end{equation}
If we neglect charge quantization, this is a continuous symmetry,
otherwise $S$ has to be integral. This descends from the usual gauge
symmetry which simultaneously shifts $B$ and the worldvolume flux
$F$. Note that \emph{if there are no walls of marginal stability
between $B+iJ$ and $B+S+iJ$}, then $B \to B+S$ with fixed $\Gamma$
is a symmetry of the BPS spectrum, and hence because of the above
gauge symmetry, likewise $\Gamma \to e^S \Gamma$ with fixed $B$ is a
symmetry of the BPS spectrum. This is the case for D4-D2-D0 systems
in the large $J$ limit. As we will see though, this is \emph{not} so
in general for D6-D4-D2-D0 systems, not even at $J \to \infty$.

\subsection{A class of examples} \label{sec:classexamples}

We will now give a class of explicit examples relevant to our
analysis below. Consider the charges
\begin{eqnarray}
 \Gamma_1 &=& r \, e^{\frac{P}{2r}} \, (1- \tb \, \frac{P^2}{r^2} - \tn \, \frac{P^3}{r^3}) \\
 &=& r + \frac{P}{2} + (\frac{1}{8} - \tb)
 \frac{P^2}{r} + (\frac{1}{48} - \frac{\tb}{2} -\tn) \frac{P^3}{r^2} \\
 \Gamma_2 &=& -r \, e^{-\frac{P}{2r}} \, (1- \tb \, \frac{P^2}{r^2} + \tn \,
 \frac{P^3}{r^3})\\
 &=& -r + \frac{P}{2} - (\frac{1}{8} - \tb)
 \frac{P^2}{r} + (\frac{1}{48} - \frac{\tb}{2} -\tn) \frac{P^3}{r^2} \\
 \Gamma&:=&\Gamma_1+\Gamma_2 = P + (\frac{1}{24} - \tb -
 2\tn)\frac{P^3}{r^2}.
\end{eqnarray}
Here $r>0$ is a D6-charge, $P=P^A D_A \in H^2(X)$ a D4-charge which
we take to be inside the K\"ahler cone (i.e.\ $P>0$), the terms
proportional to $P^2 =D_{ABC}P^A P^B \widetilde{D}^C \in H^4(X)$ are
D2-charges and those proportional to $P^3$ are
D0-charges.\footnote{We consider $P^3$ to be an element of $H^6(X)$
or a real number, depending on context.} We will work in the large
radius and large charge approximation, i.e.\ we will retain only the
cubic part of the prepotential.

We choose this parametrization such that $\tb$ and $\tn$ are
invariant under the rescalings discussed in the previous subsection,
and to simplify the entropy formulas of the two constituents as much
as possible. Note that $\Gamma_2$ is the conjugate dual charge to
$\Gamma_1$ (i.e.\ the $\Gamma_2$ is the image of $\Gamma_1$ under
the combined action of the two discrete symmetries described in the
previous subsection). This makes this class of examples particulary
symmetric. The case $\tb=\tn=0$ corresponds to the bound state of a
pure D6 with flux $F=\frac{P}{2 r} \, {\bf 1}_r$ and the anti-brane
of a pure D6 with flux $F=-\frac{P}{2 r} \, {\bf
1}_r$.\footnote{Since we work in the supergravity approximation in
this subsection, we ignore flux quantization.}

In what follows it will also be convenient to use the variables
\begin{equation}
 \nu := \frac{1}{24} - \tb -2 \tn, \qquad \mu := \frac{1}{8} - \tb,
\end{equation}
which are proportional to the D0 resp.\ D2 charges of the
constituents.

When the total D0-charge is negative, i.e. $\nu<0$, there exists a
regular attractor point for the total charge $\Gamma$, at
\begin{equation} \label{singattrpoint}
 B=0, \quad J = \sqrt{-6\nu} \, \frac{P}{r}, \quad g_{\rm top}=\frac{\pi \sqrt{-48 \nu}}{r}
 \quad \mbox{with entropy } S = 2 \pi
 \sqrt{-\nu/6} \, \frac{P^3}{r}.
\end{equation}
On the other hand, as we saw before, when $\nu > 0$, $Z(\Gamma)$ has
a zero locus; for example $Z(\Gamma)=0$ at $B=0$, $J = \sqrt{2\nu}
\frac{P}{r}$. Therefore the attractor flow associated to $\Gamma$
will crash on a regular zero, and no single centered BPS solution
exists.

\EPSFIGURE{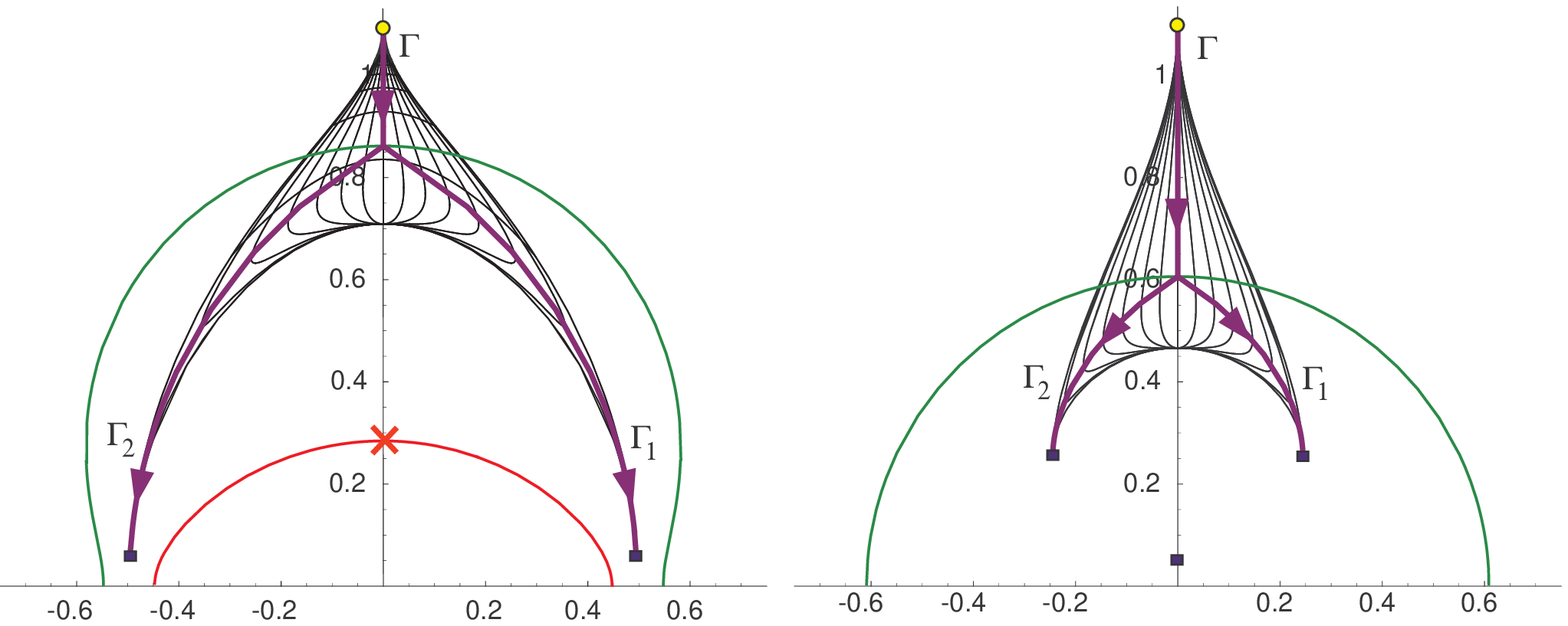,height=7cm,angle=0,trim=0 0 0 0}%
{{\bf Left:} Bound state features of the $z$-plane for
parametrization $B+i J=z P/r$, for a polar case $\tb=1.25 \times
10^{-3}$, $\tn=0$, $z_\infty=1.1 \, i$. The green (upper) line is
the line of marginal stability, where the phases of $Z_1$ and $Z_2$
align. On the red (lower) line, the phases anti-align. The red cross
is the zero of $Z(\Gamma)$. The fat split purple line is the
attractor flow tree, and the black lines forming a pair of pants
around this skeleton are the image of the moduli field $z(\IR^3)$,
following radial lines (and a few $r=$ constant lines) out of the
midpoint between the centers $\vec x=0$. {\bf Right:} Analogous plot
for a nonpolar case $24 \nu = -.01$, $8 \mu \approx 0.49$ ($\tb
\approx 0.064$, $\tn \approx -0.01$). The blue square on the
imaginary axis is the attractor point of the single flow for
$\Gamma$ which exists for this value of $\nu$.
  \label{pants}}

Thus, to verify if $\Gamma$ exists as a BPS bound state of
$\Gamma_1$ and $\Gamma_2$ when $\nu>0$, we first need to check if a
wall of $\Gamma \to \Gamma_1 + \Gamma_2$ marginal stability exists
between the zero and the value of the moduli at spatial infinity.
Taking $B=0$ at spatial infinity, we can follow a path from the
moduli there to the zero locus parametrized by $B(y)=0$, $J(y)=y
P/r$, where $y$ goes from $y=y_\infty$ (which we can think of as
very large, since we are primarily interested in BPS states in the
large radius limit) to $y=\sqrt{2\nu}$. Along this path
\begin{equation}
 Z(\Gamma_1) = \left( - \frac{i y^3}{6} + \frac{y^2}{4} + i \mu y - \frac{\nu}{2} \right)
 \frac{P^3}{r^2}, \qquad Z(\Gamma_2) = \overline{Z(\Gamma_1)}.
\end{equation}
Note that the phases of $Z(\Gamma_1)$ and $Z(\Gamma_2)$ align iff
they are both real, which is the case at
\begin{equation}
 y = y_{\rm ms} = \sqrt{6 \mu}.
\end{equation}
Thus for this to happen along the path when $y_\infty \to \infty$,
we need
\begin{equation} \label{condnupos}
 \mu \geq \frac{\nu}{3} \quad ({\rm for} \, \, \nu \geq 0), \qquad
 \mbox{i.e.} \quad \tb-\tn \leq \frac{1}{6}  \quad \mbox{(for $ \tb+ 2\tn
 \leq
 \frac{1}{24}$)}.
\end{equation}

In the nonpolar case $\nu<0$, although there are single centered
black hole solutions, there might still be 2-centered solutions as
well. In other words it might happen that both a single flow and a
split flow exists for a given charge. For this to happen, the wall
of marginal stability must separate the attractor point from the
value of the moduli at infinity. This leads to
\begin{equation} \label{condnuneg}
 \mu \geq - \nu \quad ({\rm for} \, \, \nu \leq 0), \qquad \mbox{i.e.} \quad
 \tb+\tn \leq \frac{1}{12} \quad \mbox{(for $\tb+ 2 \tn \geq \frac{1}{24}$)}.
\end{equation}

It is instructive to evaluate the necessary condition for existence
(\ref{eqsepcond}), which in the case at hand gives
\begin{equation}
 \langle \Gamma_1,\Gamma_2 \rangle \, \Im(Z_1 \overline{Z_2})_{\infty}
 \sim (\mu + \nu) (y_\infty^2 - 2 \nu) (y_\infty^2 - 6 \mu) > 0.
\end{equation}
Note that although this is always positive in the limit $y_\infty
\to \infty$ when $\mu > - \nu$, this is not enough to guarantee the
flow splits, as we just saw. Although the stable side of the
marginal stability line $y= \sqrt{6 \mu}$ \emph{near} this line is
characterized by a positive value of the left hand side, the latter
becomes also positive when we continue into the unstable side and
cross the line of \emph{anti}-marginal stability $y=\sqrt{2 \nu}$,
where the phases \emph{anti}-align. To guarantee that $y_\infty$
does not lie in this region, we need that the marginal stability
line lies above the anti-marginal stability line, i.e.\ $6 \mu> 2
\nu$.

Finally, note that when the background moduli are chosen to be at
the attractor point for $\Gamma$, the stability condition
(\ref{eqsepcond}) is not satisfied, so there will in any case be no
2-centered bound state for this value of the moduli. This is true in
general, being a direct consequence of (\ref{stabchangeflow}).

Of course, (\ref{condnupos}) or (\ref{condnuneg}) are not sufficient
to guarantee the existence of a   2-centered black hole solution
based on the split flow with two endpoints. In addition we need both
$\Gamma_1$ and $\Gamma_2$ to have attractor points, i.e.\
(\ref{attrexistence}) needs to be satisfied. In the case at hand
this reduces for both centers to
\begin{equation} \label{bhbound}
 \CD= 8 \tb^3 - 9 \tn^2 \geq 0.
\end{equation}
The attractor point is then $B+iJ = z_*(\Gamma_i) \, \frac{P}{r}$
with
\begin{equation}
 z_*(\Gamma_1) = \frac{1}{2} + \frac{3 \tn + i \sqrt{\CD}}{2 \tb},
 \qquad z_*(\Gamma_2) = -\frac{1}{2} + \frac{-3 \tn + i \sqrt{\CD}}{2
 \tb}.
\end{equation}
The corresponding entropy is
\begin{equation} \label{D6D2D0entropy}
 S_1 = S_2 = \frac{\pi}{3} \sqrt{\CD} \, \frac{P^3}{r}.
\end{equation}

\EPSFIGURE{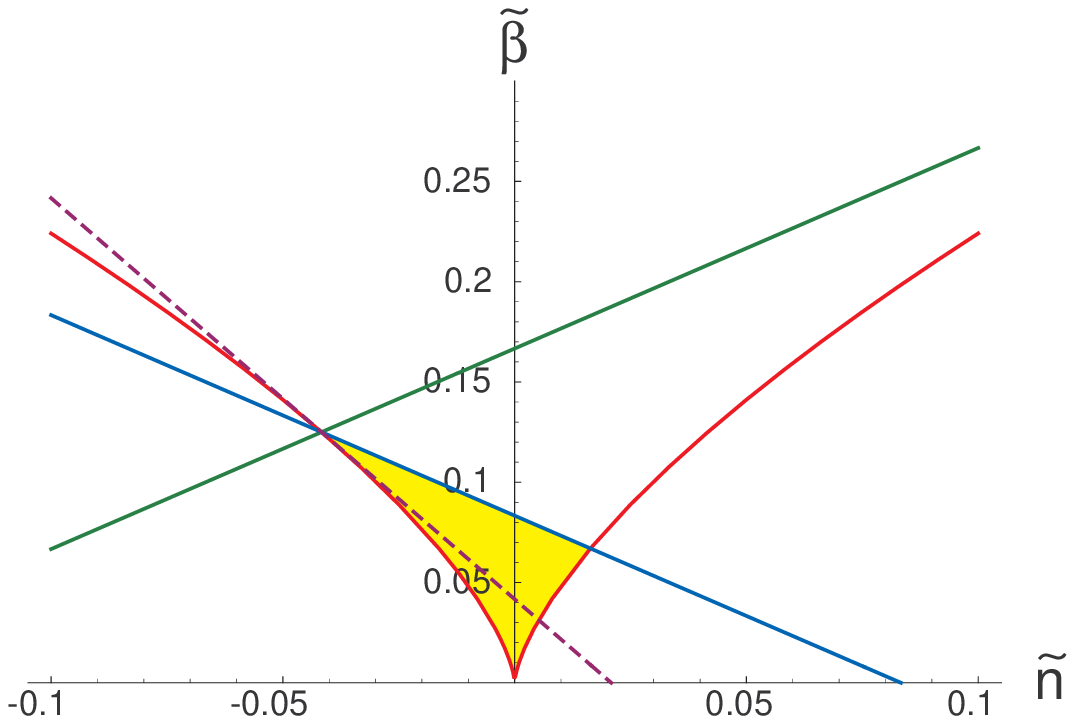,height=7.6cm,angle=0,trim=0 0 0 0}%
{Region in $(\tn,\tb)$-space supporting two-centered black hole
bound states (yellow shaded triangle). The red curved line is the
black hole bound (\ref{bhbound}), the upper green line represents
the split bound (\ref{condnupos}) for the polar case and the lower
blue line is the split bound (\ref{condnuneg}) for the nonpolar
case. The purple dotted line, corresponding to $\nu=0$,
 separates polar from nonpolar charges;
the polar region lies below. The blue boundary corresponds to the
limit in which the intersection product and hence the separation
between the two centers vanishes. The rightmost boundary vertex
gives the most negative value of the total D0-charge $\nu$,
$\nu_{\rm min} = (3-2 \sqrt{3})/8 \approx -0.058$. The origin gives
the most positive value, $\nu_{\rm max} = 1/24$. The vertex on the
left through which all lines pass has $(\tn,\tb)=(-1/24,1/8)$,
$(\mu,\nu) = (0,0)$.
  \label{stabregion}}

Note that when $r=1$, the total charges in the limiting case
$\tb=\tn=0$ are exactly those of a pure D4-brane wrapped on $P$. In
particular the D0 charge is $\widehat{q}_0=q_0=P^3/24$, which as we
saw in section \ref{sec:fareytail}, eq. (\ref{q0hatH}) is indeed the
highest possible value of $\widehat{q}_0$.\footnote{at least in the
large charge supergravity approximation in which we are working in
this subsection, which drops the subleading $c_2 P/24$ correction.
It is not hard to check that this correction is also correctly
reproduced after taking into account the $c_2$ corrections to the
$\Gamma_i$.} This suggests that the pure D4 is in fact a bound state
of a pure $D6$ plus flux and a pure anti-D6 plus flux. We will
confirm this picture in detail in the next section, both
microscopically and macroscopically. The cases with $\tb,\tn$ small
then correspond to adding a ``dilute gas'' of D2 and D0 branes to
these D6 and anti-D6 branes.

For $r>1$, the total D0-charge is not that of a single smooth
D4-brane wrapping the class $P$, but rather that of $r$ D4-branes
each wrapping the class $P/r$, as might have been expected from a
bound state of rank $r$ D6 and anti-D6 branes. In particular the
D0-charge is $r \, (P/r)^3/24$, which when $P$ is very large has a
large gap to the most polar charge $P^3/24$ obtained at $r=1$. Hence
such configurations enter far from the most polar terms in the
partition sum $\CZ$. This will be important for the derivation of
the OSV conjecture later on.

\subsection{The Entropy Enigma} \label{sec:entropyenigma}

The example of the previous section
 illustrates a remarkable phenomenon, namely, when we
scale up a nonpolar total charge $\Gamma \to \Lambda \Gamma$, with
$\Lambda \to \infty$, \emph{the 2-centered BH entropy dominates over
the single centered BH entropy, scaling as $\Lambda^3$ as opposed to
the single centered $\Lambda^2$!} Here we define the 2-centered
Bekenstein-Hawking entropy as the sum of the Bekenstein-Hawking
entropies of the two individual constituent black holes.

First, let us recall from section \ref{sec:symmetries} that in the
large $\Lambda$ limit, the single centered entropy always scales as
\begin{equation}
 S_{\rm 1 \, center}(\Lambda \Gamma) = \Lambda^2 S(\Gamma).
\end{equation}
This is easily seen to be true for (\ref{singattrpoint}), but
extends beyond the large radius approximation in which that
expression is valid.

Now let us compare this to the two centered case. To illustrate our
point, consider the case $r=1$, $\tn=0$, $\tb=1/24$, which is
clearly inside the stability domain of fig.\ \ref{stabregion}, and
corresponds to $\nu=0$, $\mu=1/12$. The total charge is then simply
$\Gamma = P$, so we achieve uniform scaling by $P \to \Lambda P$.
But then from (\ref{D6D2D0entropy})
\begin{equation}
 S_{\rm 2 \, centers} = S_1 + S_2 \to \frac{\pi}{36 \sqrt{3}}
 \Lambda^3 P^3.
\end{equation}
Thus we get the claimed $\Lambda^3$ scaling. Note that this is not
the only configuration with total charge $\Gamma=P$. Other
configurations will have different numerical prefactors replacing
$\frac{\pi}{36 \sqrt{3}}$.

This behavior is completely generic and valid for \emph{any}
D4-D2-D0 charge $\Gamma$ which is uniformly scaled up (if $P>0$). To
see this, first note that because of the shift symmetry discussed at
the end of section \ref{sec:symmetries}, we can assume the D2-charge
to be zero without loss of generality. Then we can as above split
$\Gamma=(0,\Lambda P,0, \Lambda q_0)$ into $\Gamma_{1,2} = \pm r +
\Lambda \frac{P}{2} \pm \Lambda^2 \mu \frac{P^2}{r} + \Lambda^3
\frac{q_0}{2 \Lambda^2}$ for some suitably chosen $\mu$. Because of
the second scaling symmetry discussed in section
\ref{sec:symmetries}, this exists as a 2-centered solution iff the
split into $\Gamma_{1,2} = \pm r + \frac{P}{2} \pm \mu \frac{P^2}{r}
+ \frac{q_0}{2 \Lambda^2}$ exists. In the limit $\Lambda \to
\infty$, the last term (the D0-charge) can be neglected, so this
boils down to existence of a 2-centered configuration in the class
of examples given above with $\nu = 0$. Obviously there are plenty
of such configurations; the example given in the previous paragraph
is one possibility, but it is easy to see that there is a whole
family of more general choices of $r$, $\tb$ and $\tn$ leading to a
2-centered configuration.

For any such choice, the total BH entropy is nonzero and scales as
$\Lambda^3$, because this is how $S$ scales under the second scaling
symmetry of section \ref{sec:symmetries}.

This establishes the existence of 2-centered BPS black hole bound
states at $J_{\infty}$ sufficiently large ($>\CO(\Lambda)$) for
\emph{any} charge $\Lambda \Gamma$ where  $\Gamma=(0,P>0,Q,q_0)$ and
$\Lambda \to \infty$, with entropy scaling as $\Lambda^3$ rather
than the single centered scaling $\Lambda^2$. Note however that when
$J_{\infty}$ is kept fixed at some finite, $\Lambda$-independent
value, eventually, the 2 centered solutions will cease to exist. The
reason is that the wall of marginal stability for the configuration
lies at a value of $J$ of order $\Lambda$, which when $\Lambda \to
\infty$ runs off to infinity, moving our background point out of the
stability domain.

Note also that this is not in contradiction with the microscopic
computation of the entropy of D4-D2-D0 systems in
\cite{Maldacena:1997de,Vafa:1997gr} and its successful matching with
the single centered entropy, since the regime of validity of this
computation is $\vert \hat{q}_0 \vert \gg P^3$, precisely the regime
in which there are \emph{no} multicentered solutions, and a regime
from which one automatically exits when all charges are uniformly
scaled up.

Nevertheless, since this $\Lambda^3$ scaling is surprising, to say
the least, in the remainder of this section we will justify
carefully the validity of these solutions, and of the entropy
computed from them.

Let us fix a particular two centered solution and denote the fields
and parameters associated to this by a subscript 0, e.g.\ $\Gamma_0$
is the total charge, $B_0 + i J_0$ the moduli fields and so on. For
simplicity, let us more concretely consider some case with $\nu=0$
in our class of examples (so that $\Gamma_0 = P_0$ and scaling $P_0$
is equivalent to scaling $\Gamma_0$), in some asymptotic background
$(B_0 +i J_0)|_{\infty} = z_0|_{\infty} \frac{P}{r}$ with
$z_0|_{\infty}$ above the line of marginal stability. We can scale
up
\begin{equation}
 r \to \xi \, r_0, \qquad P \to \xi \, \Lambda \, P_0
\end{equation}
without affecting the split attractor flow in rescaled coordinates
$z$ defined by $B + i J =: z \frac{P}{r}$. Since $\Im \, z$ stays
bounded away from zero, we thus see that when $\Lambda \to \infty$,
we have $J = J_0 \Lambda \to \infty$ and the large CY radius
approximation (dropping instanton corrections) is justified.

Note that the $\xi$-scaling  implements the symmetry
(\ref{unifscaling}), while the $\Lambda$-scaling implements
(\ref{lefshetzscaling}). Consequently all characteristic length
scales $L$ of the four dimensional solutions can be expressed in the
form
\begin{equation}
 L = c_0 \, \xi \, \Lambda^{3/2} \, \ell_4,
\end{equation}
where $c_0$ depends only on $r_0$, $P_0$, $\tn_0$, $\tb_0$ and
$t_0|_{\infty}$. Hence all curvature radii in 4d Planck units go to
infinity when $\Lambda \to \infty$. Note that this scaling also
implies the $\Lambda^3$ scaling of the entropy is consistent with
holography, since the area in Planck units of any surface enclosing
the centers will scale as $\Lambda^3$.

To express $L$ in string units, we use $\ell_4 = g_{4d} \ell_s$
where $g_{4d}$ is the four dimensional IIA string coupling constant,
related to the ten dimensional $g_{\rm IIA}$ by $g_{4d}^2 = g_{\rm
IIA}^2 / V_{\rm IIA}$, with $V_{IIA} = J^3/6$ the IIA CY volume in
string units. Considered as a field, $g_{4d}(\vec x)$ sits in a
hypermultiplet and does not vary over space (so $g_{\rm IIA}(\vec
x)$ does vary, since $J(\vec x)$ does). Hence, keeping the
asymptotic value of $g_{\rm IIA}$ fixed at $g_{IIA,0}$, we have the
scaling
\begin{equation}\label{Ltoell}
 L= \frac{c_0 \, g_{\rm IIA,0}}{\sqrt{V_{\rm IIA,0}}} \, \xi \,
 \ell_s,
\end{equation}
where $g_{\rm IIA,0}$ and $V_{\rm IIA,0}$ should be thought of as
asymptotic values at spatial infinity. Note that equation
(\ref{Ltoell}) no longer scales with $\Lambda$, but we can still
make it as large as we wish by scaling up $\xi$, i.e.\ by
considering large $r$, so at least in this regime higher order
curvature corrections are certainly under control, and we have no
reason left to doubt our solutions. Related to this, note that the
effective topological string coupling constant as given by
(\ref{osvsubst}), which controls $F$-term $R^2$ corrections, does
not scale with $\Lambda$ though it does scale as $1/\xi$, according
to (\ref{unifscaling}) and (\ref{lefshetzscaling}).

One could worry about cases with small $r$, since in this case at
small $g_{\rm IIA,0}$ the characteristic distance scales are small
in string units, so one might fear that $R^2$ corrections will get
out of control and we cannot trust the entropy formula we found.
However in this case we can switch to the M-theory description using
the 4d-5d correspondence of
\cite{Gaiotto:2005gf,Gaiotto:2005xt,Bena:2005ni,Elvang:2005sa} to
get a reliable picture. To achieve this, instead of keeping the
asymptotic value of $g_{\rm IIA}$ fixed, we let it scale with
$\Lambda$ to keep the M-theory CY volume in 11d Planck units $V_M =
g_{4d}^{-1}$ fixed, which amounts to taking $g_{\rm IIA}(\Lambda) =
\Lambda^{3/2} g_{\rm IIA,0}$. Since $g_{\rm IIA} = R_M^{3/2}$ where
$R_M$ is the radius of the M-theory circle in 11d Planck units, this
means we have
\begin{equation}
 R_M = \Lambda R_{M,0}, \qquad L =
 \frac{c_0}{\sqrt{R_{M,0} V_{M,0}}} \, \xi \, \Lambda^{3/2} \, \ell_{11}
\end{equation}
where we used $g_{\rm IIA,0} \ell_s = R_{M,0} \ell_{11}$. Hence we
see that all characteristic length scales of the solution (including
$R_M$) go to infinity in 11d Planck units when $\Lambda \to \infty$.
Moreover the M-theory CY volume $V_M = V_{M,0} = R_{M,0}^{-3}
J_0^3/6$ is constant over space, remains constant under the scalings
and can be taken as large as we wish (as it is a hypermultiplet
scalar). Finally we can also take $R_{M,0}$ as large as we wish, by
taking the IIA asymptotic K\"ahler class $J_0$ large (although this
changes a vector multiplet scalar, we saw this preserves the
2-centered solution).

The 4d solution near the D6D4D2D0 centers lifts up to a 5d BMPV
spinning M2 black hole with $q_{M2} = \pm \tb \frac{P^2}{r} \sim \xi
\Lambda^2$, $J^3_L = \frac{1}{2} \tn \frac{P^3}{r^2} \sim \xi
\Lambda^3$ located at the center of a $\IZ_r$ quotient of Taub-NUT
\cite{Gaiotto:2005gf}, which was shown in \cite{Gaiotto:2005gf} to
have exactly the same Bekenstein-Hawking entropy as the
corresponding 4d black hole, scaling as $S \sim \xi^2 \Lambda^3$ in
the case at hand.

Thus we conclude that even for small $r$, the enigmatic $\Lambda^3$
entropy growth we find is reliable.

From the point of view of the topological string, this is perhaps
more surprising. Solving (\ref{legendreattreq}) for the $\Gamma_1$
attractor point and substituting this in (\ref{osvsubst}), we find
\begin{equation}
 g_{\rm top} = \frac{4 \pi}{\frac{- 3 r \tn}{\sqrt{\CD}}+ i r}
\end{equation}
with $\CD$ as in (\ref{bhbound}). This does not scale with $\Lambda$
and is generically of order 1 for $r$ of order 1. How can it be then
that $R^2$ corrections to the entropy (obtained by replacing the
genus zero $\CF_0$ by the all genus $\CF=\log |\CZ_{\rm top}|^2$ in
(\ref{legendreattreq}) and (\ref{Slegform})) can be neglected in our
$\Lambda \to \infty$ scaling limit?

The puzzle is resolved by considering the product representation
(\ref{gvprodexp}) of $\CZ_{\rm top}$, and noting that all
contributions of curves with nonvanishing charge $q$ are
exponentially suppressed as $e^{-\Lambda}$ since $J \sim \Lambda$.
This leaves only the MacMahon function (\ref{ZDT0MacMahon}), which
since $g_{\rm top}$ does not scale with $\Lambda$ gives only a
finite contribution to the entropy, independent of $\Lambda$. Hence
for $\Lambda \to \infty$, this contribution can indeed be neglected.

Incidentally, the same kind of reasoning resolves the puzzle why the
D4D2D0 entropy in the large D0-charge limit does not receive
enormous $R^2$ corrections, despite the fact that $g_{\rm top} \to
\infty$ when $|q_0| \to \infty$. Again, this this system becomes
weakly curved in the M-theory description, and again all corrections
are manifestly suppressed when using the product formula for
$\CZ_{\rm top}$.

Now, having convinced ourselves that our solutions and the entropy
computed from them are reliable, we face a puzzle. The OSV
conjecture is supposed to be valid precisely at large $\Lambda$. But
its prediction for the leading asymptotic of $\ln \Omega$ is, by
construction, the single centered black hole entropy, which scales
as $\Lambda^2$, not $\Lambda^3$. So how can this be compatible with
what we find here?

Despite the obvious tension this creates, this does not immediately
mean the OSV conjecture is wrong. There are two important
subtleties. The first one is that $\Omega(\Gamma)$ is an
\emph{index}, the second is that the $\Lambda^3$ scaling holds at
$J=i \infty$ but for example not at the attractor point of $\Gamma$,
where two centered solutions do not exist.

To address these subtleties, we need a better understanding of
various types of composite BPS states, as well as the computation of
their contributions to the index, which we do in the following
sections. We postpone further discussion to section
\ref{sec:discussion}.

\subsection{D6-D0 bound states}
\label{sec:D6D0}

At large volume and zero B-field, D6 and D0 branes do not form BPS
bound states. However this can change when the B-field is
sufficiently large
\cite{Kachru:1999vj,Douglas:2000ah,Witten:2000mf}, or equivalently
when a sufficiently large $U(1)$ flux is turned on on the D6. Let
\begin{equation}
 \Gamma_1 = (p^0,0,0,0), \quad \Gamma_2 = (0,0,0,q_0), \quad
 \Gamma=\Gamma_1+\Gamma_2.
\end{equation}
Then $Z_{\rm hol}(\Gamma) = p^0 (B+iJ)^3/6 - q_0$, which clearly has
a zero locus in the interior of moduli space, so no single centered
BPS black hole solutions exist. To be more explicit, let us take for
example as in the previous subsection
\begin{equation}\label{zeetotee}
B+iJ = z P/|p^0|
\end{equation}
 with $z=x+iy \in \IC$ and $P$ some
positive class in $H^2(X,\IZ)$, and write $q_0=\rho \, P^3/(p^0)^2$.
Then up to an overall constant positive factor $P^3/(p^0)^2$ we have
$Z_1={\rm sign}(p^0) \, z^3/6$, $Z_2=-\rho$, and $Z = {\rm
sign}(p^0) \, z^3/6 - \rho$, which has a zero in the upper half
$z$-plane.

\EPSFIGURE{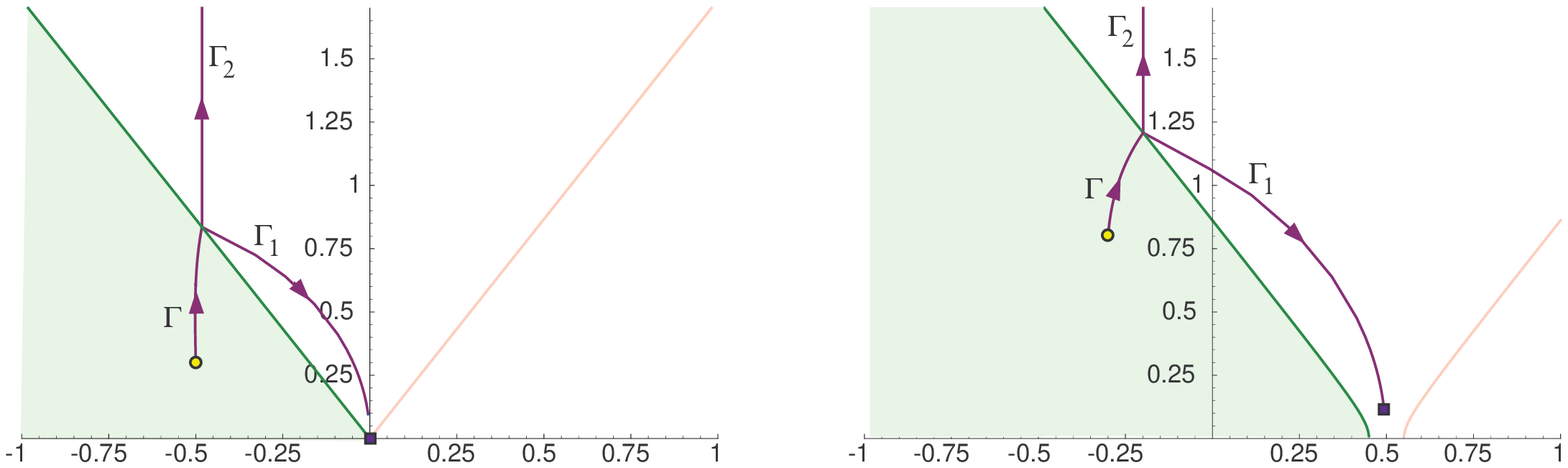,height=5.4cm,angle=0,trim=0 0 0 0}%
{{\bf Left}: attractor flow tree in $z$-plane for $p^0>0$,
$\rho=-1$, $z_{\infty}=-0.5 + 0.3 \, i$. The shaded region on the
left is the stable region, in which the BPS state exists. It is
bounded by the marginal stability line (green line). The light pink
line on the right is the line of anti-marginal stability (where the
phases anti-align). The D0-attractor flow ($\Gamma_2$) continues up
to $\Im\,z = \infty$. {\bf Right}: Analogous plot for $\Gamma_1$
defined as for fig.\ \ref{pants}a and $\Gamma_2=(0,0,0,-1)P^3/r^2$
with $P$ and $r$ as for fig.\ \ref{pants}.
  \label{D6D0}}

To see if there is a bound state of $\Gamma_1$ and $\Gamma_2$ in
some region of moduli space, it is sufficient to check if a marginal
stability wall exists, since we know that the constituents
themselves, the D0 and the D6, exist everywhere in moduli space (at
least in the large volume region we are considering). It is easy to
see that $\Im(Z_1 \bar{Z}_2)=0$ when $y=\sqrt{3}|x|$. To have the
phases align rather than anti-align on this line, we moreover need
$\Re(Z_1 \bar{Z}_2)>0$, i.e.\ ${\rm sign}(p^0) \, \rho \, x > 0$. To
see which side of this line is stable, we can use (\ref{eqsepcond}),
which gives $|x|>|y|/\sqrt{3}$. Taking into account that only a true
line of marginal stability can bound a stability domain, we get as
our final result for the zone in the upper half $z$-plane where
there exists a stable D6-D0 BPS bound state:
\begin{equation} \label{D6D0existencecond}
  |\Re \, z| > \Im\, z/\sqrt{3}, \qquad {\rm sign}(\Re \, z) = {\rm sign}(p^0 q_0).
\end{equation}
This is illustrated in fig.\ \ref{D6D0}a. Similar (but
mathematically slightly more complicated) considerations hold when
the D6 is replaced by a more general D6-D4-D2-D0 brane; an example
is shown in fig.\ \ref{D6D0}b. Note that when the D6-D4-D2-D0 has a
nonzero entropy, it can also ``absorb'' some of the D0 inside its
horizon. The amount of D0-brane charge which can be absorbed in this
way is always bounded however, as can be seen for example from
(\ref{attrexistence}), which always goes negative when $n \to
\infty$. This is another example of multiple BPS realizations of the
same charge. Again, recall that according the the split attractor flow
conjecture the number of possibilities is bounded.

\subsection{Sun-Earth-Moon systems} \label{sec:SEM}

\EPSFIGURE{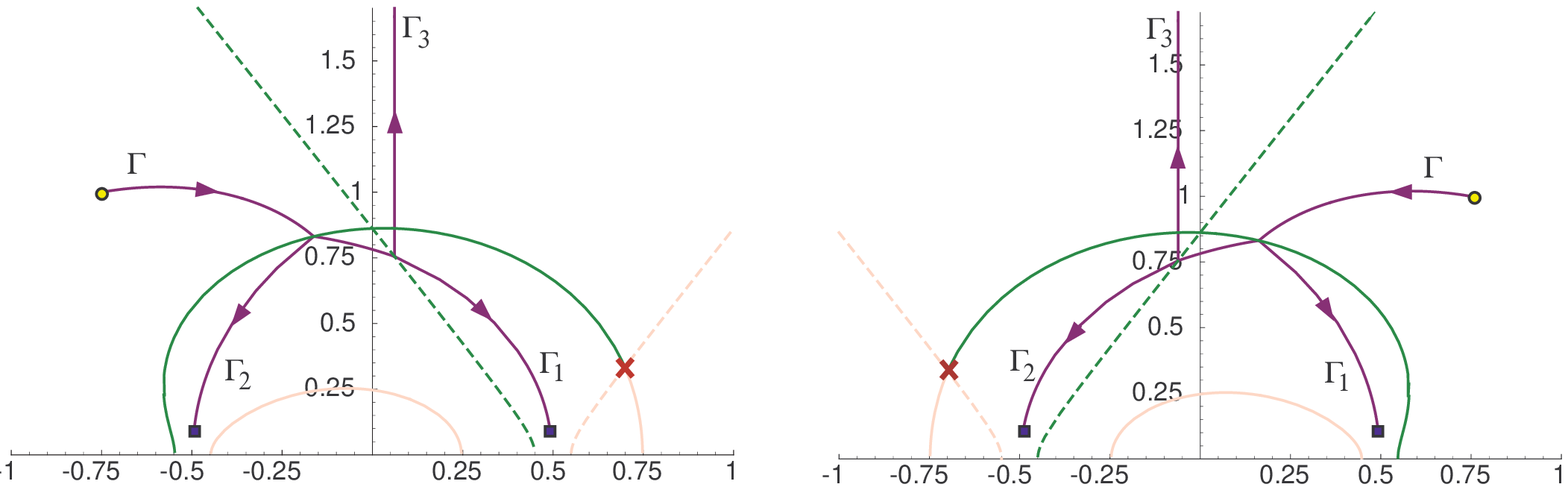,height=6cm,angle=0,trim=0 0 0 0}%
{Three-legged flow trees with $\Gamma_1$ and $\Gamma_2$ as for fig.\
\ref{pants}a and $\Gamma_3=(0,0,0,-0.01) P^3/r^2$. On the left
$z_{\infty}=-0.75 + i$, on the right $z_{\infty}=0.75+i$. Note that
this choice affects the order of the splittings: on the left we have
$\Gamma \to (\Gamma_2,\Gamma_1+\Gamma_3)$ followed by $\Gamma_1 +
\Gamma_3 \to (\Gamma_1,\Gamma_3)$, while on the right $\Gamma_1$
splits off first. The solid green line indicates the marginal
stability line for the first split, the dotted green line for the
second. The light pink lines are the corresponding anti-marginal
stability lines. The red cross indicates a zero of the central
charge of the intermediate charge ($\Gamma_1+\Gamma_3$ on the left),
implying that this charge does not have a single centered
realization.
  \label{triplesplit}}

We can also combine this sort of bound state with another state to
produce a state with overall D6 charge zero.   For example we can
dress the $(\Gamma_1,\Gamma_2)$ solutions of section
\ref{sec:classexamples} with a $\overline{D0}$-brane bound to one of
the centers. The corresponding attractor flow trees are illustrated
in fig.\ \ref{triplesplit}. Note that the charge to which the
$\overline{D0}$ binds depends on the choice of $B$-field at
infinity, i.e.\ $\Re\, z_\infty$. The transition between the two
occurs when the initial attractor flow (for charge $\Gamma$) hits
the point where all three phases of the $Z(\Gamma_i)$ align, that is
at the intersection point of the dotted and solid green lines in the
figure. In the case at hand, this happens when $z_{\infty}$ crosses
the imaginary axis. When $z_{\infty}$ is exactly on the imaginary
axis, $\Gamma_3$ (a $\overline{D0}$) is at most marginally bound:
its phase lines up there with the phase of $\Gamma_1+\Gamma_2$ (a
$D4+D0$), and there is no energetic obstruction to taking away
$\Gamma_3$ from $\Gamma_1+\Gamma_2$ as far as one wishes.

For positive D4-charge $P$, it is not possible to construct such
bound states with $\Gamma_3$ a $D0$-brane rather than
$\overline{D0}$-brane; in the figure above, this would essentially
flip the marginal and anti-marginal stability lines involving
$\Gamma_3$, so after the first split one would be outside of the
stable region for the remaining bound state involving $\Gamma_3$.
This corresponds to the fact that only $\overline{D0}$-branes, not
$D0$-branes, form bound states with D4-branes in our sign
conventions.

\EPSFIGURE{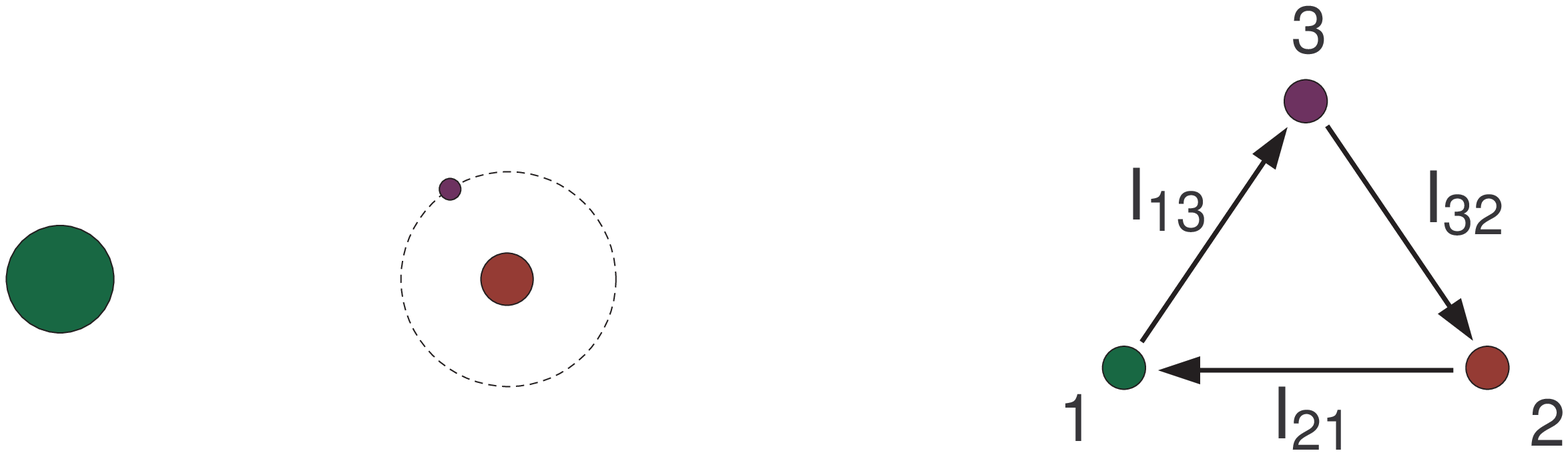,height=3.5cm,angle=0,trim=0 0 0 0}%
{{\bf Left:} Sketch of a BPS Sun - Earth - Moon configuration in
space. {\bf Right:} Quiver diagram representing intersection
products $I_{ij}=\langle \Gamma_i,\Gamma_j \rangle$ between the
centers. In the case at hand ${I_{21}},I_{13},I_{32}>0$. For the
$D6$-$\overline{D6}$-$\overline{D0}$ system, the microscopic quiver
would look identical except for an additional multiplicity 3 arrow
going from the $\overline{D0}$ node to itself, representing the
three moduli corresponding to the $\overline{D0}$ moving around in
$X$.
  \label{orbits2}}

The supergravity solutions representing these bound states are
Sun-Earth-Moon configurations, as shown in fig.\ \ref{orbits2}a. The
positions of the centers $\vec{x}_i$, $i=1,2,3$ are constrained by
the integrability conditions (\ref{centerconstraints}):
\begin{equation} \label{eqposeq}
 \frac{I_{13}}{R_{13}} - \frac{{I_{21}}}{R_{21}} = \theta_1  \qquad +
 \mbox{ cycl. perm.}
\end{equation}
where $I_{ij} = \langle \Gamma_i,\Gamma_j \rangle$,
$R_{ij}=|\vec{x}_i-\vec{x}_j|$, $\theta_i = 2 \Im(e^{-i \alpha}
Z_i)_{\infty}$. Note that $\theta_1+\theta_2+\theta_3=0$ and the
third equation is just the sum of the first two. More concretely in
the case at hand we can take say $\Gamma_1$ to be a pure D6 with
$U(1)$ flux $F=S_1$ turned on, $\Gamma_2$ the anti-brane of a pure
D6 with flux $F=S_2$ turned on, and $\Gamma_3$ a charge $-n$ anti-D0
brane, so according to (\ref{chargeformula}) we have
\begin{equation} \label{SEMcharges}
 \Gamma_1 = e^{S_1}(1+\frac{c_2}{24}), \quad \Gamma_2 =
 -e^{S_2}(1+\frac{c_2}{24}), \quad \Gamma_3 = -n \, \omega.
\end{equation}
In this case, $I_{13} = n$, $I_{32} = n$, and ${I_{21}} =
\frac{P^3}{6} + \frac{c_2 \cdot P}{12} = I_P$, where $P = S_1 - S_2$
is the total D4-brane charge. \footnote{ If we wanted to establish
the existence of these multicentered solutions directly in
supergravity without invoking the split attractor flow conjecture,
we would have to check that $H(\vec x)$ lies in ${\rm dom} \, S$ for
all $\vec x$. This is difficult. One can show (dropping $c_2$
corrections) that when the integrability conditions are satisfied,
the discriminant of $H(\vec x)$ goes to a positive constant at
infinity and goes to $+\infty$ near each of the three centers, and
therefore takes on its minimal value at some finite point in
$\IR^3$. If this point is on an axis of symmetry then one can
further show rigorously that $\CD(H(\vec x))$ is bounded below by a
positive constant.}

Note that depending on the sign of $\theta_3 \sim - \sin \alpha$,
which is determined by the value of the B-field, $R_{23}$ is smaller
or larger than $R_{31}$, corresponding to the anti-D0 binding to the
D6 or to the anti-D6. When $\theta_3=0$, the anti-D0 moves on a
plane equidistant from the D6 and the anti-D6 center, so it can
escape to infinity. Indeed, at this locus in moduli space (which
includes zero B-field in the case of zero total D2-charge), the
bound state between an anti-D0 and a D4 (=$\Gamma_1+\Gamma_2$) is
only marginal.

\subsection{Scaling solutions} \label{sec:scalingsol}

Considering the $R_{ij}$ as independent variables, the equations
(\ref{eqposeq}) always have a scaling solution
\begin{equation} \label{scalingsol}
 R_{ij} \to \lambda I_{ij}, \qquad \lambda \to 0,
\end{equation}
independent of the $\theta_i$. In the limit $\lambda=0$, the
coordinates of the 3 centers coincide and hence the solution becomes
indistinguishable from a single centered black hole solution to a
distant observer. (However,   for an observer remaining close to the
centers, they actually stay at finite distance: Within a coordinate
distance of order $\lambda$ from the centers $H(\vec x) \sim
\lambda^{-2}$ and hence $e^{-2U(\vec x)} \sim \lambda^{-2}$, so the
presence of the warp factor in (\ref{metric}) implies that the
observer remains at an order one geodesic distance. What is
happening is that a throat is developing and the observer disappears
down the throat.)

However, the $R_{ij}$ are actually not quite independent: they equal
the lengths of the edges of a triangle in flat space, and as such
must satisfy the triangle inequality. Thus a necessary and
sufficient condition for the scaling solution to (\ref{eqposeq}) to
exist is
\begin{equation} \label{trineq}
 I_{21} + I_{13} \geq I_{32} \qquad + \mbox{ cycl. perm.}
\end{equation}
In the case at hand this reduces to $n \geq \frac{1}{2}I_P =
\frac{1}{12} P^3 + \frac{1}{24} c_2 P$. In particular, this implies
that the total charge is necessarily nonpolar, since $\hat{q}_0 =
\frac{1}{24} P^3 + \frac{1}{24} c_2 P - n < 0$. This is compatible
with general expectations, as only nonpolar states should be able to
form black holes.

Since the scaling solution is independent of the $\theta_i$, the
branch of the solution moduli space to (\ref{eqposeq}) continuously
connected to the black hole in this way will never decay when the
$\theta_i$ (in other words the background K\"ahler moduli) are
varied. It is therefore represented by a single centered attractor
flow rather than an attractor flow tree.

 Conversely, if the triangle inequalities (\ref{trineq}) are
\emph{not} satisfied, then the solutions can always be forced to
decay by varying the $\theta_i$ (so the solution is described by an
attractor flow tree).

To see this, let us assume that one of the triangle inequalities is
violated. Without loss of generality we can take
$I_{21}>I_{13}+I_{32}$. Then we claim that when $\theta_3 > 0$,
taking $\theta_1$ to zero will necessarily force $\vec{x}_1$ to
separate infinitely far from $\vec{x}_2$ and $\vec{x}_3$, and when
$\theta_3 < 0$, taking $\theta_2$ to zero will similarly separate
$\vec{x}_2$ from $\vec{x}_1$ and $\vec{x}_3$. Let us consider the
$\theta_3>0$ case, the other case is analogous. When $\theta_1=0$,
we then have $\theta_2 = - \theta_1 - \theta_3 = - \theta_3 < 0$ and
the equilibrium conditions (\ref{eqposeq}) imply \emph{either}
\begin{equation} \label{splitsol}
 R_{13} = R_{21} = \infty, \quad R_{23} = - \frac{I_{32}}{\theta_2}
\end{equation}
which corresponds to the claimed infinite separation, \emph{or}
\begin{equation}
 R_{13} = \lambda I_{13}, \quad R_{21} = \lambda I_{21}, \quad
 R_{32} = \lambda' I_{32}, \quad \mbox{ where } \lambda' <
 \lambda < \infty.
\end{equation}
However, since we must have $R_{21} \leq R_{13} + R_{32}$ to have an
actual solution, the above gives $I_{21} < I_{13} + I_{32}$,
contradicting the initial assumption. Hence (\ref{splitsol}) remains
the only possibility; the solution is forced to split at
$\theta_1=0$ (with the stable side being $\theta_1<0$, as a slight
extension of the analysis shows).

All this is of course in perfect agreement with what we expect from
the attractor flow picture, as well as with general expectations for
polar states.

In section \ref{sec:quivfact}, we will study similar bound states
both in the spacetime picture and in the microscopic quiver picture.
We will see that the BPS index factorizes precisely when the
inequalities (\ref{trineq}) are violated. Moreover,   the BPS index
undergoes some sort of phase transition --- no longer factorizing
and starting to grow exponentially --- as soon as (\ref{trineq}) are satisfied.
Note this is exactly where  the black hole branch opens up.
Hence, this qualitative change  is physically expected from
the spacetime picture, but highly nontrivial from the microscopic
point of view.

\EPSFIGURE{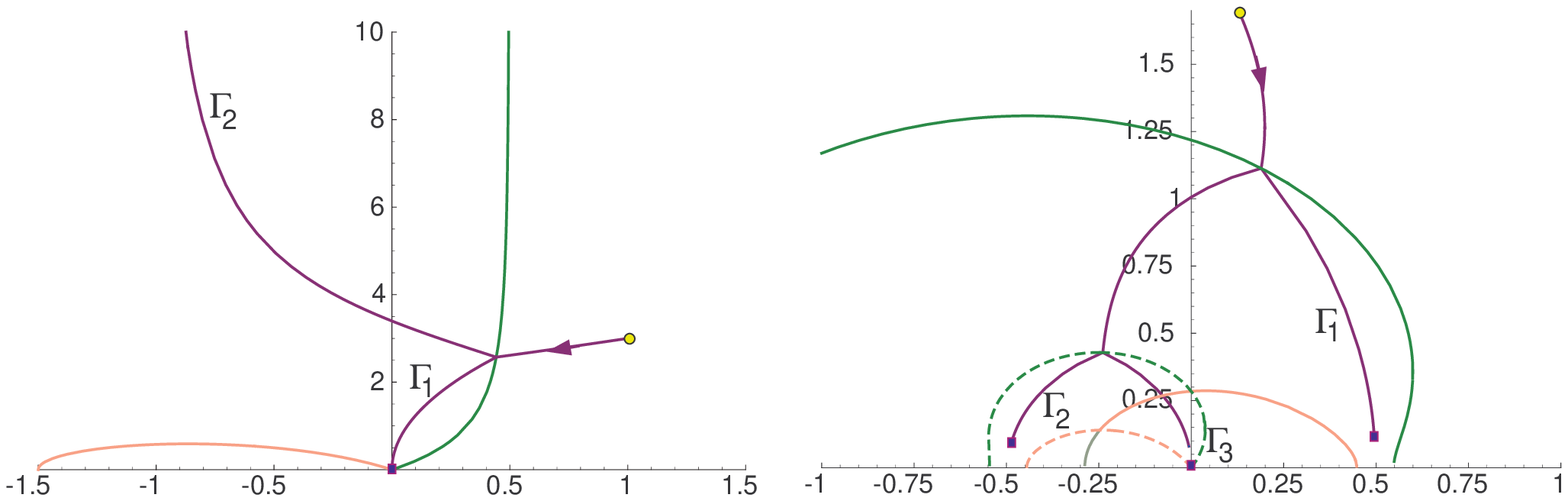,height=5.5cm,angle=0,trim=0 0 0 0}%
{Two more bound states with total D6-brane charge equal to 1. {\bf
Left:} $\Gamma_1 = D6$, $\Gamma_2 = -D2 + D0$. Note that the line of
marginal stability goes up along a vertical asymptote all the way to
infinite radius. {\bf Right:} $\Gamma_1$ and $\Gamma_2$ chosen as in
fig.\ \ref{pants} (carrying $1$ resp. $-1$ unit of D6 charge), and
$\Gamma_3=D6$.
  \label{moretrees}}

\subsection{Even more complicated multicentered bound states}

One can imagine many other multicentered configurations involving
various charges, for example we can add more anti-D0 ``moons'', or
replace the D0 particles by D2-D0 particles. Some examples with net
D6 charge 1 are shown in fig.\ \ref{moretrees}. These can in turn be
used as building blocks for the D4-D2-D0 bound states of interest,
and so on, even leading to fractal-like flow-trees, as shown in
fig.\ \ref{fr}.

\EPSFIGURE{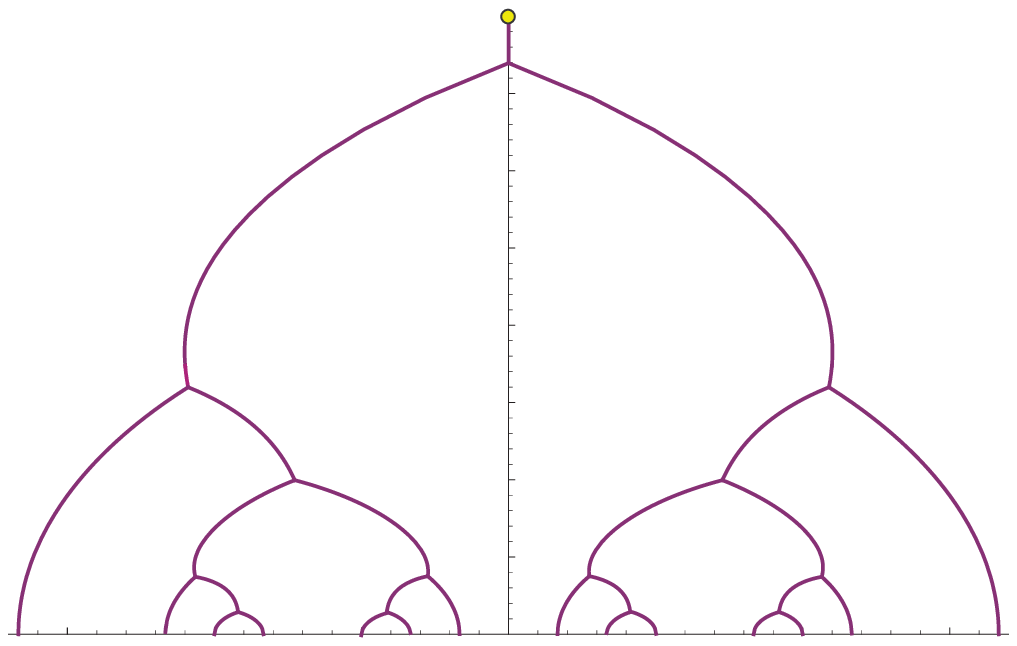,height=5cm,angle=0,trim=0 0 0 0}%
{The D6 split flows of fig.\ \ref{moretrees}$b$ (and their
conjugates) can be iteratively combined to form fractal-like flow
trees with zero total D6 charge. An example is shown with 14 pure
(fluxed) D6 / anti-D6 centers. It is possible to write compact
analytic formulae describing these fractal flow families.
  \label{fr}}

Enumerating this zoo and taking into account all existence
conditions, let alone computing their BPS ground state degeneracies,
would appear very hard, to say the least. However, to count
degeneracies of polar states, we can use the fact that these are
guaranteed to split in two clusters at a wall of marginal stability
somewhere in moduli space. The problem is then reduced to computing
the degeneracies of the two individual clusters. This might still be
complicated if one wants to compute exact expression for the the
degeneracies based on enumerating all possible further splits
corresponding to the structure of the clusters (although one could
imagine working recursively), but in suitable cases, it is possible
to circumvent this problem. The idea is to go to a regime in  which
the most significant contributions to the fareytail series come from
the polar terms corresponding to flow trees which initially split in
branes with charges $\Gamma_1$ and $\Gamma_2$ with D6-charge $1$
resp. $-1$. The indices of BPS states for such branes turn out to be
more or less given by rank one DT invariants. (The precise relation
is explained in section \ref{sec:D6D2D0} and section
\ref{innocuouscoredump}.)  This will allow us to express the BPS
indices of the relevant polar terms in the fareytail series in terms
of the DT invariants, along the lines of section
\ref{sec:basicidea}. There is no need to consider further splits of
the flow tree, since the DT invariants already count all BPS states
of the two initial brane constituents. Using the relation between DT
and GW invariants reviewed in section \ref{sec:prelim}, we will thus
be led to an expression of $\CZ_{\rm BH}$ in terms of the
topological string partition function, and to the OSV conjecture.

Finally, we note here that although a priori we should also consider
splits in two charges both of which have $p^0=0$, those are easily
shown to be absent for flows coming from large $J$. Consider a
charge $\Gamma=(0,P,0,q_0)$ and a candidate split in
$\Gamma_1=(0,P_1,Q_1,q_{0,1})$, $\Gamma_2=(0,P_2,-Q_1,q_{0,1})$, and
let us take $B_\infty=0$, $J_\infty = y \, P$, $y \to \infty$. Then
\begin{equation} \label{NoD4splitoff}
 \langle \Gamma_1,\Gamma_2 \rangle \, \Im(Z_1 \bar{Z}_2) = -(P\cdot Q_1) \left(\frac{1}{2}
 P^3 (P\cdot Q_1) y^3 + O(y) \right) \leq 0,
\end{equation}
so (\ref{eqsepcond}) is not satisfied. 
(When $J$ is not proportional to $P$ it is perfectly possible to
have a $D4D2D0$ split into a pair of $D4D2D0$ states at $B_\infty=0$
and large $J_\infty$.) 

\section{Microscopic description} \label{sec:microdescription}

States corresponding to flow trees have a microscopic description as
well. This will be the subject of the present section.

\subsection{D-brane picture at $g_s = 0$} \label{sec:microDbrane}

The microscopic D-brane picture is valid at $g_s |\Gamma| \ll 1$
with $|\Gamma|$ some appropriate measure of the ``size'' of the
charge $\Gamma$. It describes the state as an object sitting at a
single point in the noncompact space. The macroscopic picture, valid
in the opposite regime $g_s |\Gamma| \gg 1$, at first sight looks
very different, with bound states looking like atoms or molecules
rather than D-branes geometrically glued together. Nevertheless, the
two pictures can be shown to transform smoothly into each other when
varying $g_s$; for a detailed analysis see \cite{Denef:2002ru}.

In the large radius limit, IIA D-branes are well described by
holomorphic geometrical objects wrapped around various even
dimensional cycles. The F-term constraints determining the moduli
spaces of these objects do not receive $\alpha'$ corrections
\cite{Brunner:1999jq}. On the other hand, the D-term constraints,
which govern stability and decay, \emph{do} receive important
$\alpha'$ corrections \cite{Brunner:1999jq}. As a result, phenomena
such as decay at marginal stability at some finite value of the
K\"ahler moduli tend to be invisible in the IIA large radius
geometrical description. However, there is a simple universal
microscopic   picture which does capture this phenomenon
accurately. This is   originally due to \cite{Kachru:1999vj} and
has been extended in
many works on the categorical description of D-branes (as reviewed
in \cite{Douglas:2000gi}).

\EPSFIGURE{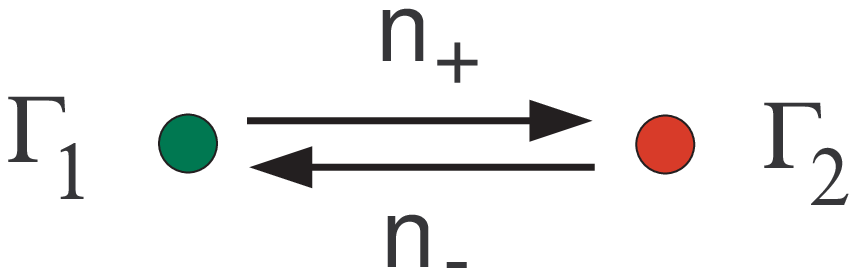,height=2.3cm,angle=0,trim=0 0 0 0}%
{Bound state quiver 
  \label{quiver}}

This goes roughly as follows. Let us consider a bound state of two
D-branes with charges $\Gamma_1$ and $\Gamma_2$. When near the wall
of marginal stability (i.e.\ when the phases $\alpha_i$ of the
central charges $Z(\Gamma_i)$ are almost identical), there are light
bosonic open string states stretching between the D-branes
corresponding to $\Gamma_1$ and $\Gamma_2$, whose mass squared
equals \cite{Berkooz:1996km}
\begin{equation} \label{masssquared}
 m^2 \sim q \, (\alpha_2 -
 \alpha_1), \qquad q=\pm 1.
\end{equation}
More precisely this is the tree level mass squared in the mirror
intersecting D3-brane picture when the D3-branes are at the same
point in $\IR^3$, the light strings corresponding to string
localized at the intersection points.\footnote{If the branes are not
at the same point in $\IR^3$ there is an additional mass term $m^2
\sim |\vec{x}_1-\vec{x}_2|^2$ and supersymmetry is generically
broken at $g_s = 0$.} In the low energy description of the D-branes
as a supersymmetric quantum mechanical system with 4 supercharges,
the light strings appear as chiral multiplets $\Phi_i$
(dimensionally reduced to $d=1$), represented by the arrows of a
quiver\footnote{This quiver should be interpreted in a loose sense
in the present discussion. In particular we allow the branes
correspondng to the nodes to have nontrivial moduli spaces here, not
necessarily realized by simple adjoint fields as in the proper
definition of a quiver. These moduli spaces might for example arise
from lumping together several standard quiver nodes into one.} with
two nodes as in fig.\ \ref{quiver}, and (\ref{masssquared}) can be
understood as being induced by a D-term potential
\cite{Kachru:1999vj}
\begin{equation}\label{deeterm}
 V(\phi) = \frac{1}{2 \mu} D^2, \qquad D = \sum_a  q_a |\phi_a|^2 -
 \mu (\alpha_2-\alpha_1).
\end{equation}
where the $q_a = \pm 1$ are charges with respect to the relative
$U(1)$ between the branes and $\mu$ is some constant (specified
below in section \ref{sec:quivdescription}). The fermionic
superpartners of the $\phi_i$ remain massless at tree level, but
when both positive and negative $q_i$ are present, disc instantons
ending on the D3-branes can produce a nontrivial superpotential
depending on the $\phi^i$, lifting pairs of massless fermions of
opposite charges, but leaving the difference of $q=\pm 1$ massless
fermions invariant. If we denote the number of stretched strings
with $q = \pm 1$ by $n_{\pm}$, then we have for the the index
\begin{equation}
 n_+ - n_- = \langle \Gamma_1,\Gamma_2 \rangle
\end{equation}
with $\langle \Gamma_1,\Gamma_2 \rangle$ the symplectic intersection
product between the charges. In type IIB this is just the geometric
intersection product between the D3-branes, $n_+$ ($n_-$) being the
number of positive (negative) intersection points. For IIA we define
this product in appendix \ref{app:defconv}.

From (\ref{masssquared}) we see that when we are close to the
marginal stability wall, on the side where
\begin{equation} \label{tachyoncond}
 \langle \Gamma_1, \Gamma_2 \rangle (\alpha_1 - \alpha_2) > 0,
\end{equation}
there will always be tachyonic strings present stretching between
the constituent branes. Condensation of these tachyons produces a
BPS bound state of total charge $\Gamma$. Since we assumed the state
decays when crossing the wall, no such tachyons exist on the other
side of the wall, where the above expression becomes negative. This
will indeed be the case when either $n_+=0$ or $n_-=0$ (possibly
effectively after lifting pairs by F-term masses). Note that this
stability condition is identical to the supergravity condition
(\ref{eqsepcond}) for small $\alpha_1-\alpha_2$.

When we have two single D-branes, hence a gauge group $U(1) \times
U(1)$, with respective deformation moduli spaces $\CM_1$ and
$\CM_2$, and if say $n_- = 0$ over all of $\CM_1 \times \CM_2$, then
the moduli space $\CM$ of the bound state will be a $\IC
\IP^{\vert\langle \Gamma_1,\Gamma_2 \rangle\vert-1}$ fibration over
$\CM_1 \times \CM_2$, with the $\IC \IP^{|\langle \Gamma_1,\Gamma_2
\rangle|-1}$ fiber coming from solving the D-flatness condition
$D=0$ and modding out by $U(1)$. If the fiber does not degenerate
anywhere, the Euler characteristic factorizes as
\begin{equation} \label{chifact}
 \chi(\CM) = \chi(\IC \IP^{|\langle \Gamma_1,\Gamma_2 \rangle|-1}) \,
 \chi(\CM_1) \, \chi(\CM_2) = |\langle \Gamma_1,\Gamma_2 \rangle| \,
 \chi(\CM_1) \, \chi(\CM_2).
\end{equation}
Identifying the Euler characteristic (up to a sign) with the index
of supersymmetric states $\Omega$, this gives a corresponding
factorization of $\Omega$.


 As described in section \ref{sec:quantumflowtrees}, attractor flow
trees provide a useful canonical prescription for an iterated
assembly or decay process of multicentered configurations, in
particular because stability is guaranteed to be preserved when
moving upstream along an attractor flow, and decay, whenever
possible, is guaranteed to occur at some point when flowing down.
Instead of splitting or joining (clusters of) centers in
supergravity, we may equally well think microscopically at $g_s=0$
and split or glue branes together through tachyon condensation as
described above, while following the same flow trees. This makes
sense since flow trees are determined entirely by central charges,
which are universal, exact data, independent of the picture in which
one is working. In this way attractor flow trees continue to be
meaningful even microscopically.

Microscopic counterparts of (\ref{eqsepcond}) exist in the framework
of the derived category as well. For a nice discussion of how it
appears for bound states of holomorphic vector bundles and its
relevance to the question of existence of stable vector bundles, see
\cite{Douglas:2006jp}.

\subsection{Quiver description of bound states}
\label{sec:quivdescription}

Building on the reasoning outlined above, one finds that quivers
give a low energy, weak string coupling description of bound states
of simple, rigid objects (such as D6 or anti-D6 branes carrying
$U(1)$ flux), near a locus in moduli space where the central charges
of the objects all line up. Let us quickly review some useful facts
about this representation, referring to \cite{Denef:2002ru} for more
details.

In a region where the phases almost line up, the objects are almost
mutually supersymmetric, and there will be open strings stretched
between them whose lightest fermionic modes are massless and whose
lightest bosonic modes have squared masses proportional to the phase
differences of the central charges, along the lines sketched above
(this is assuming the objects coincide in the noncompact space). The
system can be modeled at low energies by quiver quantum mechanics,
obtained by dimensionally reducing the corresponding $\CN=1$, $d=4$
quiver gauge theory. The multiplicities of the objects associated to
the nodes $i$ are given by the dimension vector $d_i$. The degrees
of freedom of the quantum mechanics are the (possibly nonabelian)
positions of the nodes in the noncompact space and the $U(d_i)
\times U(d_j)$ complex bifundamental scalars $\phi^a_{ij}$, $a =
1,\ldots,K_{ij}$, associated to the light open strings from node $i$
to node $j$, plus their fermionic partners.    When $g_s \to 0$
keeping other parameters fixed, the supersymmetric ground state wave
functions live on the Higgs branch, with all node positions
coincident and the bifundamental vevs subject to the D-term
constraints
\begin{equation} \label{quivDtermeq}
 \sum_j \sum_a (\phi^a_{ij})^\dagger \phi^a_{ij} -
 \sum_j \sum_a \phi^a_{ji} (\phi^a_{ji})^\dagger = \vartheta_i \,
 {\bf 1}_{d_i} \quad \forall i.
\end{equation}
The Fayet-Iliopoulos parameters $\vartheta_i$ are given by the
background moduli as
\begin{equation} \label{varthetai}
 \vartheta_i = 2 m_i (\alpha_i - \alpha_0)
\end{equation}
where $\alpha_i = \arg Z_i$, $m_i = |Z_i|$ and $\alpha_0 = \sum_i
d_i m_i \alpha_i/\sum_i d_i m_i$, with $Z_i$ the normalized central
charge of the $i$th node. Note that $\sum_i d_i \vartheta_i = 0$,
and that the condition for all the phases to almost line up is
$\vartheta_i \ll 1$. If closed oriented loops are present, there can
be a superpotential $W(\phi)$ as well. If there are no such closed
loops, gauge invariance prohibits a nonzero $W$. The quiver moduli
space is thus given by
\begin{equation} \label{genquivermodulispace}
 \CM = \{ \phi \, | \, (\ref{quivDtermeq}) \mbox{ satisfied, and } \partial W = 0
 \} \, / \, U(d_1) \times \cdots \times U(d_n).
\end{equation}

So far we kept the branes at the same point in the noncompact space.
However when one takes the objects apart (including splitting the
nodes with multiplicity $d_i > 1$ in $d_i$ separate branes, labeled
by an index $\alpha=1,\ldots,d_i$), the stretched strings become
massive and can be integrated out. At one loop this produces a
potential on position moduli space, with supersymmetric minima at
\begin{equation} \label{quivposeq}
 \sum_{j,\beta} \frac{I_{ij}}{|\vec{x}_{i\alpha} - \vec{x}_{j\beta}|} = \vartheta_i \quad
 \forall i,
\end{equation}
where $I_{ij}= K_{ij} - K_{ji} = \langle \Gamma_i,\Gamma_j \rangle$.
When the solutions to this equation have separations $|\vec{x}_i -
\vec{x}_j|$ which are sufficiently large, the procedure of
integrating out  the stretched strings is self-consistent. Depending
on the parameter regime, the supersymmetric ground state wave
functions will peak on the ``Coulomb branch'' ($\phi = 0$, $\Delta
\vec x\not=0$) or on the ``Higgs branch'' ($\phi\not= 0 $, $\Delta
\vec x =0$), thus interpolating between the two pictures of bound
states \cite{Denef:2002ru}. Note that equation (\ref{quivposeq}) is
almost exactly the same as the supergravity position constraint
equations (\ref{centerconstraints}):
\begin{equation} \label{sugraposeq}
 \sum_{j,\beta} \frac{I_{ij}}{|\vec{x}_{i\alpha} - \vec{x}_{j\beta}|} = \theta_i \quad
 \forall i,
\end{equation}
where $\theta_i = 2 \, \Im(e^{-i \alpha} Z_i) = 2 m_i
\sin(\alpha_i-\alpha)$, $\alpha = \arg (\sum_i m_i e^{i \alpha_i})$,
$\sum_i d_i \theta_i = 0$. The identity of the form of these
equations despite being in very different regimes is due to a
non-renormalization theorem. Note furthermore that in the strict
physical domain of validity of the quiver picture,  we have
$\vartheta_i \ll 1$, so $\alpha_0 \approx \alpha$ and $\vartheta_i
\approx \theta_i$. When moving away from the locus where all phases
line up, $\vartheta_i$ and $\theta_i$ start to deviate; this should
not come as a surprise, since the value of constants on the right
hand side of the one loop result (\ref{quivposeq}) are not protected
and will receive corrections.

Thus we see that in the quiver description of bound states, the
correspondence between multicentered solutions and microscopic bound
states is rather explicit.

\subsection{Geometrical relations between D4 and D6-anti-D6 bound
states} \label{sec:microrel}

IIA D-brane bound states are rather well understood in the $J \to
\infty$, $g_{\rm IIA} \to 0$ limit, where they are essentially given
by holomorphic vector bundles, or more generally coherent sheaves.
In this geometric description, F-term constraints do not receive
$\alpha'$ corrections, but D-term constraints do. D-terms govern
stability, and as a result many decay phenomena are completely
invisible at large radius from the microscopic point of view.  This
is not universally  true, since $\mu$-stability can be seen at large
radius.  However,   decays of the kind we have investigated such as
a D4 splitting into a D6 and anti-D6 are not detectable if one
limits one's attention to holomorphic vector bundles on holomorphic
4-cycles.

In spite of all this, in this section   we will nevertheless arrive
at a picture for (sufficiently polar) D4-D2-D0 brane states in the
language of holomorphic sheaves which tantalizingly hints at the
``split'' nature of the corresponding BPS states. In particular,
although in the geometrical regime we cannot literally see those
states split in the D6 and anti-D6 branes which are their building
blocks according to the split flow picture, a lot of the structure
of their moduli spaces is suggestive of this structure.

The picture we develop here was first proposed in
\cite{Gaiotto:2005rp} and exploited further in
\cite{Gaiotto:2006wm}. We review it here for completeness and add a
number of observations. The picture we arrive at is heuristic and will not
be used in the proof of the OSV formula. It is nevertheless a source
of very useful intuition.

If the divisor $\Sigma$ in the class $P$ is \emph{frozen} at
$\Sigma=\Sigma_0$, the moduli space of BPS configurations reduces to
${\rm Hilb}^N \Sigma_0$
\cite{Vafa:1995zh,Harvey:1996gc,Witten:1997yu,
Vafa:1997gr,Dijkgraaf:1998zd}, i.e.\ the Hilbert scheme of $N$
points on $\Sigma_0$, and by (\ref{indextoeuler}), since $\dim {\rm
Hilb}^N \Sigma_0 = N \dim \Sigma_0$ is always even,
\begin{equation}
 d_{\Sigma_0}(F,N) =  \chi({\rm Hilb}^N \Sigma_0).
\end{equation}
The generating function for these Euler characteristics is given by
G\"ottsche's formula \cite{goettsche} (see \cite{Dijkgraaf:1998zd}
for a pedagogical review)
\begin{equation} \label{gottsche}
 \sum_N \chi({\rm Hilb}^N \Sigma_0) \, q^N = \prod_{n \geq 1}
 (1-q^n)^{-\chi(\Sigma_0)}.
\end{equation}

However, in reality, the divisor $\Sigma$ is not some fixed
$\Sigma_0$, but has a deformation moduli space, and even when a
sufficiently generic flux is turned on such that all deformation
degrees of freedom are frozen by the condition $F^{2,0}=0$, there
might be several such isolated points in the divisor moduli space.
Moreover, we need to sum over different fluxes giving the same total
charge. In the limit $N \to \infty$, all those extra degrees of
freedom only give subleading contributions to the entropy, but at
smaller $N$, in particular for the polar states, this is not the
case.

One could try to correct this by considering the full moduli space,
say for $F$ a flux pulled back from $H^2(X)$ (such that none of the
deformation moduli of $P$ are obstructed), as a fibration over
$\CM_P=\IC\IP^{I_P-1}$ with fiber given by ${\rm Hilb}^N P$. If the
fibration has no singular fibers, the orbifold Euler characteristic
of the total space would just be the product of $\chi(\CM_P) = I_P$
and $\chi({\rm Hilb}^N P)$, and the generating function would be
obtained simply by multiplying (\ref{gottsche}) by $I_P$. A simple
example shows this idea to be too naive: Consider the moduli space
with one pointlike instanton. This fibers over $X$ with fiber $\IC
P^{I_P-2}$, and hence the Euler character is $\chi(X) (I_P-1)$. The
reason for the discrepancy is the presence of a complicated,
self-intersecting locus in $\CM_P$ where the fiber $P$ becomes
singular, so the simple factorization formula does not hold.
Figuring out the correct formula in this picture appears very hard.

\EPSFIGURE{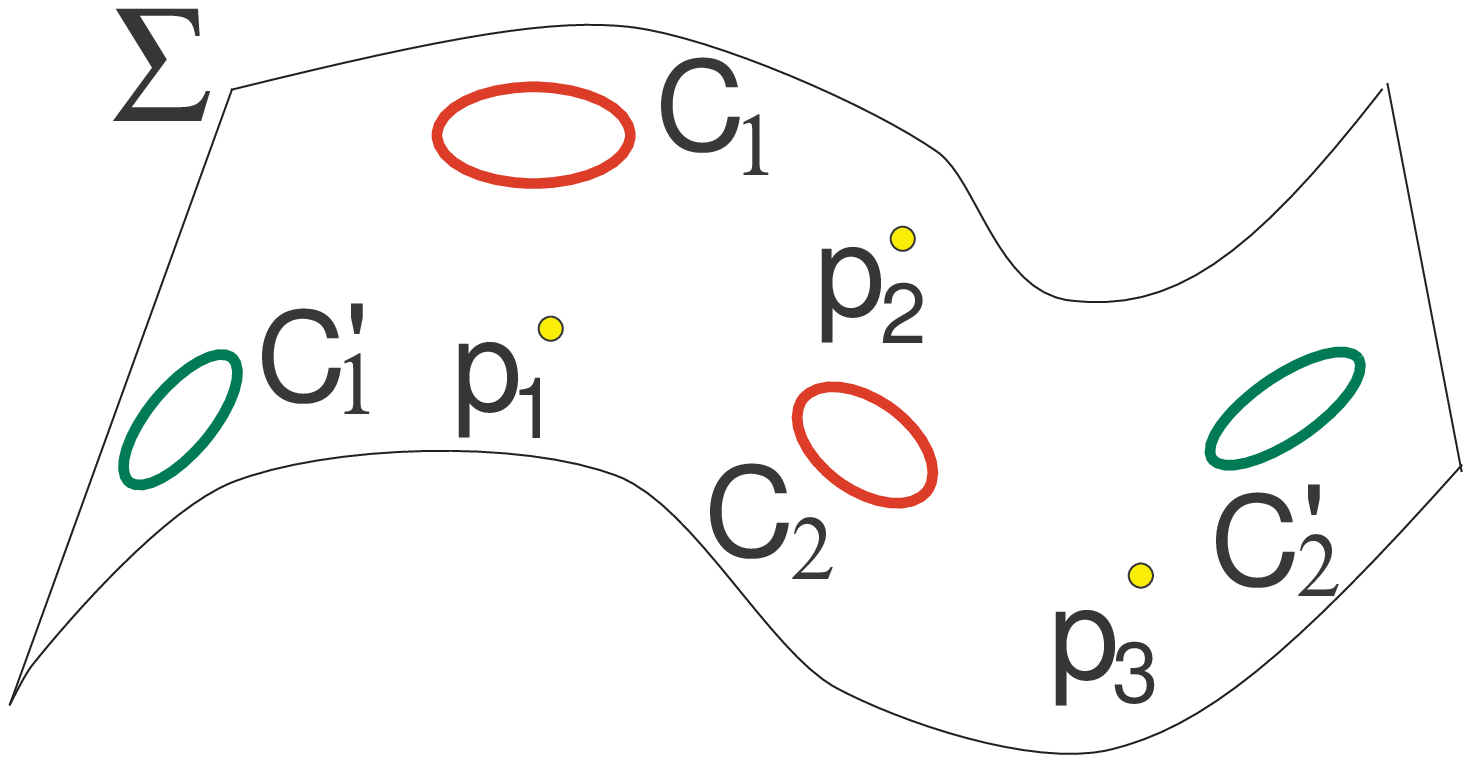,height=4cm,angle=0,trim=0 0 0 0}%
{Irreducible curves $C_k$, $C'_{k'}$ and points $p_i$ contained in
divisor $\Sigma$.
  \label{sigmas} }

An alternative way of thinking about the moduli space, at least for
sufficiently large $P$ and sufficiently small $N$ and $F$ (i.e.\
sufficiently polar states) is as follows (see also
\cite{Gaiotto:2005rp,Gaiotto:2006wm}). Write as in (\ref{Fdecomp})
$F=\frac{P}{2} + f^\| + \gamma + f^\bot$. Recall that supersymmetry
requires $F^{0,2}=(\gamma + f^\bot)^{0,2}=0$, which is equivalent to
the statement that $\gamma+f^{\bot}$ is Poincar\'e dual to a
collection of holomorphic 2-cycles on $\Sigma$. Note that this puts
restrictions on the divisor deformation moduli, since generically
the divisor will not contain curves other than those obtained by
intersecting other divisors (which correspond to $f^\|$). More
precisely we have
\begin{equation} \label{Fsplitsigmasigma}
 F = \iota_{\Sigma}^* S + [C]_{\Sigma} -
 [C']_{\Sigma}
\end{equation}
where $S\in \frac{P}{2} + H^2(X,\IZ)$, $\iota_{\Sigma}^* S =
\frac{P}{2}+f^\|$, $C$ and $C'$ are collections of holomorphic
curves, and $[\cdot]_{\Sigma}$ denotes the corresponding
(co)homology class on $\Sigma$. (Note that this formula suffers from
the ambiguity $C \to C+C'', C' \to C' +C''$, which is one of the
reasons why the picture developed here is rather heuristic.) Hence
we can build supersymmetric configurations by first picking a set of
points $p_i$, $i=1,\ldots,N$, and a collections of holomorphic
curves $C$, $C'$ in $X$, and require our divisor $\Sigma$ to contain
all of those (see fig.\ \ref{sigmas}). This is possible when the
number of points and curves (and their degrees) is sufficiently
small compared to the number of deformation moduli of $\Sigma$. For
example for the hyperplane $\sum_n a_n x_n=0$ in the Fermat Quintic
$Q:=\sum_n x_n^5=0$, requiring the curve $x_1=-x_2$, $x_3=-x_4$,
$x_5=0$ to lie in the hyperplane puts $a_1=a_2$, $a_3=a_4$, reducing
the moduli space from $\IC\IP^4$ to $\IC\IP^2$.

The adjunction formula for irreducible holomorphic curves $C$ on
$\Sigma$ gives $2 \, \chi_h(C) = -C^2 - K_\Sigma \cdot C$ where
$\chi_h$ is the holomorphic Euler characteristic, i.e.\ one minus
the genus of the curve. We can also write $K_\Sigma\cdot C = \int_C
P:= P \cdot [C]$ where the first intersection product is on $\Sigma$ and
the last on $X$. By additivity of the Euler characteristic, this
formula extends to collections of holomorphic curves. Using this,
the charges (\ref{q2val}) and (\ref{q0val}) can be computed as
\begin{eqnarray}
 q_A &=& D_A \cdot ([C] - [C'] + P \cdot S) \\
 q_0 &=& \frac{P^3 + c_2 \cdot P}{24} -N + \frac{1}{2} PS^2 + S \cdot ([C]-[C']) \\
 && -\chi_h(C) - \frac{P}{2} \cdot [C]
 - \chi_h(C') - \frac{P}{2} \cdot [C'] - [C]\cdot [C']
 \label{chargessigmasystem}
\end{eqnarray}
All intersection products are on $X$, except for the last term,
which is an intersection of two curves within $\Sigma$.

Now let us compare this to the charges of a bound state of the kind
described in the previous sections.

Start with a single D6 brane containing a BPS ``gas'' of D2- and
D0-branes, with D2-charge $-\beta_1 \in H_2(X,\IZ)$ where $\beta_1$
is an effective curve class,\footnote{In our conventions, D6-branes
form BPS states with \emph{anti}-D2 branes.} and D0-charge $n_1 \in
\IZ$. Now add D4-brane charge by turning on a flux $S_1$, giving
according to (\ref{chargeformula}) a total charge
\begin{eqnarray} \label{eSparcharge}
 \Gamma_1 &=& e^{S_1}(1-\beta_1 + n_1 \,
 \omega)(1+\frac{c_2(X)}{24})\\
 &=&\left(1,\,S_1,\,\frac{S_1^2}{2} -
 \beta_1 + \frac{c_2}{24},\,\frac{S_1^3}{6} - \beta_1 S_1 + \frac{c_2}{24}S_1 +
 n_1\right).
\end{eqnarray}
Do the same for a second D6-brane and take its charge conjugate, so
\begin{eqnarray} \label{eSparcharge2}
 \Gamma_2 &=& -e^{S_2}(1-\beta_2 + n_2 \, \omega)(1+\frac{c_2(X)}{24})\\
 &=&\left(-1,\,-S_2,\,-\frac{S_2^2}{2}
 + \beta_2 - \frac{c_2}{24},\,-\frac{S_2^3}{6}
 + \beta_2 S_2 - \frac{c_2}{24}S_2 - n_2 \right).
\end{eqnarray}
Defining
\begin{equation}
 \tilde{P}:=S_1-S_2, \qquad \tilde{S}:=\frac{S_1+S_2}{2},
\end{equation}
the total charge $\Gamma=\Gamma_1+\Gamma_2$ can be written as
\begin{equation}\label{totalddd}
 \Gamma = \left(0,\,\tilde{P},\,\beta_2-\beta_1
 + \tilde{P} \tilde{S},\,\frac{\tilde{P}^3+c_2 \tilde{P}}{24}
 + \frac{1}{2} \tilde{P} \tilde{S}^2 + \tilde{S} (\beta_2-\beta_1)
 -n_2 - \frac{\tilde{P}}{2} \beta_2
  + n_1 - \frac{\tilde{P}}{2} \beta_1 \right).
\end{equation}
So we see that if we identify
\begin{equation}\label{identifycharges}
 \tilde{P}=P, \quad \tilde{S}=S, \quad \beta_2=[C], \quad
 \beta_1=[C'], \quad n_2 = \chi_h(C)+N_2, \quad
 n_1=-\chi_h(C')-N_1
\end{equation}
with $N=N_1+N_2$, this almost matches exactly with
(\ref{chargessigmasystem}), including the correct quantization
condition on $S$.

This match is so good that we expect that generically $[C]\cdot
[C']=0$. This is certainly true for two generic homology classes in
$X$, and we will assume that for those classes sitting inside a
common holomorphic surface $\Sigma$ it is still generically true.
Granted this point, the
identifications are interpreted as follows:

 At large volume, rank 1 D6-D2-D0 bound
states are described by ideal sheaves $\CI$
\cite{Harvey:1996gc,MNOP1,MNOP2,Iqbal:2003ds} or their duals
$\CI^*$.\footnote{We do not mean the sheaf-theoretic dual here. If
we identify the objects in the category of topological B-branes with
the stable objects in the derived category of coherent sheaves then
we should take the derived dual. This will be a complex whose
cohomology is not supported in a single degree, and hence will not
be a sheaf. We thank Paul Aspinwall for pointing this out to us.}
More precisely $\CI$ corresponds to a collection of curves $C_{\CI}$
and points $\pi_{\CI}$, where the D2-charge is given by
$-\beta+c_2(X)/24$ and the D0-charge by $n$, where
\begin{equation} \label{idealsheafcharges}
 \beta=-{\rm ch}_2(\CI)=[C_{\CI}], \qquad
 n={\rm ch}_3(\CI)=\chi_h(C_{\CI} \cup
\pi_{\CI})=\chi_h(C_{\CI})+N_{\CI},
\end{equation}
where $N_{\CI}$ is the number of points in $\pi_{\CI}$ (counted with
multiplicities). Taking the dual  inverts the odd Chern characters,
so the D2-charge of $\CI^*$ is given by $-\beta+c_2(X)/24$ where
$\beta=[C_{\CI}]$ and the D0-charge is $n=-\chi_h(C_{\CI})-N_{\CI}$.
Hence the above expressions for the charges suggest we identify the
$\Gamma_1$ system with a D6-D4-D2-D0 bound state described as the
dual $\CI_1^*$ of an ideal sheaf $\CI_1$ ``shifted'' by a $U(1)$
flux $S_1$, and $\Gamma_2$ similarly as the anti-brane of a
D6-D4-D2-D0 bound state described as an ideal sheaf $\CI_2$, shifted
by $S_2$. Under this identification, we simply have
\begin{equation}
 C_{\CI_1} = C', \quad C_{\CI_2} = C, \quad
 N_{\CI_1}+N_{\CI_2} = N.
\end{equation}
Since the $N_\CI$ are nonnegative note that $n_1$ is bounded above, and not
below, while $n_2$ is bounded below, and not above. This will be important
in keeping certain signs straight in the derivation of the OSV formula.

Thus we arrive at the following heuristic picture for polar BPS
states: The curve
collections which are dual to  $\gamma+f^\bot$ in $\Sigma$
are  the remnants of gases of D2-branes inside  D6 and anti-D6 branes
with fluxes turned on. The D6-antiD6 condense producing a D4 brane
which has captured a gas of D2 and D0 branes.

We can further strengthen this picture by computing moduli degrees
of freedom. As explained in section \ref{sec:microDbrane},
 a bound
state of the two branes under consideration is expected to be a
cohomology class of a moduli space which is a  $\IC\IP^{k}$
fibration over the product of the moduli spaces of the two
constituent branes. Here $k+1$ equals the intersection product of
the constituents
\begin{equation} \label{kplus1}
 k+1 = \langle \Gamma_2,\Gamma_1 \rangle = \frac{P^3}{6} + \frac{P \cdot
 c_2(X)}{12}-P \cdot (\beta_1 + \beta_2) + n_1 - n_2.
\end{equation}
In particular in the case at hand, after freezing the curves and
points in the D6 and anti-D6 branes representing the D2-D0 gases, we
expect $k$ residual degrees of freedom, coming from light open
string modes stretching between the branes.

If the proposed picture is correct, at large radius, these $k$
residual degrees of freedom should correspond to the divisor moduli
that remain unfixed in the generic case after requiring the curve
collections $C$ and $C'$ and the set of $N$ points $p_i$ to be
contained in it. To verify this, rewrite (\ref{kplus1}) using the
above identifications as
\begin{eqnarray}\label{indexworks}
 k+1 &=& \frac{P^3}{6} + \frac{P \cdot
 c_2(X)}{12}-P \cdot C - \chi_h(C) -
 P \cdot C' - \chi_h(C') - N \\
 &=& \int_X e^P \, {\rm Td} \, X - \int_{C} e^P \, {\rm Td} \, C
 - \int_{C'} e^P \, {\rm Td} \, C' - N. \label{kplus1Todd}
\end{eqnarray}
We claim that for $P$ sufficiently large, this agrees exactly with
the generic number of deformations of a divisor constrained to
contain $N$ points and the curves $C$ and $C'$. As a simple first
check, note that when $C=C'=0$, $N=0$, i.e.\ the pure D4 with at
most flux pulled back from $H^2(X)$ turned on, this formula
reproduces precisely the dimension $I_P-1$ of the linear system $P$.
Furthermore, if we think of the divisor for example as a
hypersurface given by some homogeneous polynomial equation, then it
is clear that if we fix $N$ generic points in $X$ and require the
divisor to pass through it, this will give $N$ linear constraints on
the polynomial coefficients and thus generically reduce the residual
divisor moduli space from $\IC\IP^{I_P-1}$ to $\IC\IP^{I_P-N-1}$.

 We now give a proof for
the general case for $P$ sufficiently ample. The basic ideas are
$(i)$ for $P$ sufficiently ample, we can use index formulas to
compute the actual number of deformations, and $(ii)$ the first term
in (\ref{kplus1Todd}) is the index counting the number of
holomorphic sections of the line bundle describing $P$, and the
second and third terms are the indices counting the number of those
sections which when restricted to $C$ resp.\ $C'$ are nontrivial.
Subtracting these terms from the first one thus gives the number of
sections of the divisor line bundle which are zero on $C$ and $C'$,
i.e.\ one plus the number of divisor deformations fixing $C$ and
$C'$.

More precisely, this goes as follows.\footnote{We thank E.
Diaconescu and T. Pantev for helpful discussoins about this.} Define
the ideal sheaf:
\begin{equation}
0 \rightarrow I_C \rightarrow \CO_X \rightarrow \CO_C \rightarrow 0
\end{equation}
Our problem is to compute the dimension of $H^0(\CO(P)\otimes I_C)$.

Tensor the exact sequence with $\CO(P)$. This preserves exact
sequences since $\CO(P)$ is a line bundle. We write the
corresponding long exact sequence
\begin{equation}
0 \rightarrow H^0(I_C\otimes \CO(P)) \rightarrow H^0(\CO_X\otimes
\CO(P)) \rightarrow H^0(\CO_C\otimes \CO(P))\rightarrow
H^1(I_C\otimes \CO(P))\rightarrow 0
\end{equation}

For $P$ sufficiently ample $H^1(I_C\otimes \CO(P))=0$ and moreover
$h^1(\CO_C(P)) =0$, but now we can use Riemann-Roch to compute
\begin{eqnarray}
h^0(\CO_C(P)) - h^1(\CO_C(P)) &=& {\rm deg}(P\vert_C) - g(C) +1\\
& = & \int_C e^P {\rm Td}(TC)
\end{eqnarray}
Now compare with equation (\ref{indexworks}).

A closely related, but alternative argument proceeds as follows. For
concreteness let us take the example of the quintic in $\ICP^4$. Let
$W$ be the space of homogeneous polynomials in $X_0, X_1, X_2, X_3,
X_4$, and $W_d$ those of degree $d$. The quintic Calabi-Yau is given
by the polynomial equation $Q=0$, with $Q \in W_5$. Let $\langle Q
\rangle$
 be the ideal generated by $Q$ and define $W' := W / \langle Q \rangle$, and let
 $W'_d$ be the restriction of $W'$ to degree $d$ polynomials. Then
 $W'_d$ (projectivized) can be identified with the moduli
 space of divisors of degree $d$ on the quintic.

Fix a curve $C$ in the quintic described as the vanishing locus of
some homogeneous polynomial ideal $I(C)$ (which includes $Q$). Then
the moduli space of degree $d$ divisors on the quintic which contain
$C$
 can be identified with the (projectivization of ) $I'(C)_d$, the degree $d$ part of
 $I'(C) := I(C) / \langle Q \rangle$.
  So we are interested in
 computing $\dim I'(C)_d$. This is almost directly given
 by the Hilbert polynomial of $C$. Define $M(C) := W' / I'(C) = W / I(C)$, i.e.
 the homogeneous polynomial module associated
 to $C$. Then the Hilbert function of $C$ is by definition $f_h(d) := \dim M(C)_d$,
 and the Hilbert-Serre theorem says that this becomes a
 polynomial $p_h(d)$ for sufficiently large $d$. Moreover, that
 polynomial can be computed from the index theorem.

Since by construction $\dim M(C)_d + \dim I'(C)_d = \dim W'_d$, this
gives the expression for $\dim I'(C)_d$:
\begin{equation}
\dim I'(C)_d = \dim W'_d - p_h(d)
\end{equation}
for sufficiently large $d$. Now  we have
\begin{equation}
p_h(d) = \int_C e^{d H} Td(C) = \int_C (1+d H) (1 + c_1(C)/2) = d H
\cdot C + \chi_h(C) = P \cdot C + \chi_h(C)
\end{equation}
where $H$ is the hyperplane class and $\chi_h(C)$ the holomorphic
Euler characteristic of $C$. Therefore
\begin{equation}
\dim I'(C)_d = I_P - P \cdot C - \chi_h(C)
\end{equation}
in agreement with the above general proof.

These observations give rather strong evidence for the proposed
correspondence, although considerably more work would be needed to
make things more precise. There is some ambiguity in the
identifications in the two pictures, and constructing an exact map
between moduli spaces is presumably too much to hope for. In
particular we have not analyzed situations in which points or curves
coincide so the $\ICP^k$ fiber dimension jumps. It appears that here
the naive geometrical D4-D2-D0 picture and the D6-anti-D6 bound
state picture start to differ, with the latter apparently giving
some sort of regularization and stratification of these singular
loci. Indeed the considerations of section \ref{sec:quivfact}
strongly suggest that in the D6-anti-D6 picture the relevant
$\ICP^k$ fibrations are always regular for polar states.

We will not attempt to make this map more precise here, but instead
proceed by taking the physical D6-anti-D6 picture as a starting
point for computing the polar degeneracies. The degree to which the
heuristic picture sketched above is accurate will therefore not be
essential for the remainder of this paper.

\section{Wall-crossing formulae and factorization of indices}
\label{sec:quivfact}

In this section we derive wall crossing and factorization formulae
for indices, which among other applications will lead to (a refined
version of) (\ref{Omfact}) and eventually in section
\ref{sec:counting} to the factorization $\CZ_{\rm top} \sim \CZ_{\rm
top} \overline{\CZ_{\rm top}}$.

\subsection{Physical derivation}
\label{sec:physargfact}

Let $\CH'(\Gamma)_{t_\infty}$ be the (reduced) Hilbert space of BPS
states of charge $\Gamma$ for background moduli $t_\infty$. Then
\footnote{Elsewhere in the paper we also use the notation
$\CH'(\Gamma;t_\infty)$ and $\Omega(\Gamma;t_\infty) $.}
\begin{equation} \label{BPSindexoncemore}
 \Omega(\Gamma)|_{t_\infty} := {\rm Tr}_{\CH'(\Gamma)_{t_\infty}} (-1)^{2 J_3'}
\end{equation}
with $J_3'$ the angular momentum with center of mass degrees of
freedom factored out.

In the four dimensional supergravity picture, the index
(\ref{BPSindexoncemore}) can get contributions from several distinct
multicentered configurations, with different constituent charges
summing up to the same total charge $\Gamma$, or equivalently from
several different topologically distinct attractor flow trees. Apart
from the trivial flow tree (i.e.\ the single flow), all of these
will decay when the initial flow tree point $t_{\infty}$ passes
through the wall of marginal stability on which the first split
$\Gamma \to \Gamma_1 + \Gamma_2$ of that tree occurs (this will be a
different wall for every tree in general). Therefore as soon as
there are nontrivial tree contributions to the index, the index can
be expected to jump at these walls of marginal stability.

To derive the amount by which the index jumps at a $\Gamma \to
\Gamma_1 + \Gamma_2$ MS wall, we will first assume $\Gamma_1$ and
$\Gamma_2$ are both primitive. In that case all states decaying at
this wall will necessarily look like two clusters of bound particles
of charge $\Gamma_1$ resp.\ $\Gamma_2$, which get infinitely far
separated from each other when the wall is approached (recall eq.
(\ref{eqsep})). Denote the part of $\CH'(\Gamma)_{t_{\infty}}$
corresponding to these nearly decaying states by $\CH'(\Gamma \to
\Gamma_1 + \Gamma_2)_{t_{\infty}}$, where we let $t_{\infty} \to
t_{\rm ms}$, $t_{\rm ms}$ being a point on the marginal stability
wall under consideration. One expects this Hilbert space to
factorize as
\begin{equation} \label{Hilbfact}
 \CH'(\Gamma \to \Gamma_1 + \Gamma_2)_{t_{\rm ms}}
 = \mbox{$\left( {\bf \frac{|I_{12}|-1}{2}} \right)$}
 \otimes \CH'(\Gamma_1)_{t_{\rm ms}}
 \otimes \CH'(\Gamma_2)_{t_{\rm ms}}.
\end{equation}
The first factor comes from the quantization of the centers of mass
of the two clusters and their associated fermionic degrees of
freedom, which as reviewed in section \ref{sec:reviewtrees} yields a
spin $J_3'=\frac{|I_{12}|-1}{2}$ multiplet, where $I_{12} \equiv
\langle \Gamma_1,\Gamma_2 \rangle$. Thus, one expects a jump in the
index given by
\begin{equation} \label{wcf}
 \Delta \Omega|_{t_{\rm ms}} = (-1)^{I_{12}-1} \, |I_{12}| \, \Omega(\Gamma_1)|_{t_{\rm
 ms}} \, \Omega(\Gamma_2)|_{t_{\rm ms}}
\end{equation}
when going from the unstable to the stable side of the marginal
stability wall.\footnote{Of course, (\ref{Hilbfact}) implies
something stronger than (\ref{wcf}).  We could for example state an
analogous wall-crossing formula for the full character ${\rm Tr}
y^{2J_3'}$ implying a wall-crossing formula for the Hodge
polynomials of the relevant moduli spaces.}

The main physical input that went into this derivation is the
factorization of Hilbert spaces (\ref{Hilbfact}) for infinitely
separated clusters. Although plausible, this is not completely
obvious, since one could imagine interactions e.g.\ between the spin
of one cluster and the magnetic field produced by the other cluster,
which could spoil supersymmetry by a tiny but nonzero bit. The spin
of the clusters depends on the relative positions of the centers
(see (\ref{spinformula})), so translated to these degrees of freedom
one should check if there are non-infinitesimal effects on the
relative BPS position constraints of the centers within one cluster,
coming from the presence of the second cluster.  If the
integrability constraints admit solutions with a cluster of  centers
$\vec x_{\alpha}$ going to infinity, while other centers $\vec
x'_{i}$ remain finite then clearly the effect on the remaining
centers $\vec x'_{i}$ in (\ref{centerconstraints}) is negligible and
amounts effectively merely to an infinitesimal shift of the constant
term on the right hand side of the constraint equations. (The cases
where this constant term is zero are nongeneric and can be
eliminated by slightly perturbing $t_\infty$, which for the sake of
this argument we are free to choose anywhere as long as $t_\infty$
stays very near the wall of marginal stability on the stable side).
To strengthen our confidence in these arguments, we will give
several mathematical tests of the wall crossing formula in the
following subsections.

The wall crossing formula can be used to derive a refined version of
(\ref{Omfact}). Fix some $t_{\infty}=t_i$ and consider all splits
$\Gamma \to \Gamma_1 + \Gamma_2$ encountered along the single
$\Gamma$ attractor flow starting at $t=t_i$ and ending at $t=t_f$,
where $t_f$ is either the attractor point or a zero of $Z(\Gamma)$.
Note that by the time this endpoint is reached, all configurations
contributing to the index $\Omega(\Gamma;t_{\infty})$ that could
decay, have decayed; there are no nontrivial trees left at this
point. Repeating the wall crossing formula (\ref{wcf}) for each jump
encountered along the attractor flow gives the formula
\begin{equation} \label{BPSindfact2}
 \Omega(\Gamma)|_{t_i} =  \Omega(\Gamma)|_{t_f} + \sum_{\Gamma \to \Gamma_1 + \Gamma_2}
 (-1)^{\langle \Gamma_1,\Gamma_2
 \rangle-1} |\langle
 \Gamma_1,\Gamma_2
 \rangle| \,\, \Omega(\Gamma_1)|_{t_{\rm ms}(\Gamma_1,\Gamma_2,t_i)} \,\,
 \Omega(\Gamma_2)|_{t_{\rm ms}(\Gamma_1,\Gamma_2,t_i)}
\end{equation}
where the sum is over all $\Gamma \to \Gamma_1 + \Gamma_2$
splittings along the attractor flow and $t_{\rm
ms}(\Gamma_1,\Gamma_2,t_i)$ is the point where the flow crosses the
corresponding $\Gamma \to \Gamma_1 + \Gamma_2$ marginal stability
wall. When the final point corresponds to a zero, as is the case for
polar D4-D2-D0 states, we moreover have $\Omega(\Gamma)|_{t_f} = 0$,
and all contributions have a factorized form.

Iteratively repeating this for each of the
$\Omega(\Gamma_i)|_{t_{\rm ms}}$ eventually gives the expression
\begin{equation} \label{treerecind}
 \Omega(\Gamma)|_{t_{\infty}} = \sum_{T \in {\cal T}(\Gamma,t_{\infty})}
 \prod_{\Gamma_a \to \Gamma_b + \Gamma_c \in {\rm Vert}(T)}
 (-1)^{\langle \Gamma_b,\Gamma_c \rangle-1} |\langle
 \Gamma_b,\Gamma_c \rangle| \,
 \prod_{\Gamma_i \in {\rm Term}(T)}
 \Omega(\Gamma_i,t_*(\Gamma_i))
\end{equation}
where ${\cal T}(\Gamma,t_{\infty})$ is the set of all attractor flow
trees of total charge $\Gamma$ starting at  $t_{\infty}$, ${\rm
Vert}(T)$ is the set of vertices of the flow tree $T$, characterized
as splits $\Gamma_a \to \Gamma_b + \Gamma_c$, ${\rm Term}(T)$ is the
set of terminal charges of the flow tree $T$, and $t_*(\Gamma_i)$
the attractor point of $\Gamma_i$. Thus we see that flow trees give
a canonical way of reducing indices in general backgrounds to
irreducible\footnote{irreducible in the sense that they correspond
to states which cannot be made to decay --- there might be more
refined reductions which further factorize even these irreducible
indices.} indices $\Omega(\Gamma_i,t_*(\Gamma_i))$ associated to
black holes or simple particles.

In the wall crossing formula (\ref{wcf}) and the subsequent formulae
we have assumed that all splits $\Gamma \to \Gamma_1 + \Gamma_2$ are
primitive, i.e.\ no integral $\Gamma'_1$ and integer $N_1 > 1$ exist
such that $\Gamma_1 = N_1 \Gamma_1'$, and similarly for $\Gamma_2$.
In general this need not be the case. It is possible to extend the
wall crossing formula (\ref{wcf}) to some nonprimitive cases as
well. This is done most efficiently by using generating functions;
examples will be analyzed in detail in the section
\ref{sec:D6D4D2D0}, but for completeness we give a more general wall
crossing formula here already, for arbitrary splits $\Gamma \to
\Gamma_1 + N \Gamma_2$, $N \in \IZ^+$ (which is a nonprimitive split
when $N>1$):
\begin{equation}\label{degenwc}
\Omega(\Gamma_1)|_{t_{\rm ms}}+  \sum_{N>0} \Delta \Omega(\Gamma_1 +
N \Gamma_2)|_{t_{\rm ms}} \, q^N =
 \Omega(\Gamma_1)|_{t_{\rm ms}} \prod_{k>0} \biggl( 1-(-1)^{k \langle \Gamma_1,\Gamma_2
 \rangle}\,  q^k \biggr)^{k |\langle \Gamma_1,\Gamma_2 \rangle| \, \Omega(k \Gamma_2)|_{t_{\rm
 ms}}}
\end{equation}
where $\Delta \Omega$ denotes the index jump at the appropriate
marginal stability point $t_{\rm ms}(\Gamma_1,\Gamma_2,t_{\infty})$,
going from unstable to stable side. These splits correspond to
``halo'' states, consisting of $N$ $\Gamma_2$ particles moving on a
sphere around $\Gamma_1$. We will see several special cases in
section \ref{sec:D6D4D2D0}, after which it will be clear that this
formula is the correct generalization. Note that it reduces to
(\ref{wcf}) for $N=1$. It would be interesting to generalize this
formula further to splits $\Gamma \to N_1 \Gamma_1 + N_2 \Gamma_2$,
but we expect this to be significantly more complicated. In this
case   the configuration space will be much more complicated.
Moreover, from  our arguments above $\Delta\Omega(N_1\Gamma_1 + N_2
\Gamma_2)$ is related to the Euler character of a quiver with 2
nodes with dimension vector $(N_1,N_2)$ and $k=\langle
\Gamma_1,\Gamma_2\rangle$ arrows. However, the known expressions for
these Euler characters are very complicated \cite{reineke}.

We will however primarily use (\ref{BPSindfact2}) to factorize the
polar part of the D4-partition function, and will argue that for the
purpose of deriving the OSV conjecture it is sufficient to restrict
to the contribution from splits in two clusters with a single D6 and
a single anti-D6 brane charge, which are of course automatically
primitive.

 Finally, note that we could have tried to derive the wall crossing
formula microscopically from (\ref{chifact}) (adding the proper
signs obtained from the identification of $J_3'$ with Lefshetz spin
as explained above (\ref{indextoeuler})). However, at least to be
able to use this in a straightforward fashion, this would have
required us to assume that whenever a D-brane is close to decaying
into two branes, its connected moduli space component $\CM$ has the
structure of a $\ICP^{|\langle \Gamma_1,\Gamma_2 \rangle|-1}$
fibration over the product of the moduli spaces $\CM_1$ and $\CM_2$
of the consitutent branes, \emph{without} any degenerations of the
fiber. The latter is  not clear \emph{a priori}. Turning things
around, the physical arguments given above give a prediction that
the fibration will indeed be regular in these cases, or at least
that this can be effectively assumed for the purpose of computing
the jump of the index.

In the following we will test our wall-crossing formula
microscopically, and we will see that indeed this regular fibration
structure arises in association to decaying states, often in rather
nontrivial ways.

\subsection{Mathematical tests and applications} \label{sec:mathtests}

We have verified our wall crossing formulae, and the index
factorizations derived from it, both microscopically by comparing to
large radius geometrical results as well as by studying  examples of
quiver moduli spaces. The latter often are under good mathematical
control \cite{reineke}, and their relation to multicentered
configurations is well understood in a number of cases
\cite{Denef:2002ru}.

The simplest example is a pure (very ample) D4 of charge $P$, which
as we saw in the previous sections corresponds to a bound state of a
D6 and an anti-D6 with suitable fluxes turned on, with corresponding
charges $\Gamma_1 = e^{S_1}(1+c_2/24)$, $\Gamma_2 = -
e^{S_2}(1+c_2/24)$ such that $P=S_1-S_2$. The intersection product
between the two constituents equals $I_P = P^3/6 + c_2 \cdot P/12$.
For a single D6 brane with flux   the attractor point lies on the
boundary of moduli space. However,   the low energy gauge theory is
free Maxwell theory and hence,  on a Calabi-Yau $X$ with proper
$SU(3)$ holonomy, there is a unique ground state of the Maxwell
theory in a fixed flux sector. \footnote{In this paper we have
ignored torsion in the cohomology groups. However, at this point
torsion in $H^2(X;\IZ)$ plays an interesting role. In this case one
cannot simultaneously specify the electric and magnetic flux sectors
on the D6 brane, and in fact, the ground states of the theory form a
representation of the Heisenberg group extension of
$H^2_{tors}(X;\IZ)\times H^5_{tors}(X;\IZ)$ defined by the torsion
pairing \cite{Freed:2006ya,Freed:2006yc}. Thus, it is more
appropriate to take $\Omega(\Gamma_i;t_*(\Gamma_i))=\vert
H^2_{tors}(X;\IZ)\vert$. Since different attractor flow trees
terminate on different numbers of pure six branes the torsion
effects will modify the indices in interesting ways. We have not
systematically investigated these consequences of nonzero torsion. }
Therefore, $\CH'(\Gamma;t_*(\Gamma))$ is one-dimensional, and hence
$\Omega(\Gamma_i;t_*(\Gamma_i))=1$. We will assume $\Gamma \to
\Gamma_1 + \Gamma_2$ is the only flow tree (an assertion very well
supported by our numerical and analytical searches), and hence
equation (\ref{treerecind}) immediately gives $\Omega(P;t_{\infty})
= (-1)^{I_P-1} I_P$. This is in exact agreement with the microscopic
index computed as the euler characteristic of the linear system
corresponding to the divisor $P$, which is $\ICP^{I_P-1}$.  Note
that this is also the moduli space of a two-node quiver with $I_P$
arrows and dimension vector $(1,1)$, in accordance with the
discussion in section \ref{sec:quivdescription}, and in particular
(\ref{genquivermodulispace}).

Further tests along these lines can be extracted from the discussion
towards the end of section \ref{sec:microrel}.

In the following subsections we will consider a number of more
complicated examples, some of which are of independent interest.
Finally, the results for halo degeneracies we will obtain in the
next sections can also be considered as further tests of these
ideas.

\subsubsection{Four node quiver without closed loops}
\label{sec:fournodequiv}

\EPSFIGURE{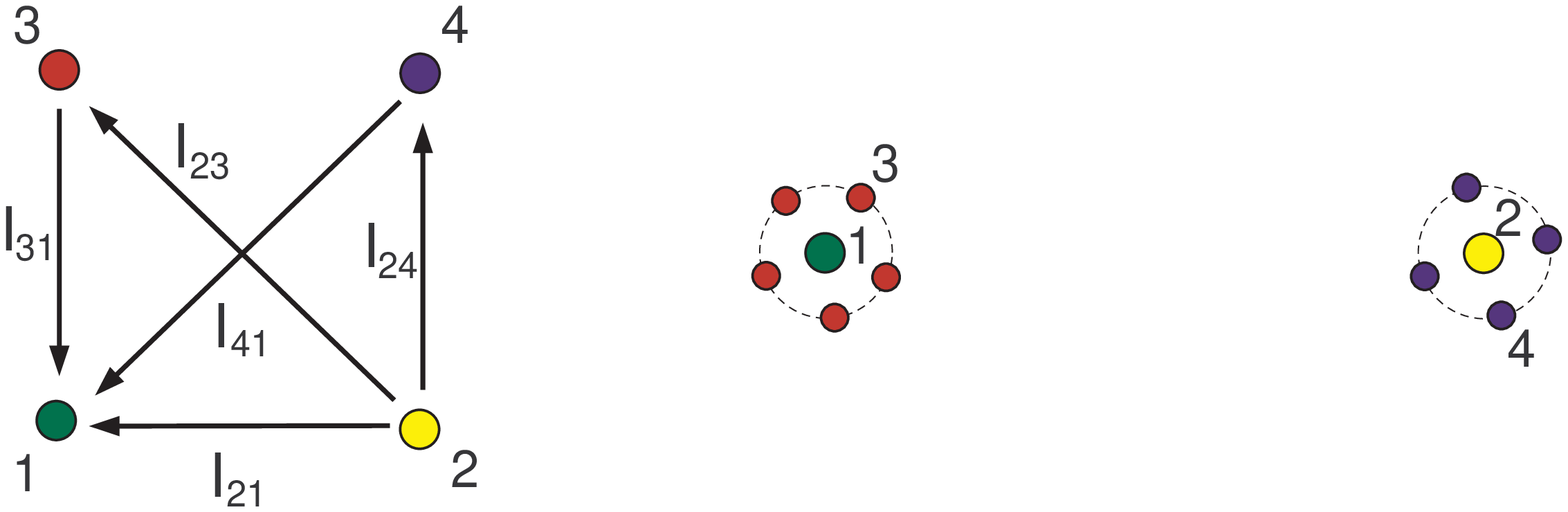,height=5cm,angle=0,trim=0 0 0 0}%
{{\bf Left:} Four node quiver without closed loops. The indicated
arrows are $I_{ij}$-fold degenerate, with $I_{ij}\geq 0$. The
dimension vector is taken to be $(1,1,d_3,d_4)$, with $d_3,d_4\geq
1$. {\bf Right:} Corresponding splitting multicentered configuration
when $\vartheta_1 + d_3\vartheta_3$ is approaching zero.
\label{factorquiver}}

To verify our physical arguments for the absence of long distance
spin-spin interactions spoiling factorization, we consider a system
described by the quiver of fig.\ \ref{factorquiver}, close to a
locus where the two nodes on the left hand side split off from those
on the right hand side.  More precisely we will go to a locus of FI
parameter space where the multicentered solutions of
(\ref{quivposeq}) split in these two clusters and we thus expect the
index to factorize accordingly. We  test this by showing that the
quiver moduli space (\ref{genquivermodulispace}) is a $\ICP^{I-1}$
fibration over the product of the moduli spaces of the two
subsectors, with $I$ the intersection product between the two
clusters, and that its cohomology factorizes as physically expected.
As discussed at the end of section \ref{sec:physargfact}, this is
sufficient to reproduce the wall crossing formula for the index.
However establishing this regular fibration structure turns out to
be rather nontrivial.

We will not try to give an actual physical or flow tree realization
of this quiver, which is not necessary for the kind of comparison we
are trying to make here.

The morphisms (or bosonic stretched open string modes) are described
by
\begin{eqnarray}
\phi_{21} \in &  &   \IC^{I_{21}}\\
\phi_{31} \in &  & {\rm Mat}(1,d_3) \times \IC^{I_{31}} \\
\phi_{23}  \in & & {\rm Mat}(d_3,1) \times \IC^{I_{23}}\\
\phi_{41} \in &  & {\rm Mat}(1,d_4) \times \IC^{I_{41}} \\
\phi_{24}  \in & & {\rm Mat}(d_4,1) \times \IC^{I_{24}}
\end{eqnarray}
denote $\phi_{31}^{\alpha,j}$ with $\alpha=1,\dots, d_3$,
$j=1,\dots, I_{31}$ and $\phi_{23}^{j,\alpha}$ with $j=1,\dots,
I_{23}$, and similarly for $\phi_{41},\phi_{24}$. The D-term
equations are given by (\ref{quivDtermeq}):
\begin{eqnarray}
  -\vert \phi_{21}\vert^2 - \vert \phi_{31}\vert^2 - \vert
  \phi_{41}\vert^2 & = & \vartheta_1\\
  \vert\phi_{21}\vert^2 + \vert \phi_{23}\vert^2 + \vert
  \phi_{24}\vert^2 & = & \vartheta_2 \\
\sum_{j=1}^{I_{31}} \phi_{31}^{\alpha,j}
(\phi_{31}^{\beta,j})^*-\sum_{j=1}^{I_{23}} (\phi_{23}^{j,
\alpha})^* \phi_{23}^{j,\beta } &= & \vartheta_3 \delta^{\alpha,\beta}\\
\sum_{j=1}^{I_{41}} \phi_{41}^{\alpha,j}
(\phi_{41}^{\beta,j})^*-\sum_{j=1}^{I_{24}} (\phi_{24}^{j,
\alpha})^* \phi_{24}^{j,\beta } &=& \vartheta_4
\delta^{\alpha,\beta}
\end{eqnarray}
with $\vartheta_i$ as in (\ref{varthetai}) and as usual $\vartheta_1
+ \vartheta_2 + d_3 \vartheta_3 + d_4 \vartheta_4 =0$. The
corresponding supersymmetric particle configuration constraints
(\ref{quivposeq}) are
\begin{eqnarray} \label{posconstrfournode}
- \frac{I_{21}}{\vd{1}{2}} - \sum_{\beta=1}^{d_3}
\frac{I_{31}}{\vd{1}{3\beta}} - \sum_{\beta=1}^{d_4} \frac{I_{41}}{\vd{1}{4\beta}}& = &  \vartheta_1 \\
 \frac{I_{21}}{\vd{2}{1}}  +\sum_{\beta=1}^{d_3} \frac{I_{23}}{\vd{2}{3\beta}}
 + \sum_{\beta=1}^{d_4} \frac{I_{24}}{\vd{2}{4 \beta}} & = &  \vartheta_2 \\
\frac{I_{31}}{\vd{3\alpha}{1}}  - \frac{I_{23}}{\vd{3}{2\alpha}} & =
& \vartheta_3\\
\frac{I_{41}}{\vd{4\alpha}{1}}  - \frac{I_{24}}{\vd{4\alpha}{2}} & =
& \vartheta_4,
\end{eqnarray}
which as mentioned in section (\ref{sec:quivdescription}) coincides
with the supergravity position constraints (\ref{sugraposeq}) in the
regime of validity of the quiver quantum mechanics, $\theta_i \ll
1$, since $\theta_i \approx \vartheta_i$ in this regime.

 Now, we want $\vec x_1, \vec x_{3,\alpha} \to \infty$ with $|\vec
x_1-\vec x_{3,\alpha}|$ held finite. Therefore, in the quiver
picture we should send $\vartheta_1+ d_3 \vartheta_3 \to 0$, holding
$\vartheta_1, \vartheta_3, \vartheta_2, \vartheta_4 $ all nonzero.
The system then splits in two clusters with charges $\Gamma_1+d_3
\Gamma_3$ and $\Gamma_2 + d_4\Gamma_4$, respectively, with mutual
intersection product
\begin{equation}\label{clusterip}
I = \langle \Gamma_2 + d_4 \Gamma_4, \Gamma_1 + d_3 \Gamma_3\rangle
= I_{21} + d_3 I_{23} + d_4 I_{41}.
\end{equation}
Clearly we must have $\vartheta_1<0$, and therefore $\vartheta_3>0$.
Similarly, since $\vartheta_1+d_3\vartheta_3 = -
(\vartheta_2+d_4\vartheta_4)$ and since $\vartheta_2>0$ we must have
$\vartheta_4<0$.

It follows from the third and fourth D-term  equations that there is
a well-defined projection to a product of  Grassmannians:
$[\phi_{31}^{\alpha j}] \in Gr(d_3, I_{31})$ and
$[\phi_{24}^{j\alpha}]\in Gr(d_4, I_{24})$. This leaves $\phi_{23}$
and $\phi_{41}$ undetermined, and the remaining equations determines
the fiber of the map to be a complex projective space so that the
moduli space is a smooth fibration \footnote{We omit many details in
the argument here.}
\begin{equation}
\IC P^{I_{21}+ d_3 I_{23} + d_4 I_{41} -1} \to\CM\to Gr(d_3,I_{31})
\times Gr(d_4, I_{24}).
\end{equation}
Note that the Grassmannians are the moduli spaces $\CM_1$, $\CM_2$
of the two 2-node sub-quivers in which our 4-node quiver splits.
Restricting the gauge invariant form $d\phi_{21}\wedge d
\overline{\phi_{21}}+ d\phi_{23}\wedge d
\overline{\phi_{23}}+d\phi_{41}\wedge d \overline{\phi_{41}}$ to the
fibers gives a generator of the cohomology of the fibers, so by the
Leray-Hirsch theorem the cohomology factorizes:
\begin{equation}\label{factorcoho}
H^*(\CM) = H^*(\IC P^{I_{21}+ d_3 I_{23} + d_4 I_{41} -1})\otimes
H^*(\CM_1) \otimes H^*(\CM_2)
\end{equation}
Comparing with (\ref{clusterip}) we see that   the factorization
(\ref{factorcoho}) is precisely that predicted by the physics, and
rather nontrivially so.

\subsubsection{A D6-D2-D0 as a 3 centered
$D6-D6-\overline{D6}$ bound state} \label{sec:D6D2D0asthreenode}

\EPSFIGURE{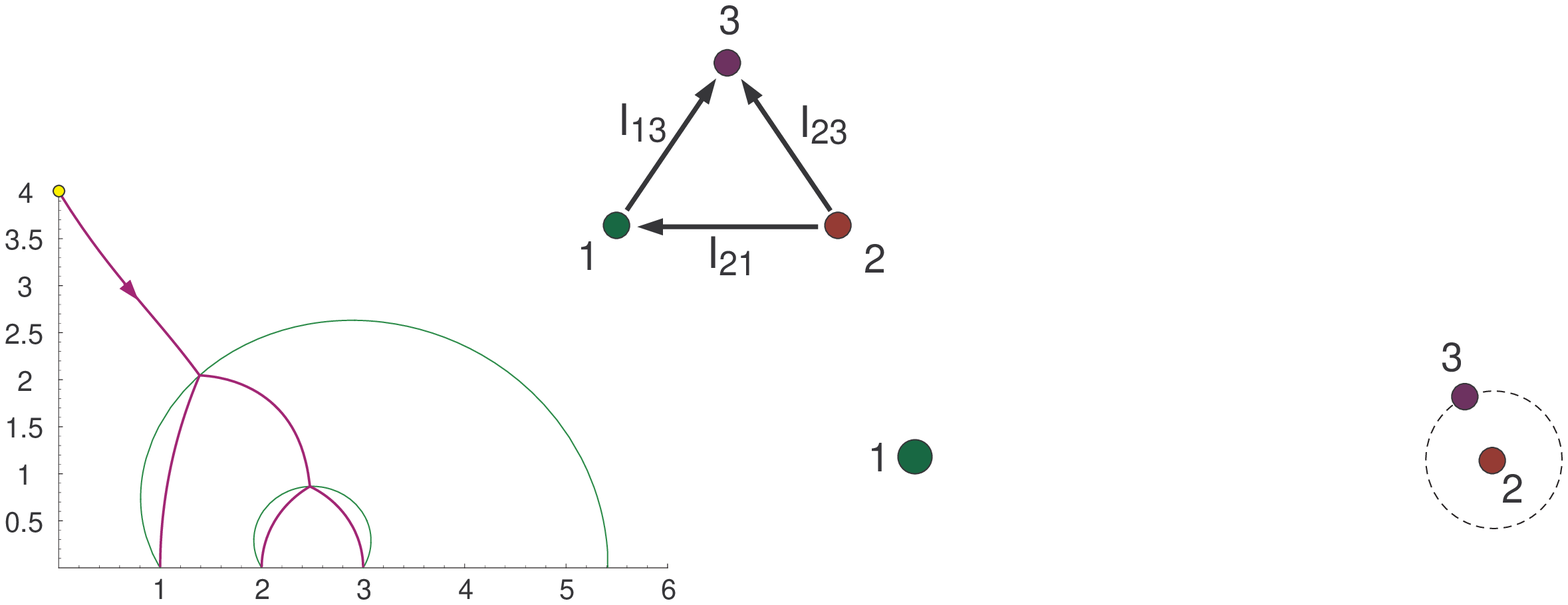,height=6cm,angle=0,trim=0 0 0 0}%
{{\bf Left:} Flow tree corresponding to the D6-D2-D0 system
described in the text, in the 1-modulus case with $U=D_1$, $V=2 \,
D_1$. The attractor points of $\Gamma_1,\Gamma_2,\Gamma_3$ are at
$x=1,2,3$ respectively. {\bf Up:} Corresponding quiver formally
associated to   this system, but in fact \emph{not} describing the
system. {\bf Right:} Corresponding splitting multicentered
Sun-Earth-Moon type configuration when the MS wall $\arg Z_1 = \arg
(Z_2 + Z_3)$ is approached.
  \label{D6D2D0triple} }

Next we give a 3-centered example with an actual flow tree
realization and a microscopic description as a geometric D-brane in
the IIA large radius limit. Consider a bound state of the following
three charges
\begin{eqnarray} \label{threenodequivcharges}
 \Gamma_1 = e^{U}(1+\frac{c_2}{24}), \quad
 \Gamma_2 = e^V(1+\frac{c_2}{24}), \quad
 \Gamma_3 = -e^{U+V}(1+\frac{c_2}{24})
\end{eqnarray}
with $U,V,V-U$ positive divisors (i.e.\ inside the K\"ahler cone).
The total charge is
\begin{eqnarray} \label{totalchargeD6D2D0triple}
 \Gamma =  1 \, - \, UV +
 \frac{c_2}{24} \, - \, \frac{1}{2}(UV^2+U^2V)
\end{eqnarray}
so this is a D6-D2-D0 bound state. Denoting $I_{ij} \equiv \langle
\Gamma_i,\Gamma_j \rangle$, we have
\begin{eqnarray}
 I_{21} = \frac{(V-U)^3}{6} + \frac{c_2 \cdot (V-U)}{12}, \quad I_{23} =
 \frac{U^3}{6} + \frac{c_2 \cdot U}{12},
 \quad I_{13} = \frac{V^3}{6} + \frac{c_2 \cdot V}{12}.
\end{eqnarray}
Because $U$, $V$ and $V-U$ are all positive, all of these
intersection numbers are positive. An example of a corresponding
attractor flow tree in the one modulus case is shown in fig.\
\ref{D6D2D0triple}, as well as the quiver encoding the intersection
products and the corresponding multicentered configuration when
approaching a $(1,2+3)$ line of marginal stability. Crucial is that
the sequence of splits is $(123) \to (1,23) \to (1,2,3)$. At least
in the one modulus case, it can be checked that this is the only
possible flow tree for the given charges starting from large $\Im \,
t$. Let us assume this is true in the general case as well.

Our factorization arguments immediately yield the following index of
BPS states associated to this flow tree:
\begin{equation} \label{indexD6D2D0triple}
 \Omega = (-1)^{I_{21}+I_{23}+I_{13}} \, |I_{13}-I_{21}| \, I_{23}
  = (-1)^{I_U + I_V + I_{V-U}} |I_V-I_{V-U}| I_U,
\end{equation}
where we used the notation $I_U \equiv \frac{U^3}{6} + \frac{c_2
\cdot U}{12}$ and so on, and the fact that for proper $SU(3)$
holonomy Calabi-Yau manifolds, the pure D6 has a unique ground state
after factoring out the center of mass hypermultiplet, i.e.\
$\Omega(e^S \, \Gamma(0,0))|_t = 1$.

Now let us compare this to the microscopic large radius geometrical
picture of this D6-D2-D0 as the ideal sheaf $\CI_C$ given by the
curve $C = U \cap V$. The charges are given by
(\ref{idealsheafcharges}), which yields, using the adjunction
formula,
\begin{equation}
 q_{D2} = - U \cdot V + \frac{c_2}{24}, \qquad q_0 = \chi_h(C) = -\frac{1}{2}
 \left( C^2|_V + C \cdot V \right) = -\frac{1}{2} (U^2V + UV^2),
\end{equation}
in agreement with (\ref{totalchargeD6D2D0triple}). We can
parametrize the moduli space of this ideal sheaf as follows. First
recall that the moduli space of very ample divisors $D$ (= $U$, $V$
and $V-U$ here) is parametrized by the vector space of holomorphic
sections $s_D$ of the associated line bundles $\CL_D$, modulo
overall rescaling of the section. That is,
\begin{equation}
 \CM_D = \ICP^{I_D-1}.
\end{equation}
Now pick a divisor representative $U_0$ in the class $U$, described
by the vanishing locus of a section $s_{U_0}$ of $\CL_U$. Note that
we can write any section of $\CL_V$ as
\begin{equation}
 s_V = s_{U_0} \, s_{V-U} + \tilde{s}_V,
\end{equation}
where $s_{V-U}$ is some section of $\CL_{V-U}$ and $\tilde{s}_V$ a
section of $\CL_V$. Conversely, any such expression gives a
holomorphic section of $\CL_V$. Now changing $s_{V-U}$ in this
expression will not change $C_0 := U_0 \cap V=\{ s_{U_0}=0 \} \cap
\{ s_V =0 \}$, and thus the moduli space of curves $C_0$ is
described by the vector space of sections $\tilde{s}_V$ of $\CL_V$
modulo products of sections of $\CL_{V-U}$ with $s_{U_0}$, and
modulo overall rescalings, that is
\begin{equation}
 \CM_{C_0} = \ICP^{I_V-I_{V-U}-1}.
\end{equation}
In conclusion, the sheaf moduli space $\CM_C$ is a regular
$\ICP^{I_V-I_{V-U}-1}$ fibration over $\ICP^{I_U-1}$. Computing the
Euler characteristic of this space immediately reproduces
(\ref{indexD6D2D0triple}).


Note that for this example, there is no region in moduli space where
all phases of the constituents line up, as can be seen directly in
fig.\ \ref{D6D2D0triple} since the two marginal stability lines do
not intersect. As a result, the quiver quantum mechanics picture as
reviewed in section \ref{sec:quivdescription} is not reliable in
this case. And indeed, if one tries to compute the index as the
euler characteristic of the moduli space $\CM$ for the quiver of
fig.\ \ref{D6D2D0triple} with dimension vector $(1,1,1)$, as given
by (\ref{genquivermodulispace}) (with $W=0$ since there is no closed
loop), one finds the \emph{wrong} result $\Omega =
(-1)^{I_{21}+I_{23}+I_{13}} \, (I_{21}+I_{23}) \, I_{13}$ or
$(-1)^{I_{21}+I_{23}+I_{13}} \, (I_{13}+I_{23}) \, I_{21}$,
depending on the sign of $\vartheta_1$. (This can be computed with
the methods described in section \ref{sec:fournodequiv}.) This
illustrates that the split flow picture is more general than the
quiver picture.

It can be shown that the correct index is that of the part of the
quiver BPS Hilbert space which jumps at the MS wall $\vartheta_1=0$,
i.e.\ the difference of the Hilbert spaces for $\vartheta_1>0$ and
$\vartheta_1<0$, suggesting an identification of the ideal sheaf
cohomology with this part of the quiver cohomology. It would be
interesting to clarify this point further.

When we invert $U \to -U$ in (\ref{threenodequivcharges}) the
situation changes significantly. In this case, there \emph{is} a
region of moduli space where all phases line up, and the quiver
description becomes accurate. The relevant quiver now has a closed
loop however, allowing a nontrivial superpotential. We now turn to
this case.

\subsubsection{Three node quiver with closed loop} \label{sec:threenodeclosedloop}

\EPSFIGURE{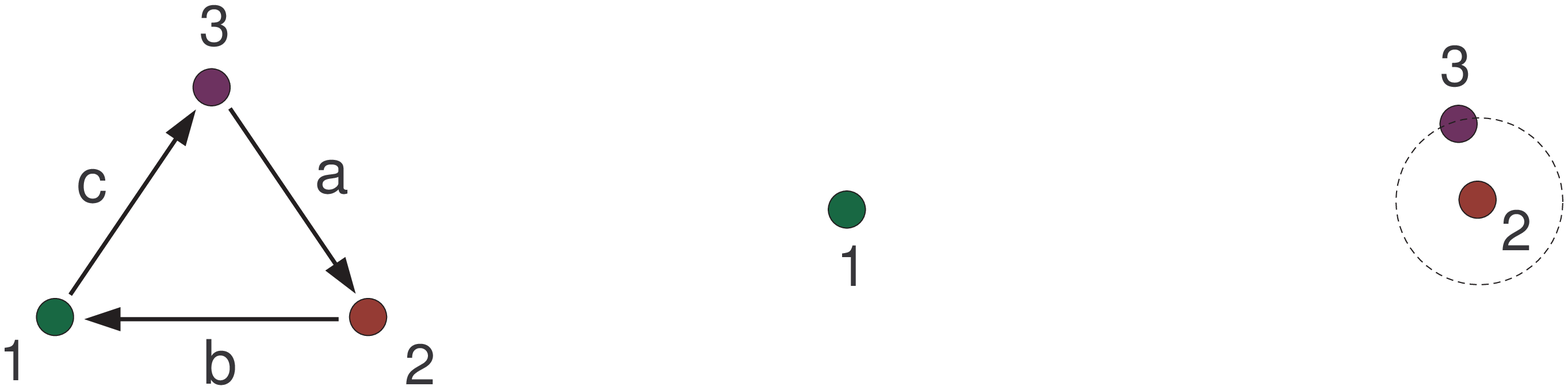,height=3.5cm,angle=0,trim=0 0 0 0}%
{{\bf Left:} Three node quiver with closed loop. {\bf Right:}
Corresponding splitting multicentered configuration when $b>a+c$ and
$\theta_1$ approaches zero.
  \label{threenodequiver}}

Our third nontrivial example is given by the quiver given in fig.\
\ref{orbits2}$b$, which has a closed loop. The latter brings in some
qualitatively new features, such as the presence of a superpotential
and the possibility of scaling solutions.

For simplicity we take the dimension vector to be $(1,1,1)$, and
assume no internal moduli associated to the vertices. This could
correspond for instance to a bound state of three single D6 or
anti-D6 branes with suitable $U(1)$ fluxes turned on on their
worldvolumes. In particular our previous example
(\ref{threenodequivcharges}) with $U$ inverted, i.e.,
\begin{eqnarray} \label{threenodequivchargesinv}
 \Gamma_1 = e^{-U}, \quad
 \Gamma_2 = e^V, \quad
 \Gamma_3 = -e^{V-U}
\end{eqnarray}
realizes this, where we take $U,V$ to be positive divisors and we
are dropping
  $c_2$ corrections for simplicity.  We have $a=U^3/6$, $b=(U+V)^3/6$,
$c=V^3/6$. Since $U,V$ are positive divisors we have $b>a+c$.

\EPSFIGURE{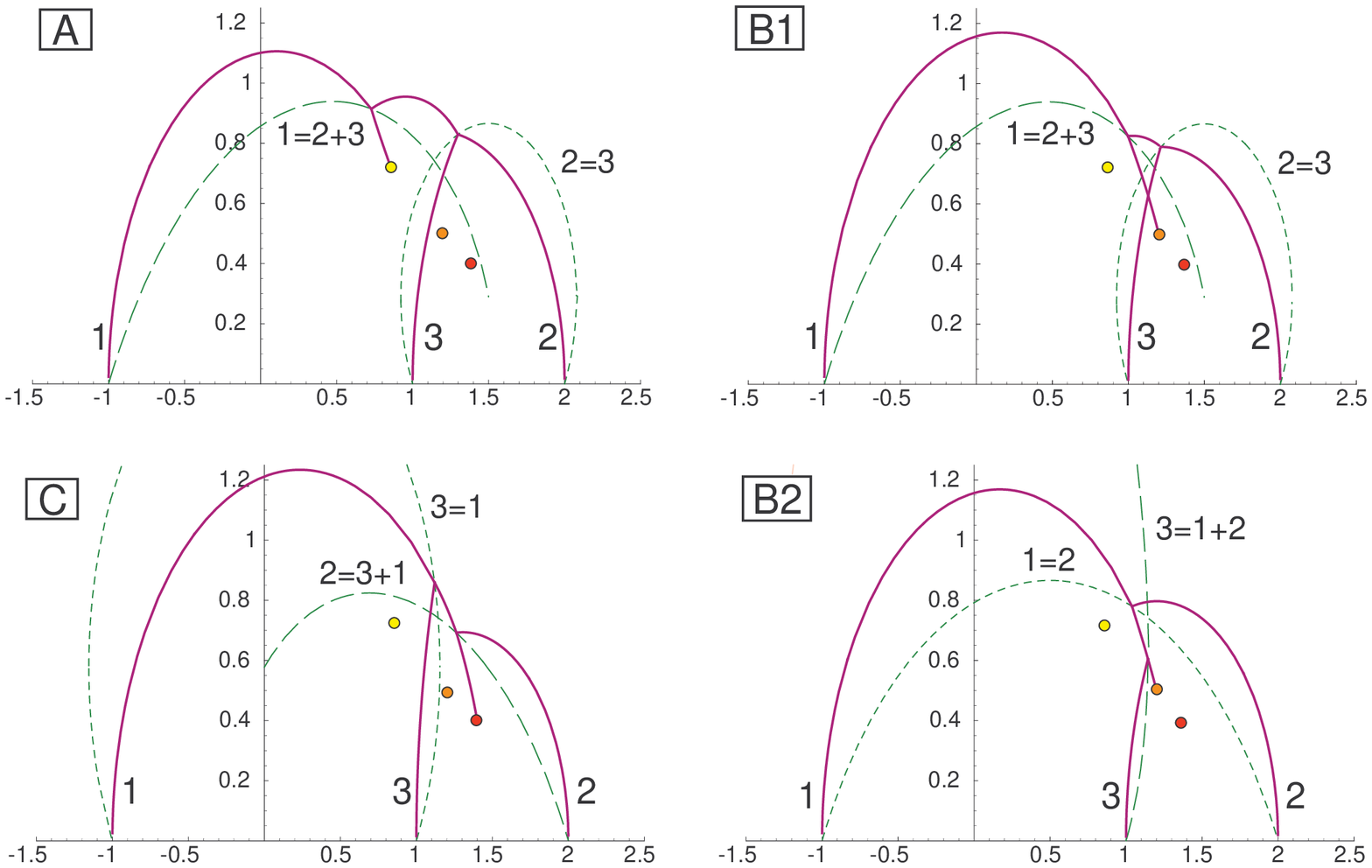,height=10cm,angle=0,trim=0 0 0 0}%
{One modulus example of a realization of the quiver in fig.
\ref{threenodequiver}. The three terminal charges $\Gamma_i$ are
given by (\ref{threenodequivchargesinv}) (dropping $c_2$
corrections) with $U=D_1$, $V = 2 D_2$, and the corresponding
terminal flows labeled by 1, 2 and 3. The green dotted lines are
lines of marginal stability. The ms line ``$2=3$'' corresponds to
$Z_2$ and $Z_3$ lining up, ``$1=2+3$'' to $Z_1$ and $Z_2 + Z_3$
lining up, and so on. For each flow tree, we only show the ms lines
on which the tree has vertices; this is different for each tree.
Three different initial points are considered, corresponding to the
labels $A$, $B$ and $C$. We consider the three kinds of flow
patterns $(1(23))$, $(2(13))$ and $(3(12))$ at each of the initial
points. The initial (yellow) point in case $A$ only supports the
flow $(1(23))$. The initial point in  case $B$ supports  the two
flows $(1(23))$ and $(3(12))$, illustrated in  $B1$ and $B2$.
Finally, the initial point in   case $C$ again supports only the
flow $(2(13))$. There are also regions where no flow tree exists,
for example the large $\Im \, t$ region. Here the attractor flow is
a single centered flow for the total charge which crashes on a zero
of $Z(\Gamma;t)$. \label{3quivsplitflows}}

 Fig.\ \ref{3quivsplitflows} shows a one modulus example of
 flows associated with (\ref{threenodequivchargesinv})
 with
$U=D_1$, $V=2 D_1$ and three different initial points. No flow trees
exist in the large radius regime. In other regions, one or more of
the three possible tree topologies $(1,(2,3))$, $(2,(3,1))$,
$(3,(1,2))$ is realized. For initial point $A$, we have only
$(1,(2,3))$. When moving to point $B$, we pass through the marginal
stability line where $Z_3$ and $Z_1+Z_2$ line up, i.e.\ $\theta_3 =
0$. The $(1,(2,3))$ remains alive ($B1$), but a new tree, of
topology $(3,(1,2))$ ($B2$) comes into existence. So in this case
the total Hilbert space $\CH(\Gamma;t_{\infty})$ will be partitioned
by two trees. Finally, when moving to point $C$, we do not pass
through any relevant marginal stability line, but along the way a
tree-topology-changing transition takes place: at some point, both
flow trees we had in $B$ become degenerate (with one 4-valent vertex
instead of two 3-valent) and identical, and going beyond that, we
are left with one new tree, of topology $(2,(3,1))$.

According to our general framework, the index should jump between
$A$ and $B$, but not between $B$ and $C$. This is confirmed by the
explicit expressions obtained from our factorization formulae
\begin{eqnarray}
 \Omega(A) &=& (-1)^{\langle \Gamma_1,\Gamma_2 + \Gamma_3
 \rangle + \langle \Gamma_2,\Gamma_3 \rangle} \, |\langle \Gamma_1,\Gamma_2 + \Gamma_3
 \rangle| |\langle \Gamma_2,\Gamma_3 \rangle| = (-1)^{a+b+c} (b-c) a
 \\
 \Omega(B) &=& \Omega(B1) + \Omega(B2) = (-1)^{a+b+c} \left( (b-c)a + (c-a)b \right) \\
 & =&
 (-1)^{a+b+c} (b-a) c \\
 \Omega(C) &=& (-1)^{a+b+c} (b-a) c,
\end{eqnarray}
where we used $b>a+c$ and $c > a$. Note that the two $B$-trees
indeed nicely combine to give the same index of the single $C$-tree!

Let us now turn   to the microscopic description. Since there are no
flow trees starting from the large radius regime, there won't be a
geometrical large radius D-brane realization. Indeed, we now have a
D6-D2-D0 charge with \emph{positive} D2 charge, which never exists
as a BPS state in the large radius regime. Fortunately however, we
see that there is a region in moduli space where all phases of the
nodes line up, so we can use quiver quantum mechanics to verify our
results.

As in section \ref{sec:fournodequiv}, we can actually verify our
results independently  of the split flow picture by just comparing
the results obtained from the 3-particle quantum mechanics to those
from the microscopic quiver moduli space.

We label the stretched open string scalars by $z_i, i=1,\dots,
I_{13}:=c$, $x_j, j=1\dots, I_{32}:=a$, $y_k, k=1,\dots, I_{21}:=b$.
The D-term constraints are
\begin{eqnarray}
 \sum_i |z_i|^2 - \sum_k |y_k|^2 &=& \vartheta_1 \label{Dflat1}\\
 \sum_k |y_k|^2 - \sum_j |x_j|^2 &=& \vartheta_2 \label{Dflat2}\\
 \sum_j |x_j|^2 - \sum_i |z_i|^2 &=& \vartheta_3,\label{Dflat3}
\end{eqnarray}
Since the quiver has a closed loop, a nontrivial gauge invariant
superpotential is possible. We assume this to have a generic cubic
form
\begin{equation}
 W(x,y,z) = \sum_{ijk} c_{ijk} z_i x_j y_k.
\end{equation}
Higher order terms can self-consistently be neglected as long as the
$(x,y,z)$ are small. As we will see, this can be enforced by making
$\vert \vartheta_i\vert$ sufficiently small.

From the discussion in section \ref{sec:scalingsol} we expect
factorization when at least one of the triangle inequalities
(\ref{trineq}) is violated, say, as in our concrete realization
described above,
\begin{equation} \label{assumption}
 b > a + c.
\end{equation}
As we saw in section \ref{sec:scalingsol}, for $\theta_3 > 0$, the
configuration will split for $\theta_1$ approaching 0 (from below)
by separating $\vec{x}_1$ infinitely far from $\vec{x}_2$ and
$\vec{x}_3$, so our general physical arguments lead to an index for
this particular system given by
\begin{equation} \label{Omegamacro}
 \Omega_{\rm macro} = (-1)^{c-b-1} \, (b-c) \, \Omega(1) \, \Omega(2+3)
 = (-1)^{c+b+a} \, (b-c)\, a.
\end{equation}
This corresponds to our case $A$.

From the point of view of the microscopic quiver moduli space
\begin{equation}
 \CM := \{ (x,y,z) | (\ref{Dflat1})-(\ref{Dflat3}) \mbox{ satisfied and } \partial W = 0
 \}/U(1)^3
\end{equation}
the result (\ref{Omegamacro}) is not at all obvious. For instance it
appears rather mysterious why the microscopic quiver description
should care about triangle inequalities.

Let us therefore compute the index directly from the quiver moduli
space, and check if factorization holds when expected. The index of
this system is given by
\begin{equation} \label{Omegamicro}
 \Omega_{\rm micro} = (-1)^{\dim \CM} \chi(\CM) = (-1)^{c+a+b}\,
 \chi(\CM).
\end{equation}
(Recall the origin of the sign factor is the identification of
$J_3'$ with Lefshetz spin, as explained above (\ref{indextoeuler}).)
For generic $c_{ijk}$, the solutions to $\partial W = 0$ split in
three branches, one with $x=0$, one with $y=0$ and one with $z=0$.
This can be seen as follows.\footnote{We thank Davide Gaiotto for
providing this argument.} Assume there are other solutions, i.e.\
with $x \neq 0$, $y \neq 0$, $z \neq 0$. Relabeling indices we can
assume say $x_1 \neq 0$, $y_1 \neq 0$, $z_1 \neq 0$. Now note that
the equations have scaling symmetries $x \to \lambda_1 x$, $y \to
\lambda_2 y$, $z \to \lambda_3 z$, so without loss of generality we
can assume $x_1=y_1=z_1=1$. The equations corresponding to partial
derivatives with respect to the other variables are then a nice set
$S$ of equations that can be solved with a finite set of solutions.
The equations corresponding to partial derivatives with respect to
$x_1$, $y_1$, $z_1$ are an extra set of constraints. From the
homogeneity of $W(x,y,z)$ it follows that these three extra
equations are satisfied iff the superpotential evaluated on the
solutions to $S$ is zero. On the other hand the coefficient
$c_{111}$ does not enter the first set of equations $S$. Picking
different values for $c_{111}$ one can get any possible value for
the superpotential. It follows that branches with $x,y,z$ all
different from zero can exist only for a codimension one set of
coefficients $c_{ijk}$, and are generically absent, proving our
claim.

 Which of the three branches is turned on depends on the signs
of the Fayet-Iliopoulos parameters $\vartheta_i$:

\begin{itemize}

\item $x_j=0$ corresponds to $\vartheta_3<0, \vartheta_2>0$, with
$\vartheta_1$ of either sign

\item $y_k=0$ corresponds to $\vartheta_1>0$, $\vartheta_2<0$,
with $\vartheta_3$ of either sign.

\item $z_i=0$ corresponds to $\vartheta_3>0$, $\vartheta_1<0$,
with $\vartheta_2$ of either sign.

\end{itemize}

Note that on any branch, eqs. (\ref{Dflat1}-\ref{Dflat3}) show that
the nonzero $\vert x_j\vert^2, \vert y_k\vert^2, \vert z_i\vert^2$
are bounded by $\vert \vartheta_i\vert$ and hence can be made small,
justifying our use of the cubic superpotential.

When (\ref{assumption}) is satisfied, there is an even simpler
argument for the absence of branches with all $x,y,z$ nonzero: in
this case $\partial_y W = 0$ imposes more equations than there are
unknowns $x_j$, $z_i$, so this equation will generically not have
any solutions apart from the trivial ones $x=0$ or $z=0$. Assuming
$\vartheta_3
> 0$ (the case $\vartheta_3 < 0$ can be dealt with analogously) then implies
using (\ref{Dflat3}) that $x \neq 0$, so we must put $z=0$. From
(\ref{Dflat3}) it then follows that there can only be solutions for
$\vartheta_1 < 0$, decaying at marginal stability $\vartheta_1=0$ by
splitting off $\Gamma_1$, in accordance with what we found in the
spacetime picture in section \ref{sec:scalingsol}.

Since $z=0$, the D- and F-constraints reduce to
\begin{equation} \label{modspacetriquiv}
 \CM = \{ (x,y) \in \IC \IP^{{a}-1} \times \IC
\IP^{b-1} \, | \, \sum_{j,k} c_{ijk} x_j y_k =0, \quad i=1,\dots, c
\},
\end{equation}
with $x$ and $y$ now interpreted as homogeneous coordinates. Now for
any fixed $x \in \ICP^{a-1}$, the above equations cut out a
$\ICP^{b-c-1}$ in $\ICP^{b-1}$. It is clear that this is true for
\emph{generic} $x$, but when (\ref{assumption}) is satisfied (and
$c_{ijk}$ is generic, as we assume throughout), it will in fact be
true for \emph{any} $x$. This is easily seen for example by taking
the $c_{ijk}$ such that $M_{ik}(x) \equiv \sum_j c_{ijk} x_j \equiv
x_{k-i+1}$ (with $x_j \equiv 0$ if $j$ is outside the range
$1,\ldots,a$), and noting that $M(x)$ manifestly has always maximal
rank on $\ICP^{a-1}$.

Therefore in this regime $\CM$ is a $\ICP^{b-c-1}$ fibration over
$\ICP^{a-1}$, without any fibers degenerating. Therefore
\begin{equation} \label{threenodefactchi}
 \chi(\CM) = (b-c)\, a,
\end{equation}
which brings (\ref{Omegamicro}) in exact agreement with
(\ref{Omegamacro}).

Note that this is again an explicit realization of the picture
outlined in section \ref{sec:microDbrane}, and of the assumptions
made there. In particular we see explicitly that the F-constraints
effectively put the net intersection product between the two custers
$\langle \Gamma_2 + \Gamma_3,\Gamma_1 \rangle = b-c$ equal to the
total number of nonzero bifundamental scalars between theses two
constituents, and that the fibration is regular.

It would be very interesting --- even within the context of the
present example --- to do a more systematic study of the various
flow trees that can arise and in which regions they do arise. It
would also be interesting to compare more systematically with the
quiver picture. This picture will be accurate in an open region
around the point where all three charges $Z_i$ are aligned, but it
is easy to see that it must fail at some distance of order one from
this point, because the lines $\vartheta_i=0$ and $\theta_i=0$ will
not coincide. However, these matters lie outside the scope of this
short note, so we will leave them for future work.

\subsection{Entropy of the three node quiver in the scaling regime}
\label{sec:3quiver}

The analysis of the three node closed loop quiver changes
significantly when all three triangle inequalities are satisfied.
For one thing, fiber jumps now become possible, with the potential
of drastically increasing the complexity and Euler characteristic of
$\CM$. This is expected physically: as we saw in section
\ref{sec:scalingsol}, in this case there is no obstruction to
letting the centers approach each other arbitrarily closely,
asymptotically forming a black hole. Such solutions can no longer be
forced to split. Thus they are not described by a flow tree but
rather by a single flow, so our factorization arguments no longer
apply, and the emergence of a horizon in the asymptotic limit
suggests instead an exponential black hole type ground state
degeneracy.

Note that our realization of the quiver described above and
exemplified in figure \ref{3quivsplitflows} is never in this regime,
since it always violates the triangle inequalities by $b>a+c$. It is
not extremely easy to find realizations of scaling solutions in
terms of rigid constituents.  However, we expect that a simple
realization with rigid nodes satisfying the triangle inequalities
and corresponding to a scaling solution can be obtained by adding a
rigid D2-brane to the charge $\Gamma_3$ in eq.(\ref{SEMcharges}).
(We say ``expect'' because we have not verified that the
discriminant $\CD(H(\vec x))$ is everywhere positive.) We assume
there are such realizations and proceed.

To compute the Euler characteristic in this case requires more
sophisticated machinery. According to the general formulae of
\cite{Hosono:1994ax} the Euler characteristic is given in the
general case   by
\begin{eqnarray}
\chi(\CM)  & = &
\frac{\p_{J_1}^{{a}-1}}{({a}-1)!}\frac{\p_{J_2}^{{b}-1}}{({b}-1)!}
\frac{(1+J_1)^{a}(1+J_2)^{b}}{(1+J_1+J_2)^{c}}
(J_1+J_2)^{c}\vert_{J_1=J_2=0}\\
& = & \oint dJ_1 \oint dJ_2 J_1^{-{a}}
J_2^{-{b}}\frac{(1+J_1)^{a}(1+J_2)^{b}}{(1+J_1+J_2)^{c}}
(J_1+J_2)^{c}. \label{threenodeint}
\end{eqnarray}
The contour integrals are on small contours $J_i = \epsilon_i e^{i
\theta_i}$ and the relative sizes of $\epsilon_i$ do not matter.

The evaluation of these integrals is nontrivial and given in
appendix \ref{app:eulerint}. We find the following elegant exact
expression:
\begin{equation} \label{triplelaguerreint}
\chi(\CM) = ab - \int_0^\infty ds \, e^{-s} \, L^1_{a-1}(s) \,
L^1_{b-1}(s) \, L^1_{c-1}(s),
\end{equation}
where the $L^1_*$ are Laguerre polynomials. Amusingly, these kinds
of integrals arise in atomic physics, since the Laguerre polynomials
are the radial eigenfunctions of an electron in a Coulomb potential.
We don't know if this is a coincidence or has a deeper physical
explanation. Recall that the small asymmetry between $(a,b)$ and $c$
arises from the fact that we are considering the case $\theta_1<0$,
$\theta_3>0$, putting us on the branch $z=0$. The other cases give
rise to expressions with obvious modifications.

Equation (\ref{triplelaguerreint}) reproduces
(\ref{threenodefactchi}) when $b+1\geq a+c$, so in particular also
when (\ref{assumption}) is satisfied, as expected. More
interestingly, when all triangle inequalities are satisfied, we find
that the degeneracies start increasing exponentially. This is in
beautiful agreement with the fact that in this regime, the state no
longer splits and a black hole can be formed, as we saw in section
\ref{sec:scalingsol}. More precisely, we find the remarkably simple
and suggestive result
\begin{equation}\label{abcentropy}
\chi(\CM) \sim (abc)^{-1/3} \, 2^{a+b+c}
\end{equation}
in the regime in which $(a,b,c)$ are not too different from each
other. Note that this amounts to a macroscopic entropy, since the
intersection products $(a,b,c)$ scale as $\Lambda^2$ when scaling up
uniformly all charges by $\Lambda$, just like a large black hole.

The  formula (\ref{abcentropy}) suggests an interpretation in terms
of fermionic degrees of freedom stretched between the centers, one
per unit of intersection product, at least at large $a$, $b$, $c$.
It is an interesting open problem to explain this result. Clearly,
one would have a hard time getting such exponential degeneracies
from a simple three particle quantum mechanics, unless new degrees
of freedom appear in the scaling regime.


\section{Counting BPS degeneracies} \label{sec:counting}

In this section we specialize the general tools developed so far to
our actual problem of counting D4-D2-D0 state degeneracies. The idea
is roughly as follows. We use the fareytail expansion to reduce the
counting problem to computing polar D4-D2-D0 indices. We show, using
our index factorization formulae, that at least for the ``extreme''
polar states, these indices factorize into D6 and anti-D6 indices.
This leads to an approximate factorization of the leading term of
the fareytail series into a D6 and an anti-D6 partition function. We
argue that the latter can be identified with the topological string
partition function. Putting the pieces together, we obtain an
expression of the form $\CZ_{\rm BH} \sim \CZ_{\rm top}
\overline{\CZ_{\rm top}}$.

In section \ref{sec:D6D4D2D0}, we explore the relation of
D6-D4-D2-D0 indices and DT invariants. This is nontrivial because of
the background dependence of the former. We also show how quantizing
particle halos leads to MacMahon and Gopakumar-Vafa-type generating
functions. In section \ref{sec:D6antiD6deg}, we compute indices of
D6-anti-D6 bound states using index factorization, and in section
\ref{sec:dilutegaspf} we show that this leads to a suitably
factorized generating function. In section \ref{sec:D4D2D0} we show
to what extent this can be used to compute polar D4-D2-D0 indices,
and based on this we give a derivation of a refined version of the
OSV conjecture (\ref{osvwithmf}).

\subsection{D6-D4-D2-D0 degeneracies} \label{sec:D6D4D2D0}

We now turn more specifically to degeneracies of bound states of a
single D6 with lower dimensional branes. Donaldson-Thomas
invariants, which ``count'' ideal sheaves, count in some sense BPS
bound states of a single D6 with D2- and D0-branes. However, there
is an immediate problem with this interpretation: the actual BPS
indices of such bound states depends strongly on the background
moduli, due to jumping phenomena at marginal stability walls, while
DT invariants are insensitive to the background. We saw examples of
the background dependence of such BPS states in sections
\ref{sec:D6D0} and \ref{sec:quivfact}. Examples of supergravity
realizations of such states with a limited domain of stability are
halos of D0-branes around some D6-D4-D2-D0 core. More general such
halo-configurations exist, for example replacing the D0-particles by
D2-D0 particles, as shown in fig.\ \ref{moretrees}$a$. The latter
have an even richer structure of marginal stability walls, and they
extend all the way to the infinite radius limit, so unlike for
D4-D2-D0 systems, one cannot avoid these issues by restricting to
the large radius limit. Moreover, as is manifest e.g.\ in equation
(\ref{BPSindfact2}), when these D6-D4-D2-D0 bound states are used as
building blocks for more complicated configurations or flow trees,
the relevant moduli are determined by the split points of the
attractor flow trees, so we cannot just pick some values we happen
to like.

\subsubsection{D6 + D0-halos}\label{sec:mcmahonsection}

Let us begin by considering the degree zero part $\CZ_{DT}^0$ of the
DT partition function, introduced in (\ref{ZDT0MacMahon}). This
supposedly counts D6-D0 bound states, but we know that at $B=0$ for
example, there are no such bound states. On the other hand, for $B$
sufficiently large, these bound states do exist, and then
$\CZ_{DT}^0$ correctly counts them.

A simple way to derive this is through the D0-halo picture in
supergravity. The following gives a sketch of how this is done,
based on the detailed study of analogous systems in
\cite{Denef:2002ru}, to which we refer for more details.

Since D0-branes can form bound states with each other of arbitrary
charge, the particles in the halo can have arbitrary D0-charge. To
have a BPS configuration, all charges have to be of the same sign
though, determined by the sign of the $B$-field, as in
(\ref{D6D0existencecond}). Let us assume this is positive.

As reviewed under (\ref{spinformula}), the contribution to the
angular momentum of a particle of D0-charge $n$ moving in the
magnetic field of a D6, arising from its position degrees of freedom
and the intrinsic monopole-electron type angular momentum stored in
the electromagnetic field, equals $j=\frac{1}{2} \langle D6, n D0
\rangle - \frac{1}{2} = (n-1)/2$, hence the contribution to the
degeneracy of BPS ground states from quantizing these degrees of
freedom equals $2 j + 1=n$.

In addition, the particle has a number of ``internal'' BPS ground
states, obtained by quantizing its position moduli (super)space
inside the Calabi-Yau threefold $X$. These are simply given in the
usual way by the cohomology of $X$, and their spin is determined by
the Lefshetz SU(2) action on $H^*(X)$; in particular a $p$-form has
spin $j_3 = (p-3)/2$, which is half-integral when $p$ is even. Since as we saw,
the $\IR^3$ position hypermultiplet is forced by the radial magnetic
field to be in a spin 1/2 state, the even cohomology will thus
correspond to bosonic particles, and the odd cohomology to fermionic
particles.\footnote{Here  ``bosonic'' and ``fermionic'' refers to
the nature of the individual particles (the electrons). Whether the
bound state with the D6 as a whole (the atom) will be fermionic or
bosonic also depends on the spin $(n-1)/2$ coming from the
quantization of the monopole-electron system as discussed above.}

To summarize, one particle ground states are labeled by their
D0-charge $n \geq 1$, an integer $m \in \{0,\ldots,n-1\}$ labeling
the lowest Landau levels, and an element $\omega$ of $H^*(X)$. Since
the particles in the halo are mutually BPS, classically they do not
exert static forces on each other, and hence in the $\hbar\to 0$
limit the multiparticle ground states are simply labeled by
occupation numbers of single particle states: $|\{ k_{n,m,\omega}
\}_{n,m,\omega} \rangle$, where $\omega$ runs over a basis for the
cohomology.  Since we are considering a Fock space of D0 particles
in a fixed D6 background we form a Fock space of bosonic one
particle states corresponding to even degree cohomology classes and
of fermionic one particle states corresponding to odd degree
cohomology classes. Thus we have bosonic occupation numbers
$k_{n,m,\omega_e} \in \IN$, where $\omega_e$ runs over a basis for
the even-degree cohomology and fermionic occupation numbers
$k_{n,m,\omega_o} \in \{0,1 \}$, where $\omega_o$ runs over a basis
for odd-degree cohomology. While the fermi/bose nature of the
individual particles is governed by the degree of the cohomology
class, the total spin $J_3'$ (as usual with the contribution of the
center of mass degrees of freedom factored out) is $J_3'=\sum
k_{n,m,\omega} (m-\frac{n-1}{2} + \frac{\deg(\omega)-3}{2})$.
Putting all this together, and letting $b_e, b_o$ denote the
dimension of the even, respectively odd cohomology,  the generating
function for the index $d_N$ of D6-D0 BPS bound states of total
D0-charge $N$ (defined following (\ref{indexdef})-(\ref{indexred}))
is
\begin{eqnarray}
 \sum_{N} d_N \, u^N &=& {\rm Tr} \, (-1)^{2 J_3'} \, u^N \nonumber \\
 &=& \sum_{k_{n,m,\omega_e}}
 \sum_{k_{n,m,\omega_o}} (-1)^{\sum k_{n,m,\omega_o} + \sum n (k_{n,m,\omega_e} + k_{n,m,\omega_o})}
 \,
 u^{\sum n (k_{n,m,\omega_e}+ k_{n,m,\omega_o})} \nonumber \\
 &=& \prod_{n=1}^\infty \left( \sum_{k=0}^\infty (-u)^{nk}
 \right)^{n b_e} \left( \sum_{k=0}^1 (-1)^k (-u)^{nk}
 \right)^{n b_o} \nonumber \\
 &=& \prod_{n=1}^\infty (1-(-u)^n)^{n(-b_e+b_o)} = \prod_{n=1}^\infty
 (1-(-u)^n)^{-n\chi(X)} = M(-u)^{\chi(X)} \label{macmahonfactor}
\end{eqnarray}
where $M$ is the MacMahon function. This exactly reproduces the
expression (\ref{ZDT0MacMahon}) for $\CZ_{DT}^0$, including all
signs.

If the value of the $B$ field is such that the BPS condition
requires negative $D0$ charge, the generating function is obtained
from the one above by substituting $u \to u^{-1}$.

Incidentally, from the weak coupling expansion (\ref{hzergw}) of the
MacMahon function --- in particular from  the $\exp[\frac{
\chi(X)\zeta(3)}{2g^2}]$ singularity --- we can extract the $n\to
\infty$ asymptotics
\begin{equation} \label{mmgrr}
 d_n = N_{DT}(0, n) \sim  \left\{
  \begin{array}{l}
   \exp[\frac{3}{2}(\chi \zeta(3)n^2)^{1/3} + (\half - \frac{\chi}{72})\log n]
    \qquad\qquad\qquad\ \mbox{ if } \quad\  \chi(X)>0 \\
      {\rm Re}\biggl[ e^{i\phi}
    \exp[e^{i\pi/3}\frac{3}{2}(\vert \chi \vert\zeta(3)n^2)^{1/3}
    + (\half - \frac{\chi}{72})\log n] \biggr] \mbox{ if }    \quad\ \chi(X)<0
  \end{array}
\right.
\end{equation}
where  $\phi$ is a real constant and we dropped the $1$-loop
prefactor. Hence the large $n$ entropy of the D0-halo goes roughly
like $S \sim n^{2/3}$, but with large oscillatory fluctuations when
$\chi(X)<0$.

As we saw in section \ref{sec:SEM}, D0-halos appear as part of
multicentered D4D2D0 states, and thus the MacMahon function
naturally appears in generating functions determining black hole
entropies. In this way the above derivation resolves an old puzzle.
It seemed mysterious how microstate counting could account for the
strange term $\chi(X)\zeta(3)/g^2$ in the expansion of the
topological free energy $F_{\rm top}$ and therefore in the black
hole entropy formula. Now we see where it comes from: the MacMahon
function arises from counting D0-halo states, and this in turn gives
rise to the term $\chi(X)\zeta(3)/g^2$ from the small $g$ asymptotic
expansion (\ref{hzergw}).

\subsubsection{D6 + D2-D0-halos and relation
between BPS indices and DT invariants} \label{sec:D6D2D0}

Before we get to counting bound states of D6-branes with D2-D0 halos
around them, let us briefly review the counting of D2-D0 BPS states
and the definitions of BPS and DT invariants.

At zero string coupling, single D2-D0 particle states of $(D2,D0)$
charge $(Q,n)$ are given by cohomology classes of D2 moduli space.
The $(D2,D0)$ moduli space is a   torus fibration over the
deformation moduli space of a holomorphic curve in homology class
$Q$ in $X$. The curve represents the supersymmetric cycle on which
$D2$ is wrapped while the torus fiber accounts for Wilson line
moduli. The cohomology can be decomposed according to
representations of the $SU(2)_L \times SU(2)_R$ Lefshetz action on
moduli space
\cite{Gopakumar:1998ii,Gopakumar:1998jq,Katz:1999xq,Hosono:2001gf}
where roughly $SU(2)_R$ acts on the base cohomology while $SU(2)_L$
acts on the fiber cohomology. After uplifting to M-theory, the
$SU(2)$'s can be identified with the factors of the 5d little group
$SO(4)=SU(2)_L \times SU(2)_R$ \cite{Witten:1996qb}.

Let $N_{Q}^{m_L,m_R}$ be the dimension of the cohomology group   of
the moduli space D2 of branes of charge $Q$ and $(J_L^3,J_R^3) =
(m_L,m_R)$.  One can construct a well-behaved index from this by
tracing over the $SU(2)_R$ factor:
\begin{equation} \label{Omjldef}
 N_Q^{m_L}:= \sum_{m_R} (-1)^{2 m_R} N_Q^{m_L,m_R}.
\end{equation}
The usual Witten index of all BPS states, which up to a sign is the
Euler characteristic of the moduli space, is obtained by tracing
this index in turn over the $SU(2)_L$ factor:
\begin{equation} \label{Nqdef}
 N_Q := \sum_{m_L} (-1)^{2 m_L} N_Q^{m_L}.
\end{equation}
%
%
These indices are related to the BPS invariants $n^r_Q$ by
\cite{Katz:1999xq}
\begin{equation} \label{NGVrel}
 N_Q^{m_L} = \sum_{r \geq |2 m_L|}
 {2r \choose r+2 m_L} \, n^r_{Q} \, , \qquad N_Q = n^0_Q.
\end{equation}
Note the interesting cancelation leading to the last expression, due
to the binomial formula $\sum_{m=0}^n (-1)^m {n \choose m} = (1-1)^n
= 0$. More fundamentally this follows from the fact that a D2-brane
of genus greater than 0 comes with a moduli space containing a torus
factor from the Wilson lines, which has zero total Euler
characteristic when excluding degeneration limits. Only a complete
degeneration to $g=0$ components eliminates the torus fiber and
gives rise to a nonzero Euler characteristic.

With these invariants and using the GW-GV-DT correspondence outlined
in section \ref{sec:prelim}, one can build up a generating function
for DT invariants, as follows
\begin{equation}
\CZ_{DT}(u,v) := \sum_{\beta,n} N_{DT}(\beta,n) \, u^n \, v^{\beta}
= \prod_{Q > 0,m_L,k>0} (1-(-u)^{k + 2 m_L} v^{Q})^{k (-1)^{2 m_L}
N_Q^{m_L}}.
\end{equation}
After some manipulations starting from (\ref{NGVrel}) and involving
binomial identities and changing product variables, this can also be
written as \cite{Klemm:2004km}
\begin{eqnarray}
 \CZ_{DT}(u,v) &=& \CZ_{DT}^0(u,v) \, \CZ_{DT}'(u,v) \label{ZDTZDT1} \\
 \CZ_{DT}^0(u,v) &=& \prod_{k>0} (1-(-u)^k)^{-k \chi(X)} \label{ZDTZDT2}\\
 \CZ_{DT}'(u,v) &=& \CZ_{DT}^{\prime,r=0}(u,v) \, \CZ_{DT}^{\prime,r>0}(u,v) \label{ZDTZDT3}\\
 \CZ_{DT}^{\prime,r=0}(u,v) &=& \prod_{Q>0,k>0} (1-(-u)^{k} v^{Q})^{k
 n_Q^0} \\
 \CZ_{DT}^{\prime,r>0}(u,v) &=& \prod_{Q>0,r>0} \prod_{\ell=0}^{2r-2}
 \left( 1 - (-u)^{r-\ell-1} v^Q \right)^{(-1)^{r+\ell} {2r-2 \choose \ell} \,
 n^r_Q}. \label{ZDTZDT5}
\end{eqnarray}

The DT invariants count in some sense D6-D2-D0 bound states, as we
will make precise below, but we can also use them to count more
general D6-D4-D2-D0 bound states (with $p^0=1$), by parametrizing
the charge $\Gamma$ as in (\ref{eSparcharge}):
\begin{equation} \label{Sparrep}
 \Gamma = e^S \, \Gamma(\beta,n) := e^S(1-\beta + n \,
 \omega)(1+\mbox{$\frac{c_2}{24}$}) = e^S(1-\beta + \mbox{$\frac{1}{24}$}
 c_2
 +n \, \omega).
\end{equation}
Multiplying by $e^S$ amounts to tensoring by a line bundle with
field strength $F=S$, or gauge equivalently, to shifting $B \to B -
S$. The gauge invariant statement is that this transformation turns
on a nonzero $\CF:=F-B$.   It is well known that tensoring by a line
bundle this does not affect $\mu$-stability. Essentially this
argument was used in \cite{Gaiotto:2005rp,Dijkgraaf:2006um} to
conclude that $\Omega(\Gamma)\vert_t$ does not depend on $S$.
However, $\mu$-stability does not precisely coincide with physical
stability, not even at infinite radius,\footnote{Thus disproving a
conjecture made in \cite{Moore:1998pn}.} and indeed as we saw (and
are going to elaborate on in what follows) the large radius BPS
spectrum of D6-D2-D0 bound states is in fact \emph{not} invariant
under arbitrary shifts of $\CF$, so the physical
$\Omega(\Gamma)\vert_t$ will actually depend on $S$. Thus, the
question arises then in which regime, if any, the DT invariants do
count physical D6-D2-D0 BPS states.

Another issue, already mentioned in section \ref{sec:microrel}, is
that not only ideal sheaves $\CI$ are suitable to model D6-D2-D0
bound states with $p^0=1$, but their duals $\CI^*$  are as well. They
differ for example in that ideal sheaves have D0 charge bounded from
below at any fixed D2 charge, whereas their duals have D0 charge
bounded above. This leads to the puzzle which of the two we should
consider.

Both of these conundrums are resolved if the DT invariants
correspond to BPS invariants only for suitable limits of the $B$
field. In particular, we will consider limits in which $B$ is taken
proportional to $J$ and taken to plus or minus infinity. The
dichotomy between ideal sheaves and their (derived) duals then
depends on the sign of the $B$-field.

To be specific, let us assume $P$ is some arbitrary auxiliary class
inside the K\"ahler cone, and
\begin{equation} \label{BJSF}
 B+i J = (x+i y) P, \qquad F = S = s \, P, \qquad \CF:= F -B = (s-x) P =: f \, P.
\end{equation}
Then we claim that any ideal sheaf with fixed $(\beta,n)$ specified
as in (\ref{Sparrep}) will become stable for sufficiently large
negative $f$, and their degeneracies $\Omega(\Gamma)|_{(x+iy)P}$
counted by the DT invariants $N_{DT}(\beta,n)$, while for
sufficiently large positive $f$, the duals of ideal sheaves are
stabilized, and their degeneracies counted by $N_{DT}(\beta,-n)$.

At the end of this subsection, we will outline an argument for the
correctness of this claim by uplifting to M-theory, refining the
analysis of \cite{Dijkgraaf:2006um}. Before we get to this, we will
make our claim more precise and elaborate on its consequences in the
IIA picture.

First note that from the discussion of D6-D0 bound states above, it
follows immediately that this proposal is correct for $\beta=0$.
When $\beta \neq 0$, there are in general various possible
configurations with the same charge, consisting of a core which
could for example be a simple D6-D4-D2-D0 black hole, surrounded by
halos of D2-D0 particles at radii fixed by the D2-D0 charges and the
background K\"ahler moduli. The typical state will thus look like an
onion with many different layers of D2-D0 halos, as illustrated in
fig.\ \ref{onion}$a$.

\EPSFIGURE{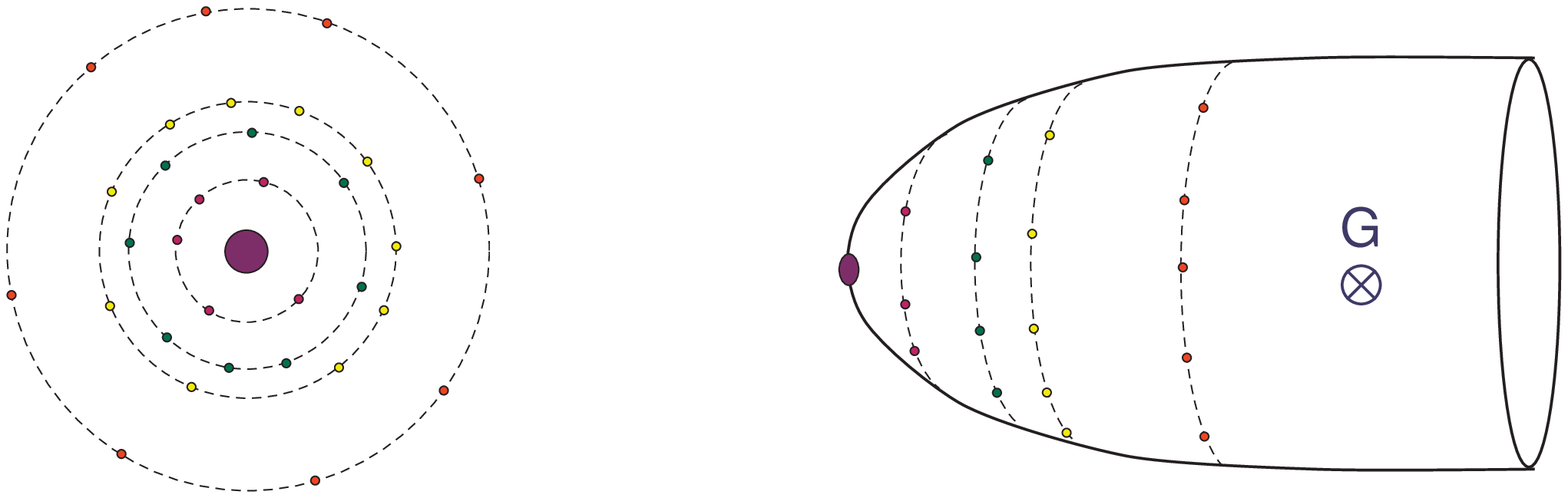,height=5cm,angle=0,trim=0 0 0 0}%
{{\bf Left:} Sketch of a typical D6-D4-D2-D0 bound state in 4d,
consisting of layers of D2-D0 halos around a D6-D4-D2-D0 core (e.g.\
a black hole). The larger $\CF=F-B$ is, the more layers can be
added. Conversely, by decreasing $\CF$, layers get peeled off one by
one, moving out to $r=\infty$. {\bf Right:} Uplift to 5d: M2 branes
in Taub-NUT + flux (see below).
  \label{onion} }

Halo configurations have walls of marginal stability and only exist
for a certain range of values of $\CF$. Applying the stability
condition (\ref{eqsepcond}) to a two-particle system of total charge
$\Gamma=e^S \Gamma(\beta,n)$, consisting of a core of charge
$\Gamma_c=e^S \Gamma(\beta_c,n_c)$ around which a D2-D0 halo of
charge $\Gamma_h = e^S(-\beta_h + n_h \omega)$ is orbiting, with
$S$, $B$ and $J$ as in (\ref{BJSF}), we find
\begin{equation} \label{cloudstab}
 - n_h \left( 2 (P \cdot \beta_h) \left( f(f^2 + y^2) - \frac{3 \, n}{P^3} \right)
 +n_h \left( y^2-3 f^2  + \frac{6 \, P \cdot \beta}{P^3} \right) \right) > 0.
\end{equation}
Asymptotically for $y \to \infty$ this becomes
\begin{equation} \label{largeradcloudstab}
 - n_h \left( 2 (P \cdot \beta_h) f
 +n_h \right) > 0,
\end{equation}
so there is a marginal stability line running all the way to
infinity, at $x = s + n_h / 2 (P \cdot \beta_h)$. Actually for this
to be a true marginal stability line where the phases of the halo
and core central charges align, we also need $\beta_h \cdot P > 0$,
as can be seen by examining the asymptotic behavior of the central
charges for $J \to \infty$.

Note in particular that (\ref{largeradcloudstab}) implies that at
$f=0$ (and $y \to \infty$), there are \emph{never} such BPS states,
while for $f \to - \infty$, all $n_h > 0$ states become stable,
while all $n_h < 0$ states become stable in the opposite regime $f
\to + \infty$. In fact the latter is also true at finite values of
$y$, as is easily seen from (\ref{cloudstab}).

When we add more D2-D0 particles, we should in principle use the
more general multicentered stability conditions
(\ref{centerconstraints}). Since the mutual intersection products
between the D2-D0 particles are zero, this effectively boils down
again to the 2-centered stability condition we used to obtain
(\ref{cloudstab}) for each individual particle in the halo, with
$e^S(-\beta_h+n_h \omega)$ the halo particle charge considered and
$e^S\Gamma(\beta,n)$ the total charge of the system.

Note though that for $y\to \infty$ the $(\beta,n)$ dependence in
 (\ref{largeradcloudstab}) drops out so the
  stability conditions are given by a set of simple,
independent constraints. Similarly, independent of which halo
particles are present, it will always remain true that for $x \to
+\infty$, $n_h > 0$ halos are stabilized, while for $x \to -
\infty$, $n_h<0$ halos are stabilized. Hence we see that the large
radius spectrum of D6-D2-D0 bound states has total D0-brane charge
$n$ unbounded above at $x \to + \infty$, while $n$ is unbounded
below at $x \to - \infty$. The former is characteristic for ideal
sheaves, while the latter is characteristic for their duals. (Recall
eq. \ref{idealsheafcharges}.) This supports our claim above.

One could ask how exactly the index of BPS states
$\Omega(\Gamma)|_{B+i J}$ changes when we change $\CF$. This is most
easily described by considering  the generating function
\begin{equation} \label{ZD6D2D0def}
 \CZ_{\rm D6-D2-D0}(u,v;B+iJ) := \sum_{\beta,n}
 \Omega(\Gamma(\beta,n))|_{B+i J}
 \,
 \, u^n \, v^{\beta}
\end{equation}
where $v^\beta \equiv \prod_A (v_A)^{\beta_A}$ and we take
$\Gamma(\beta,n)$ as in (\ref{Sparrep}) (so here $\CF = -B$). We
will in particular be interested in the case $B + i J = (x+ i y) P$,
with $P$ at this point an arbitrary auxiliary class inside the
K\"ahler cone.

The generating function $\CZ_{\rm D6-D2-D0}(u,v;(x+i y)P)$ will jump
whenever $x+i y$ is changed such that (\ref{cloudstab}) goes from
not being satisfied to being satisfied (or vice versa) for some
$(\beta_h,n_h)$. This adds states consisting of an
arbitrary\footnote{Note that (\ref{cloudstab}) is invariant under
$(\beta,n) \to (\beta,n) + k (\beta_h,n_h)$, as it should, so the
number of particles of charge $(\beta_h,n_h)$ we add does not matter
for stability.} number of D2-D0 particles (consistent with the
exclusion principle of course if the particles are fermions),
leading, following a reasoning similar to the derivation of the
D0-halo degeneracy to a jump
\begin{equation}
 \CZ_{D6-D2-D0} \to (1-(-u)^{n_h} v^{\beta_h})^{|n_h| N_{\beta_h}} \,
 \CZ_{D6-D2-D0},
\end{equation}
with $N_{\beta_h} = n^0_{\beta_h}$ as in (\ref{Nqdef}) and
(\ref{NGVrel}). Analogous to the D6-D0 system, the factor $|n_h|$
comes from the Landau degeneracy of the D2-D0 particle in the D6
background, since the intersection product equals $|n_h|$. Using
this, and recalling that at $x \to + \infty$ all $n_h>0$ halos are
stabilized, while at $x=0$, $y = \infty$ none are, we can write
\begin{eqnarray}
 \lim_{x \to +\infty} \lim_{y \to +\infty} \frac{\CZ_{D6-D2-D0}(u,v;(x+iy)P)}{\CZ_{D6-D2-D0}(u,v;i y P)}
 &=&  \prod_{\beta_h > 0,n_h>0}
 (1-(-u)^{n_h} v^{\beta_h})^{n_h n^0_{\beta_h}}
 \label{prodformdeg} \\
 &=& \CZ_{DT}^{\prime,r=0}(u,v).
\end{eqnarray}
When the limits are interchanged, there is an additional factor for
$\beta_h=0$ given by (\ref{macmahonfactor}), corresponding to
D0-halos.

Comparing to (\ref{ZDTZDT1})-(\ref{ZDTZDT5}), we see that our
proposed identification of
$$\lim_{x \to + \infty} \lim_{y \to
+\infty} \CZ_{D6-D2-D0}(u,v;(x+iy)P)$$ with $\CZ'_{DT}(u,v)$ is
valid provided
\begin{equation} \label{coredeg}
 \lim_{y \to \infty} \CZ_{D6-D2-D0}(u,v;i y P) =
 \CZ^{\prime,r>0}_{DT}(u,v).
\end{equation}
Note that this is manifestly invariant under $u \to u^{-1}$ (i.e.\
inversion of D0-charge) and has a finite range of D0-charge for
fixed $\beta_h$.  Similarly, if we require our proposed
identification in the opposite regime, namely $\lim_{x \to -\infty}
\lim_{y \to +\infty} \CZ_{D6-D2-D0}(u,v;(x+iy)P) =
\CZ'_{DT}(u^{-1},v)$, to hold, we find again (\ref{coredeg}).

In fact, for our analysis below, we will need to refine these
statements. There are four distinct ways of taking a limit to
infinity, due to the fact that D6D0-type lines of marginal stability
go all the way to infinity, being asymptotically of the form $z :=
x+iy= \lambda e^{2\pi i/3}$ and $z=\lambda e^{\pi i/3}$ with
$\lambda \to  + \infty$. Correspondingly, we distinguish the limits
$z \to \CL^+$, along a line infinitesimally above $z= \lambda
e^{2\pi i/3}$, and $z\to \CL^-$, along a line infinitesimally below
$z= \lambda e^{2\pi i/3}$. Similarly, $z\to \CR^\pm$ is the limit
going infinitesimally above (below) the line $z=\lambda e^{\pi
i/3}$. Now we have
\begin{eqnarray}
\lim_{z \to \CL^- } \CZ_{D6-D2-D0}(u,v;z P) &  = &  \CZ^{
}_{DT}(u^{-1},v) \label{fourinfinity1}\\
\lim_{z \to \CL^+ } \CZ_{D6-D2-D0}(u,v;z P) & = &  \CZ^{\prime
}_{DT}(u^{-1},v) \label{fourinfinity2} \\
\lim_{z \to \CR^- } \CZ_{D6-D2-D0}(u,v;z P) &  = &  \CZ^{
}_{DT}(u,v) \label{fourinfinity3} \\
\lim_{z \to \CR^+} \CZ_{D6-D2-D0}(u,v;z P) &  = &  \CZ^{\prime
}_{DT}(u,v) \label{fourinfinity4}
\end{eqnarray}

Finally, we sketch how to establish the proposed identification by
uplifting to M-theory, refining the analysis of
\cite{Dijkgraaf:2006um}. The lift of D6-D4-D2-D0 bound states to
M-theory is given by M2 branes in Taub-NUT (deformed by the flux)
times the Calabi-Yau, with the D0-charge corresponding to the
$U(1)_L$ isometry along the Taub-NUT circle. This is sketched in
fig.\ \ref{onion}$b$. Turning on wordvolume flux $F=S=s P>0$ on the
D6 corresponds to turning on an M-theory 4-form flux $G = s P \wedge
\omega_{TN}$, where $\omega_{TN}$ is the harmonic self-dual 2-form
on Taub-NUT. In the absence of this flux, the M2 branes must sit at
the center of Taub-NUT to be BPS. This corrseponds to the core
states. Now when we turn on the magnetic flux, the M2 branes get
access to a number of lowest Landau level states, carrying
increasing $U(1)_L$ charge. These will in general be localized at a
nonzero distance from the center of Taub-NUT, and thus correspond to
the halo states in four dimensions. However, for a finite total
integrated flux, there is a bound on the number of lowest Landau
levels that fit in the Taub-NUT space --- beyond this cutoff, the
equilibrium location of the M2 branes runs off to radial infinity.
The number of lowest Landau levels that do fit in the Taub-NUT space
is proportional to the flux, and this leads to the dependence of the
BPS spectrum on the flux derived above in the IIA picture. The limit
$|s| \to \infty$ corresponds to infinite total integrated flux,
which removes the bound on allowed lowest Landau levels. If at the
same time we let the Taub-NUT radius go to infinity, we end up with
a constant, arbitrarily small flux density $G \sim i(dz_1 \wedge
d\bar{z}_1 + dz_2 \wedge d\bar{z}_2) \wedge P$ on flat $\IC^2 \times
X$, which is indeed the background implicitly used in
\cite{Dijkgraaf:2006um} to show that M2 BPS states are counted by
$\CZ_{top} = \CZ_{GV} = \CZ_{DT}$. Therefore, in the $|s| \to
\infty$ regime --- and only in this regime --- the derivation of
\cite{Dijkgraaf:2006um} goes through, proving our claim.

\subsubsection{Core states   }
\label{sec:corestates}

In the previous section we were  led to interpret the factor
$\CZ^{\prime,r=0}_{DT}(u,v)$ as counting halo states. It is natural
to wonder about the physical interpretation of the remaining,
non-halo states counted by $\CZ^{\prime,r>0}_{DT}(u,v)$ in
(\ref{coredeg}). We will refer to these states as \textit{core
states}. Core states are characterized by the absence of marginal
stability walls extending to infinite radius.

One immediate consequence is that at sufficiently large background
$J$, these states are stable for all values of the $B$-field. Single
centered black holes are of course the simplest example, but
multicentered configurations are also possible as we will show
below. In this case the centers can be squeezed arbitrarily close
together (at least in coordinate distance) by taking $J$
sufficiently large. These states can subsequently be ``dressed''
with the D2-D0 halos described in the previous subsection, which
even at infinite $J$ can be given arbitrarily large radius by tuning
the $B$-field close to the wall of marginal stability. This
justifies the names core and halo states.

Another consequence is that core states, unlike halo states, have
degeneracies at $J \to \infty$ symmetric under inversion of
D0-charge. This is a result of combining the $\Gamma \to \Gamma^*$,
$B \to - B$ symmetry described in section \ref{sec:symmetries} with
the absence of walls of marginal stability for $J \to \infty$. This
is in agreement with the fact that $\CZ_{\rm DT}^{\prime,r>0}$ is
invariant under $u \to u^{-1}$, while $\CZ_{\rm DT}^{\prime,r=0}$ is
not.

Note that multicentered bound states of charges which all have
nonzero magnetic (D4 or D6) charges are always core states. In other
words halos with lines of marginal stability extending to infinite
radius can never contain magnetic charge. For halo particles with
nonzero D4 but zero D6 charge this follows from the fact that since
the total charge $\Gamma$ must have D6-charge 1 (by assumption), at
$J=\infty$ we have $Z(\Gamma) \sim -i J^3$ imaginary while $Z(D4)
\sim P \cdot J^2$ is real, so there cannot possibly be a wall of
marginal stability for splitting off the D4 extending all the way to
$J=\infty$. For halo particles with negative D6 charge centers the
reasoning is similar: now the central charges both are imaginary at
infinity, but with opposite phases. For halo particles with positive
D6 charge the complement has necessarily negative or zero D6 charge,
so the previous reasoning can be applied to the central charge of
the complement. Only when the complement has nothing but D2-D0
charge (with $P\cdot Q_{D2} < 0$) is there  a wall of marginal
stability which extends to infinity, but in this case the original
center of course corresponds again to a core state, the complement
being a D2-D0 particle or halo orbiting around it.

An example of a nontrivial core state was in fact already discussed
above in section \ref{sec:D6D2D0asthreenode}. We will now examine
another class of core states, which we will use to construct ``swing
states'' in section \ref{innocuouscoredump}. These will play an
important role in delimiting the region of validity of the OSV
conjecture. For this reason, we will give a detailed stability
analysis of this class.

We consider bound states of a pure fluxed D6 with a D4-D2-D0 black
hole, such that the total charge $\Gamma$ has no D4 charge, that is
we take (neglecting the $c_2/24$ correction)
\begin{equation}
 \Gamma=\Gamma(\beta_1,n_1)=1-\beta_1   + n_1 \, \omega
\end{equation}
constructed as a bound state
\begin{equation}\label{newcore}
\Gamma = \tilde \Gamma + \tilde \Gamma'
\end{equation}
with
\begin{eqnarray}
\tilde \Gamma = e^{-U-V}(U + q_0 \, \omega) \qquad &\& & \qquad
\tilde \Gamma' = e^{-U}
\end{eqnarray}
so
\begin{eqnarray}
\beta_1 = UV + \half U^2 \qquad & \&  & \qquad  n_1 =  \half U
(U+V)^2 - \frac{1}{6}U^3 + q_0 \, \omega.
\end{eqnarray}

To be more specific, we assume that $U,V$ are positive divisors,
which for simplicity we take to be proportional to $P$. This allows
us to restrict the moduli to the complex plane $B+i J = z \, P$,
rendering the problem effectively one dimensional. It is furthermore
convenient to define the notation $U= u P, V=v P , q_0 = \tilde q_0
P^3$. We will assume that $\tilde \Gamma$ is realized as a single
centered black hole, which amounts to taking $\tilde q_0<0$. We will
more specifically be interested in cases with small $u$ and $v$ of
order 1, with $|\tilde q_0|$ sufficiently small so
\begin{equation}\label{stc2}
\half u v^2 + \tilde q_0>0.
\end{equation}
An example is shown in fig.\ \ref{core_examples}$(a)$.

\EPSFIGURE{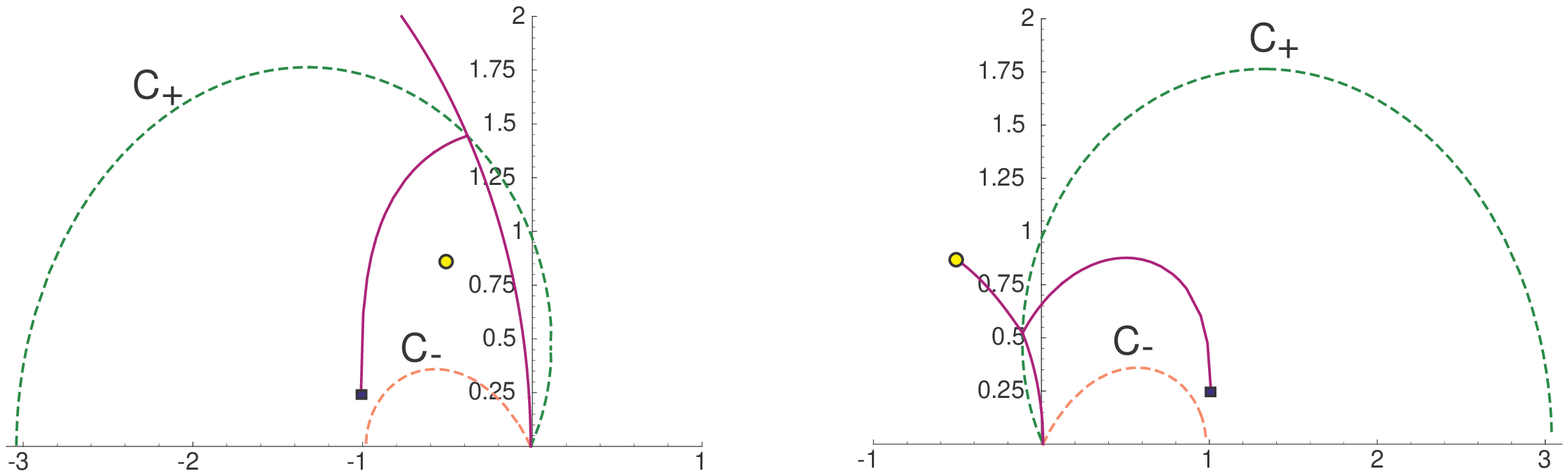,height=5cm,angle=0,trim=0 0 0 0}%
{{\bf Left:} Plot in the $z$-plane of the split flow described in
the text with $u=10^{-2}$, $v=1$, $q_0=-10^{-4}$, $z_{\infty} = 3
e^{2 i \pi/3}$. The green dotted line labeled $C_+$ is the ms line,
the pink dotted line labeled $C_-$ the anti-ms line. Note that at
$z_{\infty}=e^{2 i \pi/3}$ (yellow dot), the bound state does not
exist. {\bf Right:} Split flow for dual charges obtained by setting
$u=-10^{-2}$, $v=-1$, $q_0=10^{-4}$, with $z_{\infty} = e^{2 \pi
i/3}$. We do not analyze such negative $u,v$ cases in detail in the
text because unlike the positive $(u,v)$ cases, their stability for
$z_{\infty}$ near $e^{2 i \pi/3}$ is guaranteed, making them less of
an issue in the analysis in the following sections.
  \label{core_examples}}

Let us analyze for what values of $z_\infty$  this state exists.
Since $\tilde \Gamma$ is a single centered black hole and $\tilde
\Gamma'$ is just a pure $D6$ with flux, the two constituents are
guaranteed to exist everywhere in moduli space. Therefore the
stability analysis reduces to an analysis of the existence and
location of a line of marginal stability for the split flow $\Gamma
\to \tilde \Gamma + \tilde \Gamma'$.

The total discriminant is
\begin{equation}
 8 u^3 (v + u/2)^3- 9 (uv^2/2+u^2 v +
u^3/3+\tilde q_0)^2
 \end{equation}
up to a positive coefficient of proportionality. Hence, when $v$ is
of order $1$ and $u$ is small, $\CD(\Gamma)$ will be negative. This
means there will be a zero of the central charge, which should be
reached by the $\Gamma$ attractor flow only after crossing a
marginal stability line if we want the split flow to exist.

Now we consider the stability condition (\ref{eqsepcond}). Define
$\hat z := z + u$  so that the central charges are:
\begin{eqnarray}
Z(\tilde \Gamma;B+i J) = - \half u ( (\hat z+v)^2 + \frac{2\tilde
q_0}{u}) P^3\qquad & \& & \qquad  Z(\tilde \Gamma';B+i J) =
\frac{\hat z^3}{6}P^3
\end{eqnarray}
so  the stability condition becomes
\begin{equation}\label{stc1}
- \bigl( \half u v^2 + \tilde q_0\bigr) \, {\rm Im}\biggl[
\bigl(\half u (\hat z+v)^2 + \tilde q_0\bigr) \bar{\hat
z}^3\biggr]>0
\end{equation}
The marginal stability curve
\begin{equation}
{\rm Im}\biggl[(\half u (\hat z+v)^2 + \tilde q_0) \bar {\hat
z}^3\biggr]=0
\end{equation}
can be written as:
$$
(x^2+y^2)^2 + 4 v (x^2+y^2)x + (v^2+ \frac{2\tilde q_0}{u})
(3x^2-y^2) =0 $$
where $\hat{z}=:x+iy$.  The solution is the $x$-axis together with
two bounded components roughly of the shape of a cardiode with the
tip at the origin $\hat z=0$. Writing $\hat z=re^{i \theta}$ they
are given by
\begin{equation}\label{twobr}
r = - 2 v \cos(\theta) \pm \sqrt{v^2 + \frac{2 \tilde
q_0}{u}(1-4\cos^2(\theta))}
\end{equation}
Call the plus branch $\CC_+$ and the minus branch $\CC_-$.

Under the condition (\ref{stc2}) we find that (in the upper half
plane) $\CC_+$ is swept out   $\frac{\pi}{3}\leq \theta \leq \pi$
and $\CC_-$ is swept out   for $\frac{2\pi}{3}\leq \theta \leq \pi$.
Clearly $\CC_+$ encloses $\CC_-$ and they only intersect at the
origin.   The region at $r \to \infty$ is a region of stability.

Under our conditions $\CC_+$  is indeed a line of positive marginal
stability and $\CC_-$  is a line of anti-marginal stability, and in
fact the entire inside region of $\CC_+$ is a region of instability.

(To prove the above statement, we note that the lines ${\rm Im }
Z(\tilde \Gamma) \overline{ Z(\tilde \Gamma')}$ and  ${\rm Re }
Z(\tilde \Gamma) \overline{ Z(\tilde \Gamma')}$ can only intersect
when the product of central charges is zero, that is, at $\hat z=0$
or at $\hat z = -v\pm \sqrt{\frac{-2\tilde q_0}{u}}$. However the
curve ${\rm Re } Z(\tilde \Gamma) \overline{ Z(\tilde \Gamma')}=0$
intersects the $x$ axis only at $x=y=0$ and at   $\hat z = -v\pm
\sqrt{\frac{-2\tilde q_0}{u}}$. Now   $\CC_+$  intersects the $x$
axis at $x_+ =-2v - \sqrt{v^2- \frac{6\tilde q_0}{u}}$ and since
\begin{equation}
x_+ <-v - \sqrt{\frac{-2\tilde q_0}{u}}
\end{equation}
it follows that the change of sign of ${\rm Re } Z(\tilde \Gamma)
\overline{ Z(\tilde \Gamma')}$ happens inside the region enclosed by
$\CC_+$. Finally,   note that up to a positive coefficient ${\rm Re
} Z(\tilde \Gamma) \overline{ Z(\tilde \Gamma')}$ is given by
\begin{equation}
- \biggl[ (x^2+y^2)^2 x + 2v (x^4-y^4) + (v^2+ \frac{2\tilde
q_0}{u})(x^3-3xy^2) \biggr]
\end{equation}
and hence equals
$$
- \half u r^5 \cos(\theta) +\CO(r^4)
$$
for large $r$,  so clearly if $\theta> \pi/2$, $r\to \infty$ the
quantity is positive.)

It follows that if $z_\infty$ is outside the compact region enclosed
by $\CC_+$ then the split state does exist: The zero of $Z(\Gamma;z
P)$ lies on the antimarginal stability curve $\CC_-$ which is
contained within $\CC_+$. The attractor flow heads toward this zero,
and splits on the line $\CC_+$. On the other hand, if $z_\infty$ is
inside the curve $\CC_+$ then the split state does not exist.

Note that the M-theory uplift of such a 2-centered configuration is
a black ring orbiting the center of a Taub-NUT space with flux,
obtained from wrapping an M5 around a divisor $U$ and the Taub-NUT
circle and giving it some momentum $q_0$ around the Taub-NUT circle
\cite{Gaiotto:2005xt,Bena:2005ni,Elvang:2005sa}. The fact that the
discriminant of the total charge is negative means that this charge
cannot be realized as a BMPV black hole in Taub-NUT.

\subsection{D6-anti-D6 degeneracies} \label{sec:D6antiD6deg}

\subsubsection{Spectrum and flow trees} \label{sec:specandflow}

We now turn to our main goal, namely computing degeneracies of polar
D4-D2-D0 BPS states represented as D6-anti-D6 bound states. The
attractor flow trees corresponding to those can be as simple as
fig.\ \ref{pants}$a$ or as complex as fig.\ \ref{fr}, but in any
case, the first split will be into a pair of charges with D6-charges
$r$ and $-r$ with $r>0$.\footnote{Recall that the case $r=0$ was
excluded by (\ref{NoD4splitoff}).} The case $r=1$ will turn out to
be the most important one, so let us consider pairs of charges
$\Gamma_1$ and $\Gamma_2$, parametrized as in
(\ref{eSparcharge})-(\ref{eSparcharge2}), i.e.\
\begin{eqnarray}
 \Gamma_1 &=& e^{S_1} \Gamma(\beta_1,n_1) = e^{S_1}(1-\beta_1' + n_1 \, \omega),
 \qquad \beta_1' := \beta_1 - \mbox{$\frac{c_2}{24}$} \label{Gammaione} \\
 \Gamma_2 &=& - e^{S_2} \Gamma(\beta_2,n_2) = - e^{S_2}(1-\beta_2' + n_2 \,
 \omega), \qquad \beta_2' := \beta_2 - \mbox{$\frac{c_2}{24}$} \label{Gammaitwo}
\end{eqnarray}
chosen such that the total magnetic charge $P=S_1-S_2$ is fixed at
some large value inside the K\"ahler cone. The intersection product
is $\langle \Gamma_2,\Gamma_1 \rangle = P^3/6 -P \cdot (\beta_1' +
\beta_2') + n_1 - n_2 =  I_P -P \cdot (\beta_1 + \beta_2) + n_1 -
n_2$. In the large radius limit $J \to \infty$, the stability
condition (\ref{eqsepcond}) simply amounts to
\begin{equation} \label{D6antiD6stabcond1}
 \langle \Gamma_2,\Gamma_1 \rangle  = \frac{P^3}{6} -P \cdot (\beta_1' +
\beta_2') + n_1 - n_2 > 0.
\end{equation}
In the limit $P \to \infty$, $n_i$, $\beta_i$ fixed, this is
automatically satisfied. Recall however that this is only a
necessary, not a sufficient condition for existence.

When $n_i,\beta_i'=0$, we have essentially the extremal case
$\tn=\tb=0$ in the class of examples studied section
\ref{sec:classexamples}. Indeed we saw there that in this case these
always exist as  bound states, and that this remains true for small
perturbations away from $\tn=\tb=0$ as long as the charges
$\Gamma_1$ and $\Gamma_2$ support BPS states (see fig.\
\ref{stabregion}).

Let us be more precise. To establish the existence of a D4-D2-D0
bound state at large radius, it is sufficient to establish this at a
conveniently chosen value of the $B$-field, since we know the
infinite radius limit of the D4-D2-D0 spectrum is invariant under
shifts of $B$.

We will take this value to be $\tilde{B}=0$ after making the
uniformizing change of variables $B \to \tilde{B}$
(\ref{covBtrick}). In the case at hand
\begin{equation} \label{Bshifvalue}
 B = \frac{S_1+S_2}{2} + D_A D^{AB} \Delta \beta_B + \tilde{B}, \qquad
 \Delta \beta \equiv \beta_2 - \beta_1
\end{equation}
which  puts the total central charge in the form
(\ref{Zholaftercov}). Using this it is straightforward to show that
the attractor flow of the total charge $\Gamma$ starting at
$\tilde{B}=0$ will remain at $\tilde{B}=0$ and run straight down
till it crashes on a zero of the central charge at $J_0 = \sqrt{2
\widehat{q}_0/P^3} \, P$, where
\begin{eqnarray}
 \widehat{q}_0 &:=& q_0 - \frac{1}{2} D^{AB} q_A q_B =
 \frac{P^3}{24} - \frac{1}{2} P (\beta_1' + \beta_2') + n_1 - n_2 -
 \frac{1}{2} (\Delta \beta)^2 > 0, \label{hatq0expr}
\end{eqnarray}
and $(\Delta \beta)^2$ is defined with the $D^{AB}$ metric. If we
moreover take the initial point of the flow at $J_\infty = y_\infty
P$, the flow will simply be given by $J = y \, P$, where $y$ runs
down from $y_\infty$ to $y_0 = \sqrt{2 \widehat{q}_0/P^3}$. Hence
our choice of $\tilde{B}=0$ corresponds to a line to which attractor
flows coming in from large radius converge, thus making it a
particularly natural choice to make.

In order for the first split $\Gamma \to \Gamma_1 + \Gamma_2$ of the
flow tree to exist, a wall of marginal stability must be met before
the attractor flow hits $y_0$. Since the total central charge is
real along the flow, this wall must be at a solution $y$ of $\Im
Z(\Gamma_1)|_y = \Im Z(\Gamma_2)|_y = 0$. Thus the split point is
given by
\begin{equation} \label{splitpointvalue}
  t_{\rm ms}=(B+i J)_{\rm ms} = \frac{S_1+S_2}{2} +D_A D^{AB} \Delta \beta_B +  iy_{\rm ms} P.
\end{equation}
where we choose the unique positive root:
\begin{equation} \label{ymsval}
 y_{\rm ms} = \frac{1}{\sqrt{P^3}} \sqrt{\frac{3 P^3}{4} - 3 P(\beta_1'+\beta_2') + 3 (\Delta
 \beta)^2}.
\end{equation}
Note that   the argument of the square root is positive. To have
$y_{\rm ms} > y_0$, we thus need
\begin{equation}  \label{D6antiD6stabcond2}
 \frac{3 P^3}{8} - \frac{3}{2} P(\beta_1'+\beta_2') + \frac{3}{2} (\Delta
 \beta)^2 > \widehat{q}_0.
\end{equation}
Note that this again automatically satisfied when $P \to \infty$ at
fixed $n_i$, $\beta_i$.

This condition is still not quite enough however, since it is not
enough for the $Z(\Gamma_i)$ to be real to have a true marginal
stability wall at $y=y_{\rm ms}$ --- they must have the same sign as
well. In fact this sign must be positive since the total central
charge $Z$ is positive in the limit $y \to \infty$ and remains so
till it hits zero. This gives the somewhat complicated conditions
\begin{eqnarray}
 Z_1|_{\rm ms} &=& \frac{P^3}{6} + \frac{3}{2} \frac{P \Delta \beta}{P^3} \left(
  P(\beta_1'+\beta_2') - (\Delta \beta)^2
 \right) \nonumber \\
 && - P \beta_2' + (\Delta \beta)^2 - \frac{1}{2} (\beta_1'+\beta_2')
 \Delta \beta + \frac{(\Delta \beta)^3}{6} - n_1 > 0 \label{D6antiD6stabcond3} \\
 Z_2|_{\rm ms} &=& \frac{P^3}{6} - \frac{3}{2} \frac{P \Delta \beta}{P^3} \left(
  P(\beta_1'+\beta_2') - (\Delta \beta)^2
 \right) \nonumber \\
 && - P \beta_1' + (\Delta \beta)^2 + \frac{1}{2} (\beta_1'+\beta_2')
 \Delta \beta - \frac{(\Delta \beta)^3}{6} + n_2 > 0. \label{D6antiD6stabcond4}
\end{eqnarray}
Here $(\Delta \beta)^3 := D_{ABC} (\Delta \beta)^A (\Delta \beta)^B
(\Delta \beta)^C$, with $(\Delta \beta)^A := D^{AB} (\Delta
\beta)_B$. Again, these conditions are automatically satisfied when
$P \to \infty$ at fixed $\beta_i$, $n_i$.

In summary, when $\widehat{q}_0>0$ the conditions for the split flow
$\Gamma \to \Gamma_1 + \Gamma_2$ to exist are given by the
inequalities (\ref{D6antiD6stabcond2}), (\ref{D6antiD6stabcond3})
and (\ref{D6antiD6stabcond4}).\footnote{It can be checked easily
that when $\widehat{q}_0 > 0$, (\ref{D6antiD6stabcond2}) actually
implies (\ref{D6antiD6stabcond1}). Given the other two inequalities,
one can also replace $\widehat{q}_0$ by 0 on the right hand side of
(\ref{D6antiD6stabcond2}), since existence of $y_{\rm ms}$ and
positivity of $\Re \, Z$ there imply that $y_{\rm ms} > y_0$, as
$\Re \, Z$ is positive at $y=\infty$ and changes sign at $y=y_0$.}
To correspond to an actual BPS bound state, we furthermore need that
$\Gamma_1$ and $\Gamma_2$ each support BPS states at $y=y_{\rm ms}$.
This is straightforward if $\Gamma_1$ and $\Gamma_2$ are realized as
single attractor flows, but becomes again nontrivial when these
charges themselves correspond to split flows: this is one of the
main technical difficulties we face.

\EPSFIGURE{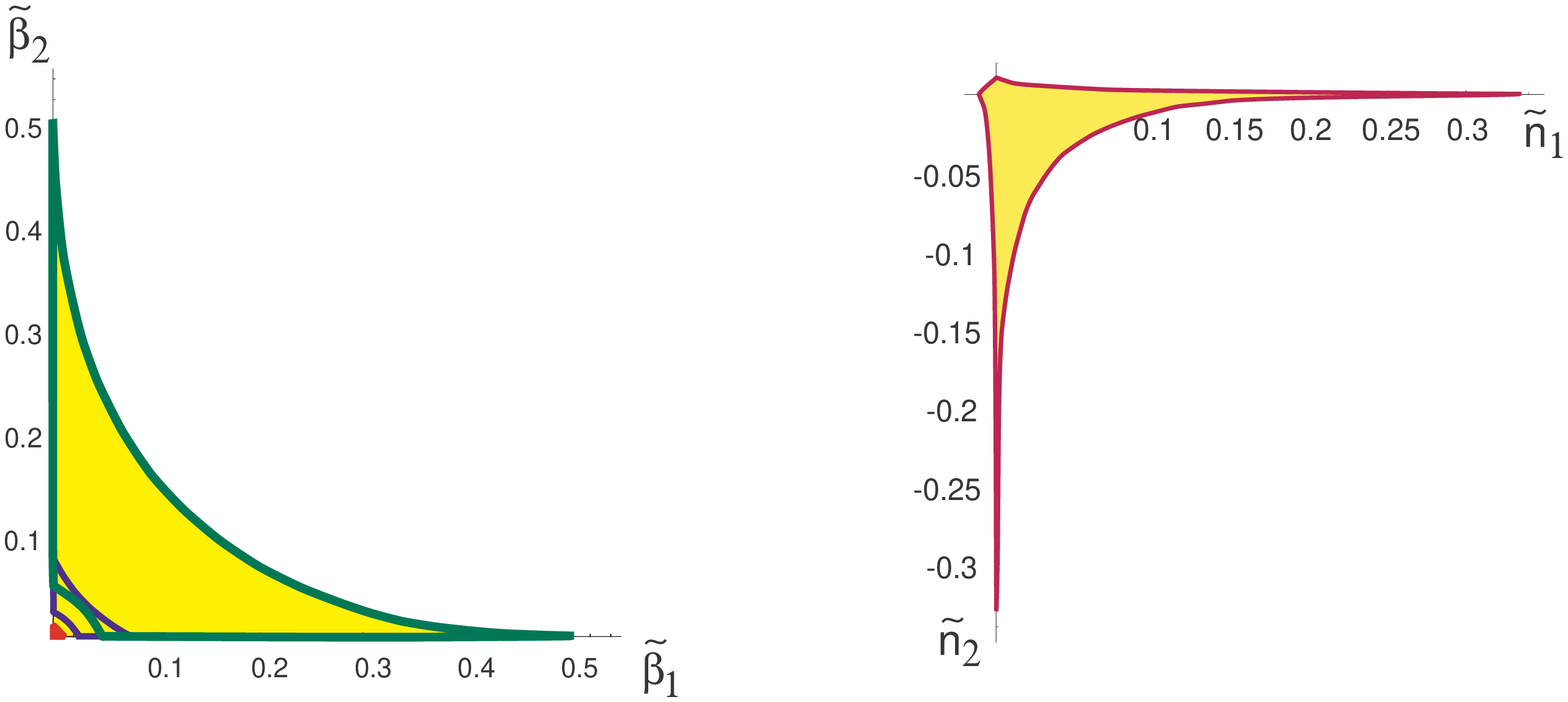,height=6cm,angle=0,trim=0 0 0 0}%
{{\bf Left:} The yellow shaded area is the projection of $\CS[0,1]$
into the $(\tb_1,\tb_2)$-plane. The green outline is the projection
of $\CS[.9,1]$, the (smaller) blue one of $\CS[.4,.5]$ and the
(smallest) red one of $\CS[0,.1]$. {\bf Right:} Projection of
$\CS[0,1]$ into the $(\tn_1,\tn_2)$-plane.
  \label{b1b2range}}

To get a feeling for the implications of these conditions let us
consider the simplest example: We take a CY with one K\"ahler
modulus (for example the quintic CY) and suppose that $\Gamma_1$ and
$\Gamma_2$ support single center attractor flows.   Parametrizing
$\beta'_i =: \tb_i P^2$, $n_i \omega=: \tn_i P^3$, $\widehat{q}_0 =:
(1-\eta) P^3/24$, the $P$-dependence scales out of all inequalities
(\ref{D6antiD6stabcond1})-(\ref{D6antiD6stabcond4}), while the
condition (\ref{attrexistence}) for existence of the regular
attractor points for $\Gamma_i$
  becomes
\begin{equation} \label{BHstabcondition}
 \tb_i \geq 0, \qquad 8 \tb_i^3 - 9 \tn_i^2 \geq 0.
\end{equation}
Since the system of inequalities is rather complicated we scanned the solution
spaces
\begin{equation}
 \CS[a,b] := \{ (\tb_1,\tb_2,\tn_1,\tn_2) | \mbox{ 2-centered solution exists with }
  a \leq 1-\eta \leq b \}
\end{equation}
numerically for various intervals $[a,b] \subseteq [0,1]$. Fig.\
\ref{b1b2range} shows the corresponding projections to the
$(\tb_1,\tb_2)$- and $(\tn_1,\tn_2)$-planes.

One thing that transpires from this analysis which is not
immediately obvious from the inequalities, although expected
physically, is that, taking into account charge quantization, the
solution space is finite. It is furthermore clear from the plots
that the solution space does not factorize, in the sense that the
choice of $(\tb_1,\tn_1)$ influences the stability domain of
$(\tb_2,\tn_2)$. Another distinct feature is the correlation between
the size of $(\tb_i,\tn_i)$ and $\eta$: the more polar the state is,
i.e.\ the closer $\eta$ approaches 0, the smaller $\tb_i$ and
$\tn_i$ are forced to be. For $\tb_i, \tn_i \ll 1$ this is not hard
to deduce analytically in the case at hand. When $\tb_i, \tn_i \ll
1$, all the required inequalities are automatically satisfied,
except (\ref{BHstabcondition}), which remains nontrivial. The
relation between $\eta$ and the $(\tb_i,\tn_i)$ is given by
\begin{equation}
 \eta = \frac{1}{2} (\tb_1 + \tb_2) - \tn_1 + \tn_2 + \frac{1}{2} (\tb_1 -
 \tb_2)^2. \label{etarelation}
\end{equation}
When $\tb_i \ll 1$, the term quadratic in the $\tb_i$ is negligible
compared the term linear in the $\tb_i$, and $\tn_1 - \tn_2$ as well
because of (\ref{BHstabcondition}). Thus we get the simple relation
$\tb_1 + \tb_2 = 2\eta$, $\tb_i > 0$, $3 \tn_i < (2 \tb_i)^{3/2}$,
making it obvious that $\tb_i$ and $\tn_i$ get smaller when
$\widehat{q}_0$ approaches its maximum, i.e.\ $\eta \to
0$.\footnote{Actually the maximal value of $\widehat{q}_0$ is
$(P^3+c_2 \cdot P)/24$, but in the $P \to \infty$ supergravity
regime we have in mind here, the linear correction is negligible.}

In section \ref{sec:extremepolar} we will conjecture that the
behavior exhibited in this example   persists in the general case as
well, namely, that the most polar states correspond to $\beta_i,n_i$
which are in some sense small compared to the scales set by $P$.

We are thus interested in charges in which the $\beta_i,n_i$ are
``small'' compared to the scales set by $P$. In the language of
\cite{Gaiotto:2006ns,deBoer:2006vg}, these are ``dilute gas''
states, which in our setup can  be thought of microscopically  as
D2-D0 branes sparsely floating around inside the D6 and the anti-D6
branes. Let us now make this notion of dilute gas more precise.

Define  the   scale of $P$ by
\begin{equation}
 |P| := (P^3)^{1/3}.
\end{equation}
We will take it to be large, and in the OSV conjecture it will scale
to infinity. Next,  define a set of ``small'' $(\beta,n)$ as
follows\footnote{By $\epsilon_{|P|}$ we mean to indicate that
$\epsilon$ can be taken to depend on $|P|$, e.g.\ $\epsilon_{|P|}
\sim |P|^{-\xi}$, for $|P| \to \infty$. We will usually just write
$\epsilon$ though.}
\begin{equation} \label{CCdef}
 \CC(P,\epsilon) := \{ (\beta,n) \, |
  \, \beta \mbox{ effective}, \, \, \beta \cdot P
  < \epsilon_{|P|} \, |P|^3, \, \, |n| < \epsilon_{|P|} \, |P|^3
  \}.
\end{equation}
  Note that because $P$ is very ample and $\beta$
effective, we have in components with respect to a basis of the
K\"ahler cone that $\beta_A \geq 0$, $P^A > 0$, so the bound on
$\beta$ implies for each component
\begin{equation}
 \beta_A < \epsilon \, \frac{|P|}{P^A} \, |P|^2 \sim \CO(\epsilon
 |P|^2),
\end{equation}
where we used that $\frac{|P|}{P^A} \sim \CO(|P|^0)$ when we scale
up $P$ uniformly. Using this, it is clear that for
sufficiently small $\epsilon$ we have
\begin{equation}
  (\beta_1,n_1)\ \& \ (\beta_2,n_2) \in \CC(P,\epsilon) \Rightarrow  (\ref{D6antiD6stabcond2}),
(\ref{D6antiD6stabcond3}) \mbox{ and } (\ref{D6antiD6stabcond4})
\mbox{ are satisfied.} \label{outoflabelinspiration}
\end{equation}
and hence the split $\Gamma \to \Gamma_1 + \Gamma_2$ exists.

\subsubsection{The extreme polar state conjecture} \label{sec:extremepolar}

In the previous section we have examined a particular class of
examples in which a D4-D2-D0 BPS state is accounted for as a
D6-antiD6 split state. We are particularly, interested in polar
states. As we will see, the  ``more polar'' a state is - that is,
the larger the value of $\hat q_0$ -  the more important is the
contribution of that polar state to the OSV formula. Clearly, the
polar states analyzed in the previous section do not account for all
polar states, since they only involve D6 branes with $r=1$. In this
section  we will  state a conjecture which claims that nevertheless,
if we restrict attention to sufficiently polar states, then the
examples of the previous section are indeed the most general
examples. We will give some evidence for this conjecture.

We know that any polar state splits into charges
\begin{equation}
 \Gamma_1 = r e^{S_1}(1-\beta_1 + n_1 \, \omega), \quad \Gamma_2 = -r
e^{S_2}(1-\beta_2+n_2\,\omega),
\end{equation}
where $r(S_1-S_2)=P$. Subsequent splits can also occur, but here we
are only interested in the first split. Splits in charges with zero
D6 charge were excluded by (\ref{NoD4splitoff}).

For such a split, we have
\begin{equation}
 \widehat{q}_0 = r \left( \frac{\widehat{P}^3}{24} - \frac{1}{2}\widehat{P} \cdot
 (\beta_1+ \beta_2) + n_1 - n_2 - \frac{1}{2} (\Delta \beta)^2
 \right), \qquad \widehat{P}:=\frac{P}{r}
\end{equation}
where $(\Delta \beta)^2 = (D_{ABC}
 \widehat{P}^C)^{-1} \, (\beta_{1,A} - \beta_{2,A}) (\beta_{1,B} -
 \beta_{2,B})$.

We can introduce a measure of the degree of polarity of a  D4-D2-D0
BPS state by defining
\begin{equation}
\eta:=\frac{(\hat{q}_0)_{\rm max} - \hat{q}_0}{(\hat{q}_0)_{\rm
max}},
\end{equation}
where $(\hat{q}_0)_{\rm max} = \frac{P^3 + c_2P}{24}$. Throughout,
we will think of $|P|$ as being very large, though finite. Therefore
to good approximation, we can drop $c_2$ corrections, which for
simplicity we will do in what follows. We will define \emph{extreme
polar states} as those for which $\eta\ll 1$.  Then, we conjecture
that for sufficiently small $\eta<1$ there
 exists an $\epsilon(\eta)$ sufficiently small so
  that the   restricted class of D6-anti-D6
 bound states with charges drawn from the set  $\CC(P;\epsilon(\eta))$
 defined in (\ref{CCdef})  indeed
 account for all such extremely polar D4-D2-D0 BPS states.

It follows easily from the above formulae that  $r=1$ D6-anti-D6
bound states with $(\beta_i,n_i) \in \CC(P,\epsilon)$ have $\eta <
\CO(\epsilon)$, and hence are extreme polar for small $\epsilon$.
What we would like to know is the converse, namely that extreme
polar states are \emph{only} realized by splits with $r=1$ and small
$\beta_i,n_i$. More  precisely, we would like to prove the:

\bigskip \noindent {\bf Extreme polar state conjecture:} \vskip2mm

\noindent

a.) For any $\eta_* \ll 1$, there exists an $\epsilon(\eta_*) \ll 1$
such that every D4-D2-D0 BPS state with $\eta < \eta_*$ corresponds
to a split $(\Gamma_1,\Gamma_2)$ as in
(\ref{Gammaione})-(\ref{Gammaitwo}), with $(\beta_i,n_i) \in
\CC(P,\epsilon(\eta_*))$ as defined in (\ref{CCdef}).

b.) Moreover, there is a $P$-independent constant, $\mu$  so that we
may take $ \epsilon= \mu \eta_*$.

\bigskip

Some heuristic intuition for the absence of ``large''
$(\beta_i,n_i)$ contributions to extreme polar state realizations is
$(i)$ the complexity and entropy of D4-D2-D0 states increases with
$\eta$, and $(ii)$ the complexity and entropy of D6-D4-D2-D0 states
increases with the scale of $(\beta_i,n_i)$, so for $(\beta_i,n_i)$
too large, we would get a contribution with too much entropy. In the
one modulus examples in section \ref{sec:D6antiD6deg}, in particular
the discussion around (\ref{etarelation}), we saw this
proportionality relation between $\eta$ and the scale of
$\beta_i,n_i$ explicitly.

The absence of $r>1$ splits from the extreme polar spectrum is
perhaps more surprising at first sight, but becomes less so when one
notes that for $\beta_i=0$, $n_i=0$, $\hat{q}_0 = P^3/(24 r^2)$, so
$\eta = 1-1/r^2 \geq 3/4$ for $r \geq 2$. We also conjecture that
the latter is the maximal possible value for $\hat{q}_0$ at any
given value of $r$, reached iff $\beta_i=0$, $n_i=0$, i.e.\ for a
bound state of a pure $U(1)$ fluxed stack of D6-branes and a stack
of anti-D6 branes. The truth of this latter conjecture
  is not necessary for our derivation
of the OSV conjecture.

Unfortunately, we have not been able to find a full, general proof
of the extreme polar state conjecture. Within the class of $r=1$
bound states, the main problem is to find suitable bounds on the
positive contributions to $\hat{q}_0$ in (\ref{hatq0expr}), such
that cancelations between ``large'' values of the $\beta_i$, $n_i$
are avoided. It seems reasonable that such large cancelations are
absent, since just a slight change of such canceling large
parameters would transform an extreme polar state ($0<\eta \ll 1$)
into a super-polar state ($\eta<0$), which we know are absent.
However when looking in more detail, one finds that the bounds come
from many different existence criteria, and all of them, including
complicated inequalities like (\ref{D6antiD6stabcond3}) and
(\ref{D6antiD6stabcond4}) as well as various stability conditions
for constituent D6-D2-D0 bound states must be taken into account to
prevent such cancelations from happening.

 Let us nevertheless have a closer look at the conjecture. First
note that $\CC(P,\epsilon)$ is defined such that at fixed $\epsilon$
and ignoring charge quantization, $(\beta,n) \in \CC(P,\epsilon)$
iff $(\lambda^2 \beta,\lambda^3 n) \in \CC(\lambda P,\epsilon)$,
i.e.\ it respects the scaling symmetry (\ref{lefshetzscaling}), as
it should for the extreme polar state conjecture to make sense
(since this relates $\epsilon$ to $\eta_*$ which similarly is
invariant under the above rescalings).

It is not hard to show that the conjecture is indeed nontrivially
true
 for the important
special case of polar states splitting in two single centered black
holes with charges
\begin{equation} \label{ttttansz}
 \Gamma_1=r e^{S_1}(1-\beta_1 + n_1 \, \omega), \qquad \Gamma_2=- r
e^{S_2}(1-\beta_2 + n_2 \, \omega),
\end{equation}
where we take $\beta_1 = \beta_2 =: \beta$, such that we don't have
to worry about possible positive contributions from the $(\Delta
\beta)^2$ term in (\ref{hatq0expr}). For simplicity we will also
take $-n_1 = n_2 =: n$, but this can be easily generalized to $n_1
\neq n_2$.

We then get
\begin{equation}
 \widehat{q}_0 = r \left( \frac{\widehat{P}^3}{24} - \widehat{P} \cdot \beta -
 2n
 \right), \qquad \widehat{P}:=\frac{P}{r},
\end{equation}
and
\begin{equation}\label{etawithphat}
 \eta = (1-\frac{1}{r^2}) + \frac{24}{r^2} \frac{\widehat{P}\cdot\beta + 2 n}{\widehat{P}^3}.
\end{equation}
Furthermore the split existence conditions
(\ref{D6antiD6stabcond2}), (\ref{D6antiD6stabcond3}) and
(\ref{D6antiD6stabcond4}) reduce simply to\footnote{Although we
derived these stability inequalities only for the case $r=1$, this
is trivially extended to general $r$ by using the uniform scaling
symmetry (\ref{unifscaling}), under which $r \to \mu r$, $r(S_1-S_2)
= P \to \mu P$, and $\beta$, $n$ are invariant.}
\begin{equation}
 \frac{\widehat{P}^3}{6} - \widehat{P} \cdot \beta + n >0,
\end{equation}
while for the black hole constituents to exist,
(\ref{attrexistence}) must be satisfied, i.e.\
\begin{equation}
 8(Y^3)^2 - 9 \, n^2 \geq 0, \quad Y^2:=\beta, \quad Y > 0.
\end{equation}
Now a general inequality \footnote{Recently used in
\cite{Douglas:2006jp}.} for any triplet of divisors $X,Y,Z$ inside
the K\"ahler cone is \cite{lazarsfeld} (Theorem 1.6.1)  $X^3 \, Y^3
\, Z^3 \leq (X \cdot Y \cdot Z)^3$, so in particular we have
$(Y^3)^2 \widehat{P}^3 \leq (Y^2 \widehat{P})^3 = (\beta \cdot
\widehat{P})^3$, and from this
\begin{equation}
 |n| \leq \sqrt{\frac{8(\beta \cdot \widehat{P})^3}{9 \, \widehat{P}^3}}.
\end{equation}
Denoting $\widehat{P} \cdot \beta =: \tb \widehat{P}^3$, $n =: \tn
\widehat{P}^3$, the above expressions become
\begin{equation} \label{existenzconditions}
 \eta = 1 - \frac{1}{r^2} + \frac{24}{r^2}(\tb+2 \tn), \qquad
 \frac{1}{6} - \tb + \tn >0, \qquad |\tn| \leq \sqrt{\frac{8
 \tb^3}{9}}.
\end{equation}
Note that the inequalities are of exactly the same form as the
existence conditions in our class of examples studied in section
\ref{sec:classexamples}. From the analysis there, we can therefore
immediately conclude that $\tb+ 2 \tn \geq 0$. This implies $\eta
\geq 1 - \frac{1}{r^2}$, which immediately excludes all $r>1$
configurations, since we are considering extreme polar states here,
which by definition have $\eta \ll 1$. For $r=1$, we have
furthermore $\eta = 1 - 24 \, \nu$ in the notation of section
\ref{sec:classexamples}, and lines of constant $\eta$ in fig.\
\ref{stabregion} are given by translations of the purple dotted
line, with $\eta = 1$ corresponding to the original line and $\eta =
0$ to its translation to the left such that it contains the origin
$(\tb,\tn)=(0,0)$. It is then clear from the plot that taking $\eta$
smaller and smaller will also cause $\tb$ and $\tn$ to become
smaller and smaller.

A precise bound is easily obtained by using $\tb \leq \frac{1}{8}$
(as can be read off from fig.\ \ref{stabregion}) together with $\eta
= 24(\tb + 2 \tn) \geq 24(\tb - \frac{4 \sqrt{2}}{3} \tb^{3/2})$,
which gives $\tb \leq \frac{\eta}{8}$, $|\tn| \leq
\frac{\eta^{3/2}}{24}$. This shows that all extreme polar 2-centered
configurations with charges given by (\ref{ttttansz}) have $r=1$ and
have $(\beta_i,n_i) \in \CC(P,\epsilon)$, where we can take
$\epsilon = \frac{\eta}{8}$, thus establishing the extreme polar
state conjecture for this case. \footnote{Recently, E. Andriyash has
extended the argument to allow $\beta_1\not=\beta_2$ and
$n_1\not=n_2$.}

It is relatively straightforward to extend this proof to the case
where we add D2-D0 halos around the black hole centers while still
keeping $\beta_1 = \beta_2$ and $-n_1=n_2=n$. The conditions for
having a split point remain unchanged, since these do not care about
the composition of $\Gamma_1$ and $\Gamma_2$. Evaluating
(\ref{cloudstab})  leads to the simple stability condition $n_h(1+6
\tn - 6 \tb) > 0$ for halos around the second center, and the
opposite inequality for the first center. Now the quantity within
brackets is actually positive because of the split conditions given
above, so the halo stability condition simply becomes $n_h > 0$ for
$\Gamma_2$ and $n_h < 0$ for $\Gamma_1$. This means that the only
effect of adding these halos will be to remove the red curved
boundary on the right in fig.\ \ref{stabregion}, extending the
stable (=yellow shaded) region to the downwards sloping blue line on
the far right. It is then again clear from the plot that the extreme
polar state conjecture holds in this case.

Things become more complicated when we allow more general
multicentered core configurations (such as those described  in
section \ref{sec:corestates}), or core configurations with $\beta_1
\neq \beta_2$. We analyzed in depth a number of examples, and always
found the extreme polar state conjecture to be true. However the
detailed arguments we found vary from case to case, tend to be
messy, and are not particularly illuminating as to why the
conjecture should be true in general, so we will not report  the
details here.

On the other hand, it is possible to give a more general (albeit
incomplete) scaling argument for why only $r=1$ splits contribute to
the extreme polar states. Say we start with a multicentered BPS
configuration with initial split having $r \geq 2$. Then we can
produce from this a multicentered BPS configuration with the same
total $P$ but $r=1$ by applying the scaling symmetries of section
\ref{sec:symmetries} with $\mu = r^{-1}$, $\lambda = r$. This scales
all $(p^0,p,q,q_0) \to (p^0/r,p,r q,r^2 q_0)$ and in particular
$\hat{q}_0 \to r^2 \hat{q}_0$. However we know that the maximal
possible $\hat{q}_0$ equals $(\hat{q}_0)_{\rm max} = P^3/24$, so in
particular we have $r^2 \hat{q}_0 < (\hat{q}_0)_{\rm max}$, hence
for our original configuration $\eta
> 1-\frac{1}{r^2} \geq \frac{3}{4}$, implying it is not extreme
polar.

Regrettably, this  argument has a flaw: if some of the centers have
D6-charge $p^0$ not divisible by $r$, the rescaled configuration
violates charge quantization and is therefore unphysical, so we
cannot use the physical bound on $\hat{q}_0$ (if we allowed
fractional $p^0$ we could produce super-polar states, so we should
be strict as far as D6-charge quantization is concerned here). One
could therefore worry that, for example, by splitting up the $p^0 =
\pm r$ centers of the class of 2-centered solutions analyzed below
(\ref{ttttansz}), we could make $\hat{q}_0$ greater than the bound
derived there, or equivalently $\eta$ smaller than $1-1/r^2$.

A full analytical analysis of such multicentered configurations with
smaller D6-charges becomes rather cumbersome. Instead we performed a
numerical analysis of the four centered ${\rm
D6-D6-\overline{D6}-\overline{D6}}$, $r=2$ case, searching through
ensembles of flow trees by a simple adaptive random walk
optimization method, trying to maximize $\hat{q}_0$. The results we
obtained are fully consistent with the extreme polar state
conjecture. We refer to appendix \ref{app:attnumerics} for more
details.

In what follows we will assume the extreme polar state conjecture is
true.

\subsection{The dilute gas D6-anti-D6 partition
function }\label{sec:dilutegaspf}

\subsubsection{Definition and factorization}\label{sec:deffact}

Let us define the following generating function
\begin{equation} \label{ZD6aD6def}
 \CZ^{\epsilon}_{D6-\overline{D6}}(u,v,w)
 := \sum_{\Gamma_1,\Gamma_2} \Omega(\Gamma_1)_{\rm ms} \,
 \Omega(\Gamma_2)_{\rm ms} \, u^{q_0} \, v^Q \, w^{\langle \Gamma_2,\Gamma_1
 \rangle},
\end{equation}
where $\Gamma_1, \Gamma_2$ are parametrized as in
(\ref{Gammaione})-(\ref{Gammaitwo}) with $P$ (but not $S$) fixed and
with  $(\beta_i,n_i) \in \CC(P,\epsilon)$ as defined in
(\ref{CCdef}). Here  $q_0$ and $Q$ are the total D0- resp.\ D2-brane
charges (recall eqs. (\ref{totalddd})):
\begin{eqnarray}
 Q &=& \beta_2-\beta_1 + P S, \qquad S:=\frac{S_1+S_2}{2} \label{qSbetan} \\
 q_0 &=& \frac{P^3+c_2 P}{24}
 + \frac{1}{2} P S^2 - S \beta_1 - \frac{P}{2} \beta_1 + n_1
 + S \beta_2 - \frac{P}{2} \beta_2 -n_2, \label{nSbetan}
\end{eqnarray}
and the subscript ``ms'' as before means that the indices have to be
evaluated at the split point of the attractor flow $\Gamma \to
\Gamma_1 + \Gamma_2$, as given by (\ref{splitpointvalue}).

Two remarks on this definition are in order:

\begin{enumerate}

\item Note that
\begin{equation}\label{derivzeedsix}
 \left. \frac{\partial}{\partial w} \CZ^{\epsilon}_{D6-\overline{D6}}(u,v,w)
 \right|_{w=-1}
 = \sum_{\Gamma_1,\Gamma_2} (-1)^{\langle \Gamma_1,\Gamma_2
 \rangle-1} |\langle \Gamma_1,\Gamma_2 \rangle| \,
 \Omega(\Gamma_1)_{\rm ms} \, \Omega(\Gamma_2)_{\rm ms}
  \, u^{q_0} \, v^Q,
\end{equation}
so, comparing to (\ref{BPSindfact2}), we see that the coefficients
of this derivative count the indices of our D6-anti-D6 BPS bound
states for given total charge $(Q,q_0)$.

\item The sum over $\beta_i,n_i$ is (most likely) a finite sum, but
the sum over $S$ is definitely an infinite sum. The sum on $S$ is
always divergent  because the quadratic form defined by $PS^2$
(which appears through $q_0$ ) has signature $(1,h-1)$. However
(\ref{ZD6aD6def}) does make good sense as a formal power series, in
the sense that only a finite number of terms contributes to the
coefficient of any monomial $u^n v^\beta w^\ell$. This is true
simply because the   map $S \to PS$ is invertible.

\end{enumerate}

%

We now aim to write (\ref{ZD6aD6def}) as a sum over $S$ of
factorized expressions depending only on $(\beta_1,n_1)$ and
$(\beta_2,n_2)$, respectively.  To this end let us evaluate the
degeneracy factors $\Omega$. We are instructed by
(\ref{BPSindfact2}) to compute $\Omega(\Gamma_i)$ evaluated at the
split point (\ref{splitpointvalue}). It is convenient at this point
to use the gauge invariance under shifting the $B$-field to say:
\begin{eqnarray}\label{Ommsform}
 \Omega(\Gamma_1; t_{\rm ms} )
 & =& \Omega(\Gamma(\beta_1,n_1); t^1_{\rm ms}  )\\
 \Omega(\Gamma_2; t_{\rm ms} )
 & =& \Omega(\Gamma(\beta_2,n_2); t^2_{\rm ms}).
 \end{eqnarray}
If  $\Gamma_1, \Gamma_2\in \CC(P,\epsilon)$ then note that
 for the gauge invariant quantity $\CF_i = S_i - B$
  we have, up to $\CO(\epsilon |P|)$ corrections, $\CF_1 = \frac{P}{2}$,
$\CF_2 = -\frac{P}{2}$. Accounting for the imaginary part  we have:
\begin{eqnarray}
t^1_{\rm ms} = -\half P + D_A D^{AB}\Delta \beta_B + i y_{\rm ms}P &
= & e^{2\pi i/3} P + \CO(\epsilon|P|)\label{shiftedtms1}\\
t^2_{\rm ms} = \half P + D_A D^{AB}\Delta \beta_B + i y_{\rm ms}P &
= & e^{\pi i/3} P + \CO(\epsilon|P|)\label{shiftedtms2}.
\end{eqnarray}
Thus the degeneracies are counted by the generating function
(\ref{ZD6D2D0def}) evaluated at $B+iJ = (\pm \half + i
\frac{\sqrt{3}}{2}) P + \CO(\epsilon |P|)$, so up to $\epsilon$
corrections the degeneracies are indeed independent of each other!

Using (\ref{D6antiD6stabcond1}), (\ref{qSbetan}) and
(\ref{nSbetan}), we can write
\begin{eqnarray}
 \CZ_{D6-\overline{D6}}^{\epsilon}(u,v,w) &=& \sum_{S,\beta_i,n_i}
 \Omega(\Gamma(\beta_1,n_1);t_{\rm ms}^1) \,
 \Omega(\Gamma(\beta_2,n_2);t_{\rm ms}^2) \nonumber \\
 && \qquad \times \, w^{I_P-P \beta_1 - P \beta_2 + n_1 - n_2}
 \nonumber \\
 && \qquad \times \, u^{\frac{P^3+c_2 P}{24}
 + \frac{1}{2} P S^2 - S \beta_1 - \frac{P}{2} \beta_1 + n_1
 + S \beta_2 - \frac{P}{2} \beta_2 -n_2} \, v^{\beta_2-\beta_1 + P
 S} \\
 &=& u^{\frac{P^3+c_2 P}{24}} \, w^{I_P} \sum_S u^{\frac{1}{2} P
 S^2} v^{PS}
 \sum_{\beta_1,n_1} \sum_{\beta_2,n_2} \nonumber \\
 && \times \, \Omega(\Gamma(\beta_1,n_1);t_{\rm ms}^1) \,
 (w u)^{n_1} \, (w^{-P} u^{-S - \frac{P}{2}} v^{-1})^{\beta_1}
 \nonumber \\
 && \times \,
 \Omega(\Gamma(\beta_2,n_2));t_{\rm ms}^2) \,
 (w u)^{-n_2} \, (w^{-P} u^{S - \frac{P}{2}} v)^{\beta_2} \label{almostfact}
\end{eqnarray}
where the sums are over $S \in \frac{P}{2} + H^2(X,\IZ)$,
$(\beta_i,n_i) \in \CC(P,\epsilon)$, and $t_{\rm ms}^{1,2}$  refers
to the shifted marginal stability points
(\ref{shiftedtms1}),(\ref{shiftedtms2}).

Note that the sum is almost factorized. For the next step we  would
like to rewrite (\ref{almostfact}) as  a sum of products of DT
partition functions. We will eventually achieve this, under suitable
conditions, in equation (\ref{foefoebal2}) below, but first, in view
of the identifications (\ref{fourinfinity1})-(\ref{fourinfinity4}),
we need to compare
\begin{equation}\label{finpointa}
\Omega(\Gamma(\beta_1,n_1); t_{\rm ms}^1)
\end{equation}
with
\begin{equation}\label{infpointa}
\lim_{z\to \CL^\pm} \Omega(\Gamma(\beta_1,n_1); z P).
\end{equation}
where $\lim_{z \to \CL^\pm} $ indicates that  $z$ goes to infinity
in the left-half plane as explained just above eq.
(\ref{fourinfinity1}). Similarly we need to compare
\begin{equation}\label{finpointb}
\Omega(\Gamma(\beta_2,n_2); t_{\rm ms}^2)
\end{equation}
with
\begin{equation}\label{infpointb}
\lim_{z\to \CR^\pm} \Omega(\Gamma(\beta_2,n_2); z P).
\end{equation}

The relation between (\ref{finpointa}) and (\ref{infpointa}) and
between (\ref{finpointb}) and (\ref{infpointb}) is not at all
trivial, and in fact they \emph{will} differ in general, due to
jumps at marginal stability. In the next section we compare these
two degeneracies.

Note that, roughly, the M-theory equivalent to this is that there
can be BPS states which exist in Taub-NUT when the Taub-NUT radius
is taken to infinity, but not necessarily at arbitrary finite radii,
and similarly they do not necessarily exist when the Taub-NUT is
combined with an anti-Taub-NUT to produce the finite size $AdS_3
\times S^2$ setup of \cite{Gaiotto:2006ns,deBoer:2006vg}. In
particular this implies that it is not true that BPS states in these
finite size cases are exactly counted by the GV / DT partition
function (which counts BPS states in infinite radius Taub-NUT).
Figuring out to what extent the spectrum is truncated is a difficult
problem, and the absence of a systematic way to do this is what
prevented \cite{Gaiotto:2006ns,deBoer:2006vg} from arriving at any
error estimates in their derivation of the OSV conjecture.

Happily, the tools we have developed in this paper are exactly
designed to do this, so let us now turn to this analysis.

\subsubsection{Harmless Halos and Catastrophic Cores}\label{innocuouscoredump}

We now focus on the difference
\begin{equation}\label{diffO}
\Delta \Omega(\beta_1,n_1;\beta_2,n_2) = \Omega(\Gamma(\beta_1,n_1);
t_{\rm ms}^1(\beta_1,n_1;\beta_2,n_2))- \lim_{z\to \CL^\pm}
\Omega(\Gamma(\beta_1,n_1); z P).
\end{equation}
Nonzero contributions to $\Delta \Omega$ will lead to corrections in
the OSV-like relation we wish to derive. The reason $\Delta \Omega$
can be nonzero  is that there can be BPS states of charge
$\Gamma(\beta_1,n_1)$ which are stable at $t^1_{\rm ms}$ but
unstable at infinity and vice versa. States which make a nonzero
contribution to (\ref{diffO}) will be called  \emph{swing states}.

 To be more precise, we will call a D6-D2-D0 BPS state of charge
$\Gamma(\beta_1,n_1)$ a \emph{swing state} if $(\beta_1,n_1)\in
\CC(P,\epsilon)$ and there exists $(\beta_2,n_2)\in \CC(P,\epsilon)$
such that either the state is contained in
$\CH(\Gamma(\beta_1,n_1);t)$ with $t= z P$, $z\to \CL^\pm$, but
decays along the way to $t=t_{\rm ms}^1(\beta_1,n_1;\beta_2,n_2)$,
or vice versa, i.e.\ it exists at $t = t_{\rm ms}^1$ but not at $t
\to \CL^{\pm} P$. Recall that $t^1_{\rm ms} = e^{2 i \pi/3} P$ up to
order $\epsilon$ corrections, so the definition basically says that
a D6-D2-D0 state is a swing state when it exists at infinity but not
in an order $\epsilon$ neighborhood of $t=e^{2 i \pi/3} P$, or vice
versa.

A very useful simplification in the analysis of swing states arises
when we recall that we are evaluating the stability condition at a
special point, $t^1_{\rm ms}$. Since $Z(\Gamma_1;t_{\rm ms})>0$ the
stability condition for
\begin{equation}
\Gamma(\beta_1,n_1) \to \tilde \Gamma + \tilde \Gamma'
\end{equation}
simplifies to:
\begin{equation}\label{Simplerstability}
\langle \tilde \Gamma, \Gamma(\beta_1,n_1)\rangle  \, {\rm Im}\,
Z(\tilde \Gamma; t^1_{\rm ms})
> 0
\end{equation}

\EPSFIGURE{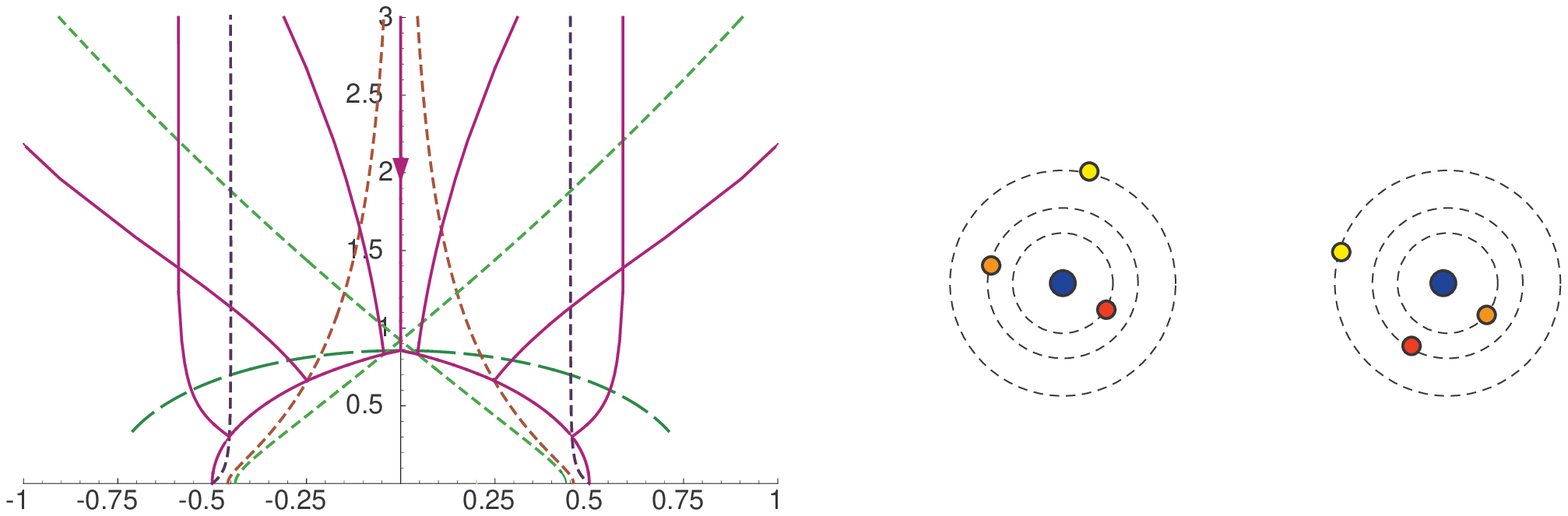,height=5.5cm,angle=0,trim=0 0 0 0}%
{{\bf Left:} Flow tree corresponding to bound state of a fluxed D6
with three halos, and its conjugate. The charges for the $\Gamma_1$
half of the tree are $\Gamma_{1,c} = e^{P/2}$, $\Gamma_{1,h,i} =
e^{P/2}(\tilde q_{2,i} P^2 + \tilde q_{0,i} P^3 \, \omega)$ with
$\tilde q_2=(-10^{-3},-10^{-3},-10^{-3})$ and $\tilde q_0 =
(-10^{-4},-10^{-3},-10^{-2})$. The larger the D0-charge (at fixed
D2), the sooner the D2D0 particles split off in the tree (so the
first to split off is $\Gamma_{1,h,3}$). The charges for the
$\Gamma_2$ half are obtained by taking the conjugates $\Gamma \to -
\Gamma^*$. The dotted lines are the MS lines corresponding to the
various splits. {\bf Right:} Sketch of a corresponding multicentered
configuration.
  \label{bluered}}

 We will now show that halos do \emph{not} cause configurations to
become swing states. Halo states have lines of MS going to infinity
and hence one might imagine these wall crossings would make a
significant contribution to the difference $\Delta \Omega$. But that
turns out not to be the case: The marginal stability curves for
$\Gamma(\beta_1,n_1) \to \Gamma(\beta_c,n_c) + \Gamma_h$ with
$\Gamma_h =  (-\beta_h + n_h \omega)$ bend over quickly from a line
of slope $- \sqrt{3}$ to a vertical line when $\beta_h \cdot P \ll
n_h$, comfortably keeping $t^1_{\rm ms}$ on their stable side, while
when $\beta_h \cdot P \gg n_h$, they come close to $t^1_{\rm ms}$
but still bend over just in time. This behavior can be observed in
fig.\ \ref{bluered}. There is an entirely analogous story for
$\Gamma_2$ --- one just reflects the picture in the $y$ axis using
the symmetry (\ref{zeedualcharge}).  This special behavior
translates into a particularly simple stability condition and
(\ref{Simplerstability}) becomes:
\begin{equation}\label{Simplerstability2}
 \mp n_h \, \beta_h \cdot P > 0
\end{equation}
where  the minus sign is for $\Gamma_1$ and the plus sign for
$\Gamma_2$. Hence for $\Gamma_2$ all halos with $n_h$ positive  are
stable and similarly for $\Gamma_1$   all halos with  $n_h$ negative
are stable.

When the D2-charge of the halo particle vanishes, the situation is
more subtle, since in this case  the split point lies exactly on the
wall of marginal stability for the D0-halo, making the indices
ambiguous. To lift the ambiguity, it suffices to take $\tilde{B}$
(defined in (\ref{Bshifvalue})) slightly different from zero. Then
we are essentially in the situation described in section
\ref{sec:D6D0}: stability requires $n_h < 0$, and depending on the
chosen sign of $\tilde{B}$, \emph{either} $\Gamma_1$ \emph{or}
$\Gamma_2$ can support D0-halos, but not both at the same time. An
important consequence of this is that for counting D6-anti-D6 bound
states, we should only include \emph{ one}  MacMahon factor
(\ref{macmahonfactor}) in the dilute gas partition function,  and
not two as one might have thought naively. We will return to this
point below (\ref{foefoebal}).

%
%

Now let us consider the possibility that there are walls for
splitting off other kinds of constituents as we move $t_\infty$ from
$t^1_{\rm ms}$ to infinity in the left-half plane. The first
observation to make is that if we consider any $P$-independent
finite set of pairs $\{( \beta_1,n_1),(\beta_2,n_2)\}$ then for
suffciently large $P$ we have $\Delta \Omega=0$. We can justify this
as follows. According to the split attractor flow conjecture,  for
any $t$, $\Gamma(\beta_1,n_1)$ only supports a finite set of split
flows. These will begin with some splitting $\Gamma(\beta_1,n_1) =
\Gamma_1' + \Gamma_1''$. Now, we know that $t^1_{\rm ms}$ lies  to
the left of all marginal stability
 lines where one of $\Gamma_1'$ or $\Gamma_1''$ are halo charges. Thus,
 we need only  worry about the case
where $\Gamma_1', \Gamma_1''$ both have magnetic charge. However, in
this case the marginal stability lines, which are subvarieties of
${\rm  Im}(Z(\Gamma_1';t)\overline{ Z(\Gamma_1''; t)})=0$ lie in
\textit{bounded} regions of moduli space. As long as we consider a
set of charges  that  makes no reference to $P$,   by making $P$
sufficiently large $t^1_{\rm ms}$ (which grows with $P$) will always
be outside the union of the compact regions where  splits are
allowed.
 Now, by the argument surrounding (\ref{Simplerstability2})
 we can freely take $x$ to $-\infty$
  since $t^1_{\rm ms}$ is to the left of all the walls for halo states.
A similar argument applies to $\Omega(\Gamma_2)$, where we must take
$x$ to $+\infty$.

Regrettably, the  above argument is not sufficient for our purposes
because the charges in $\CC(P;\epsilon)$ can in fact grow with $P$.
As a matter of fact, we will now exhibit a class of examples which
\emph{does} give a nonzero contribution to $\Delta \Omega$. That is,
we will show that swing states do indeed exist.

\EPSFIGURE{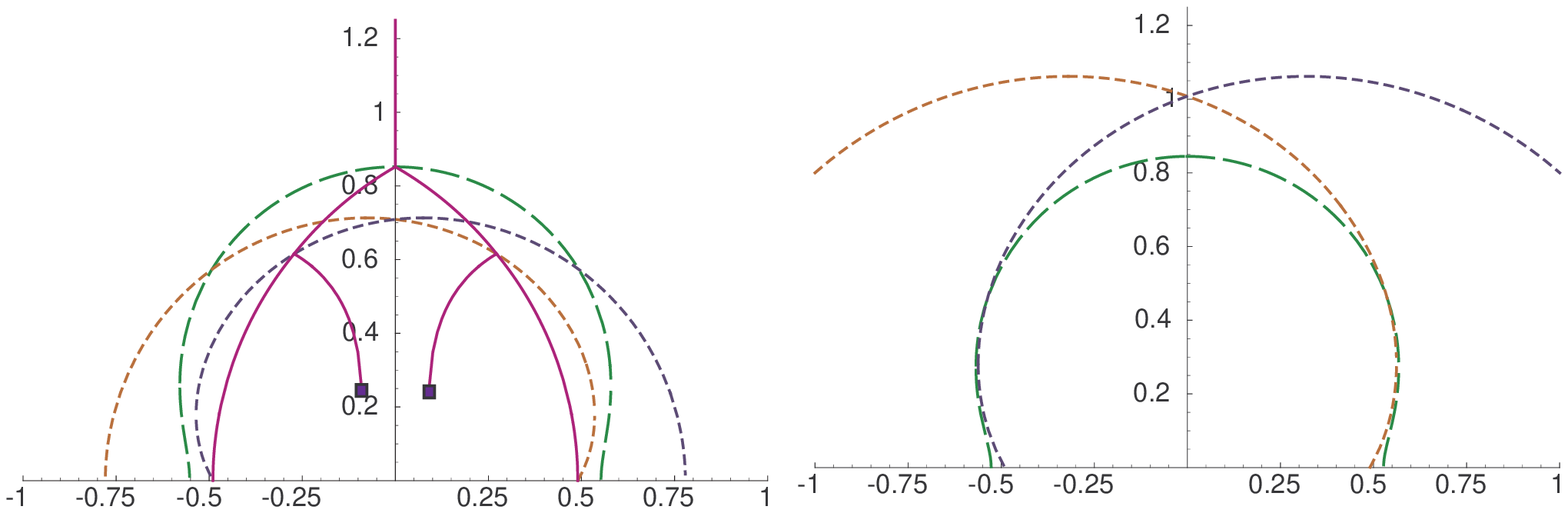,height=5.5cm,angle=0,trim=0 0 0 0}%
{{\bf Left:} Bound state of two D6-D4 core states as described in
the text, with $u=10^{-3}$, $v=0.4$, $\tilde q_0 = -10^{-4}$,
$\frac{\beta_1 \cdot P}{P^3} \approx 10^{-2}$, $\frac{n_1}{P^3}
\approx 5 \times 10^{-3}$. {\bf Right:} Failure to form a similar
bound state with $v=0.6$ instead and all other parameters the same.
The bound state cannot form because the initial $\Gamma \to \Gamma_1
+ \Gamma_2$ split point lies in the unstable region of the
constituent core states themselves: the green wide-dashed line is
the $\Gamma \to \Gamma_1 + \Gamma_2$ MS line and lies below the
short-dotted lines which are the MS lines for the states
representing the $\Gamma_i$. \label{goodcorebadcore}}

 To be concrete, we consider candidate bound states of
\begin{eqnarray}
\Gamma_1 &=& e^{P/2} \Gamma(\beta_1,n_1)\\
\Gamma_2 & = & - [e^{P/2} \Gamma(\beta_1,n_1)]^* = - e^{-P/2}
\Gamma(\beta_1,-n_1)
\end{eqnarray}
where $\Gamma(\beta_1,n_1)$ is realized as one of the core states
analyzed in section \ref{sec:corestates}, equation (\ref{newcore}).
Note that $\beta_2=\beta_1$, $n_2=-n_1$.

Recall that the split states for $\Gamma(\beta_1,n_1)$ studied in
section \ref{sec:corestates} are stable at infinity and unstable
within the curve $\CC_+$ defined in (\ref{twobr}).
 Therefore, such states will contribute to $\Delta \Omega$
if our stability condition is \emph{violated} at $t^1_{ms}$. Using
(\ref{Simplerstability}) this works out to be the simple condition
\begin{equation}\label{antistabcond}
u+v > 1/2.
\end{equation}
Thus, swing states do exist. An example where this condition is not
satisfied (so the bound state does exist) is shown in fig.\
\ref{goodcorebadcore}$(a)$, and one where it is satisfied in fig.\
\ref{goodcorebadcore}$(b)$ (so the bound state does not exist).

We are now in a position to understand why swing states are
potentially problematic. In our example we can compute
\begin{equation}
\hat q_0 = \frac{P^3+c_2\cdot P}{24} - P (UV + \half U^2) + [UV^2 +
2 U^2 V + \frac{2}{3} U^3 + q_0]
\end{equation}
and hence
\begin{equation}\label{etatoosmall}
\eta = 24 [ uv + \half u^2 - (uv^2 + 2 u^2 v+\frac{2}{3} u^3) -
\tilde q_0 ]
\end{equation}
Note that we can satisfy (\ref{antistabcond}) by taking $u\to 0$
while letting $v>1/2-u$ be order one. But then   $\eta\to 0$ and
such states are arbitrarily extreme polar. Looking ahead to the
impact on our derivation of the OSV conjecture below, we see that
such states will lead to large corrections to the OSV formula
arbitrarily close to the leading contribution, invalidating the
conjecture. How can we avoid this ``coretastrophe''?

Fortunately we can combine a choice of a suitably small $\epsilon$
with charge quantization to eliminate the contribution of this
particular example of swing states to the dilute gas partition
function. Suppose $P=p P_0$ where $P_0$ is primitive and $p$ will go
to infinity. Charge quantization implies $u= m/p$ with $m$ positive
and integral. Now, the condition  that $(\beta_1,n_1)\in
\CC(P,\epsilon)$, together with (\ref{antistabcond}) and charge
quantization  implies that
\begin{equation}\label{cdcond}
\epsilon> uv + \half u^2 > \half u(1-u)> \frac{1}{2p}(1-\frac{1}{p})
\end{equation}
Since we will be taking $p\to \infty$, if we take $\epsilon =
\delta/p$ with $\delta$ a $p$-independent constant smaller than
$1/2$ then (\ref{cdcond}) will eventually be violated, and hence
these particular swing states are eliminated from the ensemble
defined by $\CC(P,\epsilon)$.

The above argument  shows that our example of  potentially
catastrophic  \footnote{catastrophic, that is,  for the OSV
conjecture} swing states can be eliminated by making a suitable
($P$-dependent) choice of $\epsilon$. Sadly, we have no proof that
there are not other swing states which will create problems, so we
proceed as follows.

Suppose $\epsilon = \frac{\delta}{\vert P\vert^\xi}$ where $\delta$
is a $P$-independent constant. We know that if we choose $\xi=3$
then the states in $\CC(P,\epsilon)$ consist of a finite set of
$P$-independent charges $(\beta_1,n_1)$. Our argument above shows
that for such states indeed $\Delta \Omega=0$. Unfortunately, this
is not enough to prove the OSV conjecture. The reason is that, as we
show in equation (\ref{simplererror}) below, the error from the
(necessary) restriction to extreme polar states is of order
\begin{equation}
\CO\biggl(\exp[ - \frac{\pi}{12\mu}  \frac{\epsilon \vert
P\vert^3}{\phi^0}]\biggr)
\end{equation}
where $\mu$ is the constant $\epsilon= \mu \eta_*$ introduced in the
extreme polar state conjecture.  On the other hand, the worldsheet
instanton effects which make the OSV conjecture nontrivial are of
order $\exp[-2\pi \beta \cdot P/\phi^0]$. Therefore, if $\xi > 2$
the states contributing to $\Delta \Omega$ make contributions larger
than those of worldsheet instantons, eventually dominating all
worldsheet instantons in the $\vert P\vert \to \infty$ limit. If
$\xi=2$, they are of the same order, which still would not be
desirable, unless perhaps $\delta $ can be chosen to be arbitrarily
large.

This discussion motivates the following definition of the \emph{
core-dump exponent}, denoted $\xi_{cd}$:

Consider the set ${\cal S} $ of pairs $\{ \Gamma(\beta_1,n_1),
\Gamma(\beta_2,n_2)\}$ which admit flow trees   making a nonzero
contribution to $\Delta \Omega$. (Thus, the flow tree is based on
core states stable at infinity, but unstable at $t_{\rm ms}^{1,2}$,
or vice versa.)  Let $\xi_{cd}$ be the minimum of the set $\Xi$ of
numbers with the following property: For $\xi \in \Xi$, there exists
a constant $\delta$ which is independent of $\vert P\vert$ such
that, if we choose   $\epsilon = \delta \vert P \vert^{-\xi}$ then
$\CC(P,\epsilon) \times \CC(P;\epsilon)$ does not contain any of the
states in ${\cal S}$.

From the argument given above, we know $\xi_{cd} \leq 3$. From the
example (\ref{newcore}) discussed above we also know that $\xi_{cd}
\geq 1$. Then, as we have just explained, if
  $1\leq \xi_{cd} \leq 2$ we will see below that a version of the
  (strong coupling) OSV conjecture can be proven. On the other hand, if it turns out
  that  $2< \xi_{cd}$  then the OSV
  conjecture (even at strong coupling) is almost certainly not correct. We can only say
  ``almost certainly'' because we have not excluded the   possibility
  (however unlikely) that when we account for all swing states of a fixed charge
  their contributions to $\Delta \Omega$ magically sum to zero.

An independent argument sheds more light on why we must
  take $\xi_{cd}\geq 1$.  We consider, a family of $(\beta_P,n_P)$
  (we will henceforth drop the subscript) such
  that for each $P$ we have $P\cdot \beta = c_1 \vert P\vert^2, n=
  c_2 \vert P\vert^2$, where $c_1,c_2$ are constants. It turns out
  that such families exist for which $N_{DT}(\beta,-n)\not=0$
  \cite{Klemm:2004km}. On the other hand, we can attempt to build a
  boundstate using $\Gamma_1 = e^{P/2}\Gamma(\beta,n)$ and
  $\Gamma_2=-\Gamma_1^*$. One finds that such a boundstate would
  have
  \begin{equation}
  \eta=\frac{24}{P^3}(P\cdot \beta - 2n)
  \end{equation}
  and hence, the absence of superpolar states implies $\eta\geq 0$.
  On the other hand, the explicit examples of such families in \cite{Klemm:2004km} have
  constants $c_1,c_2$ such that $\eta<0$. We are thus forced to
  conclude that states with such charges $\Gamma(\beta,n)$ are
  unstable at $t_{\rm ms}^1$, and hence such states provide examples of
   swing states.
  Note that, in this case, once again we can take $\epsilon=
  \delta/\vert P\vert$ for sufficiently small $\delta$ to eliminate
  such states from the ensemble $\CC(P,\epsilon)$ thus proving once more that
  $\xi_{cd}\geq 1$.

  In fact, a generalization of the argument of the previous paragraph offers a
  hint
  that in fact  $\xi_{cd}=1$. Suppose we have a family of $(\beta_P,n_P)$
  (we henceforth drop the subscripts) such that $\beta\cdot P = c_1
  \vert P\vert^{\gamma}$ and $n= c_2 \vert P \vert^{\gamma'}$ and
  $N_{DT}(\beta,-n)\not=0$. Reasoning as above, if $\gamma'>
  \gamma$, or $\gamma'=\gamma$ and $2c_2>c_1$ then $\Gamma(\beta,n)$
  is a swing state. On the other hand, we can (following \cite{Klemm:2004km})
  use  Castelnuovo's
  inequality (see \cite{GH}, p. 252), which states that
   a curve of degree $d$ in $\IC P^n$ has its genus bounded
   above by $ g < \frac{d^2}{2(n-1)}$ for
  $d\to \infty$. Next recall from the last equation in
  (\ref{identifycharges}) that $n\leq g(\beta)-1$. Thus, we should
  have $\gamma' \leq 2(\gamma-1)$. Combining with $\gamma'\geq
  \gamma$ we see that $\gamma\geq 2$ for any such families. But then
such swing states can always be eliminated with $\xi =1 $.

In section \ref{sec:dangercore} we will give some more
circumstantial evidence for $\xi_{cd}=1$.

\subsubsection{Factorization of the dilute gas D6-anti-D6 partition function}

Let us now return to the analysis of (\ref{ZD6aD6def}). We assume
that we define this partition function with $\epsilon = \delta \vert
P\vert^{-\xi}$ with a suitable $\delta$ and $\xi$ so that we can
identify (\ref{finpointa}), (\ref{finpointb}) with
(\ref{infpointa}),(\ref{infpointb}), respectively. In this case,  we
can proceed with the derivation of   (\ref{foefoebal2}) below as
follows:

We  introduce the   $\epsilon$-dependent cut off version of
(\ref{ZD6D2D0def}):
\begin{equation}
  \CZ_{\rm D6-D2-D0}^{\epsilon}(u,v;B+i J)
  := \sum_{(\beta,n) \in \CC(P,\epsilon)} \Omega(\Gamma(\beta,n))|_{B+i J} \,
 \, u^n \, v^{\beta}.
\end{equation}
and using (\ref{almostfact}) we write:
\begin{eqnarray}
   \CZ_{D6-\overline{D6}}^{\epsilon}(u,v,w)
 &=& u^{\frac{P^3+c_2 P}{24}} \, w^{I_P} \sum_S u^{\frac{1}{2} P S^2}
 v^{PS} \nonumber \\
 && \qquad \times \, \lim_{z \to \CL^-}
 \CZ_{D6-D2-D0}^{\epsilon}(w \, u,w^{-P} \, u^{-S-\frac{P}{2}} \, v^{-1};
 z P)
 \nonumber \\
 && \qquad \times \, \lim_{z \to \CR^+ }
 \CZ_{D6-D2-D0}^{\epsilon}(w^{-1} u^{-1},w^{-P} \, u^{S-\frac{P}{2}} \, v;
 z P). \nonumber \\
 &&   \label{foefoebal}
\end{eqnarray}

  Please notice carefully the nature of the limits.
   Recall that, as we mentioned above, there is a
  subtlety when the halo particle has zero $D0$ charge, because in this
case $t_{\rm ms}$   is exactly on the wall of marginal stability for
pure D0-halos.   As discussed there, this ambiguity can be resolved
by perturbing the background $B$-field slightly, in which case there
is a D0-halo contribution \emph{either} on the first cluster
\emph{or} on the second. Therefore, as   noted in eqs.
(\ref{fourinfinity1}-\ref{fourinfinity4}), the case in which we
first take $t\to \CL^-$ does have the D0-halo MacMahon factor, while
the case $t\to \CR^+$ does not. Depending on the sign of the
perturbation of the background $B$-field we have the limits $(\CL^-,
\CR^+)$ as above or $(\CL^+, \CR^-)$. Our final answer will not
depend on this dichotomy.

Now the identifications (\ref{fourinfinity1}-\ref{fourinfinity4})
imply
\begin{eqnarray}
   \lim_{z \to \CL^-} \CZ_{D6-D2-D0}^{\epsilon}(u,v;
 z P)
 &=& \CZ_{DT}^{\epsilon}(u^{-1},v) \\
  \lim_{z \to \CR^+} \CZ_{D6-D2-D0}^{  \epsilon}(u,v;
 z P)
 &=& \CZ_{DT}^{\prime \epsilon}(u,v),
\end{eqnarray}
where the $\epsilon$-dependent $\CZ_{DT}$ is defined as the DT
partition function with sum restricted to $(\beta,n) \in
\CC(P,\epsilon)$. Similarly, for $\CZ_{DT}^{\prime \epsilon}$, we
take the infinite product $\CZ_{DT}'$ and truncate its series
expansion. In terms of these quantities  we can therefore write:
\begin{eqnarray}
 \CZ_{D6-\overline{D6}}^{\epsilon}(u,v,w) &=& u^{\frac{P^3+c_2 P}{24}}
 \, w^{I_P} \sum_S u^{\frac{1}{2} P S^2}
 v^{PS} \nonumber \\
 && \qquad \times \, \CZ_{DT}^{\epsilon}(w^{-1} u^{-1},w^{-P} \, u^{-S-\frac{P}{2}} \, v^{-1})
 \nonumber \\
 && \qquad \times \, \CZ_{DT}^{\prime \epsilon}(w^{-1} u^{-1},w^{-P} \, u^{S-\frac{P}{2}} \,
 v). \nonumber \\
 \label{foefoebal2}
\end{eqnarray}

Equation (\ref{foefoebal2})  is the main result of this section. As
noted above, the sum over $S$ is a formal sum, but the coefficients
of $u^n v^\beta w^\ell$ are well-defined.

\subsection{D4-D2-D0 degeneracies} \label{sec:D4D2D0}

 In this section we finally return to D4-D2-D0
 degeneracies and use the technology developed above to present a
 derivation of our refined OSV formula  eq. (\ref{osvwithmf}).
 We first relate the polar part
 of the D4-D2-D0 partition function to the D6-antiD6 dilute gas
 partition function. This introduces an error, but one well-controlled by
 the extreme polar state conjecture (provided it turns out that
 $\xi_{cd}\leq 2$). We then combine this with the fareytail
 expansion. At the end of the section we discuss the error terms in
the refined OSV formula.

\subsubsection{Approximate factorization of polar D4 partition function}

Now we return to the considerations of section \ref{sec:fareytail}.
The extreme polar state part of the D4 partition function
(\ref{PsiHsplitdef}) is, using the notations of section
\ref{sec:radjacobi}
\begin{eqnarray}
 \CZ^{\eta_*}(\tau,\bar{\tau},C) &:=& \sum_{\gamma}
  H_{\gamma}^{\eta_*}(\tau) \Psi_\gamma(\tau,\bar{\tau},C) \,
  \label{Zetadef} \\
 H_{\gamma}^{\eta_*}(\tau) &:=& \sum_{\eta<\eta_*}
 \Omega([\gamma,\mbox{$\frac{P^3}{24} - \eta \frac{P^3}{24}$}],t=i\infty)
 \, e^{-2 \pi i \tau \frac{P^3}{24} + 2 \pi i \tau \eta \frac{P^3}{24}}
\end{eqnarray}
where $\eta_* \ll 1$ and $\Omega([\gamma,\hat{q}_0])$ is as defined
above (\ref{HagamOm}). When $\Im \, \tau$ is sufficiently
large,\footnote{Further on we will apply the approximate
factorization we are currently deriving in the fareytail expansion
(\ref{ZPoincSer}), and will find that the dominant term for our
purposes comes from the term corresponding to $A = S$, where $\tau$
gets replaced by $-1/\tau$. Thus, in these applications, we will
need a sufficiently large $\Im \,(-1/\tau)$ to get approximate
factorization.} the extreme polar part of the partition function is
a good approximation to the full polar part $\CZ^-$ (which is
obtained by taking $\eta_*=1$):
\begin{eqnarray}
 H_{\gamma}^-(\tau) = H_{\gamma}^{\eta_*}(\tau)
 \times \left( 1 + {\cal O}( e^{-\Delta(P,\eta_*,\tau) P^3} ) \right)
 \label{Hgammapolardroperror}
\end{eqnarray}
where
\begin{eqnarray}
 \Delta(P,\eta_*,\tau) &:=& \min_{1>\eta>\eta_*}
 \left( -\Sigma(P,\eta) + \frac{\pi}{12} \Im \, \tau \, \eta \right)
 \label{minimizeta} \\
 \Sigma(P,\eta) &:=& \frac{1}{P^3} \max_\gamma
 \log \left| \Omega([\gamma,\mbox{$\frac{P^3}{24} - \eta
 \frac{P^3}{24}$}]) \right|.  \label{downunder}
\end{eqnarray}
Note that as long as $\Delta(P,\eta_*,\tau)$ is positive and doesn't
decay as $\vert P\vert^{-3}$ or faster, the error is exponentially
small when $P^3$ is large; in particular this is the case for $\Im
\, \tau$ sufficiently large.

To get an idea of the general behavior of the error term, and to
justify our notations separating out the $P^3$ factor, let us assume
for the moment that we can estimate the growth of the right hand
side of (\ref{downunder}) from the growth of the entropy of
two-centered configurations of the kind analyzed in section
\ref{sec:classexamples}, and more specifically for further
simplicity let us restrict to configurations with $\tn=0$ in the
notation used there. Then $\tb=\eta/24$ and the total
Bekenstein-Hawking entropy of the two centers is $S \sim \tb^{3/2}
P^3 \sim \eta^{3/2} P^3$. Hence this estimates $\Sigma(P,\eta)
\approx c \eta^{3/2}$ for some constant $c$ independent of $P$. In
this case we have $\Delta(P,\eta_*,\tau)>0$ if and only if $\Im \,
\tau
> 12 c/\pi$. If $\Im \, \tau$ drops below this critical value, our
error estimate blows up. Moreover, $\eta_*$ should not become too
small if we want the error to be exponentially small, since $\Delta$
is in any case smaller than $\frac{\pi}{12}\eta_* {\rm Im}\tau - c
\eta_*^{3/2}$.

Now the index $\Omega$ in (\ref{downunder}) actually receives
contributions from many other flow trees, some very complicated, so
this simple estimate might be too naive. However, it will at least
give a rough lower bound on the actual growth of $\Sigma$, unless
miraculous almost-exact cancelations occur between contributions of
different signs to the index. This implies in particular that our
approximations break down for $\Im \, \tau$ less than some order 1
critical value, unless these miraculous cancelations occur. We will
discuss this potential breakdown in detail in section
\ref{sec:discussion}, and show that it is not due to a failure of
our derivation, but intimately related to the entropy enigma of
section \ref{sec:entropyenigma}.

Combining (\ref{Hgammapolardroperror}) and (\ref{Zetadef}), we can
write
\begin{equation}
 \CZ^-(\tau,\bar{\tau},C) = \CZ^{\eta_*}(\tau,\bar{\tau},C)
 \times \left( 1 + {\cal O}( e^{-\Delta(P,\eta_*,\tau) P^3} )
 \right). \label{polartoextremepolar}
\end{equation}
If we make the OSV substitution  (\ref{osvlimit}) and  formally put
$\bar \tau=\tau$, then both sides of this equation diverge due to
the non-definiteness of the intersection product on $L_X$, so the
error estimate, strictly speaking, is not meaningful. However we can
still give it a precise meaning by considering the terms for a fixed
D2-charge (i.e.\ a fixed power of $e^{-2 \pi i C}$), or equivalently
by multiplying both sides by some $e^{2 \pi i C \cdot Q}$ and
integrating out $C$. (After a modular transformation the integral on
$C$ involves   one wrong sign Gaussian integral which is  easily
evaluated in the usual analytically continued sense). Alternatively,
we can just cut off the sum over $L_X$ in the theta-functions. Since
the error estimate provides a \emph{relative} error, it has a well
defined meaning for any such finite truncation. In the end we can
take the cutoff to infinity, which when computing any physically
meaningful quantity should give a finite result, with a well-defined
error estimate. Keeping this interpretation of the error term in
mind, we will from now on put $\bar \tau = \tau$.

The extreme polar state conjecture implies that there exists an
$\epsilon(\eta_*) \sim \eta_*$ such that all extreme polar states
(with $\eta < \eta_*$) are generated by D6-anti-D6 dilute gas pairs
with $(\beta_i,n_i) \in \CC(P,\epsilon)$. We now invoke the formula
(\ref{BPSindfact2}) for the polar degeneracies and recall eq.
(\ref{derivzeedsix}). Combining this with
(\ref{polartoextremepolar}) at $\bar \tau = \tau$ then gives
\begin{eqnarray}
 \CZ^{-}(\tau,C) &=& \left.
 \frac{1}{2 \pi} \frac{\partial}{\partial \alpha}
 \CZ_{{\rm D6 - \overline{D6}}}^{\epsilon}
 (e^{-2 \pi i \tau},e^{-2 \pi i (C + \frac{P}{2})},e^{2 \pi (\alpha-\frac{i}{2})})
 \right|_{\alpha=0} \nonumber \\
 && \times \left( 1 + {\cal O}(e^{-\Delta(P,\eta_*,\tau) P^3} ) \right)
 \label{EPfact1}
\end{eqnarray}
where $\CZ_{\rm D6-\overline{D6}}^\epsilon$ was defined in
(\ref{ZD6aD6def}). Note that although some D6-anti-D6 pairs with
$(\beta_i,n_i) \in \CC(P,\epsilon)$ will have $\eta > \eta_*$, they
will nevertheless still all be polar states (assuming $\epsilon$
sufficiently small), and therefore not affect the error any further.

Now from (\ref{foefoebal2}), we get
\begin{eqnarray}
 && \CZ_{D6-\overline{D6}}^{\epsilon}(e^{-2 \pi i \tau},e^{-2 \pi i (C + \frac{P}{2})},e^{2 \pi
 (\alpha-\frac{i}{2})}) \nonumber \\
 && = e^{-2 \pi i \tau \frac{P^3+c_2 P}{24}} \, e^{2 \pi I_P(\alpha - \frac{i}{2})}
 \sum_S e^{-\pi i \tau P S^2 -2 \pi i (C + \frac{P}{2}) P S} \nonumber \\
 && \qquad \times \, \CZ_{DT}^{\epsilon}(-e^{2 \pi i (\tau + i \alpha)},
 e^{2 \pi i [i \alpha P + \tau (S+\frac{P}{2}) + C]})
 \nonumber \\
&& \qquad \times \, \CZ_{DT}^{\prime \epsilon}(-e^{2 \pi i (\tau + i
\alpha)},
 e^{2 \pi i [i \alpha P + \tau (-S+\frac{P}{2}) - C]}). \label{EPfact2}
\end{eqnarray}

Let us make two remarks:

\begin{enumerate}

\item
 To get the factorized generating function
(\ref{foefoebal2}), we needed to restrict to $(\beta_i,n_i) \in
\CC(P,\epsilon)$, as defined in (\ref{CCdef}), and take $\epsilon =
\delta |P|^{-\xi_{cd}}$ to dump swing states. According to the
extreme polar state conjecture, we should therefore take $\eta_* =
\frac{\delta}{\mu} |P|^{-\xi_{cd}}$. The cutoff restricts the sum to
states splitting into a rank $r=1$ D6 and anti-D6; in the picture of
\cite{Gaiotto:2006ns,deBoer:2006vg}, this corresponds to leaving out
$\IZ_r$-quotients of ${\rm AdS}_3 \times S^2$ with $r>1$.
Furthermore, dumping swing states such as our example in section
\ref{innocuouscoredump} corresponds in $M$-theory to dumping certain
black M5 rings which exist in infinite radius Taub-NUT but not on
the finite size $S^2$. See section \ref{sec:comparison} below for a
more extensive discussion.

\item Again, due to the divergence of the sum over $S$, the error estimate
in (\ref{EPfact1}) is strictly speaking meaningless, but as in the
discussion under (\ref{polartoextremepolar}) we can give it a
precise meaning e.g.\ by introducing a cutoff in the sum over $S$.
Finally, putting (\ref{EPfact1}) and (\ref{EPfact2}) together thus
gives an approximate factorization formula for $\CZ^-(\tau,C)$.

\end{enumerate}

\subsubsection{Derivation of OSV}\label{sec:OSVderivation}

We are finally ready to put all our results together and derive a
refined version of the OSV conjecture. From (\ref{Opqint}), we have
\begin{equation} \label{intformwewant}
 \Omega((0,P,Q,q_0);t=i\infty) = (-i)^{h+1}\oint d \phi^0
 d \Phi \, e^{-2 \pi q_\Lambda \phi^\Lambda}
 \CZ(\tau=\bar\tau=i\phi^0,C=i \Phi - \frac{P}{2})
\end{equation}
where the $\phi^\Lambda$-integrals run over a single imaginary
period on the imaginary axis. We wish to derive an OSV-like formula
in the case $\Gamma=(0,P,Q,q_0)$ is \emph{nonpolar} and large, and
hence has a single centered black hole realization. The first step
is to substitute the fareytail expansion (\ref{ZPoincSer}) of $\CZ$
into (\ref{intformwewant}). Since we are considering nonpolar terms,
the $c=0$ part of the series will not contribute to
(\ref{intformwewant}). To leading order in the saddle point
approximation, the $c \neq 0$ terms contribute terms of order
$e^{S_{\rm sugra}/c}$, as can be seen directly from the expressions
(or see e.g.\ appendix A of \cite{Dabholkar:2005by}). Hence in the
large charge limit, the $c>1$ terms will be suppressed by a factor
$\sim e^{-k S_{\rm sugra}}$ compared to the $c=1$ terms, with $k$
some order 1 constant. Alternatively, we can say that these terms
are effectively suppressed by a factor $\sim e^{-k P^3/\phi^0}$ in
the partition function $\CZ(P,\phi)$, with $k$ some constant of
order one, and where we take $\phi^0$ to be positive (as it is at
the saddle point).

The $c=1$ terms correspond to $SL(2,\IZ)$ elements
\begin{equation}
 A = \left( \begin{array}{cc} 0 & -1 \\ 1 & d \end{array} \right) =
 S \, T^d
\end{equation}
so we can write, with $\tau=\bar \tau = i \phi^0$, $C=i \Phi -
\frac{P}{2}$,
\begin{eqnarray} \label{cis1part}
 \CZ^+(\tau,C)&=& \sum_{d \in \IZ}
  \omega_S^{-1} \omega_T^{-d} (\tau+d) \, e^{2 \pi i \frac{C^2}{2 (\tau+d)}} \,
  \CZ^-(-\frac{1}{\tau+d},\frac{C}{\tau+d}) \\
  && \times \left( 1 + \CO(e^{-k P^3/(\tau+d)})
  \right).
\end{eqnarray}
Here $\CZ^+$ is the non-polar part of $\CZ$, and the error term
comes from dropping the $c>1$ terms as discussed above. Note that
although the error is no longer exponentially small when $d \to
\infty$, the large $d$ terms themselves are exponentially suppressed
compared to the small $d$ terms, and therefore unimportant. Put
differently, the sum over $d$ can be traded for extending the
integration contour in (\ref{intformwewant}) over the entire
imaginary $\phi^0$-axis, but the large $d$ terms will correspond to
points far away from the saddle point, and are therefore
unimportant.

From (\ref{cis1part}), (\ref{EPfact1}) and (\ref{EPfact2}), we find,
after some work and substituting $\tau = \bar \tau = i\phi^0$, $C =
i \Phi - \frac{P}{2}$:
\begin{eqnarray}
\CZ_{\rm BH}^+(\phi^0,\Phi) &=& \frac{1}{2 \pi} \left.
\frac{\partial}{\partial \alpha} \right|_{\alpha=0} \biggl(
\sum_{d,\hat{S}} i (\phi^0-id)
 e^{2\pi I_P \alpha} e^{\frac{2 \pi}{\phi^0 - i d} \frac{P^3+c_2 P}{24}
 -\frac{\pi}{\phi^0-id} (\Phi + i \hat{S})^2 - 2 \pi i
\frac{P}{2}\cdot \hat{S} - 2 \pi i \frac{c_2 P}{24} d } \nonumber \\
&& \times
\CZ_{DT}^\epsilon(-e^{-2\pi(\frac{1}{\phi^0-id}+\alpha)},e^{2 \pi
i(\frac{1}{\phi^0-id}(\Phi + i \hat{S} + i \frac{P}{2}) + i \alpha P)})  \nonumber \\
&&  \times \CZ_{DT}^{\prime
\epsilon}(-e^{-2\pi(\frac{1}{\phi^0-id}+\alpha)},e^{2 \pi
i(\frac{1}{\phi^0-id}(-\Phi  - i \hat{S} + i \frac{P}{2}) + i \alpha
P)}) \biggr) \nonumber \\
&&\times \left( 1 + {\cal O} (
 e^{- \Delta(P,\eta_*,\frac{i}{\phi^0-i d}) \, P^3})
 \right)
 \label{lastffff}
\end{eqnarray}
Here $\hat{S} := S + \frac{P}{2} \in H^2(X,\IZ)$, and various
complicated phase factors have canceled in a nontrivial way.

Finally, using the identification $\CZ'_{DT}(-e^{-g},e^{2 \pi
i t}) = \CZ_{GW}'(g,t)$ discussed in section \ref{sec:prelim},
this becomes, remarkably
\begin{eqnarray}
\CZ_{\rm BH}^+(\phi^0,\Phi) &=& \frac{1}{2 \pi} \left.
\frac{\partial}{\partial \alpha} \right|_{\alpha=0} \sum_{d,\hat{S}}
 i (\phi^0-id) \, e^{\CF^\epsilon(P,\phi^0-id,\Phi+i \hat S,\alpha) - 2 \pi i
\frac{P}{2}\cdot \hat{S} - 2 \pi i \frac{c_2 P}{24} d}e^{\delta \CF}
\label{OSVresultfin}
 \end{eqnarray}
where
\begin{eqnarray}
 \CF^\epsilon(P,\phi,\alpha) &:=& F^\epsilon_{\rm top}(g,t) + \overline{F^\epsilon_{\rm
 top}(g,t)}, \label{FreeEnergy} \\
 F^\epsilon_{\rm top} &:=& \log
 \CZ_{\rm top}^\epsilon = \log \CZ_{\rm pol} +  \half\bigl( \log \CZ_{\rm
 DT}^{ \epsilon} +   \log \CZ_{\rm
 DT}^{\prime,\epsilon}\bigr) \label{FtopMod}
\end{eqnarray}
with substitutions
\begin{eqnarray}
 g \equiv \frac{2 \pi}{\phi^0} + 2 \pi \alpha, \qquad t \equiv
 \frac{1}{\phi^0} (\Phi + i \frac{P}{2}) + i \alpha P \label{derivedosvsubst}
\end{eqnarray}
and error
\begin{eqnarray}\label{deltaefferror}
 \delta \CF &=& {\cal O} \bigl(
 e^{- \Delta(P,\eta_*,\frac{i}{\phi^0}) \, P^3} \bigr).
\end{eqnarray}
 Recall that the cutoff $\epsilon$ is related to $\eta_*$ through
the extreme polar state conjecture of section \ref{sec:extremepolar}
as $\epsilon=\mu \eta_*$, and that we took $\epsilon = \delta
|P|^{-\xi_{cd}}$ to get rid of swing states.
In taking the complex conjugate in (\ref{FreeEnergy}), $\phi^0$,
$\Phi$ and $\alpha$ should formally be taken real. We also dropped
terms of quadratic and higher order in $\alpha$, since we set
$\alpha=0$ after taking the derivative. Surprisingly perhaps, the $2
\pi I_P \alpha$ in the exponent of (\ref{lastffff}) is reproduced to
this order by the $\alpha$-dependence of $F_{\rm pol}$ after
substituting (\ref{derivedosvsubst}). The inclusion of $\phi^0$ in
the measure and the $\partial_{\alpha}$ operation were absent in the
original OSV conjecture \cite{Ooguri:2004zv}. They can be traced
back respectively to the fact that $\CZ(\tau, C)$ has modular weight
$(-3/2,1/2)$ (for a proper $SU(3)$ holonomy Calabi-Yau), and to the
fact that there is a factor $\sim |\langle \Gamma_1,\Gamma_2
\rangle|$ in (\ref{BPSindfact2}). Both modifications are in
agreement with the results of section \ref{sec:smalltau} (see eq.
(\ref{resultsmallphi0})) obtained in the small $\phi^0$ limit
  by arguments independent of our D6-anti-D6 picture.

Note furthermore that the sum over $\hat{S}$ and $d$ together with
the phase factors in (\ref{OSVresultfin}) give the right hand side
precisely the same periodicity as the left hand side. These terms
also allow us to invert (\ref{OSVresultfin}) to the simple
expression
\begin{equation}\label{osvwithmf}
 \Omega(0,P,Q,q_0;t=i\infty) =  \int d \phi \, \mu(P,\phi) \, e^{-2 \pi q_\Lambda
 \phi^\Lambda} \, e^{\CF^\epsilon(P,\phi)+\delta \CF}.
\end{equation}
where the ``measure factor'' $\mu(P,\phi)$ is
\begin{equation}
 \mu(P,\phi) = \frac{(-i)^h}{2 \pi} \, \phi^0 \, \frac{\partial}{\partial \alpha}
 \CF^{\epsilon}(P,\phi,\alpha) \, |_{\alpha=0} = (-i)^h \phi^0 I_P + {\rm
 inst. \, corr.} \label{measurefactor}
\end{equation}
 Note that these instanton corrections are of order $\exp[- 2\pi
\vert P\vert/\phi^0]$ which is the same order as the other terms we
are trying to keep track of and hence are rather essential to a
correct formulation of the OSV conjecture.

Rather curiously, the measure factor can also be written as
\begin{equation}\label{mfkpot}
 \mu(P,\phi) = (-i)^h \frac{4\pi}{ g^{2}} \, e^{-K^\epsilon}
\end{equation}
where $K^\epsilon$ is a generalized K\"ahler potential, defined by
\begin{equation}\label{kpot}
e^{-K^\epsilon} = {\rm Re}\biggl[ \bar X^{\Lambda_1}
I_{\Lambda_1}^{~~\Lambda_2} \frac{\p F^{\epsilon}_{\rm top}}{\p
X^{\Lambda_2}}\biggr]
\end{equation}
with
\begin{equation}
  X^0= 2i \phi^0 \qquad\qquad X^A = (\Phi + \frac{i}{2} P)^A
\end{equation}
and we have used the property that $\phi^0$, $\Phi$ and $P$ are
real. Recall that $I_{\Lambda_1}^{~\Lambda_2}$ was defined  to be
$I_{\Lambda_1}^{~\Lambda_2}=\sigma_{\Lambda_2} \,
\delta^{\Lambda_1}_{\Lambda_2}$ where $\sigma_0=1$, $\sigma_A = -1$.
This measure factor is the same as the one  found in
\cite{Shih:2005he} for $X=T^6$ and $X=T^2 \times K3$.\footnote{There
are slight differences coming from the different power of $\phi^0$
in (\ref{measurefactor}), namely $(\phi^0)^{1-b_1}$, one gets in
those cases, and the possibility to include gravitini charges. These
factors arise in our approach for these cases as well. }

\subsubsection{Analysis of the error terms}\label{sec:erroranalysis}

Let us now consider the error term. When the saddle point lies at
sufficiently small $\phi^0$, $\Delta$ is guaranteed to be positive,
hence the first error term is exponentially small at large $P$. As
we mentioned before, the meaning of ``sufficiently small'' depends
on the growth of the polar entropies $S(\eta) \sim \Sigma(P,\eta)
P^3$ as a function of $P$. We define an exponent $\kappa$ by
$\Sigma(P,\eta) P^3 \sim   \, |P|^{\kappa}$, or, more precisely
\footnote{We take the limit supremum $\overline{\lim}$ which always
exists. We expect $\kappa$ to be at most weakly dependent on $\eta$.
}
\begin{equation}\label{precisekappa}
\kappa := 3 + \overline{\lim}_{\vert P\vert \to \infty} \frac{
\log\Sigma(P,\eta) }{\log \vert P\vert} .
\end{equation}
Then we  need
\begin{equation} \label{strongtopbound}
 g^{-1} \sim \phi^0 < \CO(\vert P\vert^{3-\kappa})
\end{equation}
for all $\eta$, $0\leq \eta \leq 1$, in order to have an
exponentially suppressed error. A simple estimate based on the BH
entropy of two-centered configurations realizations of polar states,
outlined under (\ref{downunder}), indicates $\kappa=3$, hence we
would need $g > \CO(1)$, i.e.\ strong topological string coupling.
However, since we are considering indices it might be possible in
principle that miraculous cancelations occur which effectively lower
$\kappa$. We postpone further discussion of this possibility to
section \ref{sec:cancel}.

Let us now consider the error we have when we are well within the
regime (\ref{strongtopbound}). More precisely let $\phi^0_{cr}$ be
the value at which $\Delta=0$. For $\phi^0\ll \phi^0_{cr}$, the
second term dominates in (\ref{minimizeta}), so
\begin{equation}
 \Delta \cong \frac{\pi}{12} \frac{1}{\phi^0} \eta_*
\end{equation}
and the error term becomes
\begin{equation}\label{simplererror}
 \delta \CF \sim \CO(e^{- \frac{\pi}{12}\frac{\eta_* P^3}{\phi^0}})
\end{equation}
%
For $\eta_*$ a constant independent of $P$, this agrees with the
error found in \cite{Shih:2005he} for $T^2 \times K3$ and $T^6$.
However, as discussed in section \ref{innocuouscoredump}, in general
we must take $\epsilon$, and hence $\eta_*= \epsilon/\mu$, to depend
on $\vert P\vert$. Taking $\epsilon = \delta \vert P \vert^{-
\xi_{cd}} $ these lead to corrections of order
\begin{equation} \label{corecorrections}
 \exp[- \frac{\pi \delta}{12 \mu}\frac{P^{3-\xi_{cd}}}{\phi^0}]
\end{equation}
%
%
We saw that the split configurations (\ref{newcore}) imply that we
must have $\xi_{cd}\geq 1$. On the other hand, worldsheet instantons
contribute terms of order $\exp[ - 2\pi \beta\cdot P /\phi^0]$.
Therefore if the corrections (\ref{corecorrections}) are not to
overwhelm the worldsheet instanton corrections then we must have
$\xi_{cd}\leq 2$. Unfortunately  the value of $\xi_{cd}$ is unknown.
We hasten to point out that the value $\xi_{cd}=1$ makes excellent
physical sense for reasons discussed in section \ref{sec:dangercore}
below.

%
%


\section{Discussion} \label{sec:discussion}

\subsection{Summary}

Let us summarize our final result, and discuss to what extent it
agrees with the original OSV conjecture.

We consider the index of BPS states, defined in (\ref{indexdef}),
with charge $p^0=0$, large $|P|:=(D_{ABC} P^A P^B P^C)^{1/2}$ and
$\widehat{q}_0:= q_0 - \half D^{AB} Q_A Q_B < 0$. For these charges
a  single centered black hole solutions exists. We choose the
background $t=i\infty$. Then the index  is given by
\begin{equation}\label{osvfinalfinal}
 \Omega(P,Q,q_0;t=i\infty) =  \int d \phi \, \mu(P,\phi) \, e^{-2 \pi q_\Lambda
 \phi^\Lambda} \, e^{\CF^\epsilon(P,\phi)+\delta \CF}.
\end{equation}
Where, using  the substitutions
\begin{eqnarray} \label{osvsubstoncemore}
 g \equiv \frac{2 \pi}{\phi^0}, \qquad t^A \equiv
 \frac{1}{\phi^0} (\phi^A + i \frac{P^A}{2}),
\end{eqnarray}
we have
\begin{eqnarray}
 \mu(P,\phi) &:=& (-i)^h \frac{4\pi}{ g^{2}} \, e^{-K^{\epsilon}(g,t,\bar{t})} = (-i)^{h}  \phi^0
 \mbox{$(\frac{P^3}{6}+\frac{c_2 P}{12})$}
 + {\rm inst. \, corr.} \\
 %
%
 \CF^\epsilon(P,\phi) &:=& F^\epsilon_{\rm top} + \overline{F^\epsilon_{\rm
 top}}, \label{FplusFbar} \\
 F^\epsilon_{\rm top}(g,t) &:=& \log
 \CZ_{\rm top}^\epsilon(g,t) \\
 &:=& \log \CZ_{\rm pol}(g,t) + \half\biggl( \log \CZ_{\rm DT}^{
 \epsilon}(-e^{-g},e^{2\pi i t} ) + \log \CZ_{\rm DT}^{\prime,
 \epsilon}(-e^{-g},e^{2\pi i t} )\biggr) \\
 {\cal Z}_{\rm pol}(g,t) &:=& \exp \biggl( -\frac{(2\pi i)^3}{6 g^2} D_{ABC} t^A t^B t^C
 - \frac{2 \pi i}{24} c_{2A} t^A \biggr) \\
 \CZ_{DT}^\epsilon(u,v) &:=& \sum_{|n|,\beta \cdot P < \epsilon P^3}
 N_{DT}(\beta,n) \, u^n \, v^\beta.
\end{eqnarray}
and $\CZ_{\rm DT}^{\prime,\epsilon}$ is defined analogously by
cutting off the series for $\CZ'_{\rm DT}$. The full expression for
$e^{-K^\epsilon}$ is given in (\ref{kpot}). In taking the complex
conjugate in (\ref{FplusFbar}),  $\phi$ should  be treated as real.

The error is of order
\begin{eqnarray}
 \delta \CF &=& {\cal O} \bigl(
 e^{- \Delta(P,\eta_*,\frac{i}{\phi^0}) \, P^3} \bigr)
\end{eqnarray}
where
\begin{eqnarray}
 \Delta(P,\eta_*,\frac{i}{\phi^0}) &:=&
 \min_{\eta_*<\eta<1} \left( -\Sigma(P,\eta) + \frac{\pi}{12 \, \phi^0} \, \eta \right)
 \\
 \Sigma(P,\eta) &:=& \frac{1}{P^3} \max_{[Q]}
 \log \left| \Omega \left(P,[Q],\hat{q}_0=\mbox{$(1 -
 \eta) \frac{P^3+c_2P}{24}$} \right) \right|.
\end{eqnarray}
Here $\Omega\left(P,[Q],\hat{q}_0\right)$ is the index of BPS states
of D4-charge $P$, D2-charge $[Q]$ with $[Q]$ in a fundamental domain
for the symmetry $Q_A \to Q_A + D_{ABC} P^B n^C$, $n^C \in \IZ$, and
reduced D0-charge (invariant under this symmetry) $\hat{q}_0$,
evaluated in a background $t=i\infty$. The range $0 \leq \eta < 1$
corresponds to polar charges, with $\eta=0$ being the most polar
one.

The integral in (\ref{osvfinalfinal})   runs over the imaginary
$\phi$-axes. One might worry that the error (\ref{deltaefferror})
becomes $\CO(1)$ for large $\phi^0$, but since the integral is
dominated by its saddle point, this part of the integration contour
is negligible anyway. The saddle points for $\phi$ turn out to be
real.

 For the cutoffs $\epsilon$ and $\eta$, we have the relations
\begin{equation}
 \mu \, \eta_* = \epsilon = \delta |P|^{-\xi_{cd}}.
\end{equation}
The first equality follows from the extreme polar state conjecture
of section \ref{sec:extremepolar}, and the second one is required to
get rid of swing states, as discussed in section
\ref{innocuouscoredump}. Here $\delta$ and $\mu$ are $P$-independent
constants. The core dump exponent $\xi_{cd}$ is a kind of critical
exponent, which we bounded by $1 \leq \xi_{cd} \leq 3$.


%
%

In order to clarify  the domain of validity of (\ref{osvfinalfinal})
we introduced a second critical exponent, defined by the growth of
the   polar state indices growth with $P$ at fixed $\eta$:
\begin{equation} \label{polargrowthkdef}
 \log |\Omega| \sim |P|^\kappa
\end{equation}
or, more precisely, as in (\ref{precisekappa}).
 In
terms of $\kappa$ the error term is controlled only if
\begin{equation} \label{gtopisstrongbound}
 g > \CO(|P|^{\kappa-3})
\end{equation}
 If we estimate the polar index
growth by the growth of the two-centered black hole realizations, we
get
\begin{equation}
 \kappa=3
\end{equation}
implying OSV is only valid at \emph{strong} topological string
coupling. We discuss the possibility that $\kappa$ gets miraculously
lowered by cancelations in the index in section \ref{sec:rov}.
Neglecting instanton corrections, the saddle point value of $\phi^0$
is given by $\phi^0_* \approx \sqrt{\frac{P^3}{24
|\widehat{q}_0|}}$. Hence the bound (\ref{gtopisstrongbound})
translates to a bound on the charges:
\begin{equation} \label{chargerangeov}
 |\widehat{q}_0| > \CO(|P|^{2\kappa-3}).
\end{equation}
Note that when $\kappa=3$, this bound always gets violated when
scaling up the total charge $\Gamma = (0,P,Q,q_0)$ uniformly by a
sufficiently large $\Lambda$, and that to avoid this, we need
$\kappa \leq 2$. This is not an artifact of our derivation scheme:
it is precisely as expected from the ``entropy enigma'' discussed in
section \ref{sec:entropyenigma}. The closely related two centered
configurations considered there had an entropy growing as
$\Lambda^3$, so if no miraculous cancelations occur between
contributions to the index bringing down the growth of $\log
\Omega(\Lambda \Gamma)$ to $\Lambda^2$ or lower, the OSV conjecture
at $t=i\infty$ necessarily breaks down at sufficiently large
$\Lambda$, since   it predicts $\Omega(\Lambda \Gamma) \sim
\Lambda^2$ to leading order at large $\Lambda$. We will discuss this
in more detail in section \ref{sec:rov}.

Finally, when $g$ is well inside the regime
(\ref{gtopisstrongbound}), the error simplifies to
\begin{equation} \label{corecorrections2}
 \delta \CF \sim \exp[- \frac{\pi \delta}{12 \mu}\frac{P^{3-\xi_{cd}}}{\phi^0}]
\end{equation}
For this to be negligible compared to the instanton contributions to
the free energy, which are suppressed as $e^{-\beta \cdot
P/\phi^0}$, we need $\xi_{cd} < 2$ or $\xi_{cd} = 2$ and $\delta \gg
1$. Using $\Im t \sim P/\phi^0$ and $g \sim 1/\phi^0$, we can also
write this as
\begin{equation}
 \delta \CF \sim \exp[- \frac{c}{g^{2-\xi_{cd}}} |\Im t|^{3-\xi_{cd}}],
\end{equation}
with $c$ a constant. Note that for $\xi_{cd}=1$, this is suggestive
of a D4/M5 contribution to the Schwinger computation of the
topological string free energy
\cite{Gopakumar:1998ii,Gopakumar:1998jq}. We return to the issue of
determining the value of $\xi_{cd}$ in section \ref{sec:dangercore}.

\subsection{Differences with original OSV conjecture}

We note the following differences with the original OSV conjecture:
\begin{enumerate}
 \item There is an additional measure factor $\mu(P,\phi)$, in agreement with the special cases
 studied in
 \cite{Dabholkar:2005by,Dabholkar:2005dt,Shih:2005he}.
 This does not affect the leading saddle point
 evaluation of the entropy, but does affect the inverse charge corrections to
 it. The origin of this measure factor is essentially the presence
 of  the angular momentum factor $|\langle \Gamma_1,\Gamma_2
 \rangle|$ in the factorization formula (\ref{BPSindfact2}) for degeneracies of polar
 states.

  Furthermore, if we state the OSV formula using $\vert \Psi_{\rm top}\vert^2$
 with the standard definition of $\Psi_{\rm top}$ then, since
  our definition of the topological
 string partition function in (\ref{FtopMod}) was nonstandard
 (because the degree zero terms  $\CZ_{\rm DT}^{0}$ make use of  the MacMahon function,
 which differs from the standard perturbative $F_{\rm top}$ by a term proportional to
 $\frac{\chi}{24}\log g$) one would have to include a further factor
 of $g^{\chi/24}$ in the measure.

 \item  The topological string partition function is cut off. The
 cutoff cannot be removed, since the full $\CZ_{\rm top}$ has zero
 radius of convergence and hence does not exist as a function which
 can be integrated, not even in a saddle point approximation. We were led to
 put a cutoff on the DT invariants $N_{DT}(\beta,n)$ contributing to
$\CZ_{\rm
 top}$, namely $|n|<\epsilon P^3$, $\beta \cdot P < \epsilon P^3$.
 This does not translate into a simple cutoff on the
 corresponding M2 BPS invariants in the infinite product representation.
 Physically, this happens because the
 existence of D6-anti-D6 bound states depends on the \emph{total}
 D2-D0 charge of the constituents, while the M2 BPS invariants refer to
 ``one particle'' contributions to these D2-D0 charges. As usual
 with cutoffs, there is some arbitrariness in their choice; possibly
 there are other ``regularization schemes'' than the one we used.


\item We find corrections, exponentially suppressed at large $P$.
These are due to the $c>1$ terms of the fareytail series as well as
to the non-extreme polar states we dropped. Taking into account
terms corresponding to   $SL(2,\IZ)$ transforms with $c>1$ in the
fareytail series would add $|\CZ_{\rm top}|^2$ type terms to the
integrand on the right hand side of the OSV formula, but with
substitutions different from (\ref{osvsubstoncemore}). These give
corrections $\delta \CF \sim e^{-g \, P^3}$. Taking into account
contributions of D6-anti-D6 bound states with D6 multiplicities
$r>1$ (which are necessarily non-extreme polar with $\eta \geq 3/4$)
presumably would spoil the simple relation to $\CZ_{\rm top}$. These
contributions give corrections $\delta \CF \sim e^{-g \, P^3}$.
Moreover, taking into account non-extreme polar states ($\eta >
\eta_*$), even at $r=1$, would spoil factorization, as suggested
e.g.\ already by fig.\ \ref{b1b2range}. More importantly, the
existence of the swing states of section \ref{innocuouscoredump},
which spoil factorization and the relation to DT invariants, force
us to restrict to $\eta < \eta_* \sim |P|^{-\xi_{cd}}$. This gives
corrections $\delta \CF \sim e^{-g \, |P|^{3-\xi_{cd}}}$. If
$\xi_{cd}>2$, the error actually swamps the instanton contributions
we want to keep. We only know with certainty that $1 \leq \xi_{cd}
\leq 3$, although there is some evidence that $\xi_{cd}=1$.

 \item  We find a restriction on the range of validity. We need (\ref{chargerangeov}) to be
 satisfied. In particular, if $\kappa > 2$, our formula breaks
 down when uniformly scaling up all charges by a sufficiently large
 $\Lambda$, whereas the original conjecture was meant to be valid
 precisely in this large $\Lambda$ regime. Put differently,
 since the saddle point  $g_* \sim 1/\Lambda$,
 our result is guaranteed to work in the strong topological coupling
 regime, but fails in the weak coupling regime unless there are
 miraculous cancelations between the contributions to the indices of the polar
 states. The original conjecture on the other hand was supposed to work
 at weak $g$.

\end{enumerate}

 Clearly, the last two points lead to potentially the most significant
 discrepancies with the original conjecture,
 so we will examine these points more closely and discuss the various possible
 loopholes in sections \ref{sec:rov} and \ref{sec:dangercore}.
 Before we get to this though, we will give an interpretation of our
 results in the language of the M-theory derivations \cite{Gaiotto:2006ns,deBoer:2006vg} of the OSV
 conjecture which have appeared, and demonstrate in particular
 that the discrepancies and subtleties we find are not tied to our specific picture.

\subsection{Comparison with M-theory derivations}
\label{sec:comparison}

The M-theory derivations \cite{Gaiotto:2006ns,deBoer:2006vg} of the
OSV conjecture did not detect the measure factor and did not attempt
to give bounds on the regime of validity or on the error. The
appearance of a cutoff and corrections was emphasized in
\cite{deBoer:2006vg}, but the level of analysis was insufficient to
provide explicit cutoff prescriptions or estimates of corrections.

In these derivations the polar states were represented as dilute
gasses of spinning M2 branes and anti-M2 branes orbiting the poles
of the $S^2$ in the spacetime $AdS_3 \times S^2 \times X$, the
latter carrying a $G$-flux proportional to $P$. Some of the issues
which were left open in \cite{Gaiotto:2006ns,deBoer:2006vg} were the
parameter range for which this dilute gas approximation is accurate,
what happens if the M2's and anti-M2's start ``spilling over'' into
each others hemispheres and what the effect is of other BPS states
such as M5 branes and of taking into account other geometries such
as quotients of $AdS_3 \times S^2$.

The various elements in our picture have a fairly straightforward
translation into this M-theory picture, and hence our analysis
clarifies all of these issues.\footnote{This translation was
developed in collaboration with Dieter Van den Bleeken
\cite{freddie}.} The rough idea is as follows. A D6-brane lifts in
M-theory to a Taub-NUT space, so a bound state consisting of a pure
D6 and an anti-D6 stabilized by flux lifts to a Taub-NUT-anti-Taub
NUT geometry stabilized by flux. In the M-theory limit, this
geometry becomes $AdS_3 \times S^2$ with flux proportional to the
net D4-charge $P$, with the north and south poles of the sphere
identified with the centers of Taub-NUT, projecting down to the D6
and resp.\ anti-D6. Adding D2D0 charge to the D6 and anti-D6 branes
to turn them into black holes corresponds to putting spinning M2
BMPV black holes at the north and south poles of the sphere
\cite{Gaiotto:2005gf}. Our D2D0 halos orbiting the D6 and anti-D6
lift to M2 branes orbiting the north and south poles. More
complicated D6 core states such as the swing states in section
\ref{innocuouscoredump} realized as 2-centered D6-D4 configurations
become M5 black rings in $AdS_3 \times S^2$. Higher rank $r>1$ D6
anti-D6 bound states correspond to $\IZ_r$ quotients of $AdS_3
\times S^2$ and deformations thereof. Fractal flow trees such as
fig.\ \ref{fr} lift to foamy ``bubbling'' geometries. And so on. The
upshot is that we have included \emph{all} possible contributions in
our analysis, no matter how exotic or complicated.

Now, the idea of \cite{Gaiotto:2006ns} can be summarized like this:
Cut the $AdS_3 \times S^2$ in two halves and identify one half with
part of Taub-NUT with flux and the other one with part of
anti-Taub-NUT with flux. Now complete these cut off, finite-size,
finite flux Taub-NUT spaces to complete Taub-NUT spaces of infinite
size, with infinite Taub-NUT circle radii and infinite total
integrated flux, and count BPS states on each of those. In this
infinite-size, infinite flux limit, all BPS states are
well-described by lowest Landau-levels of spinning M2 probes, and
the generating function of their indices is the Gopakumar-Vafa
partition function (as we argued in section \ref{sec:D6D2D0},
refining the analysis of \cite{Dijkgraaf:2006um}). Hence, in this
approximation, ignoring higher $r$ geometries, finite size effects
and the coupling between the two sectors, the $AdS_3 \times S^2$
elliptic genus is simply given by the product of two GV products,
leading to $\CZ \sim \CZ_{\rm top} \overline{\CZ_{\rm top}}$.

It is of course not clear to what extent this picture is justified,
and through the above dictionary, our work can be interpreted as a
thorough analysis of this problem. Let us translate a few of our
results into this M-theory picture:

\begin{itemize}

\item We found it necessary in section \ref{sec:specandflow} to cut
off the D2-D0 charges at $\beta_i \cdot P < \epsilon P^3$, $|n_i| <
\epsilon P^3$, $\epsilon \ll 1$, in order to guarantee existence at
least of the first split of the flow tree for any choice of
$(\beta_i,n_i)$ within this cut off domain. This corresponds to the
fact that when these charges grow too large, the M2's and anti-M2's
associated to the two $S^2$ poles start interacting so strongly that
they can no longer be seen as independent probes and the BPS state
can cease to exist altogether. The extreme polar state conjecture
guarantees we do not need to worry about this for sufficiently polar
states (as well as that we do not have to worry about $\IZ_r$
quotient geometries).

\item Swing states can be seen as BPS states which exist in infinite radius
Taub-NUT, but not at some smaller radius --- more precisely not in
the finite size $AdS_3 \times S^2$ --- or vice versa. The example of
the swing state of section \ref{innocuouscoredump} can be thought of
as a BPS M5 ring which would fit in a Taub-NUT of sufficiently large
radius, but not in the finite radius $AdS_3 \times S^2$, as the ring
radius becomes too large. At infinite TN radius, this should not be
counted as a separate BPS state since all BPS states can be
described by light M2 probes which are guaranteed to exist, but at
finite radius this is no longer so and BPS states might disappear
from the spectrum, hence altering the BPS free energy function which
counts BPS states.\footnote{Related to this, as was shown in
\cite{Klemm:2005pd}, in certain limits of the IIA Enriques CY moduli
space, the BPS free energy is generated by D4-branes indices rather
than D2 brane indices, as the former become the light states there.}
Somewhat surprisingly perhaps, we found that merely taking the
cutoff $\epsilon \ll 1$ is not enough to avoid this phenomenon, but
that instead one should take $\epsilon < \delta |P|^{-\xi_{cd}}$
with $\xi_{cd} \geq 1$. If $\xi_{cd}=1$ (as is the case for our
examples and for which we gave some circumstantial evidence at the
end of section \ref{innocuouscoredump}), the correction to the free
energy is of order $\delta \CF \sim e^{-g_{\rm top} |P|^2} \sim
e^{-(\Im \, t)^2/g}$, indeed suggestive of finite size
nonperturbative D4/M5 corrections to the computation of
\cite{Gopakumar:1998ii,Gopakumar:1998jq}. This can viewed as a
further physical indication for $\xi_{cd}=1$, although we will not
try to make this precise here.

\item Our restriction to sufficiently strong $g$ arose from the fact
that at weak $g$, the fareytail series ceases to be dominated by the
extreme polar terms due to entropic effects. As a result, in this
regime, it is no longer justified to disentangle the two sectors,
since the main contribution will come from complicated BPS
configurations delocalized over the sphere, which do not factorize.
The meaning of ``strong'' and ``weak'' $g$ depends on the growth of
polar indices, to be discussed in the next subsection. This is
closely related to the entropy enigma of section
\ref{sec:entropyenigma}, which in M-theory translates to the
entropic dominance of geometries containing two BMPV black holes
over the M5 black string.

\item The measure factor we find can be traced back to the fact that even
in the most dilute gas regime, the M2 and anti-M2 sectors do not
fully decouple, since there is a multiplicative contribution to the
ground state degeneracies depending on $J^3_R$ (given by $|\langle
\Gamma_1, \Gamma_2 \rangle|$ in our setup) which depends on the
charges in a non-factorized way. This factor was not taken into
account in \cite{Gaiotto:2006ns,deBoer:2006vg}, but was noted in the
M-theory context in \cite{Gaiotto:2006wm}, where it was found
necessary for modularity of the M5 elliptic genus and detailed
matching with geometrical considerations.

\end{itemize}

\subsection{Range of validity, background dependence and miraculous cancelations}
\label{sec:rov}

We now turn to the issue of weak versus strong topological string
coupling, which as we pointed out depends crucially the growth of
polar indices with $|P|$ at fixed $\eta$ where $\eta$ parametrizes
$\hat{q}_0 = (1 -
 \eta) \frac{P^3+c_2P}{24}$. Note that this is equivalent to the
growth of these indices under the scaling $(p^0,p,q,q_0) \to
(p^0,\lambda p,\lambda^2 q,\lambda^3 q_0)$ since (in the large
$\lambda$ limit) this leaves $\eta$ invariant while scaling up $|P|
\to \lambda |P|$. Recall furthermore from section
\ref{sec:symmetries} that this rescaling is a symmetry of arbitrary
multicentered BPS configurations (provided one also scales
$t_{\infty} \to \lambda t_{\infty}$), with corresponding horizon
entropy growing as $S \sim \lambda^3$. Thus, this suggests $\log
\Omega \sim P^3$ at fixed $\eta$, and hence $\kappa=3$ in
(\ref{polargrowthkdef}), implying a breakdown of the refined OSV
formula (\ref{osvfinalfinal}) at weak topological string coupling or
equivalently in the limit in which we scale up all charges
uniformly.

Now note on the other hand that when the total charge is
$(0,P,0,0)$, the same non-uniform scaling actually acts uniformly on
the \emph{total} charge, so if multicentered configurations exist
with nonzero horizon areas and such a total charge, their entropies
will scale as $\lambda^3$. We saw in section \ref{sec:entropyenigma}
that such multicentered solutions do indeed exist, and that moreover
a slight extension of the argument just given leads to the
conclusion that uniform rescaling of \emph{any} D4-D2-D0 charge
(with $P>0$) by a sufficiently large $\Lambda$ leads eventually to
multicentered configurations with horizon entropy growth $S \sim
\Lambda^3$. This suggests $\log \Omega \sim \Lambda^3$ in this
regime, in contradiction with the OSV prediction, which scales as
$\Lambda^2$. This confirms the close relation between the ``entropy
enigma'' of section \ref{sec:entropyenigma} and the breakdown of eq.
(\ref{osvfinalfinal}), and the fact that the latter is not due to a
shortcoming of our derivation itself.

There are two possible loopholes to these conclusions, which could
potentially still allow some version of the OSV conjecture to be
valid in the large $\Lambda$ limit. The first one is that we have
restricted our attention to a background $t=i \infty$, while perhaps
the OSV conjecture   should instead be taken to be valid only at
some other distinguished point.   The second one is that we are
considering an index, which gets many contributions with different
signs, so there might be miraculous cancelations bringing down the
growth of the index compared to the supergravity entropies of
individual contributing configurations.

\subsubsection{Evaluation point}

We do not have much to say about the first possibility. The reason
why we considered $t_{\infty}=i \infty$ only is that  our derivation
crucially relies on the D4 partition function being a generalized
Jacobi-form, and this is only plainly the case at $t=i \infty$. For
example, trying to construct some sort of partition function where
the indices making up the coefficients are all evaluated at the
attractor point of the charge in question would manifestly not give
a Jacobi form, since more or less by definition such a ``partition
function'' would not have a polar part (since polar charges do not
have attractor points). The necessity to restrict to $t=i \infty$ is
not only true for our derivation, but for all derivations based on
the fareytail expansion that have appeared
\cite{Gaiotto:2006ns,deBoer:2006vg}.

However one could of course contemplate   other backgrounds. A
natural choice would be to take $t_{\infty}$ at the attractor point
$t_*(p,q)$ of the charge under consideration (assuming this exists),
and postulate  a formula like
\begin{equation}\label{osvvariantone}
\Omega(p,q; t_*(p,q))\quad {\buildrel ? \over  \sim} \quad \int
d\phi \vert \CZ_{top}(p,\phi)\vert^2 e^{-2\pi \phi^\Lambda
q_\Lambda}
\end{equation}
Unlike the $t=i\infty$ case, there is certainly no evidence against
such a claim. For example, the multicentered configurations which
lead to the $\Lambda^3$ scaling of the entropy do not exist in the
attractor background --- only configurations encoded by single
centered attractor flows, i.e.\ single centered black holes as well
as multicentered ``scaling'' solutions asymptotically connected to
single centered black holes (cf.\ section \ref{sec:scalingsol}).
There are very good physical reasons to believe that, at least to
leading order, this version of the OSV conjecture should be correct,
since in this background one expects the leading order statistical
entropy to be given by the Bekenstein-Hawking-Wald entropy, which
coincides with the logarithm of the saddle point value of the right
hand side of (\ref{osvvariantone}).

More generally, one could also consider a fixed finite $t_{\infty}$
background and take $(p,q) \to \Lambda \, (p,q)$, $\Lambda \to
\infty$. From the discussion in section \ref{sec:entropyenigma}, one
can see that in such a limit, the 2-centered configurations with
$\Lambda^3$ entropy will again disappear, since the walls of
marginal stability move off to infinity when $\Lambda \to \infty$.
Note that this raises subtle order of limits issues.

Yet another alternative modifies the OSV conjecture in the form
(\ref{osvconj})  by replacing the definition of $\CZ_{\rm BH}$ by
\begin{equation} \label{Zosvnew}
 \check{\cal Z}_{\rm BH}(p,\phi) := \sum_{q}
 \Omega(p,q;t^A = \frac{\Phi^A + i P^A}{\phi^0} )
 \, e^{2 \pi \phi^\Lambda q_\Lambda }.
\end{equation}
This form of the OSV conjecture, regrettably, would seem to be
inconsistent with the wall-crossing formulae we have described, so
the right-hand side, $\vert \CZ_{\rm top}\vert^2$ would need to be
modified also in some way.

Sadly, a direct microscopic counting at $g_{IIA} = 0$ at finite
$t_{\infty}$ seems out of reach for the charges of interest, because
the microscopic description is not sufficiently understood. In the
IIA picture, the problem is that $\alpha'$ corrections to the D-term
constraints determining the moduli space (in particular determining
$\Pi$-stability \cite{Douglas:2000ah,Douglas:2000gi}) become
manifestly of crucial importance, since they are responsible for the
elimination of the ``extra'' states corresponding to multicentered
black holes with $\Lambda^3$ entropy growth existing at large
radius. These $\alpha'$ corrections are not known systematically.
One could try to use mirror symmetry to type IIB, where the relevant
D-branes are special Lagrangian 3-cycles and stability becomes just
a classical geometrical property, but the problem on this side is
$(a)$ the F-terms receive complicated disk instanton corrections and
$(b)$ even classically very little is known about special Lagrangian
3-cycles in compact manifolds.

An alternative approach to a derivation, directly at weak $g_{\rm
top}$, was suggested in \cite{Beasley:2006us}, based on AdS-CFT and
the computation of the free energy in IIA perturbation theory in a
suitable attractor background. There are several points to be
clarified in this derivation, and it is not quite a microscopic
derivation in the sense of directly counting underlying quantum
mechanical degrees of freedom of some brane model. Nevertheless, if
correct, the proposal would give a very nice explanation of why
$\CZ_{\rm top}$ should govern the corrections to the
Bekenstein-Hawking entropy.

It would clearly be desirable to know whether the index of D-brane
microstates   at $t=i\infty$, might still be governed by the OSV
formula even in the weak $g_{\rm top}$ regime, so let us next
examine the second possibility.

\subsubsection{Cancelations between index contributions?} \label{sec:cancel}

As discussed above, the only way this can happen is if the
multicentered black hole entropy for individual configurations
corresponding to polar charges grossly overestimates the actual
index for a given charge.

Recall that the D4-D2-D0 polar indices are given by the
factorization formula (\ref{BPSindfact2}):
\begin{equation} \label{splitnrone}
 \Omega(\Gamma,t_\infty) =   \sum_{\Gamma \to \Gamma_1 + \Gamma_2} (-1)^{\langle \Gamma_1,\Gamma_2
 \rangle-1} |\langle
 \Gamma_1,\Gamma_2
 \rangle| \,\, \Omega(\Gamma_1;t_{\rm ms}) \,\,
 \Omega(\Gamma_2;t_{\rm ms})
\end{equation}
where $t_{\rm ms}$ denotes the location of the MS wall for the split
$\Gamma \to \Gamma_1 + \Gamma_2$ along the $\Gamma$ attractor flow
starting at $t=i\infty$ (if this exists). As we saw, $\Gamma_1$ and
$\Gamma_2$ need to have nonzero (and of course opposite) D6-charges
for the split to exist.

Each of the $\Omega(\Gamma_i;t_{\rm ms})$ could in turn still get
contributions from different splittings and possibly also from a
single flow (corresponding to a single centered black hole
realization of $\Gamma_1$), leading to further expressions like
\begin{equation} \label{splitnrtwo}
 \Omega(\Gamma_1;t_{\rm ms}) =
 \Omega(\Gamma_1;t_*(\Gamma_1)) +
 \sum_{\Gamma_1 \to \Gamma_1' + \Gamma_2'} (-1)^{\langle
 \Gamma_1',\Gamma_2'
 \rangle-1} |\langle
 \Gamma'_1,\Gamma'_2
 \rangle| \,\, \Omega(\Gamma'_1;t'_{\rm ms}) \,\,
 \Omega(\Gamma'_2;t'_{\rm ms}),
\end{equation}
where $t_*(\Gamma_1)$ is the attractor point of $\Gamma_1$, and so
on.

Phrased in this framework, the problem we face is that there exist
splits $\Gamma \to \Gamma_1 + \Gamma_2$ into two black holes with
$S_{BH}(\Gamma_i)$ scaling as $|P|^3$ at fixed $\eta$. Identifying
$\log \Omega(\Gamma_i,t_*(\Gamma_i)) \approx S_{BH}(\Gamma_i)$, we
would thus get a contribution scaling as $e^{c |P|^3}$ to the total
index.

Hence we see there are three different ways the total index could
still grow more slowly than this:
\begin{enumerate}
 \item The identification $\log \Omega(\Gamma_i,t_*(\Gamma_i)) \approx
 S_{BH}(\Gamma_i)$ is wrong, and in fact LHS $\ll$ RHS.
 \item There are miraculous cancelations already between the contributions in
 (\ref{splitnrtwo}).
 \item There are miraculous cancelations between the contributions in
 (\ref{splitnrone}).
\end{enumerate}
Possibility (1) is extremely unlikely. In all cases in which one has
been able to compute reliably  the (proper) index of BPS states of a
large black hole at or near its attractor point (e.g.\
\cite{Maldacena:1997de,Vafa:1997gr,Diaconescu:2006qm}), its
logarithm has been found to coincide with the
Bekenstein-Hawking-Wald horizon entropy, even beyond leading order.
Although physically one expects the horizon entropy to count the
true number of BPS states at the attractor point, there is in
general no reason to expect this true number to be much larger than
the index at finite values of the string coupling, as generically
quantum tunneling effects will lift unprotected bose-fermi pairs as
soon as the coupling becomes nonzero. (See section
\ref{sec:openproblems} for more discussion about this.)

At first sight, a class of five dimensional M2 black
holes\footnote{And therefore, through the 4d-5d correspondence of
\cite{Gaiotto:2005gf}, also a class of four dimensional D6-D2-D0
black holes, for which the D6-brane charge is 1, the D2 charge
equals the M2 charge $Q$, and the D0-charge equals $2 J^3_L=2 m_L$.
In other words, precisely the kind of 4d black holes we are
interested in here. \label{4d5dfootnote}} studied in
\cite{Vafa:1997gr} seems to provide a strong counterexample to this
claim: it was found there that the black hole horizon entropy
matched the total dimension of the cohomology of the corresponding
microscopic D-brane moduli space rather than the index, which was
taken to be the Euler characteristic, i.e.\ $N_Q=n^0_Q$ as defined
in (\ref{Nqdef}) and (\ref{NGVrel}). In fact the logarithm of the
former was found to grow like $Q^{3/2}$, while the latter grows only
like $Q$, with $Q$ the M2 charge, precisely the kind of miraculous
cancelation we are after. However, upon closer inspection, one sees
that $N_Q$ is \emph{not} the proper index to compare to. Indeed, a
5d BPS black hole is characterized not only by its M2 charge $Q$,
but also $SU(2)_L$ spin $J^3_L=m_L$. Therefore, the proper index to
compare to is $N_Q^{m_L}$ defined in (\ref{Omjldef}) and in fact, we
argue in appendix \ref{app:entropy5dBH} that for this model, $\log
N_Q^{m_L} \sim \sqrt{Q^3-m_L^2}$, in perfect agreement with the
black hole horizon entropy to leading order. The origin of this huge
cancelation arising when summing over $m_L$ was explained under
(\ref{NGVrel}), and can be summarized as $(1-1)^n=0$. The
enhancement of growth going from the $n_Q^r \sim e^Q$ to the
$N_Q^{m_L} \sim e^{Q^{3/2}}$ is due to the presence of large
binomial coefficients in the relation (\ref{NGVrel}) between them,
which in turn come from degeneracies due to the Wilson line moduli.
In any case, the upshot is that the proper index again agrees with
the black hole horizon entropy.

Nevertheless, the cancelation is suggestive. Could it be that in
summing over all contributions to our total index, we are
effectively summing over $m_L$ (or, in four dimensional language as
in footnote \ref{4d5dfootnote}, over D0-charges), thus producing a
near-exact cancelation?   This brings us to possibilities (2) and
(3) in the list above. In particular, as we will see below,
possibility (2) might be related to this.

Before we go on, it is worth emphasizing that in general
cancelations changing the exponential growth behavior have to be
pretty miraculous indeed. Consider for simplicity two contributions
of nearly the same size but with opposite signs, say
\begin{equation}
 \zeta := e^{c |P|^3} - e^{(c + \epsilon) |P|^3} \approx \epsilon
 |P|^3 e^{c |P|^3}.
\end{equation}
Then to get $\zeta \sim e^{c' |P|^2}$, we need $\epsilon$ to be of
order $e^{-c |P|^3}$! So it is hard to imagine a significant
cancelation in our index unless all leading order contributions
cancel exactly, and only exponentially subleading contributions
remain.

In view of this, it seems highly unlikely that such cancelations
could occur as described in possibility (3). Moreover, even if there
were such a cancelation for contributions at $t=i\infty$, there
would not be such a cancelation at other values of $t$, since
different splits $\Gamma \to \Gamma_1 + \Gamma_2$ have different
walls of marginal stability, so if there were cancelation at one
point in moduli space, this would almost certainly not hold at some
other point, since the set of contributing splits will be different
at these two points. Thus, if cancelation is to happen, one would
expect it to take place already at the level of the contributions to
$\Omega(\Gamma_1)$ and $\Omega(\Gamma_2)$, i.e.\ possibility (2).

 Unfortunately, possibility (2) also appears  highly unlikely. In this
case, we can relate the problem, to some extent,  to a precise
mathematical question about the asymptotic growth of DT invariants.
Recall that according to (\ref{diffO}) and
(\ref{fourinfinity1})-(\ref{fourinfinity4}), we have, at least for
$\eta < \eta_* \ll 1$,
\begin{equation} \label{OMNDTrel}
 \Omega(\Gamma_i,t_{\rm ms}) = N_{DT}(\beta_i,n_i) + \Delta \Omega(\Gamma_i,t_{\rm ms})
\end{equation}
with $(\beta_i,n_i)$ defined as in
(\ref{eSparcharge})-(\ref{eSparcharge2}). As we showed in section
\ref{innocuouscoredump}, $\Delta \Omega$ can be nonzero due to swing
states. By definition, for $\eta < \delta |P|^{-\xi_{cd}}$ (with $1
\leq \xi_{cd} \leq 3$), we have $\Delta \Omega = 0$, but in the case
at hand we cannot use this since we want to keep $\eta$ fixed while
scaling up $P$, so we always exit this regime. Thus we expect
$\Delta \Omega \neq 0$.

As we will see below, the generating function for DT invariants has
some special structure which allows one to make some (very)
heuristic arguments in favor of miraculous cancelation of different
contributions to the DT invariants. If such cancelations indeed
occur, then it would become perhaps less implausible that something
like it might  happen for $\Omega(\Gamma_i,t_{\rm ms})$ as well.
However, note that this is far from obvious. First, for non-extreme
polar states, i.e.\ values of $\eta$ closer to 1, there will also be
contributions from rank $r>1$ D6-anti-D6 splits, which are not
directly related to the above DT invariants. Second, the individual
flow tree contributions to \emph{each} of the two terms on the right
hand side of (\ref{OMNDTrel}) will give contributions to the index
scaling as $e^{\lambda^3}$ under the large $\lambda$ scaling
$(p^0,p,q,q_0) \to (p^0,\lambda p,\lambda^2 q,\lambda^3 q_0)$ we are
considering (this acts on the $(\beta_i,n_i)$ parameters as
$(\beta_i,n_i) \to (\lambda^2 \beta_i,\lambda^3 n_i)$), as follows
from the general arguments of section \ref{sec:symmetries}. Hence we
would need miraculous cancelations between the contributions to
$\Delta \Omega$ separately as well. Moreover, by varying
$(\beta_2,n_2)$ while keeping $(\beta_1,n_1)$ fixed, $t_{\rm ms}$
will vary, and so $\Delta \Omega$ can change if this variation takes
$t_{\rm ms}$ over some marginal stability wall. Again individual
contributions to this variation of $\Delta \Omega$ scale as
$e^{\lambda^3}$, so cancelation should occur already within this
very reduced ensemble. It seems hard to imagine how something like
this could happen unless there is extended supersymmetry killing off
individual contributions in bose-fermi pairs.

Nevertheless, since   the problem for DT invariants can be
formulated in a mathematically precise way, and the question is of
some interest on its own, let us proceed to investigate the scaling
\begin{equation} \label{NDTscaling}
 \log N_{DT}(\lambda^2 \beta,\lambda^3 n) \sim \lambda^k,
\end{equation}
and ask whether $k=3$, as suggested by the scaling argument, or
whether cancelations occur making $k \leq 2$. (Note that the
asymptotic growth $N_{DT}(0,n) \sim e^{n^{2/3}}$, given in
(\ref{mmgrr}), suggests that we always have at least $k\geq
2$.\footnote{Except when $\chi(X)=0$. Indeed, for $T^6$ and $T^2
\times K3$, this entire discussion is superfluous: for $T^6$,
$\CZ_{\rm DT}=1$, and for $T^2 \times K3$, $\CZ_{\rm
DT}=\eta(t)^{-24}$ where $t$ is the $T^2$ K\"ahler modulus. In other
words, in these cases we \emph{do} have ``miraculous'' cancelations,
although in this case the miracle is simply extended
supersymmetry.})

To be more precise, consider the limit-supremum:
\begin{equation}\label{preciseasympts}
 k = \overline{\lim}_{\lambda\to +\infty} \frac{\log \log \vert
N_{DT}(\lambda^2 \beta, \lambda^3 n)\vert }{\log \lambda}
\end{equation}
The first question is whether this is  independent of $(\beta, n)$,
and hence equal to a constant $k$. We expect this to be the case. If
this is indeed so, then the  next and crucial question is the value
of $k$. Unfortunately, the answer to this question seems to be
unknown. Even for particular cases, we have been unable to find $k$.
The reason is that to compute the DT invariants to sufficiently high
order, one needs BPS invariants $n^r_Q$ to sufficiently high order
$Q$, and despite tremendous recent progress \cite{Huang:2006hq}, the
available data so far are not yet sufficient to get even a numerical
hint of what the correct answer could be. \footnote{There are some
examples where one could extract $k$ on compact Calabi-Yau
manifolds. These use results on $F_g$ to all orders derived from
heterotic/typeIIA duality, \cite{Marino:1998pg,Klemm:2004km,
Klemm:2005pd}. However, in precisely these cases the relevant black
holes have $P^3=0$. Thus, one should suspect that the case where
$\beta$ represents a holomorphic curve in the K3 fiber of a
K3-fibered Calabi-Yau is not representative. That is, there is in
fact some dependence of (\ref{preciseasympts}) on $\beta$ and really
we should be asking about \emph{generic} $\beta$. }

Let us therefore turn to a some heuristic arguments - two pro and
two contra - for the cancellation hypothesis.

\begin{enumerate}

\item The first heuristic argument suggesting $k=2$  goes as follows. What we want to compute is
\begin{equation} \label{NDTNDTNDT}
 N_{DT}(\lambda^2 \beta,\lambda^3 n) \sim \oint dt \,
 dg \, e^{g n \lambda^3 - 2 \pi i t \cdot \beta \lambda^2}
 \, \CZ_{DT}(-e^{-g},e^{2 \pi i t}).
\end{equation}
Now, because of the DT-GW correspondence as reviewed in section
\ref{sec:prelim}, we can formally write, at small $g$,
\begin{equation}\label{dtsingularity}
 Z_{DT}(-e^{-g},e^{2\pi i t}) \sim e^{\frac{1}{g^2} f(t)}
\end{equation}
where $f(t)$ is the generating function for genus zero Gromov-Witten
invariants. Plugging this in (\ref{NDTNDTNDT}) and doing a naive
saddle point evaluation gives saddle point equations of the form
\begin{equation}
 n \sim \frac{f(t)}{g_0^3}, \qquad \beta \sim
 \frac{f'(t)}{g_0^2}, \qquad g =: g_0/\lambda,
\end{equation}
and saddle point value
\begin{equation}
 N_{DT}(\lambda^2 \beta,\lambda^3 n) \sim e^{c \lambda^2},
\end{equation}
where $c$ is independent of $\lambda$. In other words, this indeed
suggests $k=2$, hence miraculous cancelations! On the other hand, it
is not clearly valid to use (\ref{dtsingularity}) in such a saddle
point analysis, so the argument is only heuristic.

\item To be conclusive, one would like to see the cancelations
happening directly. To see where these might come from and how they
might be related after all to the cancelations we mentioned earlier
in the context of 5d M2 black holes, recall the expression
(\ref{ZDTZDT1})-(\ref{ZDTZDT5}) for $\CZ_{DT}$. Note that when $g
\to 0$, $u \equiv -e^{-g} \to -1$ and $\CZ_{DT}^{\prime,r>0} \to
\prod_q \left( 1 - v^q \right)^{-n^1_q}$. This collapse is due to
the same kind of binomial coefficient cancelations we saw in the 5d
black hole context and as explained under (\ref{NGVrel}), killing
all $r > 1$ contributions. Of course this is as expected from the
DT-GW correspondence: only the genus zero and one contributions to
$F_{\rm top}$ survive when $g \to 0$. But it is noteworthy that
these ``miraculous cancelations'' are intimately tied together with
this correspondence, indicating that the heuristic result we found
does have something to do with the existence of cancelations.

Furthermore, on general grounds, and as suggested by the heuristic
argument, it is conceivable that the large $\lambda$ scaling of
(\ref{NDTscaling}) is governed by the $g \to 0$ behavior of
$\CZ_{DT}$, i.e.\ the $r=0$ factor and to a lesser extent the $r=1$
factor. Now if we drop the $r>1$ factors, we have effectively
dropped our reasons to expect $k=3$ in (\ref{NDTscaling}), since as
we noted in our discussion of 5d black holes above, the enhancement
of growth going from the $n_Q^r \sim e^Q$ to the $N_Q^{m_L} \sim
e^{Q^{3/2}}$ is due to the presence of large binomial coefficients
in the relation (\ref{NGVrel}) between them, but if we drop all
$r>1$ contributions, these are no longer present.

Although this fortifies the case for cancelations, it is still not
conclusive. We did some numerical experiments with very simple toy
models for which naively one could make the same reasoning as above,
but which nevertheless did not lead to any cancelations.
Unfortunately, as we noted before, not enough hard data about the
BPS invariants $n^r_Q$ is available at this time to check these
arguments by direct computation in an actual (compact) model.

\item  An argument against the cancelation hypothesis
follows if we assume the OSV conjecture holds for $p^0=1$, for some
value of the $B$-field at $J=\infty$. Then $N_{DT}(\lambda^2 \beta,
\lambda^3 n)$ should be given by the OSV formula, which for $\lambda
\to \infty$ predicts a growth $N_{DT} \sim \exp[\lambda^3
\sqrt{\beta^3-n^2}]$, leading to $k=3$. Indeed, one concrete
conclusion from these considerations is that the OSV conjecture for
$p^0 = 1$, $t_{\infty} \to i \infty$ and the weak coupling OSV
conjecture for $p^0=0$, $t_{\infty} \to i \infty$ cannot both be
true.

\item We conclude by giving a second  heuristic argument indicating
 there are no significant cancelations, so that the entropy
enigma is also an index enigma, i.e.\ that indeed $\log
\Omega(\Lambda \Gamma;t=i\infty)$ does grow as $\Lambda^3$ in the
large $\Lambda$ limit, with $\Gamma$ some D4-D2-D0
charge.\footnote{We thank D.~Gaiotto, A.~Strominger and X.~Yin for a
related suggestion leading to this argument.} For simplicity we take
$\Gamma=(0,P,0,0)$. A naive model for this is a D4-brane wrapped on
$P$ with $N = \chi(P)/24 \approx P^3/24$ pointlike
$\overline{D0}$-branes bound to it. Ignoring divisor moduli, flux
degrees of freedom and so on, the index of this system is simply the
orbifold Euler characteristic of the $N$-fold symmetric product of
$P$. This is given by the coefficient $d_N$ of $q^N$ in $\prod_n
(1-q^n)^{-\chi(P)}$. The $N \to \infty$ asymptotics are given by the
Cardy formula $\log d_N \sim 4 \pi \sqrt{(N-\frac{\chi}{24})
\frac{\chi}{24}}$, and this equals the single centered entropy for
this charge. However, the value $N=\chi/24$ of interest to us lies
outside the regime of validity of this formula --- in fact plugging
this in the formula gives zero. Numerically on the other hand, we
find $\log d_{N=\chi/24} \approx 0.17649134 * \chi \sim P^3$.
(Exponential growth of this coefficient can also be proved
analytically.) Note that this is  the same growth as suggested by
the two-centered black hole estimate without cancelations! But again
this argument is too heuristic to be taken seriously; in particular,
although this model for the D4-D0 system is fine in the limit $N \to
\infty$, it is not clearly applicable to the case $N \sim P^3$,
since there is now no justification for ignoring the divisor moduli
and fluxes.

\end{enumerate}

In conclusion, although we cannot completely exclude a miracle, it
seems very unlikely to us that a sufficient amount of cancelation
could occur to bring down the polar index growth at fixed $\eta$
from $\sim e^{P^3}$ to $\sim e^{P^2}$, and we therefore believe
$\kappa = 3$ in (\ref{polargrowthkdef}), hence a breakdown of OSV at
weak $g_{\rm top}$ (and $t_{\infty} = i \infty$). It would be
interesting to settle this question definitively.

\subsection{Dumping the dangerous swing states}\label{sec:dangercore}

We now briefly return to the second main unresolved issue, namely
the value of the core dump exponent $\xi_{cd}$ introduced in section
\ref{innocuouscoredump}.

At the end of section \ref{innocuouscoredump} we offered some
circumstantial evidence suggesting that perhaps $\xi_{cd}=1$, and
all is well. In addition to this, we  can offer the following
physical evidence that suggests that $\xi_{cd}=1$. As noted in
section \ref{sec:comparison}, in this case, the  D4-D2-D0 partition
function differs from  $\vert \CZ_{\rm top}\vert^2$ by terms whose
order indicates that they involve Schwinger pair production of
wrapped D4-brane states, giving finite size corrections to the
exactly factorized expression. Analogous states   have been seen to
play a role in topological string amplitudes before
\cite{Klemm:2005pd}. Since there are no other obvious physical
effects which would be larger, one might hope that $\xi_{cd}=1$.

Clearly it is of great interest to investigate these phenomena
further to see whether or not the core-dump exponent $\xi_{cd}$ is
larger than one. It should be relatively straightforward to come up
with further systematic analytical and numerical evidence, similar
to the evidence we accumulated in favor of the extreme polar state
conjecture, but we leave this for future work.

One of the reasons that the OSV conjecture is interesting is that it
suggests a way to give a nonperturbative definition to the
topological string. In view of this it is intriguing that the
corrections we find are indeed suggestive of nonperturbative
corrections. Therefore it might be useful to ask  how to compute the
contribution of these  nonperturbative effects   to
 F-terms in effective supergravity.

\section{Summary of open problems and potential future directions}
\label{sec:openproblems}

In this section we collect and summarize the many  issues and open
problems which arose in our derivation of the OSV conjecture. We
also suggest some potentially interesting future directions for
research.

First, as already discussed in section \ref{sec:challenges},  our
``proof'' of the OSV conjecture is really more of an outline for a
proof. The following important issues need to be settled before the
argument truly constitutes a proof, even in the strong coupling
regime:

\begin{itemize}

\item Some basic issues in the theory of split attractor flows
 and multicentered black hole solutions remain to be clarified.
 While physically very well motivated, and supported by numerous
 examples, the split attractor flow
 conjecture of section \ref{sec:flowtrees} remains to be proven mathematically. Moreover, as
 we discussed, the Hilbert space of BPS states $\CH(\Gamma;t)$ is ---
 roughly speaking --- ``graded'' by the split attractor flows associated to
 $(\Gamma;t)$, but we noted some subtleties, and hence the precise
 rule remains to be elucidated. Among other things one should understand
  better the possible
 quantum mixing between states associated to different attractor flow trees.

\item It would be desirable to have a more
 systematic derivation of the D4 partition function we used in section
 \ref{sec:fareytail}
 starting from a
path integral. Ideally, this should clarify the relation of $d(F,N)$
to general DT invariants.

\item The extreme polar state conjecture remains to be proved.

\item We have shown that certain swing (core) states
could potentially   invalidate the OSV conjecture. This led to the
definition of the core dump exponent in section
\ref{innocuouscoredump}. As discussed there, it remains to show that
$\xi_{cd} \leq 2$. If it turns out that $\xi_{cd}>2$ then the OSV
conjecture is very unlikely to be true, even at strong $g_{\rm
top}$. We outlined some indications that $\xi_{cd}=1$, but further
work is needed to test this hypothesis.

\item The equality of DT and Gromov-Witten partition
functions for compact CY remains to be proved.

\end{itemize}

Certainly, at the ``physical level of rigour'' the extreme polar
state conjecture and the claim that $\xi_{cd} \leq 2$ are the main
gaps in our argument.  We are rather confident that all the above
issues -- be they ever so challenging --  can be satisfactorily
settled, with the possible exception of $\xi_{cd}\leq 2$.  Granting
these points, there are a number of ways in which the refined  OSV
formula could be extended further.

\begin{itemize}

\item  First there is the question of the extension from strong
to weak topological string coupling. As we have demonstrated, this
is loosely related to a well-posed question regarding asymptotics of
DT invariants, namely, the evaluation of
\begin{equation}
k = \overline{\lim}_{\lambda \to \infty} \frac{\log \log \vert
N_{DT}(\lambda^2 \beta, \lambda^3 n)\vert}{\log \lambda}.
\end{equation}
However, while an anomalous value $k \leq 2$ instead of the expected
$k=3$ would certainly be suggestive, this would not immediately
imply a similar growth of the relevant indices
$\Omega(\Gamma_i,t_{\rm ms})$. Regarding the latter, we see little
hope of a cancellation and expect that the OSV conjecture fails at
weak $g_{\rm top}$.  It would however be very interesting to verify
this more directly.

\item As we saw, the ``core states'' account for the genus $r>0$
 component of the Gopakumar-Vafa product form of the
topological string partition function. While some core states are
black holes, there are also more complicated core states. It would
be very interesting to find some way to organize and classify these
core states. It would seem that this is essential to evaluating
$\xi_{cd}$.


\item  Can our methods be extended to values of $P$ which are not
in the K\"ahler cone but nevertheless support BPS states? (Consider, for example,
 a curve of resolved ADE singularities.)

\item  The original paper \cite{Ooguri:2004zv} claimed a version of the
 conjecture for \emph{all} magnetic charges, including $p^0\not=0$.
The wall-crossing formulae we have discussed would seem to pose a serious
obstacle for such a version of the conjecture, at least if it is
 based on the degeneracies at
$\Im \, t=\infty$. Is there nevertheless a version for $p^0\not=0$?

\end{itemize}

Let us now turn to various questions and potential physical
applications which our paper raises:

\begin{itemize}

\item  Our work sheds some light on the old confusion of the relevance of
absolute BPS degeneracies versus indices. On physical grounds, one
expects the total BPS Bekenstein-Hawking-Wald entropy and
refinements thereof to correspond to the absolute number of BPS
states. Naively one might therefore think one should compare the
total dimension of the cohomology of the relevant D-brane moduli
spaces rather than the euler characteristic, which is the index.
However, one should keep in mind that the quantum mechanics of
D-brane moduli spaces is only  a low energy approximation to the
true physical situation. The effective quantum mechanics  ignores
some of the degrees of freedom on the D-brane. In particular, one
should take the string coupling constant to be zero. As soon as it
is nonzero, instanton effects come into play. For example there
might be tunneling between different classical supersymmetric ground
states, i.e.\ between different components of the D-brane moduli
spaces. In particular we expect D2-instantons tunneling between
different flux sectors with the same total charge, producing effects
of order $\sim e^{-\sqrt{J^3}/g_{\rm IIA}} \sim e^{-1/g_{\rm
IIA}^{4d}}$.\footnote{Here $g_{\rm IIA}^{4d}$ is the effective four
dimensional string coupling, which sits in a hypermultiplet and is
therefore tunable without affecting any BPS indices.} These are
external to the moduli space quantum mechanics, and will generically
lift bose-fermi pairs of supersymmetric ground states of the latter.
Based on genericity, one could therefore reason that in fact, all
nonprotected states can be expected to be lifted, bringing the total
degeneracy down to the value of the index.

There is quite a bit of evidence in favor of this idea. First, the
detailed agreement we find in this paper (and the agreement found in
related work) is with the index, not with the total cohomology of
moduli space. Indeed, the BPS invariants $n_Q^r$ determining
$\CZ_{\rm DT}$ are all indices, not total betti numbers. The
simplest example of this is the power of the MacMahon factor, which
is the euler characteristic $\chi(X)$ of the Calabi-Yau, not its
total cohomology. It should furthermore be noted that these index
invariants appear already in the leading order supergravity entropy
formula, even neglecting $R^2$ and higher order curvature
corrections --- for example the contribution proportional to
$\zeta(3) \, \chi(X)$ in the IIA supergravity entropy formula is
obtained from the MacMahon factor counting indices as in
(\ref{hzergw}). This indicates that subtleties involving higher
curvature corrections (and the fact that we are comparing to the BHW
entropy taking into account only F-term $R^2$ corrections) are
largely irrelevant for this discussion.

 In fact it has been known for quite some time
\cite{LopesCardoso:1999xn} that D4-D0 black holes for $T^2 \times
K3$ and $T^6$ compactifications have, already in the large D0-charge
limit, BHW entropies which do not match the total degeneracy at
$g_{\rm IIA}=0$, while they agree with the index (this discrepancy
did not arise for $SU(3)$ holonomy Calabi-Yau compactifications
because the D4-D0 brane moduli spaces were approximated as $N$-fold
symmetric products of a very ample divisor, which in the $SU(3)$
holonomy case does not have any odd cohomology, making the index
equal to the total degeneracy). The class of 5d black holes studied
in \cite{Vafa:1997gr} seemed to go the other way, but as we show in
appendix \ref{app:entropy5dBH}, also in this case there is again
exact agreement with the index, provided the proper index is used.

 In conclusion, we see considerable evidence that the true
number of BPS states at finite coupling is in fact  given by the
microscopic index.\footnote{We should mention though that in
\cite{Dabholkar:2005by} (section 4) the BHW entropy of a class of
small black holes with untwisted sector charges in the FHSV model
was found to agree with the absolute cohomology to leading order,
but not with the index (while in the twisted sector the index did
agree). It is conceivable however that in this case there exists
again a more refined index which would also be in agreement in the
untwisted case. Alternatively, certain K\"ahler classes were set to
zero for these small black holes, and perhaps the background is too
singular. } If the total number of BPS states at zero coupling is
higher than the index, we expect a number of states with energies no
more than $\Delta E \sim e^{-1/g_{IIA}^{4d}}$ above the BPS bound.

For an in depth recent discussion of related issues in the case of
extremal nonsupersymmetric black holes see \cite{Dabholkar:2006tb}.

On the supergravity side, we have a parallel picture. Since there
are in general many alternating sign contributions to the BPS index,
possibly coming from many different multicentered configurations,
there is again ample room for tunneling effects to lift BPS states
obtained in the supergravity moduli space approximation. For example
in our D6-anti-D6 description of polar states, an anti-D2 particle
in a halo around the D6 could annihilate with a D2 particle in a
halo around the anti-D6. Following the heuristic dictionary of
section \ref{sec:microrel}, such a process should be the
supergravity analog of a D2-instanton tunneling between different
flux sectors.

Thus, on both sides, in suitable coupling regimes, this suggests a
picture of having essentially $\log |\Omega(\Gamma;t_{\infty})|$
exact BPS ground states of charge $\Gamma$ in a background
$t_{\infty}$, as well as a certain number of slightly non-BPS states
at exponentially small energies above the BPS bound. Now, this
  number might actually be quite huge: indeed if there is a
cancelation $e^{\Lambda^3} \to e^{\Lambda^2}$ in the index of the
kind discussed at length in section \ref{sec:cancel}, and if indeed
all canceling contributions in the index get lifted by tunneling
effects, then the number of near-BPS states with exponentially small
energy gaps would in fact be of order $e^{\Lambda^3}$, dwarfing the
number of exact BPS states!   Even if there is no such cancelation,
one would still expect a comparable amount of bosonic and fermionic
states, and therefore $\sim e^{\Lambda^3}$ non-BPS states with
exponentially small energy gaps.

It would be interesting to see to what extent these speculations are
correct, and if so, if perhaps there might be implications for
models of dynamical supersymmetry breaking.


\item It would be helpful to elucidate the connection between
our approach and the approach based on the MSW conformal field theory. The
MSW CFT is the effective conformal field theory with $(0,4)$ supersymmetry
describing a
string obtained  from wrapping an M5 brane on a holomorphic surface such as $P$
\cite{Maldacena:1997de,Minasian:1999qn}. The precise formulation of this
 CFT remains incomplete. Assuming  the formulation can be completed,
  it would be interesting to clarify
the relation between our D4 partition function and the elliptic genus of the
MSW CFT.

\item In equation (\ref{abcentropy}) we found a remarkably simple formula
for the entropy associated with a 3-node quiver with a loop. The challenge
remains to find a conceptual derivation of this formula. Will it extend to
other quivers with loops? Conceivably, there might be interesting applications
to the AdS/CFT correspondence.

\item  Our discussion of multicentered black holes and their moduli spaces should
have implications for the ``Mathur program,'' which aims to account
for microstate entropy from quantizing supergravity solution moduli
spaces \cite{Mathur:2005zp}. In particular in
\cite{Balasubramanian:2006gi} it was proposed that quantizing moduli
spaces of horizon-free four dimensional multicentered configurations
with given total charge might account for the corresponding black
hole entropy. The results we found for the three node quiver in
sections \ref{sec:threenodeclosedloop} and \ref{sec:3quiver} seem to
create some tension with this proposal. While in the regime in which
there are no scaling solutions to the integrability constraints, the
microscopic ($g_{\rm IIA} \to 0$) quiver degeneracies are correctly
reproduced from quantizing the associated three centered system, it
seems rather unlikely that this remains the case in the regime in
which there are scaling solutions, given the exponential growth
(\ref{abcentropy}) of the microscopic degeneracies in this regime
and the relative simplicity of the corresponding three particle
quantum mechanics. Although we can't exclude that the set of
\emph{all} multicentered configurations with the same total charge
might still add up to the same degeneracy as the set of all
microscopic realizations  with the same total charge, in our
opinion, our result is instead rather suggestive of the appearance
of new degrees of freedom and BPS configurations in the scaling
regime, qualitatively different from the multicentered
configurations considered so far. (In \cite{Balasubramanian:2006gi}
are related suggestion of the appearance of nonabelian degrees of
freedom was made, based on $SU(N)$ degrees of freedom associated
with nodes of the quiver. The striking thing about the present
example is that the nodes are associated with rank one gauge
groups.)

\item In this setting, an open problem remains. In the split attractor
flow conjecture the scaling solutions are in the same component of
moduli space as a simple single-centered flow for the total charge,
and hence are represented by the same flow. These scaling solutions
cannot be forced to decay by changing the background moduli. We do
not know however if all multicentered scaling solutions to the
integrability conditions (\ref{centerconstraints}) with the total
charge of a black hole in fact correspond to actual BPS solutions of
supergravity, and if not, whether one can significantly simplify the
rather cumbersome criterion for existence:  $\CD(H(\vec x))>0$ for
all $\vec x\in \IR^3$.

\item The appearance of the measure factor (\ref{mfkpot}) in the integral form of the
OSV formula (\ref{osvwithmf}) is striking and very reminiscent of
K\"ahler quantization. In this interpretation the black hole
degeneracies are certain kinds of Wigner functions for a
distinguished quantum state provided by the topological string. Of
course, this observation has been made before
\cite{Ooguri:2005vr,Gunaydin:2006bz}, but there have been some
difficulties making this proposal precise. (For example, the
topological string partition function is not a normalizable
wavefunction. In fact it is not even a function, since it has zero
radius of convergence.) We hope our precise version can help clarify
this conjecture.

\item Our results should have several model building applications,
since, at least classically, the conditions to have supersymmetric
brane configurations are independent of whether the branes are
space-filling or not. One issue that generally has been ignored in
phenomenological D-brane model building is stability. For example D7
branes wrapping divisors in IIB orientifold models might well decay
when the volume of the divisor gets too ``small,'' just like the
corresponding D4-branes in type IIA.  But as we saw, small is not
all that small actually if the divisor has a large Euler
characteristic. Moreover, flux  compactifications with all moduli
stabilized typically do not have parametrically large divisor
volumes. Thus, stability becomes quite relevant, and the practical
tools we developed here might be useful to get a handle on this.

\end{itemize}

Finally, we think there are mathematical implications and applications which
could be of some interest.

\begin{itemize}

\item If indeed the degeneracies $d(F,N)$ are   DT invariants then
our main claim is that these can be arranged in an interesting
modular generating function, and our factorization formulae imply
highly nontrivial polynomial relations between the DT invariants.

\item Do the physical results here, especially the picture of
section \ref{sec:microrel},  shed light on the geometry of the
Noether-Lefshetz locus?

\item In the discussion of the fareytail expansion a certain
interesting class of polynomials $h_\gamma$ arose. What are these
polynomials? Do they have a physical interpretation? What  is the analog
when $\vert w_H\vert $ is  half-integral?

\item In equation (\ref{degenwc}) we derived a wall crossing formula. However, it is
incomplete since we should like to account for all nonprimitive splits
$\Gamma \to N_1 \Gamma_1 + N_2 \Gamma_2$ across a wall of marginal stability.
Generalizing to the case where $N_1,N_2$ are both greater than one appears
to be an interesting and challenging problem.

\item
It would be interesting to clarify the relation of our framework to
the study of $\Pi$-stable objects in the derived category of
coherent sheaves on $X$. It has not escaped our attention that
D.~Joyce has also  arrived at wall-crossing formulae reminiscent of
ours from a very different point of view (see \cite{Joyce:2006pf}
and references therein). It is clearly of value to examine the
relation between these formulae.

\item  There is an old idea that there should be an
interesting algebraic structure associated with BPS states of
D-branes on Calabi-Yau manifolds
\cite{Harvey:1995fq,Harvey:1996gc,Moore:1997ar}. Indeed, infinite
products such as (\ref{ZDTZDT2})-(\ref{ZDTZDT5}) are reminiscent of
denominator products associated to generalized Kac-Moody algebras.
The algebra should be graded by $K(X)$, and should depend on the
moduli $t_\infty$. Along lines of marginal stability where
$\Gamma\to \Gamma_1 + \Gamma_2$  we should associate algebra
products  $\CH(\Gamma_1;t)\otimes \CH(\Gamma_2;t) \to
\CH(\Gamma;t)$. These algebras should generalize the geometric
construction of highest weight modules of affine Lie algebras due to
Nakajima. Perhaps some of the examples and techniques of this paper
could be usefully applied to realizing this dream.

\item As we have mentioned, in
 the dual $M$-theory viewpoint the BPS indices are computed
using the elliptic genus of the MSW CFT
\cite{Gaiotto:2006ns,Gaiotto:2006wm,deBoer:2006vg}.  It is
interesting to ask how the wall-crossing formula should arise in
this context. Studies of the $(0,4)$ elliptic genus thus far have
been limited to the region $t_\infty = i \infty$, but presumably
there is an extension of the $(0,4)$ CFT and its elliptic genus to
finite values of $t_\infty$. In this context many interesting new
issues will arise.
 First of all, the $(0,4)$ CFT is
rather subtle due to the discriminant locus in the moduli space of
the surface on which the M5 is wrapped, and hence the very
definition of the Dirac-Ramond operator will be subtle.  Next, one
may guess that as one continues the background moduli through
certain walls the Dirac-Ramond operator   fails even to be  formally
Fredholm and its character-valued index can change. It would be very
interesting to recover the wall-crossing formulae from this point of
view.

\item  In section \ref{sec:D6D2D0}  we showed that there is a difference between D6D2D0
degeneracies and DT invariants. Only in a special limit of the
$B$-field do they correspond. This raises the question of what the
general relation is, and whether other limits of the $B$-field could
be taken. (That in turn reduces to detailed applications of
wall-crossing formulae which we have not tried to sort out.)

\end{itemize}

\acknowledgments

We would like to thank Paul Aspinwall, Miranda Cheng, Emanuel
Diaconescu, Michael Douglas, Bogdan Florea, Davide Gaiotto, Albrecht
Klemm, Juan Maldacena, David Morrison, Nikita Nekrasov, Hirosi
Ooguri, Tony Pantev, Andy Strominger, Cumrun Vafa, Dieter Van den
Bleeken, Erik Verlinde and Xi Yin for useful discussions. We
acknowledge the hospitality of the Kavli Institute for Theoretical
Physics, the Aspen Center for Physics and the high energy theory
group at Harvard University, where parts of this project were done.
This work was supported in part by the National Science Foundation
under Grant No.\ PHY99-07949, and by the DOE under grant
DE-FG02-96ER40949. G.M. also thanks the Institute for Advanced Study
and the Monell foundation for support, as well as the 4$^{th}$
Simons Workshop for hospitality.

\appendix

\section{Definitions and conventions} \label{app:defconv}

In the IIA picture, we write charges $\Gamma \in H^{\rm even}(X)$ in
components as
\begin{equation}
 \Gamma = \Gamma^0 + \Gamma^A D_A + \Gamma_A \widetilde{D}^A +
 \Gamma_0 \, \omega
\end{equation}
where $\{ D_A \}$ is a basis of $H^2(X,\IZ)$, $\{ \widetilde{D}^A
\}$ is a dual basis, and $\omega$ is the unit volume element on $X$,
dual to 1. The index $A$ runs from $1$ to $h^{1,1}(X):=h$. When
$\Gamma$ is identified with $(p,q)$, we have
$p^\Lambda=\Gamma^\Lambda$, $q_\Lambda = \Gamma_\Lambda$. The index
$\Lambda$ runs from $0$ to $h$. Quantities with an $A$ index tend to
get capitalized. Note that $\Phi^A = \phi^A$, $P^A = p^A$ etc.

The Dirac-Schwinger-Zwanziger symplectic intersection product is
defined as
\begin{equation} \label{intprodform}
 \langle \Gamma,\Delta \rangle = -\Gamma^0 \Delta_{0} +
 \Gamma^A \Delta_{A} - \Gamma_{A} \Delta^A + \Gamma_0
 \Delta^0 = \int_X \Gamma \wedge \Delta^*,
\end{equation}
where $\Delta^*$ is obtained from $\Delta$ by inverting the sign of
the 2- and 6-form components. When $\Gamma_1$ and $\Gamma_2$ are
represented as sheaves $V_1$ and $V_2$, then their charges are given
by
\begin{equation} \label{chargeformula}
 \Gamma_i = {\rm ch}(V_i) \, \sqrt{\widehat{A}(X)} = {\rm ch}(V_i)
 \, (1 + \frac{c_2(X)}{24})
\end{equation}
and, by the Grothendieck-Riemann-Roch theorem,
\begin{equation}
 \sum_k (-1)^k \, \dim {\rm Ext}^k(V_1,V_2) = \langle \Gamma_1,\Gamma_2\rangle
 = \int_X {\rm ch}(V_1) \, {\rm ch}(V_2^*) \, \widehat{A}(X).
\end{equation}
The normalized period vector is defined as
\begin{equation}
\Omega_{\rm nrm}(t,\bar{t}) := e^{\frac{1}{2}K(t,\bar{t})} \,
\Omega_{\rm hol}(t) \, \, \in H^{2*}(X,\IC), \qquad K := - \ln i
\langle \Omega_{\rm hol},\overline{\Omega_{\rm hol}} \rangle,
\end{equation}
which depends on the complexified K\"ahler moduli fields $t^A = B^A
+ i J^A$ and in the large radius approximation is given by
\begin{equation}
 \Omega_{\rm hol} = -e^{B+iJ}, \qquad \Omega_{\rm nrm} = \frac{\Omega_{\rm hol}}{\sqrt{4J^3/3}}.
\end{equation}
From this, one defines the holomorphic central charge
\begin{equation}
 Z_{\rm hol}(\Gamma,t) := \langle \Gamma, \Omega_{\rm hol}(t)
 \rangle,
\end{equation}
which in the large radius approximation becomes
\begin{eqnarray} \label{centralchargedef}
 Z_{\rm hol}(p,q;t)
  &=& - \int_X (p^0+P+Q+q_0 \, \omega) \wedge
 e^{-t}\\
 &=& p^0 \frac{t^3}{6} - P \cdot \frac{t^2}{2} + Q \cdot
 t - q_0 \\
 &=& \frac{1}{6} p^0 D_{ABC} t^A t^B t^C - \frac{1}{2} p^A D_{ABC} t^B t^C
 + q_A t^A - q_0
\end{eqnarray}
where $D_{ABC}:=D_A \cdot D_B \cdot D_C :=\int_X D_A \wedge D_B
\wedge D_C$. The normalized central charge is obtained from this as
$Z_{\rm nrm} = e^{\half K} Z_{\rm hol}$. We usually drop the
subscript distinguishing holomorphic and normalized central charges
when no confusion can arise. Sometimes we abbreviate $Z(\Gamma;t)$
to $Z(\Gamma)$.

%

\section{Some algebraic geometry} \label{sec:alggeom}

We collect here some mathematical facts used in the text. Many of
these are nicely explained and were skillfully applied in the
present context in \cite{Maldacena:1997de}. Many of the mathematical
facts can be found explained in detail in  \cite{GH}, chapter 1.

Let $X$ be a projective variety. A divisor class is determined by
$P\in H^2(X,\IZ)$. If $P$ is of type $(1,1)$ it is the first Chern
class of a holomorphic line bundle $\CL_P$, and effective divisors
in the divisor class are vanishing loci of sections of $\CL_P$. The
moduli space of these divisors is a projective space $\CM_P = \IP
H^0(X, \CL_P)$, called a \textit{complete linear system}, also
denoted $\vert P \vert$. The generic divisor in  $\CM_P$ is a smooth
hypersurface in $X$, but there is a  discriminant locus $\CD$ in
$\CM_P$ of singular divisors.

For example, if $X$ is the quintic $\sum X_i^5 =0$ in $\IP^4$ then
$P = n H$ where $n>0$ is integral and $H$ is the K\"ahler class of
$\IP^4$ and  the linear system $\CM_P$  consists of the set of
divisors defined by the vanishing of a degree $n$ polynomial on
$\IP^4$ intersected with $\sum X_i^5 =0$.  The discriminant locus
$\CD$ is already quite complicated for $n=1$. In this case the
divisors are defined by $\sum \alpha_i X_i=0$ with $[\alpha_1:\cdots
:\alpha_5]\in \IP^4$ and the discriminant locus is defined by
$\alpha_i = b_i^4$ where $[b_1:\cdots : b_5]\in X$.

The dimension of the moduli space $\dim \CM_P$ can, under some
circumstances be obtained by combining the index theorem with
vanishing theorems. The index theorem says
\begin{equation}\label{indextheorem}
\sum_i (-1)^i h^i = \int_X e^P \, {\rm Td}(T^{1,0}X)
\end{equation}
where $h^i = \dim H^i(X,\CL_P)$. Now, $P$ is \textit{very ample} iff
the sections $s: X \to \IP H^0(X,\CL_P)$ define an embedding  (
\cite{GH}, p.\ 192). On the other hand, $P$ is said to be
\textit{ample} if some positive multiple of it is very ample. A criterion for
being ample is that   $P$ is positive as a (1,1)-form, i.e.\
$P_{i\bj} > 0$, which in turn is true iff $P$ lies within the
K\"ahler cone, i.e. $\beta \cdot P>0$, $D \cdot P^2 > 0$, $P^3>0$
for all effective curves $\beta$ and divisors $D$. In this case $h^i(\CL_P)=0$ for
$i>0$ (\cite{GH} p.154), and $\CM_P$ is just a projective space of
complex dimension $h^0-1$ which can be read off from
\ref{indextheorem}.

Specializing to a Calabi-Yau 3-fold $X$ we have ${\rm Td}(T^{1,0}X)
= 1+ c_2(X)/12$ and hence
\begin{equation}
\dim \CM_P = \frac{1}{6}P^3 + \frac{1}{12} Pc_2(X) -1
\end{equation}
and hence  $\chi(\CM_P) =\frac{1}{6}P^3 + \frac{1}{12} Pc_2(X)$.

 As explained in detail in
\cite{Maldacena:1997de}, using the Hirzebruch signature theorem and
$\chi(\Sigma)= \int_\Sigma c_2(T\Sigma)$ together with the
adjunction formula we get
\begin{eqnarray}
\chi(\Sigma) & = &   P^3 + P c_2(X)\label{chisigmaforma} \\
\sigma(\Sigma) & = &   - \frac{1}{3} P^3 - \frac{2}{3} P
c_2(X)\label{chisigmaformb}.
\end{eqnarray}
We often denote $\chi(\Sigma)$ by $\chi(P)$. Now eqs.
(\ref{chisigmaforma}),(\ref{chisigmaformb})  in turn imply
\begin{eqnarray}
b_2^+(\Sigma) & =  & \frac{1}{3} P^3 + \frac{1}{6} P c_2(X) + b_1(\Sigma) -1 \label{topinvtsa}\\
b_2^-(\Sigma) & = & \frac{2}{3} P^3 + \frac{5}{6} P
c_2(X)+b_1(\Sigma) -1.\label{topinvtsb}
\end{eqnarray}
Note that for ``large'' $P$ (such as we consider in this paper) the
topology of $\Sigma$ is quite complicated. For example, on the
quintic, if $P= nH$ we have $\chi( \CM_P) =\frac{ 5n^3 + 25 n}{6}$
(an integer) and $\chi(\Sigma) = 5 n^3 + 50 n$.

 In the text we use the
Lefschetz Hyperplane theorem (\cite{GH}, p. 156) which guarantees
for very ample $\Sigma$ that the pullback $H^q(X,\IQ) \to
H^q(\Sigma, \IQ)$ is an isomorphism for $q\leq \dim X -2$ and is
injective for $q =\dim X-1$. It follows that, if $X$ has generic
holonomy and $P$ is very ample, then the generic smooth surface
$\Sigma \in \CM_P$ has $b_1(\Sigma)=0$.

The lattice  $H^2(\Sigma, \IZ)$ has an intersection form. It is
embedded in the vector space $H^2(\Sigma, \IR)$, and the latter can
be decomposed orthogonally into  $H^{2,+}\oplus H^{2,-}$. As
$\Sigma$ moves in the moduli space the decomposition ``rotates''
relative to $H^2(\Sigma, \IZ)$. This is described as a variation of
weight two Hodge structures. See \cite{Minasian:1999qn} for a
discussion in the present context. There is a fixed part, $L_X=
\iota^*H^{1,1}(X,\IZ)$ which does not rotate. As explained in
\cite{Maldacena:1997de} $L_X$ has signature $(1, h^{1,1}(X)-1)$ by
the Hodge index theorem \cite{shafarevich} with the positive
direction being the K\"ahler class $J$.   Thus $H^{2,+}$ is spanned
by the $J$ and the $(2,0)+(0,2)$ forms, while $H^{2,-}$ is a
negative definite space spanned by the orthogonal $(1,1)$ forms.
Note in particular that since $L_X$ has a nondegenerate form the
matrix $D_{ABC} P^C $ is an invertible matrix --- a fact we often
use.

%
%

An important role in this paper is played by the locus $NL(F)$
defined by choosing $F\in H^2(\Sigma, \IZ)$ and considering the
locus of divisors for which $F$ is of type $(1,1)$. This is known as
the \textit{Noether-Lefschetz locus}, and, we are told, is a
somewhat mysterious object mathematically. In \cite{CDK} it is shown
that $NL(F)$ is an algebraic variety. The moduli space $\CM_{F,N}$
appearing in eq.(\ref{indextoeuler}) projects to $NL(F)$. The fiber
over a smooth element $\Sigma\in NL(F)$ is $Hilb^{N}(\Sigma)$.
Unfortunately, complicated things happen at the discriminant locus
so this is not a practical way of understanding the $d(F,N)$.


\section{Finiteness of the number of split attractor flows}\label{app:finiteness}

Throughout the paper we have assumed the following statement:

\textit{The number of distinct split attractor flows terminating on
regular attractor points,  beginning with a fixed charge $\Gamma_0$,
at a fixed initial point $t_\infty$, is finite.}

In this appendix we will prove a weaker version of this claim,
namely that the number of attractor flows terminating in any fixed compact region
of Teichm\"uller space is finite. In fact our argument proves rather more and
addresses a class of noncompact regions.  The  proof uses
some general ideas mentioned in appendix A of \cite{Denef:2001xn}.

We will be using the large K\"ahler structure formulae for the
central charges. Expansions around this point in moduli space
distinguish a duality frame of electric and magnetic charges. The
first step in the argument shows that there are a finite number of
possible collections of magnetic charges of the final regular
attractor points. To do this we consider the attractor equation for
a charge $\Gamma$ written as:
\begin{equation}
2 {\rm Im}(\bar Z(\Gamma) Z(\Gamma')) = \langle \Gamma,
\Gamma'\rangle
\end{equation}
for all charges $\Gamma'$.  It therefore follows that
\begin{equation}\label{basicineq}
\vert Z(\Gamma;t_*(\Gamma))\vert \geq \frac{1}{2}\frac{\vert \langle
\Gamma, \Gamma'\rangle\vert}{\vert Z(\Gamma';t_*(\Gamma))\vert}
\end{equation}
for all $\Gamma'$ such that $Z(\Gamma';t_*(\Gamma))\not=0$.

Let us consider first the one-dimensional case with $\Gamma = r + b
P + c P^2 + d P^3$ and $t_*(\Gamma)= (x+iy)P $.  Then applying the
inequality (\ref{basicineq}) with $\Gamma'=P^3$ we get
\begin{equation}\label{rineq}
\vert Z(\Gamma;t_*(\Gamma))\vert \geq
\frac{1}{2}\sqrt{\frac{4P^3}{3}} \vert r\vert y^{3/2}
\end{equation}
and using $\Gamma'=P^2$ we get
\begin{equation}\label{bineq}
\vert Z(\Gamma;t_*(\Gamma))\vert \geq
\frac{1}{2}\sqrt{\frac{4P^3}{3}} \vert b \vert \frac{y^{3/2}}{\vert
x+iy \vert}
\end{equation}

In order to for these inequalities to be useful we must assume that
the  final regular attractor points will be contained in a region of
Teichm\"uller space  of the form $-L \leq x \leq L$, $y\geq y_m$.
These are the noncompact regions alluded to above. The need to
restrict attention to such regions is
 the main limitation of the present argument.

Granted that the flows lie in a region of the above type,  we have
absolute lower bounds at attractor points:
\begin{equation}
\vert Z(\Gamma;t_*(\Gamma))\vert \geq
\frac{1}{2}\sqrt{\frac{4P^3}{3}} \vert r\vert y_m^{3/2}
\end{equation}
and using $\Gamma'=P^2$ we get
\begin{equation}
\vert Z(\Gamma;t_*(\Gamma))\vert \geq
\frac{1}{2}\sqrt{\frac{4P^3}{3}} \vert b \vert
\frac{y_m^{3/2}}{\vert L+iy_m \vert}
\end{equation}
Let us absorb the factor  $\sqrt{P^3/3}$ into $Z$ to define $\hat
Z$.

 Now, if we consider an attractor flow tree starting from
$(\Gamma_0,t_\infty)$ then since the flow is gradient flow for $\log
\vert Z(\Gamma;t)\vert^2$, and since at the walls of marginal
stability with vertices $\Gamma \to \Gamma_1+\Gamma_2$, we have
$\vert Z(\Gamma;t_{ms})\vert =\vert Z(\Gamma_1;t_{ms})\vert+ \vert
Z(\Gamma_2;t_{ms})\vert$, we see that  if the final regular
attractor points for the flow tree are labelled $(\Gamma_i, t_i)$,
$i=1,\dots N$ then we have
\begin{equation}
\vert \hat Z(\Gamma_0;t_\infty)\vert \frac{1}{y_m^{3/2}} \geq
\sum_{i=1}^N \vert r_i\vert
\end{equation}
\begin{equation}
\vert \hat Z(\Gamma_0;t_\infty)\vert \frac{\vert L+iy_m\vert}{y_m^{3/2}}
\geq \sum_{i=1}^N \vert b_i\vert
\end{equation}
Because charges are quantized $r_i$ are integers, and (taking $P$ to
be primitive, for simplicity) $b_i$ are integers. It follows that
there are a finite number of sets of possible final magnetic charges
$\{ (r_1,b_1), \dots, (r_{N},b_{N}) \} $. From this finite list of
charges we only keep those for which $\sum r_i = r$ and $\sum b_i =
b$. Then from the remaining list there are a finite number of
topologies of binary trees we can build up from these final charges
terminating on a single initial charge. Let us call these ``magnetic
flow trees.''

The finiteness of the number of magnetic flow trees does not yet
imply that there are a finite number of attractor flow trees because
we have not taken into account the electric charges. Now, the  the
regular attractor flows for pure electric charges, i.e. for $D2D0$
boundstates,  goes to $t=i\infty$.  For this reason the inequalities
we get taking $\Gamma'$ to be a magnetic charge are less useful and
we need to use a different kind of argument.

Suppose there were an infinite set of attractor flow trees. As we
have seen there is a finite list of magnetic flow trees  terminating
on regular attractor points at finite places in moduli space.
Therefore, there would have to be an infinite family of flow trees
with all the   $D2D0$ emissions taking place along one particular
line segment taken from one particular magnetic flow tree.  We are
not allowing splits where all three charges have zero magnetic
charge, and hence this line-segment must carry some nonzero magnetic
charge $(r_*,b_* P)\not=0$. Order the infinite set of trees with
electric emissions from this line segement and let $(c_\alpha P^2,
d_\alpha P^3)$, $\alpha = 1, \dots, \infty$ be the electric charges
emitted from this line segment in the ensemble of all trees.  Let
the point at which they are emitted be $t_\alpha = (x_\alpha + i
y_\alpha)P$. Let the charge along the line right after this emission
be $(r_*,b_* P, \hat c_\alpha P^2, \hat d_\alpha P^3)$.  Finally, if
a flow emits $(c_\alpha P^2, d_\alpha P^3)$ then there will be a set
$S_\alpha$ of numbers $\beta \leq \alpha$ accounting for all the
electric charges $(c_\beta P^2, d_\beta P^3)$ emitted up to that
point along that segment in that particular flow.

Note that we have
\begin{equation}\label{bdd2d0}
\vert \hat Z(\Gamma_0; t_\infty)\vert \geq \vert Z(r_*,b_* P, \hat
c_\alpha P^2, \hat d_\alpha P^3; t_\alpha)\vert  + \sum_{  \beta\in
S_\alpha } \vert Z(0,0,c_\beta P^2, d_\beta P^3;t_\beta)\vert
\end{equation}
for all $\alpha$.

Now, if  the set $(c_\alpha, d_\alpha)$ is not bounded in $\IZ^2$
then we clearly must have $y_\alpha \to \infty$ so that
\begin{equation}\label{interline}
\vert Z(0,0,c_\alpha P^2, d_\alpha P^3;t_\alpha)\vert = \frac{\vert
c_\alpha (x_\alpha + i y_\alpha) -d_\alpha\vert
}{y_\alpha^{3/2}}\sqrt{\frac{3 P^3}{4}}
\end{equation}
remains bounded. On the other hand, suppose the set of electric
charges  $(c_\alpha, d_\alpha)$ does remain bounded (for example,
suppose there is an infinite set of attractor trees where more and
more D2D0 lines are emitted but the charges come with alternate
signs and balance each other). Nevertheless, because the sum on
$\beta$ in (\ref{bdd2d0}) remains bounded it must be that there is a
subsequence such that
\begin{equation}
\frac{\vert (c_\alpha x_\alpha -d_\alpha) + i c_\alpha y_\alpha
\vert }{y_\alpha^{3/2}} \to 0
\end{equation}
This still implies that $y_\alpha \to \infty$. One might wonder if
we can have the numerator tend to zero while $y_\alpha $ remains
bounded. Clearly, because of the term $i c_\alpha y_\alpha$, the
$c_\alpha$ must have an infinite subsequence with all but finitely
many zero. But then we cannot have   $(c_\alpha x_\alpha
-d_\alpha)\to 0$ without $d_\alpha =0$, but then we don't have an
infinite number of nonzero D2D0 charges. Thus, we must have an
infinite subsequence with $y_\alpha \to \infty$.

Now,   because of (\ref{bdd2d0} ) we also have an upper bound
(namely, $\sqrt{\frac{4}{3P^3}}\vert \hat Z(\Gamma_0;t_\infty)\vert$) on
\begin{equation}\label{emitline}
 \vert \frac{\frac{1}{6}r_* (x_\alpha + i y_\alpha)^3 - \frac{1}{2}
b_* (x_\alpha + i y_\alpha)^2}{y_\alpha^{3/2}} + \frac{ \hat
c_\alpha(x_\alpha + i y_\alpha) - \hat d_\alpha
}{y_\alpha^{3/2}}\vert
\end{equation}
The ensemble of complex numbers in the absolute value sign is
clearly bounded. Now consider the ensemble of complex numbers
$$
\frac{ \hat c_\alpha(x_\alpha + i y_\alpha) - \hat d_\alpha
}{y_\alpha^{3/2}}
$$
We can write:
\begin{equation}\label{sumtr}
\frac{\hat c_\alpha(x_\alpha + i y_\alpha) - \hat d_\alpha
}{y_\alpha^{3/2}}= \frac{ c_*(x_\alpha + i y_\alpha) -  d_*
}{y_\alpha^{3/2}} -\frac{( \sum_{\beta \in S_\alpha} c_\beta
)(x_\alpha + i y_\alpha) - ( \sum_{\beta \in S_\alpha} d_\beta )
}{y_\alpha^{3/2}}
\end{equation}
for some fixed $c_*, d_*$.  We claim this is a bounded set of
complex numbers (as we let $\alpha \to \infty$). The first term on
the RHS of \ref{sumtr} goes to zero. For the second term  we use the
fact that the $y_\alpha$ are increasing\footnote{we might need to
choose a subsequence for this to be the case}  so, for example,
\begin{equation}
 \sum_{\beta \in S_\alpha} \vert c_\beta\vert  y_\alpha^{-1/2}
< \sum_{\beta\in S_\alpha }  \vert c_\beta \vert y_\beta^{-1/2}
\end{equation}
but the RHS of this inequality is bounded. Now,  since $y_\alpha \to
\infty$ the ensemble  of complex numbers
$$
\frac{\frac{1}{6}r_* (x_\alpha + i y_\alpha)^3 - \frac{1}{2} b_*
(x_\alpha + i y_\alpha)^2}{y_\alpha^{3/2}}
$$
in (\ref{emitline}) is unbounded.  This is a contradiction with the
boundedness of the set (\ref{emitline}) so we conclude that there
can only be a finite number of split attractor flows.

The above argument can be adapted to the general case as follows.

Let the terminal regular attractor points at finite points of moduli
space $t_i=B_i + i J_i$  have magnetic charges $(r_i, P_i)$.  Using
the basic inequality (\ref{basicineq}) with $\Gamma' = \omega$,
where $\omega$ is a unit volume form  we again find
\begin{equation}\label{rgen}
\vert Z(\Gamma_0;t_\infty)\vert \geq  \sum_i \vert
r_i\vert\sqrt{\frac{J_i^3}{3}}\geq \sqrt{\frac{J^3_{min} }{3}} \sum
\vert r_i\vert
\end{equation}
where as before there is a lower bound on the volume in the moduli
space.

Similarly, we find
\begin{equation}\label{bgen}
\vert Z(\Gamma_0;t_\infty)\vert \geq  \sum_i \frac{ \vert q_2^i\cdot
P_i\vert}{\vert q_2^i\cdot t_i\vert} \sqrt{\frac{J_i^3}{3}}
\end{equation}
Here $q_2^i$ are arbitrary charges in $H^4(X,\IZ)$ applied to each
of the terminal attractor points. The $t_i$ are in the K\"ahler
cone, as are the $P_i$ (by the attractor equation), so the most
effective choice is to take the $q_2^i$ to range over an integral
basis $\CB$ of effective curves generating $H_2(X,\IZ)$. Once again
we claim that in a region of Teichm\"uller space where the $B$
fields are bounded, and the Kahler classes are bounded below, there
is a universal lower bound for $\frac{ 1}{\vert q_2^i\cdot t_i\vert}
\sqrt{\frac{J_i^3}{3}}$ as $q_2^i$ ranges over $\CB$. It follows
that $\sum_i \vert P_i^A\vert $ is bounded above for each component
$A$ in the basis dual to $\CB$.

As before, from (\ref{rgen})(\ref{bgen}) we conclude that there is a
finite set of possible final magnetic attractor charges, and hence a
finite number of magnetic attractor trees one can make.

As before, if there are an infinite number of attractor flow trees
then there must be some line segment in some tree that supports an
infinite family of different trees emitting D2D0 charges
$(q_2^\alpha, q_0^\alpha)$. Once again,
\begin{equation}
\frac{\vert q_2^\alpha \cdot t_\alpha - q_0^\alpha
V\vert}{\sqrt{J_\alpha^3}}
\end{equation}
must go to zero for some subsequence of electric charges. Again we
conclude that $J_\alpha^3\to \infty$ is necessary, and now observe
that there is an upper bound on
\begin{equation}
\vert \frac{ \frac{1}{6} r_* t_\alpha^3- \frac{1}{2} P_*
t_\alpha^2}{\sqrt{J_\alpha^3}} + \frac{ \hat q_2^\alpha t_\alpha
-\hat q_0^\alpha V}{\sqrt{J_\alpha^3}} \vert
\end{equation}
The above is a sum of two complex numbers. The ensemble formed by
the second (as $\alpha \to \infty$ ) is bounded, but the first
cannot be, but this contradicts the fact that the norm is bounded.

Thus, there must be a finite number of split attractor flows
terminating in the regions of the type we have specified. For
physical reasons we firmly believe that the number of split
attractor flows in all of Teichm\"uller space is unconditionally
finite. Unfortunately, it appears to us that the above ideas are not
sufficiently powerful to prove this, and the proof will need a new
idea.

\section{Attractor tree numerics} \label{app:attnumerics}

As reviewed in section \ref{sec:reviewtrees}, whenever the entropy
function $S$ on charge space is known explicitly, one can explicitly
construct all solutions to the BPS equations of motion. In
particular one can in principle explicitly construct attractor flows
and the trees built from them, although explicit expressions often
become very complicated. The same explicit prescriptions can be used
though to construct highly efficient numerical algorithms, e.g.\ for
determining whether or not a tree of given topology exists in a
given background.

In this appendix, we sketch such an algorithm and explain how we
used it to check the extreme polar state conjecture.

\subsection{Existence of flow trees}

\EPSFIGURE{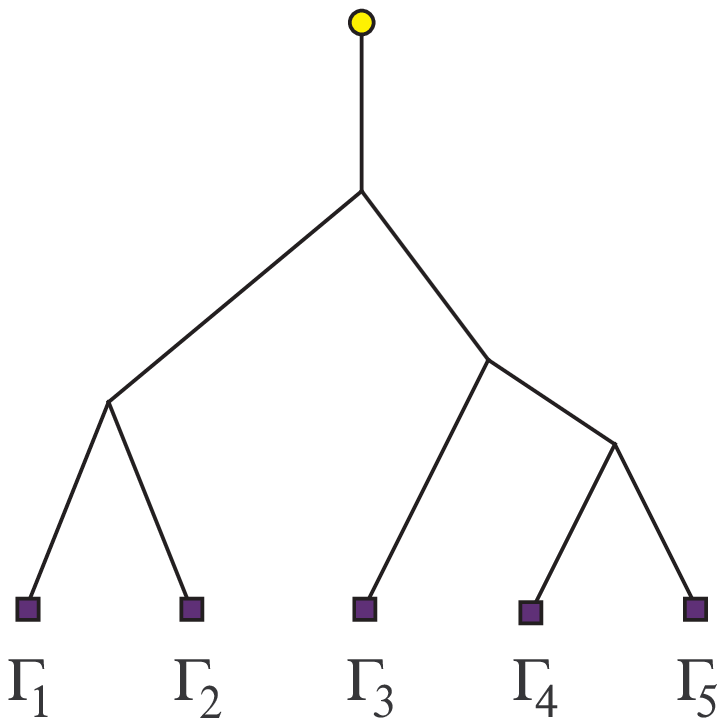,height=5cm,angle=0,trim=0 0 0 0}%
{Example of topological flow tree data. \label{treedata}}

The topological data of a flow tree can be specified as a nested
list. For example the tree sketched in fig.\ \ref{treedata} is
represented as ${\cal T} = \{ \{ \Gamma_1,\Gamma_2 \},\{ \Gamma_3,
\{ \Gamma_4, \Gamma_5 \} \}\}$.

To determine whether a split $\Gamma \to \Gamma_1 + \Gamma_2$ exists
starting from some initial point $t_{\rm in}$, i.e.\ whether a wall
of marginal stability is crossed before an attractor point or zero
of $Z(\Gamma)$ is reached, we proceed as follows. We parametrize the
attractor flow in the usual way by $\tau = 1/r$ such that the
initial point corresponds to $\tau=0$ and $U_{\tau=0} = 0$. Note
that from (\ref{stabchangeflow}) we always find a unique value of
$\tau$ where $\Im(Z_1 \bar{Z}_2) = 0$:
\begin{equation}
 \tau_0 = \frac{2 }{\langle \Gamma_1,\Gamma_2
 \rangle}\frac{ \Im(Z_1 \bar{Z}_2)}{\vert Z\vert }\biggl\vert_{\rm \tau=0}. \label{tau0}
\end{equation}
When the entropy function $S$ is known explicitly, the value of the
moduli $t^A$ and therefore the central charges $Z_1$ and $Z_2$ at
$\tau = \tau_0$ can be computed explicitly using the results of
\cite{Bates:2003vx}. For example in the (effective) one modulus,
large radius case with $(p^0,p,q,q_0)=p^0 + p D + q D^2 + q_0 D^3$
with $D \equiv D_1$ the basis divisor, we have
\begin{eqnarray*}
S(p^0,p,q,q_0) &=& \frac{\pi}{3} \,{\sqrt{3\,p^2\,q^2 - 8\,p^0\,q^3
- 6\,p^3\,q^0 +
        18\,p\,p^0\,q\,{q^0} -
        9\,{{p^0}}^2\,{{q^0}}^2}} \\
t(p^0,p,q,q_0) &=& \frac{p\,q - 3\,{p^0}\,{q^0} + i
    {\sqrt{3\,p^2\,q^2 - 6\,p^3\,{q^0} + 18\,p\,{p^0}\,q\,{q^0} -
        8\,p^0 \, q^3 - 9\,{p^0}^2\,{{q^0}}^2 }}}{p^2 -
    2\,{p^0}\,q},
\end{eqnarray*}
so the flows are given by $t(\tau)=t(H(\tau))$ with $H(\tau) = -
\Gamma \, \tau + 2 \, \Im (e^{-i \alpha} \Omega)_{\tau = 0}$.

The value of $\tau_0$ given by (\ref{tau0}) corresponds to an actual
split point if and only if $\tau_0>0$, $\Re(Z_1
\bar{Z}_2)|_{\tau_0}>0$, and $t^A|_{\tau_0}$ lies in the interior of
Teichm\"{u}ller space (in the large radius approximation in which we
work this amounts to $\Im \, t^A|_{\tau_0} > 0$).

To determine whether the full flow tree exists, it thus suffices to
check recursively through the nested list for the existence of the
subsequent splits, as outlined above, and finally whether the
$\Gamma_i$ attractor points exist for the endpoints of the tree. In
the large radius approximation, this is equivalent to having
positive discriminant $S^2(\Gamma_i)\sim \CD(\Gamma_i)$.

All of this is easily done numerically. A straightforward
implementation in Mathematica manages to check about one thousand
splits per second on a 2 GHz Pentium.

\subsection{Maximizing $\widehat{q}_0$}

\EPSFIGURE{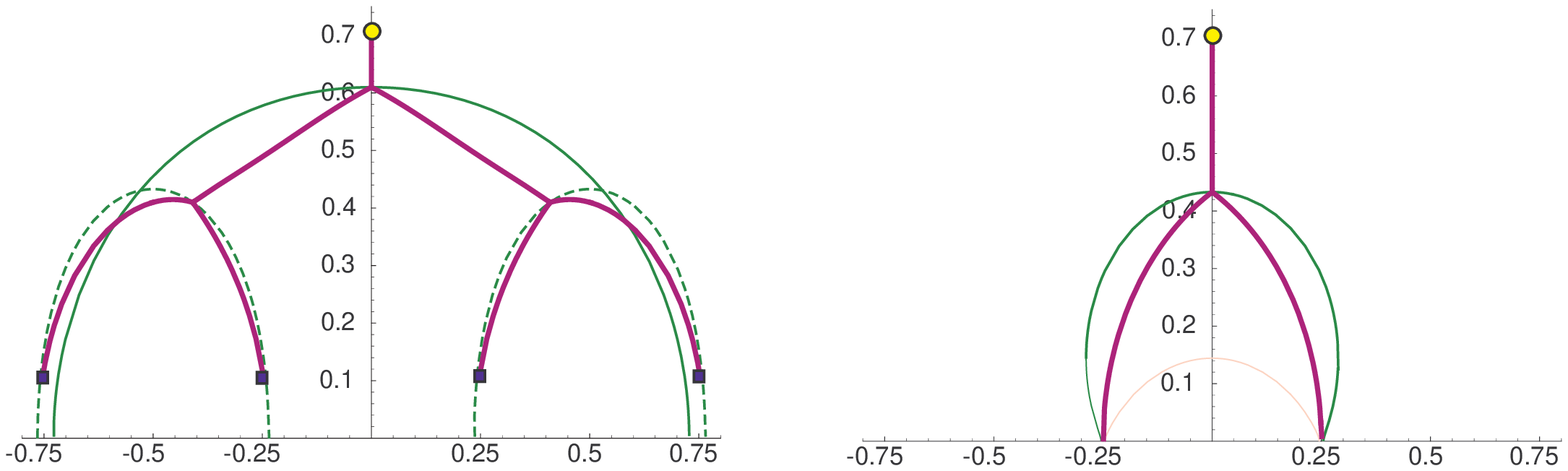,height=5cm,angle=0,trim=0 0 0 0}%
{{\bf Left:} Initial four-legged flow tree in D6-D6-anti-D6-anti-D6
optimization procedure. {\bf Right}: Endpoint of optimization. The
final two splits have become invisibly small; in spacetime this flow
tree corresponds to near-coincident ${\rm D6-D6}$ and
near-coincident ${\rm \overline{D6}-\overline{D6}}$, with nearly
pure fluxed D6 and anti-D6 branes. \label{D6D6}}

Using the procedure for checking the existence of flow trees
sketched above, we can try to find numerically   the maximally polar
states (i.e.\ maximal $\hat{q}_0$) within a specified ensemble of
flow trees, thus providing evidence for the extreme polar state
conjecture of section \ref{sec:extremepolar}.

We implemented this in Mathematica by a simple random walk
algorithm, starting from an existing attractor flow tree within an
ensemble specified by a flow tree topology and charges $\Gamma_i(u)$
depending on a set of parameters $u$. At each step random points $u$
near the latest successful point $u_{\rm prev}$ are chosen until a
value of $u$ is found which gives an actual attractor flow tree with
$\hat{q}_0$ larger than the maximal $\hat{q}_0$ so far, with some
bias in the direction of the last successful step. If the number of
trials exceeds a certain cutoff, the stepsize is decreased. This
goes on till a specified (large) number of attractor flow trees has
been evaluated. The whole process is repeated several times over,
eliminating the less successful random walks.

For example, we considered four centered ${\rm
D6-D6-\overline{D6}-\overline{D6}}$ flow trees with topology as in
fig.\ \ref{D6D6}$a$ and charges parametrized by $\Gamma_1 =
e^{P/4+S_a} (1-\tilde{\beta} D^2+\tilde{n}_1 D^3)$, $\Gamma_2 =
e^{P/4-S_a} (1-\tilde{\beta} D^2+\tilde{n}_2 D^3)$, $\Gamma_3 =
-e^{-P/4-S_b} (1-\tilde{\beta} D^2+\tilde{n}_3 D^3)$, $\Gamma_4 =
-e^{P/4+S_b} (1-\tilde{\beta} D^2+\tilde{n}_4 D^3)$. Keeping $P$
fixed at $P=1$ and starting at
$\{S_a,S_b,\beta,\tilde{n}_1,\tilde{n}_2,\tilde{n}_3,\tilde{n}_4 \}
= \{ 5 \times 10^{-2}, 5 \times 10^{-2},1.92 \times 10^{-2}, 2.03
\times 10^{-3}, -2.03 \times 10^{-3}, -2.03 \times 10^{-3}, 2.03
\times 10^{-3} \}$ (shown in fig.\ \ref{D6D6}$a$), running 100 times
at a cutoff of $100,000$ flow tree evaluations, resulted in a
maximal $\hat{q}_0$, $(\hat{q}_0)_{\rm max} = 0.0104064 \quad
\mbox{at} \quad
 \{S_a,S_b,\beta,\tilde{n}_1,\tilde{n}_2,\tilde{n}_3,\tilde{n}_4 \}
 = \{ 4.9 \times 10^{-4}, 1.5 \times 10^{-4},1.0 \times 10^{-5},
 3.1 \times {10}^{-8},-3.1 \times {10}^{-8},-3.1 \times {10}^{-8},
  7.4 \times {10}^{-9} \}$ (shown in fig.\ \ref{D6D6}$b$).
 This is fully compatible with our conjectured $(\hat{q}_0)_{\rm max}
= P^3/24 r^2 = 1/96 \approx 0.0104167$, at
$\{S_a,S_b,\beta,\tilde{n}_1,\tilde{n}_2,\tilde{n}_3,\tilde{n}_4
\}=\{0,0,0,0,0,0,0\}$, in accordance with the extreme polar state
conjecture.

\section{The three node quiver index}
\label{app:eulerint}

 In this appendix we evaluate the integral
(\ref{threenodeint}) yielding the Euler characteristic of $\CM$
defined in (\ref{modspacetriquiv}). We therefore define the
function:
\begin{equation}\label{contourone}
\chi(a,b;c):= \oint dJ_1 \oint dJ_2 J_1^{-a}
J_2^{-b}\frac{(1+J_1)^a(1+J_2)^b}{(1+J_1+J_2)^c} (J_1+J_2)^c.
\end{equation}
In this appendix we will derive the following four main properties.

First, we obviously have $\chi(a,b;c)=\chi(b,a;c)$. Second, we can
write $\chi(a,b;c)$ in terms of an integral of Laguerre polynomials:
\begin{equation}\label{laguerreform}
\chi(a,b;c) = ab - \int_0^\infty ds e^{-s} L^1_{a-1}(s) L^1_{b-1}(s)
L^1_{c-1}(s)
\end{equation}
in particular, $\chi(a,b;c) = ab - f(a,b,c)$ where $f(a,b,c)$ is
totally symmetric.

To state the third and fourth properties note that for 3 positive
integers $a,b,c$ either all three triangle inequalities are
satisfied
\begin{eqnarray}\label{tidents}
a+b  & \geq & c \\
b+c & \geq & a \\
c+a & \geq & b
\end{eqnarray}
or precisely one is violated. Our third property states that:
\begin{equation}
\chi(a,b;c) = \left\{
  \begin{array}{l}
    b (a-c) \mbox{ if } \quad a \geq b+c \\
   a(b-c)  \mbox{ if } \quad b \geq a+c\\
   0 \qquad\quad \ \mbox{if} \quad\  c\geq a+b
  \end{array}
\right.
\end{equation}
Fourth, when  all three inequalities (\ref{tidents}) are satisfied
we do not have a simple formula for $\chi(a,b;c)$, but we do have
the asymptotic formula
\begin{equation}
\chi(a,b;c) \sim k (-1)^{a+b+c}  (abc)^{-1/3} 2^a 2^b 2^c
\end{equation}
where $k$ is a constant.

\subsection{Derivation of property two }

 Write $\chi(a,b;c)$ as:
\begin{eqnarray}
\chi(a,b;c) & = & \frac{1}{(c-1)!} \oint dx_1 \oint dx_2 \left(1+
1/x_1\right)^a \left(1+1/x_2\right)^b (x_1 + x_2)^c \int_0^\infty
\frac{ds}{s} s^c
e^{-s(1+x_1 + x_2)} \\
& = & \frac{1}{(c-1)!} \int_0^\infty \frac{ds}{s} s^c e^{-s}\oint
dx_1 \oint dx_2 \left(1+ 1/x_1\right)^a \left(1+1/x_2\right)^b
\left(-\frac{\p}{\p s} \right)^c
e^{-s(x_1 + x_2)} \\
& = & \frac{1}{(c-1)!}\oint_0^\infty \frac{ds}{s} s^c e^{-s}
\left(-\frac{\p}{\p s} \right)^c\biggl[\left( \oint dx_1 \left(1+
1/x_1\right)^a  e^{-s x_1} \right) \left( \oint dx_2
\left(1+1/x_2\right)^b  e^{-s x_2}\right)\biggr]
\end{eqnarray}

Now we note that
\begin{equation}
 \oint dx_1 \left(1+
1/x_1\right)^a  e^{-s x_1} = \sum_{j=1}^{a} { a \choose j}
\frac{(-s)^{j-1}}{(j-1)!} = L_{a-1}^{1}(s)
\end{equation}
is a Laguerre polynomial. Thus we can write
\begin{equation}
\chi(a,b;c) =\frac{1}{(c-1)!}\oint_0^\infty  ds s^{c-1} e^{-s}
\left(-\frac{\p}{\p s} \right)^c\biggl( L^1_{a-1}(s)
L^1_{b-1}(s)\biggr)
\end{equation}

Now integrate by parts $c-1$ times (assuming $c-1>0$). The boundary
terms do not contribute. Next use the Rodrigues formula
\begin{equation}
\left(\frac{d}{dx}\right)^n(x^{n+a} e^{-x}) = n! x^a e^{-x} L^a_n(x)
\end{equation}
to get
\begin{equation}
\chi(a,b;c) = \int_0^\infty ds e^{-s} L^0_{c-1}(s)
\left(-\frac{\p}{\p s} \right)\biggl( L^1_{a-1}(s)
L^1_{b-1}(s)\biggr)
\end{equation}
The last integration by parts produces
\begin{equation}
\chi(a,b;c) = ab + \int_0^\infty ds \frac{d}{ds}(e^{-s}
L^0_{c-1}(s))L^1_{a-1}(s) L^1_{b-1}(s)
\end{equation}
Finally, using
\begin{equation}
\frac{d}{ds}(e^{-s} L^0_{c-1}(s))= - e^{-s} L^1_{c-1}(s)
\end{equation}
we arrive at the elegant formula (\ref{laguerreform}). Of course the
integral can be done explicitly as a triple sum:
\begin{equation}
\chi(a,b;c) = ab - \sum_{s=0}^{a-1} \sum_{t=0}^{b-1}\sum_{u=0}^{c-1}
{a \choose s+1}{b \choose t+1}{c\choose u+1} \frac{(s+t+u)!}{s! t!
u!} (-1)^{s+t+u}
\end{equation}

\subsection{Evaluation when a triangle inequality is violated}

The Laguerre form (\ref{laguerreform}) of the function does not
appear to be the most useful form for evaluating $\chi$ in this
region. Rather, in the contour integral (\ref{contourone}) it is
useful to make the change of variables
\begin{equation}
z_i := 1+ 1/J_i
\end{equation}
The contour will now be on two \textit{large} circles with radius
$\cong 1/\epsilon_i$ and we have the contour integral
\begin{equation}
\chi(a,b;c)= \CI(a,b;c) = \oint dz_1 \oint dz_2
\frac{1}{(1-z_1)^2}\frac{1}{(1-z_2)^2} z_1^{a} z_2^{{b}}
\left(\frac{z_1+z_2-2}{z_1z_2-1}\right)^{c}
\end{equation}

Let us try to do the integral by deforming the $z_1$ contour first.
Then we potentially pick up poles at $z_1=1$ and $z_1=1/z_2$. This
leads to $\CI=\CI_1 + \CI_2$ where $\CI_1$ comes from the pole at
$z_1=1$ and $\CI_2$ from the pole at $z_1=1/z_2$. We have
\begin{equation}
\CI_1= \oint dz_2 {d \over dz_1}\vert_{z_1=1} \Biggl[
\frac{z_2^{b}}{(z_2-1)^2} z_1^{a}
\left(\frac{z_1+z_2-2}{z_1z_2-1}\right)^{c}\Biggr]
\end{equation}
\begin{equation}
\CI_2 = \oint dz_2\frac{1}{(c-1)!}\left( {d \over
dz_1}\right)^{c-1}\vert_{z_1=1/z_2} \Biggl[
\frac{z_2^{{b}-c}}{(z_2-1)^2} \frac{z_1^{a}}{(z_1-1)^2} \left(
z_1+z_2-2 \right)^{c}\Biggr]
\end{equation}

It is straightforward to carry out the differentiation in $\CI_1$
and evaluate the $z_2$ integral from its pole  at $z_2=1$:
\begin{equation}
\CI_1  = \oint dz_2 (a-c) \frac{z_2^{b}}{(z_2-1)^2} =(a-c){b}
\end{equation}

In order to evaluate $\CI_2$ we expand
$$
 \left( z_1+z_2-2 \right)^{c}= \sum_{s=0}^{c}  {c \choose s}(z_1-1)^s
 (z_2-1)^{c-s}
 $$
so now we write:
\begin{equation}
\CI_2 = \CI_2^A + \CI_2^B + \CI_2^C
\end{equation}
\begin{eqnarray*}
\CI_2^A & = & \oint dz_2\frac{1}{(c-1)!}\left( {d \over
dz_1}\right)^{c-1}\vert_{z_1=1/z_2} \Biggl[
\frac{z_2^{b-c}}{(z_2-1)^2}
\frac{z_1^{a}}{(z_1-1)^2} (z_2-1)^{c}\Biggr]\\
\CI_2^B & = & \oint dz_2\frac{1}{(c-1)!}\left( {d \over
dz_1}\right)^{c-1}\vert_{z_1=1/z_2} \Biggl[
\frac{z_2^{b-c}}{(z_2-1)^2}
\frac{z_1^{a}}{(z_1-1)^2} c (z_1-1)(z_2-1)^{c-1}\Biggr]\\
\CI_2^C & = & \oint dz_2\frac{1}{(c-1)!}\left( {d \over
dz_1}\right)^{c-1}\vert_{z_1=1/z_2} \Biggl[
\frac{z_2^{b-c}}{(z_2-1)^2}
 \sum_{s=2}^{c} {c\choose s}z_1^{a} (z_1-1)^{s-2} (z_2-1)^{c-s} \Biggr]
\end{eqnarray*}

Next we write out the action of the derivative wrt $z_1$. Important
simplifications occur because after differentiating we replace
$$
z_1-1 \rightarrow - z_2^{-1} (z_2-1)
$$
The remaining $z_2$ integral will get contributions from the poles
at $z_2=1$ and, possibly, at $z_2=0$. If we replace
\begin{eqnarray*}
&& \left( {d \over dz_1}\right)^{c-1}\vert_{z_1=1/z_2}z_1^{a}
(z_1-1)^{s-2} \\
&& = \sum_{k=0}^{c-1} {c-1 \choose k} \frac{a!}{(a-(c-1-k))!}
\frac{(s-2)!}{(s-2-k)!} z_1^{a-(c-1-k)}(z_1-1)^{s-2-k}
\end{eqnarray*}
then we find after setting $z_1=1/z_2$ that the term is proportional
to
\begin{equation}\label{powers}
z_2^{{b}-a-s+1}(z_2-1)^{c-k-4}
\end{equation}
Thus, we can only get a pole for the contributions from $k =
c-3,c-2,c-1$ and then we find that only $s=c,c-1$ can contribute.

Adding up the contributions we get
\begin{equation}
\CI_2^C = a(a-1) + a({b}-a-c+1) + \oint_{z_2=0} [...]
\end{equation}
The second term arises from the contributions of the poles at
$z_2=0$. From (\ref{powers}) we see that these poles are absent if
${b}+1\geq a+c$.

In exactly the same way we find that
\begin{eqnarray*}
\CI_2^A &=& \half a (a-1)(c-2) + a(c-1)({b}-a+1) + \half c
({b}-a+1)({b}-a) + \oint_{z_2=0} [...] \\
\CI_2^B &=& -\half a (a-1)c -a c ({b}-a) - \half c ({b}-a)({b}-a-1)
+ \oint_{z_2=0} [...]
\end{eqnarray*}
where we have added up the poles at $z_2=1$. The poles at zero are
absent for ${b}+1\geq a$ for $\CI_2^A$ and ${b}\geq a$ for
$\CI_2^B$.

Thus, when ${b}+1\geq a+c$ we only have poles from $z_2=1$ and
adding up the contributions we find
\begin{equation}
\chi=\CI({b},a;c)= a ({b}-c)  \qquad \qquad a+c\leq {b}+1,
\end{equation}
in agreement with what we found in section \ref{sec:3quiver}.

What about other ranges? Clearly we have $\CI({b},a;c)=\CI(a,{b};c)$
so we also have
\begin{equation}
\chi= \CI(a,{b};c) = {b} (a-c)  \qquad \qquad {b}+c\leq a+1
\end{equation}

When $c\geq a+b$ we can use the fact that $\chi(a,b;c)=ab-f(a,b,c)$
with $f(a,b,c)$ totally symmetric to derive $\chi(a,b;c)=0$.

\subsection{Large $(a,b,c)$ asymptotics}

For estimating  asymptotics at large $a,b,c$  satisfying the
triangle inequality we return to the formula (\ref{laguerreform}).
For simplicity we consider
$$
\int_0^\infty ds e^{-s} L^1_{A }(s) L^1_{B}(s) L^1_{C}(s)
$$
For large $A,B,C$ the integrand oscillates, but has a large peak (at
least for $A=B=C$ ) and one can try to do the integral by
saddle-point approximation. The integrand is certainly very small
for $s\geq (A+B+C)$.

The appropriate asymptotic expansion  of the Laguerre polynomials
 for our needs is that given in \cite{Szego}, equation 8.22.10.
 Namely, for $x= (4n+4)\cosh^2(\phi)$, $\epsilon\leq \phi \leq \Lambda$
\begin{equation}\label{correctasymp}
L_n^1(x) \sim \frac{1}{2}(-1)^n e^{x/2}(\pi \sinh \phi)^{-1/2}
x^{-3/4} n^{1/4} \exp \biggl( (n+1)( 2\phi-\sinh 2 \phi) \biggr)(1+
\CO(n^{-1}) )
\end{equation}
This covers a region up to $(4n+4) \cosh^2 \Lambda$ for any fixed
$\Lambda$ as $n\to \infty$. We see that the integrand grows much
more slowly  than $e^{-x + 3x/2}$ in this region. Beyond this region
we will start to get exponential decay.

We consider the case where $A,B,C$ do not differ too much from some
common large integer $N$. To be more precise, define $\phi_A,
\phi_B, \phi_C$ by $s = 4(A+1) \cosh^2 \phi_A$, etc. and define also
$s=4(N+1)\cosh^2\phi$.   We define $\mu_A := \frac{A+1}{N+1}:=1 +
\delta \mu_A $ and we are considering limits where $\delta \mu_A =
(A-N)/(N+1) \sim N^{-\theta}$ with $0<\theta< 1$. We will neglect
corrections to the integral of order $1+ \CO(\delta \mu)$. These are
very complicated. But we will keep corrections to the entropy of
order $(N+1)\delta \mu^2 \sim N^{1-2\theta}$. Note that if
$1/2>\theta$ these are even dominant over the $\log N$ correction
from the one-loop prefactor.

We solve $\cosh\phi_A = (1+\delta \mu_A)^{-1/2} \cosh \phi$ by
\begin{equation}
\phi_A = \phi - \half \delta \mu_A \coth \phi + \frac{\delta
\mu_A^2}{16}\frac{(\cosh 3\phi - 3 \cosh \phi)}{(\sinh\phi)^3}
+\cdots
\end{equation}
and then expand the action to second order:
\begin{equation}
f = (N+1) \biggl(2 \cosh^2\phi + 3(2\phi -\sinh 2\phi )+
2\phi(\delta\mu_A + \delta \mu_B + \delta \mu_C)-
\frac{1}{2}(\delta\mu_A^2 + \delta\mu_B^2 + \delta \mu_C^2)\coth
\phi + \cdots \biggr)
\end{equation}
We find the stationary point for this action  is
\begin{equation}
\phi_* = \log \sqrt{2} + \frac{1}{2}(\delta\mu_A + \delta \mu_B +
\delta \mu_C)-\frac{3}{4}(\delta\mu_A + \delta \mu_B + \delta
\mu_C)^2 +(\delta\mu_A^2 + \delta\mu_B^2 + \delta \mu_C^2)
\end{equation}
and the saddle point value is
\begin{equation}
f(\phi_*)= (N+1)\biggl( \log 8 +(\delta\mu_A + \delta \mu_B + \delta
\mu_C)\log 2 + \frac{1}{2}((\delta\mu_A + \delta \mu_B + \delta
\mu_C)^2-3(\delta\mu_A^2 + \delta\mu_B^2 + \delta \mu_C^2)) +
\cdots\biggr)
\end{equation}

From this we get:
\begin{eqnarray}
I_{ABC} & \sim & \frac{2^{1/2}}{\pi 3^{7/2}}
\frac{(-1)^{A+B+C}}{(ABC)^{1/3}} 2^{A+1} 2^{B+1} 2^{C+1}\\
& & e^{(N+1) \frac{1}{2}((\delta\mu_A + \delta \mu_B + \delta
\mu_C)^2-3(\delta\mu_A^2 + \delta\mu_B^2 + \delta \mu_C^2))}
\biggl(1+\CO(\delta \mu_A, \delta \mu_B, \delta\mu_C)\biggr)
\end{eqnarray}
Note that if $A,B,C$ are not very different from each other, as we
assumed, then it is natural to take $N=(A+B+C)/3$.

Thus, translating back to our entropy we find that in this regime,
\begin{equation}
\chi \sim \frac{2^{1/2}}{\pi 3^{7/2}}\biggl( \frac{(-1)^{a}
2^{a}}{a^{1/3}}\biggr)\biggl( \frac{(-1)^{b}
2^{b}}{b^{1/3}}\biggr)\biggl( \frac{(-1)^{c}
2^{c}}{c^{1/3}}\biggr)(1+ \cdots)
\end{equation}
where the corrections in $+\cdots$ are of order
\begin{equation}  \CO(
(\frac{2a-b-c}{b+a+c}), (\frac{2b-a-c}{b+a+c}),
(\frac{2c-b-a}{b+a+c}))
\end{equation}

The leading order factorization of the answer, and especially the
factors $2^{a}$ etc.\ call for a conceptual explanation!

\section{Index vs.\ absolute cohomology and the entropy of 5d black holes} \label{app:entropy5dBH}

In \cite{Vafa:1997gr} C. Vafa adduced an example of black hole
entropy counting which appears to imply that the entropy can only be
accounted for by computing the total number of BPS states without
signs, rather than by an index of BPS states. In this appendix we
will explain that, in fact, the entropy can be correctly accounted
for using an appropriate index.

The problematic example studied in \cite{Vafa:1997gr} involves type
IIA string theory on an elliptically fibered Calabi-Yau $\pi: X \to
B$ with section. The BPS states in question are those obtained from
wrapping $D2$ branes on a curve $C \subset B$.

Let $\hat C = \pi^{-1}(C)$ be the elliptically fibered surface
covering $C$. Then, in \cite{Vafa:1997gr} it is argued that the
relevant moduli space which one should quantize to produce BPS
states is
\begin{equation}\label{productenn}
  \amalg_{n\geq 1} {\rm Sym}^n(\hat C)
\end{equation}
As usual, this quantization involves a Fock space based on
oscillators associated with the cohomology of $\hat C$. For generic
elliptic fibrations one has $h^{1,0}(\hat C) = h^{1,0}(C)$.
\footnote{Note that this explicitly excludes the case of a direct
product $\hat C = C \times T^2$. Our considerations below apply
equally well in the direct product case.} It then follows from the
adjunction formula that
\begin{equation}
  h^{1,0}(\hat C) = \half (C\cdot C + C\cdot K_B) + 1
\end{equation}
where the intersection products are taken within the surface $B$,
and $K_B$ is the canonical bundle of $B$. Now, using equations
(\ref{topinvtsa}-\ref{topinvtsb})   together with $c_2(X) = 12
\sigma \pi^*(c_1(B)) \mod \pi^*$ where $\sigma$ is the section of
the elliptic fibration, (see, for example, \cite{Friedman:1997yq},
eq. 7.28), we find \cite{Vafa:1997gr}
\begin{eqnarray}
  h^{2,0}(\hat C) & = & \half (C\cdot C - C \cdot K_B) \\
  h^{1,1}(\hat C) & = & C\cdot C - 9 C \cdot K_B + 2.
\end{eqnarray}

Now, the key to resolving the puzzle pointed out in
\cite{Vafa:1997gr} lies in considering the $SU(2) \times SU(2)$
Lefshetz decomposition of the cohomology of the moduli space. As
emphasized in \cite{Katz:1999xq} the existence of such a double
Lefshetz decomposition  follows from physical reasoning, although it
is not so obvious mathematically. The existence of an $SU(2) \times
SU(2)$ Lefshetz decomposition of the cohomology of $\hat C$ is
strongly suggested by the Leray spectral sequence, and we will
simply assume it exists. Some rigorous results along these lines
 appear in \cite{Hosono:2001gf}.

Proceeding naively, the $SU(2)_R$ raising and lowering operators are
constructed using the K\"ahler form $\omega(C)$ of the base, while
those for $SU(2)_L$ are constructed using $\omega(E):=\omega(\hat
C)-\omega(C)$ which may be regarded as the K\"ahler form of a
generic fiber. The decomposition into multiplets of the type
$(\textbf{j}_L, \textbf{j}_R)$ is then
\begin{equation}
  2 h^{1,0}(\hat C)(\half,0) \oplus  (2 h^{2,0}+h^{1,1}-2)(0,0)
  \oplus
  (\half, \half)
\end{equation}
where the last summand is the multplet  $ 1, \omega(C),\omega(E),
\omega(C)\wedge \omega(E)$.

Taking into account the symmetric products (\ref{productenn}) we
have
\begin{equation}
Z = {\rm Tr} (-1)^{2m_L + 2m_R} y^{2m_L} q^N =  \prod_{n\geq 1}
\frac{\bigl( (1-y q^n)(1-y^{-1}q^n)\bigr)^{N_f } }{(1-q^n)^{N_b} }
\end{equation}
where
\begin{eqnarray}
  N_f & = & 2h^{1,0}(\hat C)-2 \\
  N_b & = & h^{1,1}(\hat C) + 2 h^{2,0}(\hat C)-2.
\end{eqnarray}
Note that if we wish to extract the Euler character of the moduli
spaces then we set $y=1$ and study the coefficients of $q^n$. For
$y=1$ we  indeed we obtain
\begin{equation}
  \eta^{-\chi(\hat C)}
\end{equation}
where $\chi(\hat C) = - 12 C \cdot K_B$. From the Calabi-Yau condition
 $\chi(\hat C)>0$ so this
will produce exponential degeneracies  $\sim \exp[ \pi \sqrt{8 \vert
C\cdot K_B\vert n}]$ but, as stressed in \cite{Vafa:1997gr} the
growth under uniform scaling of charges $(C,n) \to (\Lambda C,
\Lambda n)$ goes as $\exp[ const. \Lambda]$ in contradiction with
the supergravity entropy which scales like $\exp[const.
\Lambda^{3/2}]$. On the other hand, since $b^{even}(\hat C)$ and
$b^{odd}(\hat C)$ each scale like $C \cdot C$ for large $C$, the
absolute cohomology will grow like $\exp[ const. \sqrt{C\cdot C n}]
\sim \exp[ const. \Lambda^{3/2}]$. This observation suggests that,
at least in this example, one needs to use the absolute cohomology
-- the sum over all BPS states without signs -- to account properly
for the entropy. Unfortunately, that proposal in turn leads to many
paradoxes.

There is an alternative however. To account for the entropy we
should work at fixed $j_L$, and compute the asymptotic growth of
$N_Q^{m_L}$ as explained in section \ref{sec:D6D2D0}. In order to do
this properly we should incorporate the Wilson line degrees of
freedom for the $D2$ wrapped on $C$. This leads to an extra torus
factor in the moduli space, and the quantization of that torus leads
to a factor $y - 2 + y^{-1} = (y^{1/2}- y^{-1/2})^2$ for each $T^2$.
Therefore, we are interested in the asymptotics of the coefficients
$D'(n,\ell)$ defined by
\begin{equation}
(y^{1/2}- y^{-1/2})^{2h^{1,0}(\hat C)} \prod_{n\geq 1} \frac{\bigl(
(1-y q^n)(1-y^{-1}q^n)\bigr)^{N_f} }{(1-q^n)^{N_b} }= \sum
D'(n,\ell) q^n y^{\ell}
\end{equation}

Setting $y= e^{2\pi i z}$ and $q=e^{2\pi i \tau}$ and using the
product formula for the theta function we see that the asymptotics
for large $n$ of $D'(n,\ell)$ are in turn governed by those in the
Fourier expansion of
\begin{equation}\label{modratwithysq}
(y^{1/2} - y^{-1/2})^2 \eta^{-\chi(\hat C)} \Biggl(
\frac{\vartheta_1(z,\tau)}{\eta^3} \Biggr)^{C\cdot C+K_B\cdot C}
\end{equation}
We are interested in the leading behavior for $(C,n) \to (\Lambda C,
\Lambda n)$ and since $\chi(\hat C)$ is linear in $C$ the  first two
factors in (\ref{modratwithysq}) lead to a subleading correction to
the entropy.

Now let us derive the asymptotics of the Fourier coefficients of
\begin{equation}\label{modrat}
 \eta^{-\chi(\hat C)} \Biggl(
\frac{\vartheta_1(z,\tau)}{\eta^3} \Biggr)^{C\cdot C+K_B\cdot C}
\end{equation}
Put $C^2 +C\cdot K_B = M $ and for simplicity assume $M$ is an even
integer, and define $k=M/2$. In this case (\ref{modrat}) is a weak
Jacobi form of index $k=M/2$ and weight $-M-\chi/2$, where $\chi =
\chi(\hat C)$. (We choose $M$ to be even to avoid certain
inconvenient phases in the modular transformations. Similarly,
strictly speaking we should take  $\chi$ to be a multiple of $24$,
but this latter point is not too essential. )

The spectral flow identity shows that (\ref{modrat}) has an
expansion of the form
\begin{equation}
\sum_{n\geq 0, \ell \in \IZ} c(2Mn-\ell^2) q^n y^\ell
\end{equation}

Decompose the sum by writing
\begin{eqnarray}
\ell & = & \mu + 2k s \\ n & = & n_0 + ks^2 + \mu s
\end{eqnarray}
and choose a fundamental domain $-k+1 \leq \mu \leq k$  so that we
can write

\begin{equation}\label{modratii}
 \eta^{-\chi} \bigl(\frac{\vartheta_1(z,\tau)}{\eta^3} \bigr)^M =
 \sum_{\mu=-k+1}^k H_\mu(\tau) \Theta_{\mu,k}(z,\tau)
\end{equation}
where

\begin{eqnarray}
H_\mu(\tau) &=& \sum_{n\in \IZ} c(4kn-\mu^2) q^{n -
\frac{\mu^2}{4k}-\chi/24}\\
& = & (-1)^\mu { 2k \choose k-\mu} q^{ -
\frac{\mu^2}{4k}-\frac{\chi}{24} }+\cdots
\end{eqnarray}

Note that the most negative power goes like
$q^{-\frac{k}{4}-\frac{\chi}{24} }$ for $\mu= \pm k$.  Also note
that by the modular transformations of level $k$ theta functions the
$H_\mu$ transform under $\tau \to -1/\tau$ by a finite fourier
transform, times the usual modular weight of $-M -\frac{\chi+1}{2}$.

Applying the Rademacher expansion we find that for $2Mn-\nu^2 \gg 1$
\begin{equation}\label{finalasymps}
c(2Mn-\nu^2)\sim   \zeta e^{\pi
\sqrt{\bigl(2Mn-\nu^2-\frac{M\chi}{12}\bigr)(1+\frac{\chi}{3M})}}
\end{equation}
 with a rather awkward prefactor
\begin{equation}
\zeta = (-1)^{\nu+M/2}\sqrt{2} \biggl(\frac{M}{2}\biggr)^{M+\half
\chi + \frac{3}{2}} \biggl(1+
\frac{\chi}{3M}\biggr)^{\frac{1}{2}(M+\chi/2 +1)} \biggl(2Mn-\nu^2-
\frac{M\chi}{12}\biggr)^{-\frac{1}{2}(M+\chi/2 +2)}
\end{equation}
Now, we can take into account the prefactor $y -2 + y^{-1}$ in
(\ref{modratwithysq}) by noting that this amounts to taking a
discrete second derivative with respect to $\nu$ of
(\ref{finalasymps}). This will leave the exponential factor and
modify the prefactor $\zeta$.

Letting $M=C^2+ C\cdot K_B$ this shows that the entropy at fixed
$m_L$ is exactly that predicted macroscopically by supergravity, at
least for large $2Mn-\nu^2$,  and we do indeed have $\exp[ const.
\Lambda^{3/2}]$ growth for an index. Note that the terms depending
on $\chi$ correct the leading supergravity result in an interesting
way.

What has happened here is that the sum over $m_L$, which corresponds
to putting $y=1$ leads to impressive cancellations. Nevertheless,
one can still  capture the entropy with an \textit{index} rather
than the absolute cohomology.

\section{A derivation of $g_{\rm top} \to \infty$ OSV using flux vacua counting techniques}
\label{app:largeq}


 Now let us return to the discussion of section
\ref{sec:Dbranemodel}. As we explained below eq. (\ref{eulercharp}),
counting BPS states involves the counting of ``open string flux
vacua.''

To make this more precise, we make use of the $\CN=1$ special
geometry structure of the D4-brane moduli space $\CM$
\cite{Lerche:2002yw}. Let $\Sigma_F$ be the Poincar\'e dual 2-cycle
to $F$,  and expand $\Sigma_F$ in a basis $\{ C_\alpha \}$ of
$H_2(P)$: $\Sigma_F=m^\alpha C_\alpha$. In a neighborhood of the
divisor moduli space, parametrized by moduli $z^i$,
$i=1,\ldots,n:=h^{2,0}$, we can define chain periods $\Pi_\alpha$ by
\begin{equation}
 \Pi_\alpha(z) := \int_{\Gamma_\alpha(z)} \Omega
\end{equation}
where $\Gamma_\alpha$ is a 3-chain with a $z$-dependent boundary
component on $P$ given by $C_\alpha$, and possibly other, fixed
boundary components, independent of $z$. With these chain periods,
we define a superpotential\footnote{This superpotential and
generalizations thereof have been discussed in
\cite{Gomis:2005wc,Martucci:2006ij}.}
\begin{equation}
 W(z) := m^\alpha \Pi_\alpha(z) = \int_{\Gamma(z)} \Omega,
\end{equation}
where $\Gamma:=m^\alpha \Gamma_\alpha$ is thus a 3-chain with
boundary $\Sigma_F$ on $P$. Critical points of $W$ precisely
correspond to points where $F^{(0,2)}=0$. To see this, note that an
infinitesimal holomorphic variation of $W$ gives
 $$
  \delta W = \int_{\delta \Gamma} \Omega = \int_{\Sigma_F} \delta n
  \cdot \Omega = \int_P F \wedge (\delta n \cdot \Omega)
 $$
where $\delta n$ is the normal holomorphic vector field
corresponding to the variation $\delta z$ of the divisor moduli and
$\delta n \cdot \Omega$ is the contraction of $\delta n$ with
$\Omega$, providing an isomorphism between the space of holomorphic
sections of the normal bundle to $P$ and $(2,0)$-forms on $P$. Thus
we see that requiring $\partial_i W=0$ is equivalent to $F^{(0,2)} =
0$ (and therefore of course also $F^{(2,0)}=0$).

For the same reason, we have that for each $i = 1,\ldots,h^{2,0}$,
$\partial_i \Pi_\alpha(z)$ is the period vector of a $(2,0)$-form
$\omega_i$ on $P$. The natural K\"ahler metric on moduli space is
given by
\begin{equation} \label{metricdef}
 g_{i\bj} := \int_P \omega_i \wedge \bar{\omega}_{\bj}
 = \partial_i \Pi_\alpha \, Q^{\alpha \beta} \, \bpartial_{\bj}
 \bPi_\beta
 = \partial_i \bpartial_{\bj} (\Pi_\alpha \, Q^{\alpha \beta} \, \bPi_\beta) ,
\end{equation}
where $Q^{\alpha\beta}$ is the inverse of the intersection form
$Q_{\alpha\beta} := C_\alpha \cdot C_\beta$.

Similar to the more familiar $\CN=2$ special geometry, acting with
further derivatives on $\partial_i \Pi$ will produce periods of
$(1,1)$- and $(0,2)$-forms on $P$, because of Griffiths
transversality \cite{Lerche:2002yw}. In particular
\begin{equation} \label{ddis11}
  \nabla_i \partial_j \Pi(z) \sim (1,1)
\end{equation}
where $\nabla_i$ is the Levi-Civita covariant derivative with
respect to the above defined metric.

Let us now compute the actual BPS partition sum. For a given flux
$F$, the number of isolated critical points of the corresponding
flux superpotential $W_F$ is given by
\begin{equation}
 \int_{\CM} d^{2n} z \, \delta^{2n}(\partial W_F) \, |\det \nabla_i
 \partial_j W_F|^2.
\end{equation}
The determinant ensures that each isolated zero of the delta
function contributes $+1$ to the integral. We are free to use
covariant derivatives instead of ordinary derivatives because the
difference is proportional to $\partial W$, which vanishes. At any
such critical point, the divisor is frozen, so the only remaining
moduli are the positions of the $N$ D0-branes bound to $P$. The
contribution to the total degeneracy or Euler characteristic from
this component of moduli space is therefore simply $\chi({\rm
Sym}^N(P))=p_{\chi}(N)$, where $p_\chi(N)$ are the partitions of $N$
into $\chi$ colors.\footnote{Because $b_1(X)=0$, we have $b_1(P)=0$
hence the Euler characteristic of the symmetric product equals the
total degeneracy.}

Thus, we get for the OSV black hole partition sum
\begin{eqnarray}
 \CZ_{\rm BH} &:=& \sum_{q} \Omega(p,q) \, e^{2 \pi \phi^0 q_0 + 2 \pi \phi^A q_A}
 \\
 &\approx& \sum_{N,F} p_\chi(N) \,
 e^{2 \pi \phi^0(-N+\frac{1}{2}F^2 + \frac{\chi}{24}) + 2 \pi \Phi \cdot
 F} \, \int_{\CM} d^{2n} z \, \delta^{2n}(\partial W_F) \, |\det \nabla_i
 \partial_j W_F|^2
\end{eqnarray}
Here $\Phi=\phi^A D_A$ viewed as an element of $H^2(P)$, and we used
(\ref{q2val})-(\ref{q0val}). Actually, the above partition sum
misses the contributions from components which have flat directions
in the divisor moduli space (since then $\det W'' = 0$), e.g.\ for
$F=0$. However, we will eventually make a continuous $F$
approximation anyway, which as we will discuss is equivalent to a
large $q_0$ or small $\phi^0$ approximation, and for generic
divisors the set of such components with flat directions has measure
zero in flux space. So we will take the above expression for
$\CZ_{\rm BH}$ as our starting point.

The sum over $N$ is easily performed and yields a factor
$1/\eta^\chi$. We furthermore expand as before $F=m^\alpha
C_\alpha$, which gives:
\begin{equation} \label{Zn}
 \CZ_{\rm BH} \approx \frac{1}{\eta^\chi(e^{-2 \pi \phi^0})} \int_{\CM} d^{2n} z \,
 \sum_m e^{\pi \phi^0 Q_{\alpha\beta} m^\alpha m^\beta + 2 \pi \Phi_\alpha m^\alpha} \,
 \delta^{2n}(m^\alpha \partial_i \Pi_\alpha) \, |\det m^\alpha \nabla_i
 \partial_j \Pi_\alpha|^2.
\end{equation}
Note that $Q_{\alpha\beta}$ is an indefinite form of signature
$(b_2^+,b_2^-)$. However, only critical points of $W$ contribute, at
which $F$ is in $H^{1,1}(P)$. Restricted to this space, $Q$ has
signature $(1,b_2^-)$. The one positive direction corresponds to the
K\"ahler form $J$ on $P$. This positive direction will cause the
black hole partition sum to diverge, but as discussed in
\cite{Dabholkar:2005dt} and at length in this paper, this divergence
is easily regularized by adding a Boltzmann factor $e^{-\beta
H(p,q)}$. To avoid cluttering of formulas, we will not do this
regularization explicitly in what follows, and use its existence
only to justify formal manipulations.

Both the delta-function and the determinant can be rewritten as
integrals of exponentials linear in $m^\alpha$:
\begin{eqnarray}
 \delta^{2n}(m^\alpha \partial_i \Pi_\alpha)
 &=& \int d^{2n} \lambda \, e^{i \pi m^\alpha (\lambda^i \partial_i \Pi_\alpha +
 \bar{\lambda}^{\bi} \bpartial_{\bi} \bPi_\alpha)} \label{deltaLM} \\
 |\det m^\alpha \nabla_i \partial_j \Pi_\alpha|^2
 &=& \frac{1}{\pi^{2n}} \int d^n \theta \, d^n \psi \, d^n \btheta \, d^n \bpsi \,
 e^{\pi m^\alpha (\nabla_i \partial_j \Pi_\alpha \, \theta^i \psi^j +
 \bnabla_{\bi} \bpartial_{\bj} \bPi_\alpha \, \btheta^{\bi}
 \bpsi^{\bj})}. \label{detLM}
\end{eqnarray}
The second integral is over fermionic variables. This recasts the
partition function (\ref{Zn}) as a Gaussian ensemble with
boson-fermion-fermion cubic interactions. To obtain the ``large
flux'' asymptotics, i.e.\ the limit of small $\phi^0$, we replace
the sum over discrete fluxes $m^\alpha$ by an integral, parallel to
\cite{Ashok:2003gk,Denef:2004ze}. The resulting integral is
Gaussian, so it can be performed exactly. This yields for the part
of (\ref{Zn}) starting at $\sum_m \approx \int d^{b_2}m$:
\begin{equation} \label{gaussint}
 \frac{1}{\pi^{2n}} (\phi^0)^{-b_2/2} e^{-\frac{\pi}{4 \phi^0} \,
 (2\Phi_\alpha + i \lambda^i \partial_i \Pi_\alpha + \nabla_i \partial_j \Pi_\alpha \psi^i \theta^j \, + \, {\rm{c.c.}})
 \,Q^{\alpha \beta}\,
 (2\Phi_\beta + i \lambda^i \partial_i \Pi_\beta + \nabla_i \partial_j \Pi_\beta \psi^i \theta^j \, + \, \rm{c.c.})
 }
\end{equation}
where $b_2:=b_2(P)$ and $+ {\rm c.c.}$ stands for the conjugate
terms in (\ref{deltaLM})-(\ref{detLM}). Crucial here is that $\det
Q_{\alpha \beta} = 1$, because the middle cohomology of a compact
manifold is always self-dual and therefore its intersection form
unimodular. In the above expression and the remainder of this
appendix, we drop overall phase factors.

We now need to work out the intersection products. At first sight,
this seems to give a lot of complicated terms. However, the
underlying $\CN=1$ special geometry structure, and in particular
Griffiths transversality, simplifies this a lot, again in parallel
to the closed string case analyzed in
\cite{Ashok:2003gk,Denef:2004ze}. First recall that $\partial_i
\Pi_\alpha \sim (2,0)$, $\nabla_i
\partial_j \Pi_\alpha \sim (1,1)$, and $\Phi_\alpha \sim (1,1)$.
Only intersection products of $(1,1)$ with $(1,1)$ or $(2,0)$ with
$(0,2)$ can be nonzero. Furthermore, the intersection product
$\Phi_\alpha \, Q^{\alpha\beta} \, \nabla_i \partial_j \Pi_\beta =
0$ because $\Phi_\alpha \, Q^{\alpha\beta} \, \partial_j \Pi_\beta =
0$ identically for all values of the moduli $z$.

The remaining nontrivial products can be computed using the Leibniz
rule and orthogonality, together with (\ref{metricdef}):
\begin{eqnarray}
 \partial_i \Pi_\alpha \, Q^{\alpha \beta} \, \bpartial_{\bj} \bPi_\beta
 &=& g_{i\bj} \\
 \nabla_i \partial_j \Pi_\alpha \, Q^{\alpha \beta} \, \bnabla_{\bk} \bpartial_{\bl} \bPi_\beta
 &=& R_{i\bk j \bl} \\
  \nabla_i \partial_j \Pi_\alpha \, Q^{\alpha \beta} \, \nabla_{k} \partial_{l} \Pi_\beta
 &=:& \CF_{ijkl} \quad \mbox{(symm.\ in $ijkl$)}
\end{eqnarray}
These are similar to (but somewhat simpler than) the closed string
expressions of \cite{Ashok:2003gk,Denef:2004ze}.

The exponential in (\ref{gaussint}) thus becomes
\begin{equation}
 e^{-\frac{\pi}{\phi^0}(\Phi^2 - \frac{1}{2} g_{i\bj} \lambda^i \bar{\lambda}^{\bj}
 + \frac{1}{2} R_{i\bk j \bl} \psi^i \bpsi^k \theta^j \btheta^l )}.
\end{equation}
The term $\CF_{ijkl} \psi^i \theta^j \psi^k \theta^l$ drops out
because $\CF_{ijkl}$ is symmetric in its indices. Doing the Gaussian
integrals over $\lambda$ and $\psi,\bpsi$ turns this in
\begin{equation}
 \pi^n \, e^{-\frac{\pi}{\phi^0} \Phi^2} (\det g_{i \bj})^{-1} \det(R_{i \bk j \bl} \theta^j \btheta^l)
\end{equation}
which is equal to
\begin{equation}
 \pi^n \, e^{-\frac{\pi}{\phi^0} \Phi^2} \det(R^{k}_{i j \bl} \theta^j
 \btheta^l).
\end{equation}
This can be combined with the measure $d^{2n} z$ in (\ref{Zn}) to
produce
\begin{equation}
 \pi^n \, e^{-\frac{\pi}{\phi^0} \Phi^2} \det R
\end{equation}
where $R$ is the curvature 2-form
\begin{equation}
 R^k_i = \frac{i}{2} R^{k}_{i j \bl} dz^j \wedge d\bz^{\bl}.
\end{equation}
We are almost ready to write down our final result. A final step is
to do a modular transformation on the $1/\eta^\chi$ factor in
(\ref{Zn}):
\begin{equation}
 \frac{1}{\eta^\chi(e^{- 2 \pi \phi^0})} =
 (\phi^0)^{\chi/2}
 \frac{1}{\eta^\chi(e^{-\frac{2 \pi}{\phi^0}})}.
\end{equation}
Putting everything together, and noting that $\chi=b_2+2$, we get
(in the continuous flux / small $\phi^0$ approximation):
\begin{eqnarray}
 \CZ_{\rm BH} &\approx& \phi^0 \frac{e^{-\frac{\pi}{\phi^0}
 \Phi^2}}{\eta^\chi(e^{-\frac{2
 \pi}{\phi^0}})} \int_{\CM} \frac{1}{\pi^n} \det R \\
 &\approx& \hat{\chi}(\CM) \, \phi^0 \, e^{\frac{2\pi}{\phi^0}
 \bigl( \frac{P^3+c_2 \cdot P}{24} - \frac{\Phi^2}{2}  \bigr)} \label{Zosvapp}
\end{eqnarray}
where
\begin{equation} \label{differentialeuler}
 \hat{\chi}(\CM) := \int_{\CM} \frac{1}{\pi^n} \det R.
\end{equation}
Alternatively
\begin{equation}
 \Omega(p,q) \approx \hat{\chi}(\CM) \int d\phi \, \phi^0 \,
 e^{-2 \pi \phi \cdot q} \, e^{\frac{2\pi}{\phi^0}
 \bigl( \frac{P^3+c_2 \cdot P}{24} - \frac{\Phi^2}{2}  \bigr)
 }. \label{Omapprox}
\end{equation}
To get to (\ref{Zosvapp}) we used the small $\phi^0$ approximation
to the $\eta$-function and $\chi=P^3 + c_2(X)\cdot P$ (the terms
dropped are exponentially suppressed). Formally
(\ref{differentialeuler}) is exactly the Euler characteristic of the
divisor moduli space $\CM$, but there might be some subtleties since
the metric on $\CM$ has singularities. Note that although this is a
natural result for counting critical points of $W$ on $\CM$, it is
not trivial: while it is true that the Euler characteristic counts
the number of zeros of a section of the cotangent bundle,
$\partial_i W$ does not give such a section because $W$ is not
single valued on $\CM$ (due to monodromies acting on the fluxes).
Indeed, for some fluxes there will be no critical points at all, for
example fluxes Poincar\'e dual to 2-cycles which are trivial on $X$,
and which moreover satisfy $F^2 > 0$, cannot satisfy $F^{(0,2)}=0$
anywhere in moduli space. Again all this has a close analog for IIB
closed string flux vacua, where the analogous index is $\int
\frac{1}{\pi^n} \det(R + \omega {\bf 1})$
\cite{Ashok:2003gk,Denef:2004ze}. The difference comes from the fact
that the relevant covariant derivatives in the closed string case
involve an additional $\partial K$ connection piece, whose curvature
is the K\"ahler form $\omega$.

The moduli space for very ample divisors $P$ is simply $\CM =
\ICP^{I_P-1}$, with $I_P:=\frac{P^3}{6} + \frac{c_2 \cdot P}{12}$.
If the ``differential Euler characteristic''
(\ref{differentialeuler}) equals the topological Euler
characteristic, we thus have
\begin{equation}
 \hat{\chi}(\CM)=\chi(\ICP^{I_P-1})=I_P.
\end{equation}
The results obtained in the bulk of this paper support this
assumption. (It might be possible to prove that $\hat{\chi}(\CM)=I_P$
directly using the estimates in \cite{Peters}. We have not
attempted to do so.)

The result obtained here is in agreement with
(\ref{resultsmallphi0}). The sum over $S$ is absent here; including
it is equivalent to extending the integration contour for $\Phi$ to
the entire imaginary axis in (\ref{Omapprox}). However, since the
saddle point of (\ref{Omapprox}) lies at
\begin{eqnarray}
 \phi_*^0 &=& \sqrt{-\frac{P^3 + c_2P}{24 \, \hat{q}_0}} \\
 \phi_*^A &=& - \phi^0 D^{AB} q_B,
\end{eqnarray}
we see that in the large $q_0$ limit at fixed $q_A$ and $p^A$,
$\phi^0_*$ and $\phi^A_*$ become small, and therefore the
contributions from the extension of the integration contour or
equivalently the $S$-shifted terms in (\ref{resultsmallphi0}) are
actually exponentially suppressed in the regime of interest here.
Hence they can be dropped consistent with our approximations.


\begin{thebibliography}{99}


\bibitem{Strominger:1996sh}
  A.~Strominger and C.~Vafa,
  ``Microscopic Origin of the Bekenstein-Hawking Entropy,''
  Phys.\ Lett.\ B {\bf 379} (1996) 99
  [arXiv:hep-th/9601029].


\bibitem{Maldacena:1997de}
  J.~M.~Maldacena, A.~Strominger and E.~Witten,
  ``Black hole entropy in M-theory,''
  JHEP {\bf 9712} (1997) 002
  [arXiv:hep-th/9711053].

\bibitem{Vafa:1997gr}
  C.~Vafa,
  ``Black holes and Calabi-Yau threefolds,''
  Adv.\ Theor.\ Math.\ Phys.\  {\bf 2}, 207 (1998)
  [arXiv:hep-th/9711067].


\bibitem{Ferrara:1995ih}
  S.~Ferrara, R.~Kallosh and A.~Strominger,
  ``N=2 extremal black holes,''
  Phys.\ Rev.\ D {\bf 52} (1995) 5412
  [arXiv:hep-th/9508072].

\bibitem{Strominger:1996kf}
  A.~Strominger,
  ``Macroscopic Entropy of $N=2$ Extremal Black Holes,''
  Phys.\ Lett.\ B {\bf 383} (1996) 39
  [arXiv:hep-th/9602111].


\bibitem{LopesCardoso:1998wt}
  G.~Lopes Cardoso, B.~de Wit and T.~Mohaupt,
  ``Corrections to macroscopic supersymmetric black-hole entropy,''
  Phys.\ Lett.\ B {\bf 451} (1999) 309
  [arXiv:hep-th/9812082].

\bibitem{LopesCardoso:1999cv}
  G.~Lopes Cardoso, B.~de Wit and T.~Mohaupt,
  ``Deviations from the area law for supersymmetric black holes,''
  Fortsch.\ Phys.\  {\bf 48} (2000) 49
  [arXiv:hep-th/9904005].

\bibitem{LopesCardoso:1999xn}
  G.~Lopes Cardoso, B.~de Wit and T.~Mohaupt,
  ``Area law corrections from state counting and supergravity,''
  Class.\ Quant.\ Grav.\  {\bf 17} (2000) 1007
  [arXiv:hep-th/9910179].


\bibitem{Mohaupt:2000mj}
  T.~Mohaupt,
  ``Black hole entropy, special geometry and strings,''
  Fortsch.\ Phys.\  {\bf 49} (2001) 3
  [arXiv:hep-th/0007195].



\bibitem{Ooguri:2004zv}
  H.~Ooguri, A.~Strominger and C.~Vafa,
  ``Black hole attractors and the topological string,''
  Phys.\ Rev.\ D {\bf 70}, 106007 (2004)
  [arXiv:hep-th/0405146].

%
%

\bibitem{Dabholkar:2005by}
  A.~Dabholkar, F.~Denef, G.~W.~Moore and B.~Pioline,
  ``Exact and asymptotic degeneracies of small black holes,''
  JHEP {\bf 0508}, 021 (2005)
  [arXiv:hep-th/0502157].

\bibitem{Dabholkar:2005dt}
  A.~Dabholkar, F.~Denef, G.~W.~Moore and B.~Pioline,
  ``Precision counting of small black holes,''
  JHEP {\bf 0510}, 096 (2005)
  [arXiv:hep-th/0507014].

\bibitem{Shih:2005he}
  D.~Shih and X.~Yin,
  ``Exact black hole degeneracies and the topological string,''
  JHEP {\bf 0604}, 034 (2006)
  [arXiv:hep-th/0508174].



\bibitem{LopesCardoso:2004xf}
  G.~Lopes Cardoso, B.~de Wit, J.~Kappeli and T.~Mohaupt,
  ``Asymptotic degeneracy of dyonic N = 4 string states and black hole
  entropy,''
  JHEP {\bf 0412} (2004) 075
  [arXiv:hep-th/0412287].


\bibitem{LopesCardoso:2006bg}
  G.~Lopes Cardoso, B.~de Wit, J.~Kappeli and T.~Mohaupt,
  ``Black hole partition functions and duality,''
  JHEP {\bf 0603} (2006) 074
  [arXiv:hep-th/0601108].




\bibitem{talks}
 F.~Denef, ``Counting D-brane ground states,'' talk given at
  \textit{Black holes, topological strings, and invariants of holomorphic submanifolds},
  Harvard,  January 31, 2006;  ``A derivation of OSV (refined),''
  talk given at workshop \textit{Black holes, black rings and topological
  strings}, Munich, April 1, 2006,
  http://www.theorie.physik.uni-muenchen.de/cosmology/seminars/talks/bhbrts/Denef1.pdf;
  ``From OSV to OSV,'' talk at \textit{Strings06},
  http://strings06.itp.ac.cn/talk-files/denef.pdf ;
  G. ~ Moore, ``Split Polar Attractors,'' Talk at the 4th Simons Workshop;
  http://insti.physics.sunysb.edu/itp/conf/simonswork4/

\bibitem{Gaiotto:2006ns}
  D.~Gaiotto, A.~Strominger and X.~Yin,
  ``From AdS(3)/CFT(2) to black holes / topological strings,''
  arXiv:hep-th/0602046;  X.~Yin, ``The M5
  brane elliptic genus,''
  http://strings06.itp.ac.cn/

\bibitem{deBoer:2006vg}
  J.~de Boer, M.~C.~N.~Cheng, R.~Dijkgraaf, J.~Manschot and E.~Verlinde,
  ``A farey tail for attractor black holes,''
  JHEP {\bf 0611}, 024 (2006)
  [arXiv:hep-th/0608059]. E. Verlinde, ``A Farey Tail for N=2 Black Holes,''
  http://strings06.itp.ac.cn/

\bibitem{Beasley:2006us}
  C.~Beasley, D.~Gaiotto, M.~Guica, L.~Huang, A.~Strominger and X.~Yin,
  ``Why Z(BH) = |Z(top)|**2,''
  arXiv:hep-th/0608021.

\bibitem{Verlinde:2004ck}
  E.~P.~Verlinde,
  ``Attractors and the holomorphic anomaly,''
  arXiv:hep-th/0412139.


\bibitem{Denef:2000nb}
  F.~Denef,
  ``Supergravity flows and D-brane stability,''
  JHEP {\bf 0008}, 050 (2000)
  [arXiv:hep-th/0005049].

\bibitem{Denef:2000ar}
  F.~Denef,
  ``On the correspondence between D-branes and stationary supergravity
  solutions of type II Calabi-Yau compactifications,''
  arXiv:hep-th/0010222.

\bibitem{Denef:2001xn}
  F.~Denef, B.~R.~Greene and M.~Raugas,
  ``Split attractor flows and the spectrum of BPS D-branes on the quintic,''
  JHEP {\bf 0105}, 012 (2001)
  [arXiv:hep-th/0101135].

\bibitem{Bates:2003vx}
  B.~Bates and F.~Denef,
  ``Exact solutions for supersymmetric stationary black hole composites,''
  arXiv:hep-th/0304094.


\bibitem{Huang:2007sb}
  M.~x.~Huang, A.~Klemm, M.~Marino and A.~Tavanfar,
  ``Black Holes and Large Order Quantum Geometry,''
  arXiv:0704.2440 [hep-th].



\bibitem{Dijkgraaf:2000fq}
  R.~Dijkgraaf, J.~M.~Maldacena, G.~W.~Moore and E.~P.~Verlinde,
  ``A black hole farey tail,''
  arXiv:hep-th/0005003.

\bibitem{Moore:2004fg}
  G.~W.~Moore,
  ``Les Houches lectures on strings and arithmetic,''
  arXiv:hep-th/0401049.

\bibitem{Kraus:2006nb}
  P.~Kraus and F.~Larsen,
  ``Partition functions and elliptic genera from supergravity,''
  arXiv:hep-th/0607138.

\bibitem{Dijkgraaf:2006um}
  R.~Dijkgraaf, C.~Vafa and E.~Verlinde,
  ``M-theory and a topological string duality,''
  arXiv:hep-th/0602087.

\bibitem{Ashok:2003gk}
  S.~Ashok and M.~R.~Douglas,
  ``Counting flux vacua,''
  JHEP {\bf 0401} (2004) 060
  [arXiv:hep-th/0307049].

\bibitem{Denef:2004ze}
  F.~Denef and M.~R.~Douglas,
  ``Distributions of flux vacua,''
  JHEP {\bf 0405}, 072 (2004)
  [arXiv:hep-th/0404116].

\bibitem{Gopakumar:1998ii}
  R.~Gopakumar and C.~Vafa,
  ``M-theory and topological strings. I,''
  arXiv:hep-th/9809187.

\bibitem{Gopakumar:1998jq}
  R.~Gopakumar and C.~Vafa,
  ``M-theory and topological strings. II,''
  arXiv:hep-th/9812127.

\bibitem{Klemm:2004km}
  A.~Klemm, M.~Kreuzer, E.~Riegler and E.~Scheidegger,
  ``Topological string amplitudes, complete intersection Calabi-Yau spaces  and
  threshold corrections,''
  JHEP {\bf 0505}, 023 (2005)
  [arXiv:hep-th/0410018].

\bibitem{Iqbal:2003ds}
  A.~Iqbal, N.~Nekrasov, A.~Okounkov and C.~Vafa,
  ``Quantum foam and topological strings,''
  arXiv:hep-th/0312022.

\bibitem{MNOP1}
 D.~Maulik, N.~Nekrasov, A.~Okounkov and R.~Pandharipande,
 ``Gromov-Witten theory and Donaldson-Thomas theory,
 I,'' arXiv:math.AG/0312059.

\bibitem{MNOP2}
 D.~Maulik, N.~Nekrasov, A.~Okounkov and R.~Pandharipande, ``Gromov-Witten theory and Donaldson-Thomas theory,
 II,'' arXiv:math.AG/0406092

\bibitem{DT1} R. Thomas, ``Gauge Theory on Calabi-Yau Manifolds,''
PhD Thesis, Oxford 1997.

\bibitem{DT2}
 S.~Donaldson and R.~Thomas, ``Gauge theory in higher dimensions,''
 in ``The geometric universe: science, geometry, and the work of
 Roger Penrose,'' S.~Huggett et.\ al.\ eds.\, Oxford Univ.\ Press,
 1998.

 \bibitem{DT3}
 R. Thomas, ``A holomorphic Casson invariant for Calabi-Yau 3-folds,
 and bundles on $K3$ fibrations,'' arXiv:math.AG/9806111.


\bibitem{Pioline:2006ni}
  B.~Pioline,
  ``Lectures on on black holes, topological strings and quantum attractors,''
  Class.\ Quant.\ Grav.\  {\bf 23} (2006) S981
  [arXiv:hep-th/0607227].



\bibitem{Brunner:1999jq}
  I.~Brunner, M.~R.~Douglas, A.~E.~Lawrence and C.~Romelsberger,
  ``D-branes on the quintic,''
  JHEP {\bf 0008}, 015 (2000)
  [arXiv:hep-th/9906200].

\bibitem{Gomis:2005wc}
  J.~Gomis, F.~Marchesano and D.~Mateos,
  ``An open string landscape,''
  JHEP {\bf 0511}, 021 (2005)
  [arXiv:hep-th/0506179].

\bibitem{Gaiotto:2005rp}
  D.~Gaiotto, M.~Guica, L.~Huang, A.~Simons, A.~Strominger and X.~Yin,
  ``D4-D0 branes on the quintic,''
  arXiv:hep-th/0509168.

\bibitem{Minasian:1997mm}
  R.~Minasian and G.~W.~Moore,
  ``K-theory and Ramond-Ramond charge,''
  JHEP {\bf 9711}, 002 (1997)
  [arXiv:hep-th/9710230].

\bibitem{Freed:1999vc}
  D.~S.~Freed and E.~Witten,
  ``Anomalies in string theory with D-branes,''
  arXiv:hep-th/9907189.

\bibitem{Gaiotto:2006wm}
  D.~Gaiotto, A.~Strominger and X.~Yin,
  ``The M5-brane elliptic genus: Modularity and BPS states,''
  arXiv:hep-th/0607010.

\bibitem{Spence:1999xb}
  B.~J.~Spence,
  ``Topological Born-Infeld actions and D-branes,''
  arXiv:hep-th/9907053.

\bibitem{Belov:2004ht}
  D.~Belov and G.~W.~Moore,
  ``Conformal blocks for AdS(5) singletons,''
  arXiv:hep-th/0412167.

\bibitem{Kapustin:2006pk}
  A.~Kapustin and E.~Witten,
  ``Electric-magnetic duality and the geometric Langlands program,''
  arXiv:hep-th/0604151.


  \bibitem{Nikita}
  N. Nekrasov, ``Localizing gauge theory,'' XIVth International
Congress on Mathematical Physics, Lisbon 2003, J.-C. Zambrini,
editor, World Scientific 2005; pp. 645-654

\bibitem{Tseytlin:1996it}
  A.~A.~Tseytlin,
  ``Self-duality of Born-Infeld action and Dirichlet 3-brane of type IIB
  superstring theory,''
  Nucl.\ Phys.\ B {\bf 469}, 51 (1996)
  [arXiv:hep-th/9602064].

\bibitem{Nikulin} V.V.~Nikulin, ``Integral symmetric
bilinear forms and some of their applications,'' Math. USSR
Izvestija Vol. 14 (1980), No. 1, p. 103

\bibitem{Gukov:2004id}
  S.~Gukov, E.~Martinec, G.~W.~Moore and A.~Strominger,
  ``Chern-Simons gauge theory and the AdS(3)/CFT(2) correspondence,''
  arXiv:hep-th/0403225.

\bibitem{Milnor} J.~Milnor and D.~Husemoller, \textit{Symmetric bilinear forms},
Springer-Verlag, New York, 1973, Ergebnisse der Mathematik und ihrer
Grenzgebiete, Band 73. MR 58 \#22129

\bibitem{Seiberg:1994rs}
  N.~Seiberg and E.~Witten,
  ``Electric - magnetic duality, monopole condensation, and confinement in N=2
  supersymmetric Yang-Mills theory,''
  Nucl.\ Phys.\ B {\bf 426} (1994) 19
  [Erratum-ibid.\ B {\bf 430} (1994) 485]
  [arXiv:hep-th/9407087].

\bibitem{Strominger:1995cz}
  A.~Strominger,
  ``Massless black holes and conifolds in string theory,''
  Nucl.\ Phys.\ B {\bf 451}, 96 (1995)
  [arXiv:hep-th/9504090].

\bibitem{Douglas:1999vm}
  M.~R.~Douglas,
  ``Topics in D-geometry,''
  Class.\ Quant.\ Grav.\  {\bf 17}, 1057 (2000)
  [arXiv:hep-th/9910170].

\bibitem{Moore:1998pn}
  G.~W.~Moore,
  ``Arithmetic and attractors,''
  arXiv:hep-th/9807087.

\bibitem{Moore:1998zu}
  G.~W.~Moore,
  ``Attractors and arithmetic,''
  arXiv:hep-th/9807056.

\bibitem{Behrndt:1997ny}
  K.~Behrndt, D.~Lust and W.~A.~Sabra,
  Nucl.\ Phys.\ B {\bf 510}, 264 (1998)
  [arXiv:hep-th/9705169].

\bibitem{LopesCardoso:2000qm}
  G.~Lopes Cardoso, B.~de Wit, J.~Kappeli and T.~Mohaupt,
``Stationary BPS solutions in N = 2 supergravity with R**2
interactions,''
  JHEP {\bf 0012}, 019 (2000)
  [arXiv:hep-th/0009234].

\bibitem{Ferrara:1997tw}
  S.~Ferrara, G.~W.~Gibbons and R.~Kallosh,
  ``Black holes and critical points in moduli space,''
  Nucl.\ Phys.\ B {\bf 500} (1997) 75
  [arXiv:hep-th/9702103].

\bibitem{Shmakova:1996nz}
  M.~Shmakova,
  ``Calabi-Yau black holes,''
  Phys.\ Rev.\ D {\bf 56} (1997) 540
  [arXiv:hep-th/9612076].

\bibitem{Denef:2002ru}
  F.~Denef,
  ``Quantum quivers and Hall/hole halos,''
  JHEP {\bf 0210}, 023 (2002)
  [arXiv:hep-th/0206072].

\bibitem{Douglas:2000gi}
  M.~R.~Douglas,
  ``D-branes, categories and N = 1 supersymmetry,''
  J.\ Math.\ Phys.\  {\bf 42}, 2818 (2001)
  [arXiv:hep-th/0011017].

\bibitem{bridgeland}
 T.~Bridgeland, ``Spaces of stability conditions,''
 arXiv:math.AG/0611510

\bibitem{Joyce:2006pf}
  D.~Joyce,
  ``Holomorphic generating functions for invariants counting coherent sheaves
  on Calabi-Yau 3-folds,''
  arXiv:hep-th/0607039.

\bibitem{Denef:2001ix}
  F.~Denef,
  ``(Dis)assembling special Lagrangians,''
  arXiv:hep-th/0107152.

\bibitem{Gaiotto:2005gf}
  D.~Gaiotto, A.~Strominger and X.~Yin,
  ``New connections between 4D and 5D black holes,''
  JHEP {\bf 0602}, 024 (2006)
  [arXiv:hep-th/0503217].

\bibitem{Gaiotto:2005xt}
  D.~Gaiotto, A.~Strominger and X.~Yin,
  ``5D black rings and 4D black holes,''
  JHEP {\bf 0602}, 023 (2006)
  [arXiv:hep-th/0504126].

\bibitem{Behrndt:2005he}
  K.~Behrndt, G.~Lopes Cardoso and S.~Mahapatra,
  ``Exploring the relation between 4D and 5D BPS solutions,''
  Nucl.\ Phys.\  B {\bf 732}, 200 (2006)
  [arXiv:hep-th/0506251].

\bibitem{Cheng:2006yq}
  M.~C.~N.~Cheng,
  ``More bubbling solutions,''
  arXiv:hep-th/0611156.

\bibitem{Bena:2004tk}
  I.~Bena and P.~Kraus,
  ``Microscopic description of black rings in AdS/CFT,''
  JHEP {\bf 0412}, 070 (2004)
  [arXiv:hep-th/0408186].

\bibitem{Bena:2005ay}
  I.~Bena and P.~Kraus,
  ``Microstates of the D1-D5-KK system,''
  Phys.\ Rev.\  D {\bf 72}, 025007 (2005)
  [arXiv:hep-th/0503053].

\bibitem{Bena:2005ni}
  I.~Bena, P.~Kraus and N.~P.~Warner,
  ``Black rings in Taub-NUT,''
  Phys.\ Rev.\  D {\bf 72}, 084019 (2005)
  [arXiv:hep-th/0504142].

\bibitem{Elvang:2005sa}
  H.~Elvang, R.~Emparan, D.~Mateos and H.~S.~Reall,
  ``Supersymmetric 4D rotating black holes from 5D black rings,''
  JHEP {\bf 0508}, 042 (2005)
  [arXiv:hep-th/0504125].

\bibitem{Berglund:2005vb}
  P.~Berglund, E.~G.~Gimon and T.~S.~Levi,
  ``Supergravity microstates for BPS black holes and black rings,''
  JHEP {\bf 0606}, 007 (2006)
  [arXiv:hep-th/0505167].

\bibitem{Balasubramanian:2006gi}
  V.~Balasubramanian, E.~G.~Gimon and T.~S.~Levi,
  ``Four dimensional black hole microstates: From D-branes to spacetime foam,''
  arXiv:hep-th/0606118.

\bibitem{Bena:2005va}
  I.~Bena and N.~P.~Warner,
  ``Bubbling supertubes and foaming black holes,''
  Phys.\ Rev.\ D {\bf 74}, 066001 (2006)
  [arXiv:hep-th/0505166].

\bibitem{Bena:2007kg}
  I.~Bena and N.~P.~Warner,
  ``Black holes, black rings and their microstates,''
  arXiv:hep-th/0701216.


\bibitem{freddie}
 F.~Denef and D.~Van den Bleeken, in preparation.

\bibitem{Kachru:1999vj}
  S.~Kachru and J.~McGreevy,
  ``Supersymmetric three-cycles and (super)symmetry breaking,''
  Phys.\ Rev.\ D {\bf 61}, 026001 (2000)
  [arXiv:hep-th/9908135].

\bibitem{Douglas:2000ah}
  M.~R.~Douglas, B.~Fiol and C.~Romelsberger,
  ``Stability and BPS branes,''
  JHEP {\bf 0509}, 006 (2005)
  [arXiv:hep-th/0002037].

\bibitem{Witten:2000mf}
  E.~Witten,
  ``BPS bound states of D0-D6 and D0-D8 systems in a B-field,''
  JHEP {\bf 0204}, 012 (2002)
  [arXiv:hep-th/0012054].

\bibitem{Berkooz:1996km}
  M.~Berkooz, M.~R.~Douglas and R.~G.~Leigh,
  ``Branes intersecting at angles,''
  Nucl.\ Phys.\ B {\bf 480}, 265 (1996)
  [arXiv:hep-th/9606139].

\bibitem{Vafa:1995zh}
  C.~Vafa,
  ``Gas of D-Branes and Hagedorn Density of BPS States,''
  Nucl.\ Phys.\ B {\bf 463} (1996) 415
  [arXiv:hep-th/9511088].

\bibitem{Harvey:1996gc}
  J.~A.~Harvey and G.~W.~Moore,
  ``On the algebras of BPS states,''
  Commun.\ Math.\ Phys.\  {\bf 197} (1998) 489
  [arXiv:hep-th/9609017].

\bibitem{Witten:1997yu}
  E.~Witten,
  ``On the conformal field theory of the Higgs branch,''
  JHEP {\bf 9707} (1997) 003
  [arXiv:hep-th/9707093].

\bibitem{Dijkgraaf:1998zd}
  R.~Dijkgraaf,
  ``Fields, strings, matrices and symmetric products,''
  arXiv:hep-th/9912104.

\bibitem{Witten:1996qb}
  E.~Witten,
  ``Phase Transitions In M-Theory And F-Theory,''
  Nucl.\ Phys.\ B {\bf 471} (1996) 195
  [arXiv:hep-th/9603150].





\bibitem{goettsche} L. G\"ottsche, ``The Betti numbers of the Hilbert
scheme of points on a smooth projective surface,'' Math. Ann. {\bf
286} 193 (1990); ``Hilbert Schemes of Zero-dimensional Subschemes of
Smooth Varieties,'' Lecture Notes in Mathematics {\bf 1572},
Springer-Verlag, 1994.



\bibitem{reineke} M.~Reineke, ``The Harder-Narasimhan system in quantum groups and
cohomology of quiver moduli,'' arXiv:math.QA/0204059.

\bibitem{Hosono:1994ax}
  S.~Hosono, A.~Klemm, S.~Theisen and S.~T.~Yau,
   ``Mirror symmetry, mirror map and applications to complete intersection
  Calabi-Yau spaces,''
  Nucl.\ Phys.\ B {\bf 433}, 501 (1995)
  [arXiv:hep-th/9406055].

\bibitem{Katz:1999xq}
  S.~H.~Katz, A.~Klemm and C.~Vafa,
  ``M-theory, topological strings and spinning black holes,''
  Adv.\ Theor.\ Math.\ Phys.\  {\bf 3}, 1445 (1999)
  [arXiv:hep-th/9910181].

\bibitem{Harvey:1996ir}
  J.~A.~Harvey and G.~W.~Moore,
  ``Fivebrane instantons and $R^2$ couplings in N = 4 string theory,''
  Phys.\ Rev.\ D {\bf 57}, 2323 (1998)
  [arXiv:hep-th/9610237].

\bibitem{Marino:1998pg}
  M.~Marino and G.~W.~Moore,
  ``Counting higher genus curves in a Calabi-Yau manifold,''
  Nucl.\ Phys.\ B {\bf 543}, 592 (1999)
  [arXiv:hep-th/9808131].


\bibitem{Klemm:2005pd}
  A.~Klemm and M.~Marino,
  ``Counting BPS states on the Enriques Calabi-Yau,''
  arXiv:hep-th/0512227.


\bibitem{lazarsfeld}
 R.~Lazarsfeld, ``Positivity in Algebraic Geometry I,''
 Ergebnisse der Mathemaik und ihrer Grenzgebiete, Vol. 48

\bibitem{Douglas:2006jp}
  M.~R.~Douglas, R.~Reinbacher and S.~T.~Yau,
  ``Branes, bundles and attractors: Bogomolov and beyond,''
  arXiv:math.ag/0604597.

\bibitem{Diaconescu:2006qm}
I.~Bena, D.~E.~Diaconescu and B.~Florea,
  ``Black hole entropy and Fourier-Mukai transform,''
  JHEP {\bf 0704}, 045 (2007)
  [arXiv:hep-th/0610068].

\bibitem{Huang:2006hq}
  M.~x.~Huang, A.~Klemm and S.~Quackenbush,
  ``Topological string theory on compact Calabi-Yau: Modularity and boundary
  conditions,''
  arXiv:hep-th/0612125.

\bibitem{Dabholkar:2006tb}
  A.~Dabholkar, A.~Sen and S.~P.~Trivedi,
  ``Black hole microstates and attractor without supersymmetry,''
  arXiv:hep-th/0611143.

\bibitem{GH}
P. Griffiths and J. Harris, \textit{Principles of Algebraic
Geometry}, Wiley, 1978.

\bibitem{Minasian:1999qn}
  R.~Minasian, G.~W.~Moore and D.~Tsimpis,
  ``Calabi-Yau black holes and (0,4) sigma models,''
  Commun.\ Math.\ Phys.\  {\bf 209} (2000) 325
  [arXiv:hep-th/9904217].

\bibitem{shafarevich} I.R.~Shafarevich, ``Basic Algebraic Geometry,'' vol. 1 Springer 1994, p. 245.

\bibitem{CDK} E. Cattani, A. Kaplan, and P. Deligne, ``On the locus of
Hodge classes,'' J. Amer. Math. Soc. \textbf{8}(1995) 483.

\bibitem{Lerche:2002yw}
  W.~Lerche, P.~Mayr and N.~Warner,
  ``N = 1 special geometry, mixed Hodge variations and toric geometry,''
  arXiv:hep-th/0208039.

\bibitem{Martucci:2006ij}
  L.~Martucci,
  ``D-branes on general N = 1 backgrounds: Superpotentials and D-terms,''
  JHEP {\bf 0606}, 033 (2006)
  [arXiv:hep-th/0602129].



\bibitem{Friedman:1997yq}
  R.~Friedman, J.~Morgan and E.~Witten,
  ``Vector bundles and F theory,''
  Commun.\ Math.\ Phys.\  {\bf 187} (1997) 679
  [arXiv:hep-th/9701162].


\bibitem{Hosono:2001gf}
  S.~Hosono, M.~H.~Saito and A.~Takahashi,
  ``Relative Lefschetz Action and BPS State Counting,''
  arXiv:math.ag/0105148.



\bibitem{Szego} G. Szeg\"o, \textit{Orthogonal Polynomials}, Fourth Edition,
Amer. Math. Soc.



\bibitem{Ooguri:2005vr}
  H.~Ooguri, C.~Vafa and E.~P.~Verlinde,
  ``Hartle-Hawking wave-function for flux compactifications,''
  Lett.\ Math.\ Phys.\  {\bf 74} (2005) 311
  [arXiv:hep-th/0502211].

\bibitem{Gunaydin:2006bz}
  M.~Gunaydin, A.~Neitzke and B.~Pioline,
  ``Topological wave functions and heat equations,''
  JHEP {\bf 0612} (2006) 070
  [arXiv:hep-th/0607200].



\bibitem{Mathur:2005zp}
  S.~D.~Mathur,
  ``The fuzzball proposal for black holes: An elementary review,''
  Fortsch.\ Phys.\  {\bf 53} (2005) 793
  [arXiv:hep-th/0502050].

\bibitem{Candelas:1990rm}
  P.~Candelas, X.~C.~De La Ossa, P.~S.~Green and L.~Parkes,
  ``A pair of Calabi-Yau manifolds as an exactly soluble superconformal
  Nucl.\ Phys.\ B {\bf 359} (1991) 21.

\bibitem{Harvey:1995fq}
  J.~A.~Harvey and G.~W.~Moore,
  ``Algebras, BPS States, and Strings,''
  Nucl.\ Phys.\ B {\bf 463} (1996) 315
  [arXiv:hep-th/9510182].





\bibitem{Moore:1997ar}
  G.~W.~Moore,
  ``String duality, automorphic forms, and generalized Kac-Moody algebras,''
  Nucl.\ Phys.\ Proc.\ Suppl.\  {\bf 67} (1998) 56
  [arXiv:hep-th/9710198].



\bibitem{Peters}
C.A.M. Peters, ``A criterion for flatness of Hodge bundles over
curves and geometric applications,'' Math. Ann. {\bf 268} (1984)1.



\bibitem{Freed:2006ya}
  D.~S.~Freed, G.~W.~Moore and G.~Segal,
  ``The uncertainty of fluxes,''
  arXiv:hep-th/0605198.

\bibitem{Freed:2006yc}
  D.~S.~Freed, G.~W.~Moore and G.~Segal,
  ``Heisenberg groups and noncommutative fluxes,''
  Annals Phys.\  {\bf 322} (2007) 236
  [arXiv:hep-th/0605200].

\bibitem{Moore:2004jv}
  G.~W.~Moore,
  ``Anomalies, Gauss laws, and page charges in M-theory,''
  Comptes Rendus Physique {\bf 6} (2005) 251
  [arXiv:hep-th/0409158].




\end{thebibliography}
\end{document}